\documentclass[a4paper,fleqn,usenatbib]{mnras}

\usepackage{newtxtext,newtxmath}

\usepackage[T1]{fontenc}
\usepackage{ae,aecompl}

\usepackage{adjustbox}
\usepackage{placeins}
\usepackage{afterpage}
\usepackage{float}
\usepackage{graphicx}	
\usepackage{amsmath}	
\usepackage{amssymb}	

\usepackage{epsfig,color,pifont,array}
\usepackage[outdir=./]{epstopdf}

\usepackage{bm}

\usepackage{etoolbox}

\makeatletter

\patchcmd{\NAT@citex}
  {\@citea\NAT@hyper@{%
     \NAT@nmfmt{\NAT@nm}%
     \hyper@natlinkbreak{\NAT@aysep\NAT@spacechar}{\@citeb\@extra@b@citeb}%
     \NAT@date}}
  {\@citea\NAT@nmfmt{\NAT@nm}%
   \NAT@aysep\NAT@spacechar\NAT@hyper@{\NAT@date}}{}{}

\patchcmd{\NAT@citex}
  {\@citea\NAT@hyper@{%
     \NAT@nmfmt{\NAT@nm}%
     \hyper@natlinkbreak{\NAT@spacechar\NAT@@open\if*#1*\else#1\NAT@spacechar\fi}%
       {\@citeb\@extra@b@citeb}%
     \NAT@date}}
  {\@citea\NAT@nmfmt{\NAT@nm}%
   \NAT@spacechar\NAT@@open\if*#1*\else#1\NAT@spacechar\fi\NAT@hyper@{\NAT@date}}
  {}{}

\makeatother

\newcommand\Tstrut{\rule{0pt}{2.6ex}}         
\newcommand\Bstrut{\rule[-1.5ex]{0pt}{2.6ex}}   
\newcommand\Dstrut{\rule[-3.0ex]{0pt}{2.6ex}}   

\defcitealias{2016MNRAS.455.1309B}{Paper I}

\title[Filaments in wind-cloud interactions (II)]{Filament formation in wind-cloud interactions. II. Clouds with turbulent density, velocity, and magnetic fields}

\author[W.~E.~Banda-Barrag\'{a}n, C.~Federrath, R.~M.~Crocker, and G.~V.~Bicknell]{
W.~E.~Banda-Barrag\'{a}n,$^{1,2}$\thanks{E-mail: wlady.bsc@gmail.com (WBB)}
C.~Federrath,$^{1}$
R.~M.~Crocker,$^{1}$
and G.~V.~Bicknell$^{1}$
\\
$^{1}$Research School of Astronomy and Astrophysics, Australian National University, Canberra, ACT 2611, Australia\\
$^{2}$Observatorio Astron\'{o}mico de Quito, Escuela Polit\'{e}cnica Nacional, Av. Gran Colombia S/N, Quito 170403, Ecuador\\
}

\date{Accepted XXX. Received YYY; in original form ZZZ}

\pubyear{2017}

\begin{document}
\label{firstpage}
\pagerange{\pageref{firstpage}--\pageref{lastpage}}
\maketitle

\begin{abstract}
{
We present a set of numerical experiments designed to systematically investigate how turbulence and magnetic fields influence the morphology, energetics, and dynamics of filaments produced in wind-cloud interactions. We cover three-dimensional, magnetohydrodynamic systems of supersonic winds impacting clouds with turbulent density, velocity, and magnetic fields. We find that log-normal density distributions aid shock propagation through clouds, increasing their velocity dispersion and producing filaments with expanded cross sections and highly-magnetised knots and sub-filaments. In self-consistently turbulent scenarios the ratio of filament to initial cloud magnetic energy densities is $\sim 1$. The effect of Gaussian velocity fields is bound to the turbulence Mach number: Supersonic velocities trigger a rapid cloud expansion; subsonic velocities only have a minor impact. The role of turbulent magnetic fields depends on their tension and is similar to the effect of radiative losses: the stronger the magnetic field or the softer the gas equation of state, the greater the magnetic shielding at wind-filament interfaces and the suppression of Kelvin-Helmholtz instabilities. Overall, we show that including turbulence and magnetic fields is crucial to understanding cold gas entrainment in multi-phase winds. While cloud porosity and supersonic turbulence enhance the acceleration of clouds, magnetic shielding protects them from ablation and causes Rayleigh-Taylor-driven sub-filamentation. Wind-swept clouds in turbulent models reach distances $\sim15-20$ times their core radius and acquire bulk speeds $\sim0.3-0.4$ of the wind speed in one cloud-crushing time, which are three times larger than in non-turbulent models. In all simulations the ratio of turbulent magnetic to kinetic energy densities asymptotes at $\sim 0.1-0.4$, and convergence of all relevant dynamical properties requires at least $64$ cells per cloud radius.
}
\end{abstract}




\section{Introduction}
\label{sec:Introduction}
Investigating the dynamics and longevity of wind-swept interstellar clouds is essential to understanding stellar- and supernova-driven, multi-phase winds and outflows, as well as the formation and evolution of filaments embedded in them. Wind-swept clouds and filaments have been observed and studied at various scales in the interstellar medium (ISM), such as in the shells of supernova remnants (e.g., see \citealt{1996ApJ...456..225H,2007ApJ...657..308K,2009ApJ...693.1883S,2009ApJ...697..535P,2013ApJ...774..120M,2015ApJ...800..119N} for observations; and \citealt{1992ApJ...390L..17S}; \citealt*{2005AA...443..495M}; \citealt{2006MNRAS.373..811M}; \citealt{2005A&A...444..505O,2006A&A...457..545O,2008ApJ...678..274O,2009MNRAS.394..157L} for models), in molecular cloud complexes (e.g., see \citealt*{2003A&A...403..399C}; \citealt*{2012ApJ...751...69S,2012ApJ...761L..21S}; \citealt{2012ApJ...746L..21W,2014ApJS..213....8T,2014ApJ...780...72E,2015MNRAS.453.2036B} for observations; and \citealt{2006ApJ...647..397M}; \citealt{2007ApJ...668..980M}; \citealt*{2010MNRAS.403..714M} for models), in tidally-disrupted clouds (e.g., see \citealt{2012Natur.481...51G,2013ApJ...763...78G} for observations; and \citealt{2012ApJ...750...58B,2012ApJ...755..155S,2015ApJ...811..155S,2013ApJ...776...13B,2016ApJ...819L..28B} for models), in the Galactic centre magnetosphere (e.g., see \citealt*{1984Natur.310..557Y}, \citealt{1985AJ.....90.2511M}, \citealt{1999ApJ...521L..41L}, \citealt{2000AJ....119..207L,2004ApJ...607..302L}, \citealt*{2004ApJS..155..421Y}, \citealt*{2014IAUS..303..369M} for observations; and \citealt{1999ApJ...521..587S}, \citealt{2002ApJ...568..220D}, \citealt*{2005PASJ...57L..39S} for models), in large-scale galactic winds, outflows, and fountains (e.g., see \citealt{1988Natur.334...43B,1998ApJ...493..129S}; \citealt*{2002ApJ...576..745C}; \citealt{2005MNRAS.363..216C,2009ApJ...701.1636M,2012ApJS..199...12M,2013ApJ...770L...4M,2015Natur.519..436T,2017arXiv170600443V} for observations; and \citealt{2000MNRAS.314..511S,2008MNRAS.388..573M,2009MNRAS.399.1089M}; \citealt{2008ApJ...674..157C,2009ApJ...703..330C}; \citealt*{2013MNRAS.430.3235M,2017ApJ...837...28S} for models), and also in ram-pressure-stripped galaxies (e.g., see \citealt{2014AJ....147...63A}; \citealt*{2015AJ....150...59K} for observations; and \citealt*{2003MNRAS.345.1329M}; \citealt{2008A&A...481..337K}; \citealt{2010NatPh...6..520P,2015MNRAS.449.2312V} for models).\par

A crucial question to be answered when studying wind-swept clouds in all the above scenarios is how long a cold and dense cloud is able to survive embedded in a hot and diffuse wind. In particular, if we consider the case of large-scale, multi-phase galactic winds, such as the one in the galaxy M82 (e.g., see \citealt*{1997A&A...320..378S,1999ApJ...523..575L}) or the one in our own Galaxy (e.g., see \citealt*{1984Natur.310..568S,2003ApJ...582..246B,2010ApJ...724.1044S}), a key problem is to explain the origin of the populations of high-latitude dense clouds and filaments observed in the star-formation-driven winds of these galaxies (e.g., see \citealt{1998ApJ...493..129S,2004A&A...416...67G,2008AJ....135.1983C} for the wind in M82, and \citealt{2012ApJS..199...12M,2013ApJ...770L...4M,2016ApJ...826..215L} for the wind in the Milky Way). Did these structures form at high latitudes as a result of wind gas cooling down to sufficiently-low temperatures to trigger radiative instabilities as discussed in \cite{1995ApJ...444..590W,2015arXiv150701951Z,2016MNRAS.455.1830T}? Were these clouds and filaments transported from low to high latitudes by the ram pressure or the radiation pressure of global outflows as discussed in \cite{2008ApJ...674..157C,2009ApJ...703..330C,2015ApJ...805..158S,2015MNRAS.449....2M,2016ApJ...822...31B} and in \cite*{2005ApJ...618..569M,2012MNRAS.424.1170Z,2012ApJ...760..155K,2013MNRAS.434.2329K}, respectively? If so, how did these entrained structures survive disruption and ablation to reach latitudes $\sim0.4-3\,\rm kpc$ above and below the galactic planes of these galaxies?\par

As shown in previous numerical studies of shock-cloud and wind-cloud systems, supersonic winds have the ability to disrupt clouds via ram pressure and dynamical instabilities in only a few shock-crossing time-scales (e.g., see \citealt*{1994ApJ...420..213K}, \citealt{1994ApJ...433..757M,1995ApJ...454..172X}; \citealt*{1996ApJ...473..365J}; \citealt{1999ApJ...527L.113G,2000ApJ...543..775G,2005ApJ...619..327F,2006ApJS..164..477N,2008ApJ...680..336S,2009MNRAS.394.1351P}; \citealt*{2010ApJ...722..412Y,2010MNRAS.405..821P}; \citealt{2011Ap&SS.336..239P,2011MNRAS.411L..41P,2013ApJ...766...45J,2013ApJ...774..133L,2015MNRAS.449....2M,2016MNRAS.457.4470P,2016MNRAS.458.1139P,2016MNRAS.461..578G,2017A&A...600A.134M}). The reader is referred to Chapter 2 in \cite{2016PhDT.......154B} for a recent review of the literature on wind-cloud and shock-cloud interactions. In \cite{2016MNRAS.455.1309B} (hereafter \citetalias{2016MNRAS.455.1309B}), for instance, we showed that an adiabatic, spherical cloud embedded in a supersonic wind can only reach distances of the order of $3-4$ times the cloud's core radius (see Figure 12 in that paper) in one cloud-crushing time, $t_{\rm cc}$, defined as the time it takes for the initially-refracted shock to cross one cloud diameter (see Section \ref{subsec:DynamicalTime-Scales} below for further details). Moreover, previous simulations of shock-cloud interactions showed that the destruction of clouds occurs in only a few cloud-crushing times, typically of the order of $t_{\rm des}/t_{\rm cc}\sim1.5-2$ in purely adiabatic, hydrodynamic (hereafter HD) models (e.g, see \citealt{1994ApJ...420..213K}; \citealt*{2002ApJ...576..832P}; \citealt{2006ApJS..164..477N}), or $t_{\rm des}/t_{\rm cc}\sim4-6$ in models that incorporate radiative cooling (e.g., see \citealt{2005AA...443..495M}), where the cloud destruction time, $t_{\rm des}$, is conventionally defined as the time when the mass of the cloud (or of its largest fragment) has dropped by $1/e$ (see Section 2.2 of \citealt{2006ApJS..164..477N}). Similarly to radiative cooling, thermal conduction has also been demonstrated to prolong the lifetime of clouds embedded in a hot wind via suppression of dynamical instabilities (e.g., see \citealt{2005A&A...444..505O,2007AA...472..141V,2008ApJ...678..274O}).\par

In all the above cases, however, explaining cloud entrainment has been difficult (e.g., see \citealt{2015arXiv150701951Z,2016ApJ...822...31B}). The modelled wind-swept clouds do survive shredding in scenarios with radiative cooling and thermal conduction to form dense cloudlets and filaments, but their cross sections become smaller than in purely HD cases as a result of efficient cooling (in radiative clouds) or evaporation (in thermally-conducting clouds). Smaller cross sections mean lower drag forces, so clouds in these models do not reach the high, asymptotic velocities of a few hundred $\rm km\,s^{-1}$ that are characteristic of the cold phases in multi-phase galactic winds (e.g., see \citealt{1998ApJ...493..129S} for M82, and \citealt{2013ApJ...770L...4M} for our Galaxy). Thus, the nature of high-latitude, dense gas entrained in the aforementioned large-scale, galactic winds is still puzzling from a theoretical point of view (see \citealt*{1985Natur.317...44C,2005ApJ...621..227M,2005ARA&A..43..769V}; \citealt{2016arXiv161001164M}; \citealt*{2017MNRAS.466.1213K,2017arXiv170109062H,2017ApJ...837...28S} for thorough reviews of galactic winds and/or discussions on wind-launching mechanisms).\par

Despite this, we show in this paper that two ingredients are key to understanding the prevalence of cold gas in hot winds. The first one is turbulence, previously considered by \cite{2008ApJ...674..157C,2009ApJ...703..330C,2017ApJ...834..144S}, and the second one is magnetic fields, whose importance has been pointed out by \cite{1999ApJ...527L.113G,2000ApJ...543..775G,2013ApJ...774..133L,2015MNRAS.449....2M}; and ourselves in \citetalias{2016MNRAS.455.1309B}. The results obtained by these studies have been inconclusive so far. On the one hand, the simulations presented by \cite{2008ApJ...674..157C,2009ApJ...703..330C} showed that fractal, radiative clouds are fragmented by a global galactic wind to form smaller cloudlets and filaments that acquire velocities of $\sim \rm  100-800\,km\,s^{-1}$, matching observational measurements of the cold phase velocity in galactic winds and suggesting that entrainment of cloudlets and filaments is actually possible. On the other hand, \cite{2017ApJ...834..144S} found that the entrainment of turbulent clouds is inefficient, owing to the smaller accelerations of their densest cores.\par

A similar situation occurs in studies with magnetic fields. On the one hand, \cite{2013ApJ...774..133L,2015MNRAS.449....2M} and ourselves (see \citetalias{2016MNRAS.455.1309B}) showed that tangled and transverse (to the direction of streaming) magnetic fields, respectively, effectively suppress Kelvin-Helmholtz (hereafter KH; e.g., see \citealt{1961hhs..book.....C,FLM:383106,2000ifd..book.....B}) instabilities at the sides of wind-swept clouds (via a magnetic shielding effect), thus reducing mixing and prolonging the lifetime of wind-swept clouds. On the other hand, \cite{1999ApJ...527L.113G,2000ApJ...543..775G,2013ApJ...774..133L} and ourselves (see \citetalias{2016MNRAS.455.1309B}) showed that transverse magnetic fields also hasten the growth of Rayleigh-Taylor (hereafter RT; e.g., see \citealt{1984PhyD...12....3S,2002ihs..book.....D,2004hyst.book.....D}) instabilities at the leading edge of the cloud (via a magnetic bumper effect), thus contributing to their break-up and sub-filamentation. The stronger the magnetic field, the more accentuated the aforementioned affects (e.g., see \citealt{2008ApJ...680..336S}). In addition, \cite{1994ApJ...433..757M,1996ApJ...473..365J,1999ApJ...510..726M,1999ApJ...517..242M}; and ourselves (see \citetalias{2016MNRAS.455.1309B}) found that aligned (to the direction of streaming) magnetic fields do not have a significant impact on the cloud dynamics when compared to purely HD models. Despite the seemingly-different results mentioned above, they are all complementary and indicate that magnetic fields have different effects on the morphology and dynamics of clouds, depending on their tension and orientation. They also motivate the study presented here with more realistic models for the clouds as most authors, including ourselves, have considered idealised systems with either spherical clouds and uniform or tangled magnetic fields, or turbulent/fractal clouds without magnetic fields. Thus, in this paper we reconcile the above results by systematically studying the evolution of initially-turbulent clouds embedded in supersonic winds and concentrate on studying both qualitatively and quantitatively the morphology, energetics, and dynamics of these clouds and their ensuing filaments. We show that the inclusion of turbulence and magnetic fields in a self-consistent manner is crucial to understanding cold gas entrainment in multi-phase winds.\par

This paper is organised as follows. In Section \ref{sec:Importance} we explain the significance of our work in the context of the literature on ISM turbulence. In Section \ref{sec:Method} we include a description of the numerical methods, initial and boundary conditions, time-scales, and diagnostics, which we employ in our study. In Section \ref{sec:Results}, we present our results, including an overall description of filament formation as well as comparisons between different initial configurations in non-turbulent and turbulent clouds. We utilise 3D volume renderings and several diagnostics to illustrate the structure, kinematics, and survival of filaments against dynamical instabilities, as well as the evolution, in the magnetotails, of the different components of the energy density. In Section \ref{sec:Summary} we summarise our findings and conclusions. At the end of the paper, we include several Appendices that contain further details on the methods that we follow in this study and a thorough discussion on the effects of numerical resolution and simulation domain size on the diagnostics presented here.

\section{Relevance of this study}
\label{sec:Importance}
In this series of papers we study wind-cloud interactions and their associated filaments with increasing complexity. In \citetalias{2016MNRAS.455.1309B} we investigated the formation of filamentary structures in systems that included spherical clouds with smoothed density profiles and supersonic winds with uniformly-distributed magnetic fields. Here we investigate wind-cloud systems in which the clouds have log-normal density distributions, Gaussian velocity fields, and turbulent magnetic fields. The significance of this study lies in three main points:

\begin{enumerate}
\item We include magnetic fields as they are ubiquitous in the ISM and should therefore be considered in any realistic models of wind/shock-swept clouds;
\item We incorporate turbulence within the clouds (i.e., turbulent distributions for the density, velocity, and magnetic fields) as this is also an intrinsic characteristic of ISM clouds;
\item We implement, for the first time, points (i) and (ii) in a systematic and self-consistent manner in the initial conditions of our wind-cloud models.
\end{enumerate}

\noindent We explain each of these points below and then pose some questions to be answered throughout this paper.

Magnetic fields and turbulence are fundamental elements of the ISM (see \citealt{1981MNRAS.194..809L}; \citealt*{1997MNRAS.288..145P,2004RvMP...76..125M,2001RvMP...73.1031F,2007EAS....23....3F,2011MmSAI..82..824F,2004ARA&A..42..211E,2004ARA&A..42..275S,2007ARA&A..45..565M}). First, magnetic fields have a reciprocal relationship with the gas in which they are frozen. Maxwell stresses act upon the magnetised gas changing its dynamics, whilst the resulting motion of the gas affects the topology of the magnetic field lines via shearing and vortical motions (e.g., see \citealt{1976magn.book.....C}, \citealt*{1999ApJ...517..242M}). As a result, enhancement and annihilation of magnetic energy occur in such environments through dynamo action (e.g., see \citealt{1999PhRvL..83.2957S,2005PhR...417....1B,2014ApJ...797L..19F}), and reconnection events (e.g., see \citealt{1999ApJ...517..700L,2014SSRv..181....1L}), respectively. Second, turbulence also plays an essential role in shaping the ISM and influencing the processes occurring in it, such as star formation (e.g., see \citealt{2004RvMP...76..125M,2010ApJ...714..442S}; \citealt*{2012ApJ...747...21S}; \citealt{2012ApJ...761..156F}; \citealt*{2013A&A...553L...8K}; \citealt{2014prpl.conf...77P,2013ApJ...777...46L}; \citealt*{2015ApJ...806L..36S}, \citealt{2015MNRAS.453..739K}), dynamo-regulated growth of magnetic fields (e.g., see \citealt{2013NJPh...15b3017S,2014ApJ...781...84S,2015PhRvE..92b3010S}; \citealt*{2016MNRAS.461..240B}), and acceleration and diffusion of cosmic rays (e.g., see \citealt{2002PhRvL..89B1102Y,2015ApJ...811....8W}).\par

Furthermore, clouds in the ISM are intrinsically turbulent as they emerge from the non-isotropic condensation of thermally-unstable gas (e.g., see \citealt{1965ApJ...142..531F,2001ApJ...547...99Y,2007A&A...471..213V}; \citealt*{2010MNRAS.406.1260V}; \citealt{2012ApJ...759...35I,2015ApJ...804..137P}), or from thin shell instabilities in colliding winds (e.g., see \citealt*{1992ApJ...386..265S}, \citealt*{1993A&A...267..155D}, \citealt{1994ApJ...428..186V}, \citealt{2011ApJ...726..105P}, \citealt{2015arXiv150707012C}). Observational and numerical studies of clumpy media show, for example, that the density profiles inside clouds are best described by log-normal distributions in supersonic, transonic, and subsonic scenarios (e.g., see \citealt{1994ApJ...423..681V,1998PhRvE..58.4501P}; \citealt{1999intu.conf..218N}; \citealt*{2008ApJ...688L..79F}; \citealt{2009A&A...508L..35K,2010A&A...512A..81F,2014Sci...344..183K}; \citealt{2012A&A...540L..11S,2013ApJ...766L..17S,2016A&A...587A..74S}). Similarly, velocity fields inside clouds are best represented by Gaussian random distributions (e.g., see \citealt{2002PhRvE..65e6304M,2002A&A...390..307O,2013MNRAS.436.1245F}). Lastly, numerical studies on compressible isothermal turbulence show that the magnetic field perturbations in such environments are well described by monotonic probability distributions (e.g., see \citealt{2009ApJ...693.1728P,2010ApJ...725..466C}). Thus, incorporating turbulent clouds constitutes a substantial improvement with respect to previous studies as: 1) the profiles of the density, velocity, and magnetic fields in turbulent clouds are correlated as a result of the coupling between density, velocity, and magnetic field governed by the MHD equations, and 2) the log-normal or Gaussian distributions in turbulent models are skewed and contain higher order moments, which manifests in the so-called intermittency (see \citealt{2010A&A...512A..81F,2013MNRAS.430.1880H}). These two realistic features of turbulence are absent in artificially-generated fractal clouds (e.g., see our group's earlier works with fractal clouds in \citealt{2008ApJ...674..157C,2009ApJ...703..330C}) and other non-uniform density profiles (e.g., see \citealt*{2005RMxAA..41...45R}).\par

Our study here probes the physics of filament formation in three-dimensional, turbulent wind-cloud interactions at high resolution. The aim of this paper is to study the morphology, kinematics, and magnetic properties of the magnetised filaments that arise as a result of wind-swept turbulent clouds. By examining the impact of turbulence on the interplay between winds and clouds, we address the following questions: 1) Are the mechanisms involved in the formation of filaments universal, i.e., the same for uniform and turbulent (non-uniform) clouds? 2) How does the internal structure of filaments change when turbulence is included in the initial conditions? 3) To what distances and velocities are wind-swept turbulent clouds ram-pressure accelerated by the wind? 4) What are the effects of varying the Mach number of turbulent velocity fields? 5) What are the effects of changing the strength of turbulent magnetic fields? 6) What is the ratio of filament magnetic field to initial magnetic field in the cloud? 7) What kind of energy densities are involved? 8) What is the fate of dense gas within turbulent clouds when entrained in the wind? \par

\section{Method}
\label{sec:Method}

\subsection{Simulation code}
\label{subsec:SimulationCode}
To perform the simulations reported in this paper we solve the time-dependent equations of ideal magnetohydrodynamics (hereafter MHD). We utilise the {\sevensize PLUTOv4.0} code (see \citealt{2007ApJS..170..228M,2012ApJS..198....7M}) in a 3D cartesian coordinate system $(X_1,X_2,X_3)$ to solve the following equations for mass, momentum, energy conservation, and magnetic induction:

\begin{equation}
\frac{\partial \rho}{\partial t}+\bm{\nabla\cdot}\left[{\rho \bm{v}}\right]=0,
\label{eq:MassConservation}
\end{equation}

\begin{equation}
\frac{\partial \left[\rho \bm{v}\right]}{\partial t}+\bm{\nabla\cdot}\left[{\rho\bm{v}\bm{v}}-{\bm{B}\bm{B}}+{\bm{I}}P\right]=0,
\label{eq:MomentumConservation}
\end{equation}

\begin{equation}
\frac{\partial E}{\partial t}+\bm{\nabla\cdot}\left[\left(E+P\right)\bm{v}-\bm{B}\left(\bm{v}\bm{\cdot B}\right)\right]=0,
\label{eq:EnergyConservation}
\end{equation}

\begin{equation}
\frac{\partial \bm{B}}{\partial t}-\bm{\nabla\times}\left(\bm{v}\bm{\times B}\right)=0,
\label{eq:induction}
\end{equation}

\noindent where $\rho$ is the mass density, $\bm{v}$ is the velocity, $\bm{B}$ is the magnetic field\footnote{Note that the factor $\frac{1}{\sqrt{4\pi}}$ is subsumed into the definition of magnetic field. The same normalisation applies henceforth.}, $P=P_{\rm th}+P_{\rm mag}$ is the total pressure (i.e., thermal plus magnetic: $P_{\rm mag}=\frac{1}{2}|\bm{B}|^2$), $E=\rho\epsilon+\frac{1}{2}\rho\bm{v^2}+\frac{1}{2}|\bm{B}|^2$ is the total energy density, and $\epsilon$ is the specific internal energy.  We use an ideal equation of state to close the above system of conservation laws: 

\begin{equation}
P_{\rm {th}}=P_{\rm th}(\rho,\epsilon)=\left(\gamma-1\right)\rho\epsilon, 
\end{equation}

\noindent where we assume a ratio of the specific heat capacities at constant pressure and volume of $\gamma=\frac{5}{3}$ for adiabatic models and $\gamma=1.1$ for the quasi-isothermal model (which approximates the effects of radiative cooling as in \citealt{1994ApJ...420..213K,2006ApJS..164..477N}). Using a softer polytropic index is an effective way of investigating the role of radiative cooling in wind-cloud systems, without having to worry about problems with unresolved cooling length scales (e.g., see \citealt{2010ApJ...722..412Y}) or "runaway" cooling effects (e.g., see \citealt*{2002AA...395L..13M}; \citealt{2004ApJ...604...74F,2013ApJ...766...45J}). This approach also allows us to keep our results scale-free and provide a thorough qualitative and quantitive analysis of wind-swept clouds in a general magneto-hydrodynamical context; this can be insightful for current and future work addressing specific wind/shock cloud systems and scales (in specific cooling regimes). Similarly to \citetalias{2016MNRAS.455.1309B}, we solve three additional advection equations of the form:

\begin{equation}
\frac{\partial\left[\rho C_{\alpha}\right]}{\partial t}+\bm{\nabla\cdot}\left[{\rho C_{\alpha} \bm{v}}\right]=0,
\label{eq:tracer}
\end{equation}

\noindent where $C_{\alpha}$ represents a set of three Lagrangian scalars used to track the evolution of gas initially contained in the cloud as a whole (when $\alpha=\rm cloud/filament$), in its core (when $\alpha=\rm core/footpoint$), and in its envelope (when $\alpha=\rm envelope/tail$). Initially we define $C_{\alpha}=1$ for the whole cloud, the cloud core, and the cloud envelope, respectively, and $C_{\alpha}=0$ everywhere else. This configuration allows us to study the evolution of each component of the cloud separately.\par

In order to solve the above system of hyperbolic conservation laws and to preserve the solenoidal condition, $\bm{\nabla\cdot B}=0$, we configure the {\sevensize PLUTO} code to use the \verb#HLLD# approximate Riemann solver of \cite{Miyoshi:2005} jointly with the constrained-transport upwind scheme of \cite{2005JCoPh.205..509G,2008JCoPh.227.4123G}. The magnetic vector potential $\bm{A}$, where $\bm{B}=\bm{\nabla\times A}$, is used to initialise the field and the Courant-Friedrichs-Lewy (CFL) number is $C_{\rm a}=0.3$ in all cases to achieve numerical stability.

\subsection{Initial and boundary conditions}
\label{subsec:Initial and Boundary Conditions}
In these simulations we consider a two-phase ISM composed of a single uniform or turbulent cloud (in a spherical volume) surrounded by a hot, tenuous, supersonic wind. Similarly to \citetalias{2016MNRAS.455.1309B} (see Figure 1 there), the cloud is initially immersed in a uniform velocity field, i.e., a wind with Mach numbers:

\begin{equation}
{\cal M_{\rm w}}=\frac{|\bm{v_{\rm w}}|}{c_{\rm w}}=4.0\:{\rm or}\:4.9,
\label{eq:MachNumber}
\end{equation}

\noindent depending on whether the model is adiabatic or quasi-isothermal, respectively. In Equation (\ref{eq:MachNumber}), $|{\bm{v_{\rm w}}}|\equiv v_{\rm w}$ and $c_{\rm w}=\sqrt{\gamma \frac{P_{\rm th}}{\rho_{\rm w}}}$ are the speed and sound speed of the wind, respectively, and $\rho_{\rm w}$ is the density of the wind.\par

We employ Cartesian $(X_1,X_2,X_3)$ coordinates for all the simulations reported here. The simulation domain consists of a rectangular prism that comes in two different configurations: M (standard, medium domain) covering the spatial range $-3\,r_{\rm c}\leq X_1\leq3\,r_{\rm c}$, $-2\,r_{\rm c}\leq X_2\leq16\,r_{\rm c}$, and $-3\,r_{\rm c}\leq X_3\leq 3\,r_{\rm c}$, and S (small domain) covering the spatial range $-2\,r_{\rm c}\leq X_1\leq2\,r_{\rm c}$, $-2\,r_{\rm c}\leq X_2\leq10\,r_{\rm c}$, and $-2\,r_{\rm c}\leq X_3\leq 2\,r_{\rm c}$, where $r_{\rm c}$ is the radius of the cloud. In the former configuration, M, the uniform grid resolution is $(N_{\rm X_{1}}\times N_{\rm X_{2}}\times N_{\rm X_{3}})=(384\times1152\times384)$, so that $64$ cells cover the cloud radius ($R_{\rm 64}$) and $32$ cells cover the core radius (defined as $r_{\rm co}=0.5\,r_{\rm c}$). In the latter configuration, S, the uniform grid resolution is $(N_{\rm X_{1}}\times N_{\rm X_{2}}\times N_{\rm X_{3}})=(512\times1536\times512)$, so that $128$ cells cover the cloud radius ($R_{\rm 128}$) and $64$ cells cover the core radius (defined as $r_{\rm co}=0.5\,r_{\rm c}$). Models with other resolutions ($R_{16-128}$) and with a larger-domain configuration, L, are also described in Appendices \ref{sec:Appendix3} and \ref{sec:Appendix4}, respectively. The cloud is initially centred in the origin $(0,0,0)$ of the simulation domain.\par

We prescribe diode boundary conditions (i.e., gas outflow is allowed while inflow is prevented) on five sides of the simulation domain and an inflow boundary condition (i.e., an injection zone) on the remaining side. A constant supply of wind material is ensured by setting the injection zone at the ghost zone (of the computational domain) that faces the leading edge of the cloud.\par

\begin{figure}\centering
\includegraphics[scale=1]{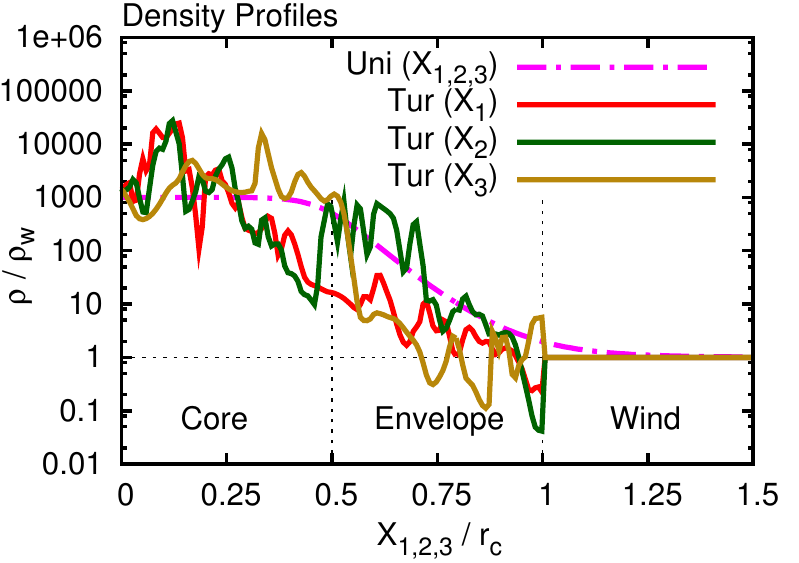}
\caption{Initial density profiles of a uniform cloud (dash-dotted line) and a turbulent cloud (solid lines) along the $X_1$, $X_2$, and $X_3$ directions. The horizontal axis shows the extent of the cloud core, cloud envelope, and ambient wind material up to $X_{1,2,3}/r_{\rm c}=1.5$, and the vertical axis shows the normalised density, $\rho/\rho_{\rm w}$, in logarithmic scale.}
\label{Figure1}
\end{figure}

For consistency, clouds with either uniform or turbulent profiles are assigned a spherical density distribution that smoothly decreases away from its centre (see \citealt{2000ApJ...531..366K,2006ApJS..164..477N}). The function describing the radial density gradient is:

\begin{equation}
\frac{\rho(r)}{\rho_{\rm w}}=1+\frac{\chi-1}{1+\left(\frac{2r}{r_{\rm c}}\right)^N},
\label{eq:DensityProfile}
\end{equation}

\noindent where $N$ is an integer that determines the steepness of the curve (see Figure \ref{Figure1}), and $\chi$ represents the density contrast between wind and cloud material, which is:

\begin{equation}
\chi=\frac{\rho_{\rm c}}{\rho_{\rm w}}=10^3
\label{eq:DensityContrast}
\end{equation}

\noindent for all models, where $\rho_{\rm c}$ is the target density at the centre of the cloud. Since the density profile of Equation (\ref{eq:DensityProfile}) extends to infinity, we impose a boundary for the cloud by selecting $N=10$ and a cut-off radius, $r_{\rm cut}$. In our model with a uniform cloud, we truncate the density function at $r_{\rm cut}=1.58\,r_{\rm c}$, at which point $\rho(r_{\rm cut})=1.01\,\rho_{\rm w}$, and we define the boundary of the cloud at $r_{\rm boundary}=1.0\,r_{\rm c}$, at which point $\rho(r_{\rm c})=2.0\,\rho_{\rm w}$. This ensures a smooth transition into the background gas. In our models with turbulent clouds, we define $r_{\rm cut}=r_{\rm boundary}=1.0\,r_{\rm c}$ for all configurations. Density gradients, similar to that described by Equation (\ref{eq:DensityProfile}), are expected in e.g., ISM atomic and molecular clouds, in which dense cores are surrounded by warm, low-density envelopes (e.g., \citealt*{1990ApJ...358..116W}; \citealt{1994ApJ...423..223C}; \citealt*{2009ApJ...698..350H}). All the clouds in our models are in thermal pressure equilibrium with the ambient medium at the beginning of the calculations.\par

Despite the smooth density transition between cloud and wind material achieved with Equation (\ref{eq:DensityProfile}), spherically-symmetric clouds with uniform densities are only ideal approximations. In reality, we know that ISM clumps are turbulent and have density profiles described by log-normal distributions (e.g., see \citealt{1999ApJ...526..279P}, \citealt{1999intu.conf..190W}, \citealt{2000AnRFM..32..203W}, \citealt{2004ASPC..323...79H}, \citealt*{2009A&A...504..883B}, \citealt{2012A&ARv..20...55H}, \citealt{2011ApJ...727L..21P}, \citealt{2012MNRAS.423.2680M}, \citealt{2012ApJ...761..149K}, \citealt{2015MNRAS.448.3297F}, \citealt{2015MNRAS.451.1380N,2015A&A...575A..79S} and references in Section \ref{sec:Importance}). Therefore, in order to initialise our simulations with more physical density profiles for the clumps, we take a $256^3$-sized data cube from a simulation snapshot of isothermal turbulence (see model 21 in Table 2 of \citealt{2012ApJ...761..156F}), and interpolate its density structure into our simulation domain. The probability density function (hereafter PDF) of the turbulent clouds studied in this paper is presented in Appendix \ref{sec:Appendix1}.\par

Before proceeding with the interpolation, we first taper the density profile in the data cube with the function given in Equation (\ref{eq:DensityProfile}), and mask regions in the cube outside a spherical volume of radius $r_{\rm c}$. The ensuing turbulent cloud is then interpolated into our simulation region and placed at the grid origin ($0,0,0$). As a final step, before initialising the simulations, we scale the density distribution in order to obtain the same initial mean density of the uniform clouds reported in \citetalias{2016MNRAS.455.1309B} and below. As a result, all models start off with clouds of the same mass and average density, $[~\rho_{\rm cloud}~]$. 

To prescribe the initial turbulent velocity field for the clouds, we use the velocity components from the aforementioned snapshot of model 21 in \cite{2012ApJ...761..156F} and scale the initial velocity dispersion by selecting an rms Mach number:

\begin{equation}
{\cal M_{\rm tu}}=\frac{|\bm{\delta v_{\rm cloud}}|}{c_{\rm cloud}}=7.2;{\rm\:8.9;\:or\:0.33},
\label{eq:CloudMachNumber}
\end{equation}

\noindent representative of adiabatic, supersonically-turbulent clouds; self-consistent, quasi-isothermal, supersonically-turbulent clouds; or adiabatic, subsonically-turbulent clouds, respectively. In Equation (\ref{eq:CloudMachNumber}), $|\bm{\delta v_{\rm cloud}}|\equiv \delta v_{\rm cloud}$ and $c_{\rm cloud}=\sqrt{\gamma\frac{P_{\rm {th}}}{\rho_{\rm c}}}$ are the initial velocity dispersion and initial sound speed of the cloud, respectively. The supersonic setups are representative of molecular (e.g., see \citealt{1981MNRAS.194..809L}) or cold, atomic (e.g., see \citealt{2003ApJ...586.1067H}) clouds, while the subsonic setups are more appropriate for warm, atomic (e.g., see \citealt{2014A&A...567A..16S}) or partially-ionised (e.g., see \citealt{2004ApJ...613.1004R,2011Natur.478..214G}) clouds in the ISM.

This study comprises nine models in total, for which we adopt a naming convention WWW-YYY-ZZZ for the models with the M configuration and Www-Yyy-Zzz for the models with the S configuration, such that WWW and Www describe the type of density field (i.e., UNI=Uni=Uniform, TUR=Tur=Turbulent), YYY and Yyy describe the type of turbulent velocity field (i.e., 0=Null, Sub=Subsonic, SUP=Sup=Supersonic), and ZZZ and Zzz describe the type of turbulent magnetic field (i.e., 0=Null, Bwk=Weak $\bm B$ field, BST=Bst=Strong $\bm B$ field). Our nine models are split into two sets of numerical MHD simulations (see Table \ref{Table1}). The first set has three models that utilise the M configuration mentioned above. Model UNI-0-0 includes a uniform (i.e., non-turbulent) cloud embedded in a uniform magnetic field, while models TUR-SUP-BST and TUR-SUP-BST-ISO (where ISO stands for quasi-isothermal) include turbulent clouds with self-consistent density, velocity, and magnetic field profiles. The second set has six models that utilise the S configuration. Model Uni-0-0 includes a uniform cloud and serves as a comparison between filament formation mechanisms in models with and without turbulence. Model Tur-0-0 is our control run and includes a turbulent cloud with the log-normal density PDF mentioned above (see Equation \ref{eq:PDF} in Appendix \ref{sec:Appendix1}). Models Tur-Sub-0 and Tur-Sup-0 include a turbulent cloud with the same density PDF of model Tur-0-0, plus the Gaussian velocity field mentioned above with subsonic and supersonic Mach numbers, respectively, as described by Equation (\ref{eq:CloudMachNumber}).\par

\begin{table*}\centering
\caption{Simulation parameters for different MHD models. In column $1$, we provide the model names. In columns $2$ and $3$, we indicate the computational domain configurations (i.e., model identifiers or IDs and sizes) and resolutions, respectively. Note that the model identifiers (M1-3 and S4-9) are also used in the panels with the 3D renderings presented in Section \ref{sec:Results}. In column $4$, we describe the type of density profile that is initialised in each cloud model. In columns $5-7$, we provide the polytropic index, the Mach number of the wind, and the wind-cloud density contrast, respectively. In columns $8$ and $9$, we provide details on the configuration and sonic Mach number, respectively, of the initial velocity field in the clouds. In columns $10-12$, we describe the initial topology, the plasma beta of the uniform component, and the mean plasma beta of the turbulent component, respectively, of the initial magnetic field in the cloud. The magnetic field in the wind is uniform (with an oblique topology) and has a plasma beta of $\beta_{\rm ob}=100$ in all cases.}
\begin{adjustbox}{max width=\textwidth}
\begin{tabular}{c c c c c c c c c c c c}
\hline
\textbf{(1)} & \textbf{(2)} & \textbf{(3)} & \textbf{(4)} & \textbf{(5)} & \textbf{(6)} & \textbf{(7)} & \textbf{(8)} & \textbf{(9)} & \textbf{(10)} & \textbf{(11)} & \textbf{(12)}\Tstrut\\
\textbf{Model} & \textbf{ID \& Domain} &\textbf{Resolution} & \textbf{Density Field} & $\gamma$ & $\cal M_{\rm w}$ & $\chi$ & \textbf{Velocity Field} & $\cal M_{\rm tu}$ & \textbf{Magnetic Field} & $\beta_{\rm ob}$ & $[~\beta_{\rm tu}~]$\Tstrut\Bstrut \\ \hline
UNI-0-0 & ${\rm M1}\equiv(6\times18\times6)\,r_{\rm c}$ & $R_{64}$ & Uniform &  $1.667$ & 4 & $10^3$ & Null & -- & Uniform Oblique & $100$ & -- \Tstrut \\
TUR-SUP-BST & ${\rm M2}\equiv(6\times18\times6)\,r_{\rm c}$ & $R_{64}$ & Turbulent &  $1.667$ & 4 & $10^3$ & Turbulent & 7.2 & Oblique + Turbulent & $100$ & $0.04$ \\
TUR-SUP-BST-ISO & ${\rm M3}\equiv(6\times18\times6)\,r_{\rm c}$ & $R_{64}$ & Turbulent &  $1.100$ & 4.9 & $10^3$ & Turbulent & 8.9 & Oblique + Turbulent & $100$ & $0.04$\Bstrut\\\hline
Uni-0-0 & ${\rm S4}\equiv(4\times12\times4)\,r_{\rm c}$ & $R_{128}$ & Uniform &  $1.667$ & 4 & $10^3$ & Null & -- & Uniform Oblique & $100$ & -- \Tstrut \\
Tur-0-0 & ${\rm S5}\equiv(4\times12\times4)\,r_{\rm c}$ & $R_{128}$ & Turbulent &  $1.667$ & 4 & $10^3$ & Null & -- & Uniform Oblique & $100$ & --\\
Tur-Sub-0 & ${\rm S6}\equiv(4\times12\times4)\,r_{\rm c}$ & $R_{128}$ & Turbulent &  $1.667$ & 4 & $10^3$ & Turbulent & $0.33$ & Uniform Oblique & $100$ & --\\
Tur-Sup-0 & ${\rm S7}\equiv(4\times12\times4)\,r_{\rm c}$ & $R_{128}$ &Turbulent &  $1.667$ & 4 & $10^3$ & Turbulent & $8.9$ & Uniform Oblique & $100$ & -- \\ 
Tur-Sub-Bwk & ${\rm S8}\equiv(4\times12\times4)\,r_{\rm c}$ & $R_{128}$ &Turbulent &  $1.667$ & 4 & $10^3$ & Turbulent & $0.33$ & Oblique + Turbulent & $100$ & $4$\\
Tur-Sub-Bst & ${\rm S9}\equiv(4\times12\times4)\,r_{\rm c}$ & $R_{128}$ &Turbulent &  $1.667$ & 4 & $10^3$ & Turbulent & $0.33$ & Oblique + Turbulent & $100$ & $0.04$\Bstrut\\\hline
\end{tabular}
\end{adjustbox}
\label{Table1}
\end{table*} 

In five of the models mentioned above (UNI-0-0, Uni-0-0, Tur-0-0, Tur-Sub-0, and Tur-Sup-0) we add an oblique magnetic field, uniformly distributed over the entire simulation domain (similar to our setup in \citetalias{2016MNRAS.455.1309B}), using the following equation:

\begin{equation}
\bm{B}=\bm{B_{\rm ob}=B_{\rm 1}+B_{\rm 2}+B_{\rm 3}},
\label{eq:ObliqueField}
\end{equation}

\noindent in which the 3D magnetic field has components along $X_1$, $X_2$, and $X_3$ of identical magnitude:

\begin{equation}
|\bm{B_{\rm 1}|=|B_{\rm 2}|=|B_{\rm 3}}|=\sqrt{\frac{2P_{\rm th}}{3\beta_{\rm ob}}}
\label{eq:BComponents}
\end{equation}

\noindent where the plasma beta, $\beta_{\rm ob}$ is a dimensionless number that relates the thermal pressure, $P_{\rm th}$, to the magnetic pressure in the oblique field, $P_{\rm mag, ob}=\frac{1}{2}|{\bf{B}}|^2=\frac{1}{2}|{\bf{B_{\rm ob}}}|^2$, and is given by:

\begin{equation}
\beta_{\rm ob}=\frac{P_{\rm th}}{P_{\rm mag, ob}}=\frac{P_{\rm th}}{\frac{1}{2}|\bm{B_{\rm ob}}|^2}=100,
\label{eq:Beta}
\end{equation}

\noindent in all of our models. In order to isolate the effects of a tangled, turbulent magnetic field on the formation of filaments, models TUR-SUP-BST, TUR-SUP-BST-ISO, Tur-Sub-Bwk, and Tur-Sub-Bst include turbulent clouds with the density PDF and Gaussian velocity fields (with supersonic Mach numbers in the first two cases and subsonic Mach numbers in the other two cases), plus a two-component magnetic field given by:

\begin{equation}
\bm{B=B_{\rm ob}+B_{\rm tu}=(B_{\rm 1}+B_{\rm 2}+B_{\rm 3})+B_{\rm tu}},
\label{eq:TurbulentField}
\end{equation}

\noindent i.e., the total magnetic field in the cloud is the sum of a 3D uniform magnetic field obliquely oriented with respect to the wind direction with components given by Equation (\ref{eq:BComponents}), and a turbulent magnetic field extracted from model 21 of \cite{2012ApJ...761..156F} with a turbulent plasma beta defined by:

\begin{equation}
\beta_{\rm tu}=\frac{P_{\rm th}}{P_{\rm mag, tu}}=\frac{P_{\rm th}}{\frac{1}{2}|\bm{B_{\rm tu}}|^2}=0.04;\rm{\:or\:4},
\label{eq:BetaTurb}
\end{equation}

\noindent for strong- and weak-field simulations, respectively (the reader is referred to Panel C of Figure 7.6 in \citealt{2016PhDT.......154B} for a 3D streamline plot of the initial topology of the turbulent magnetic field in the simulation domain). In models TUR-SUP-BST, TUR-SUP-BST-ISO, and Tur-Sub-Bst the initial turbulent magnetic field is scaled so that its average plasma beta is $[~\beta_{\rm tu}~]=0.04$ (i.e., the magnetic field is strong and consistent with the magnetic distribution in the initial turbulence data cube from \citealt{2012ApJ...761..156F}), while in model Tur-Sub-Bwk the initial turbulent magnetic field is scaled so that its average plasma beta is $[~\beta_{\rm tu}~]=4$ (i.e., the magnetic field is weak).\par

Note that the initial magnetic field in the wind is $\bm{B_{\rm ob}}$ in all cases, so if the magnetic fields given in Equation (\ref{eq:TurbulentField}) were directly interpolated into the simulation grids of models TUR-SUP-BST, TUR-SUP-BST-ISO, Tur-Sub-Bst, and Tur-Sub-Bwk, the solenoidal property would be violated at the boundaries of the clouds (due to truncation). In order to ensure that the initial magnetic fields in these models are solenoidal (i.e., that $\bm{\nabla\cdot B}=0$), we clean the divergence errors before initialising these simulations. We follow the hyperbolic, divergence-cleaning algorithm introduced by \cite{2002JCoPh.175..645D} and implemented by \cite{2010JCoPh.229.5896M} to perform this operation (see Appendix \ref{sec:Appendix2} for further details). Once the magnetic fields for models TUR-SUP-BST, TUR-SUP-BST-ISO, Tur-Sub-Bst, and Tur-Sub-Bwk satisfy the divergence-free constraint and have the desired value of $[~\beta_{\rm tu}~]$, we interpolate them into our simulation domains and update them numerically with the system of equations described in Section \ref{subsec:SimulationCode}.

\subsection{Diagnostics}
\label{subsec:Diagnostics}
Similarly to \citetalias{2016MNRAS.455.1309B}, we use the following global diagnostics to study the formation and evolution of filaments in our simulations:\par

i) The volume-averaged value of a variable $\cal F$ is denoted by square brackets as follows:

\begin{equation}
[~{\cal F}_{\alpha}~]=\frac{\int {\cal F}C_{\alpha}dV}{V_{\rm cl}}=\frac{\int {\cal F}C_{\alpha}dV}{\int C_{\alpha}dV},
\label{eq:AveragedF}
\end{equation}

\noindent where $V$ is the volume, $C_{\alpha}$ are the advected scalars defined in Section \ref{subsec:SimulationCode}, and $V_{\rm cl}$ is the total cloud volume. Using Equation (\ref{eq:AveragedF}), we define functions describing the average density, $[~\rho_{\alpha}~]$; the average plasma beta, $[~\beta_{\alpha}~]$; the average magnetic field, $[~B_{{\rm j},\alpha}~]$; and its rms along each axis, $[~B^2_{{\rm j},\alpha}~]^{\frac{1}{2}}$. The subscript $\rm j=1,2,3$ specifies the direction along $X_1$, $X_2$, and $X_3$, respectively.\par

ii) The mass-weighted volume average of the variable $\cal G$ is denoted by angle brackets as follows:

\begin{equation}
\langle~{\cal G}_{\alpha}~\rangle=\frac{\int {\cal G}\rho C_{\alpha}dV}{M_{\rm cl}}=\frac{\int {\cal G}\rho C_{\alpha}dV}{\int \rho C_{\alpha}dV},
\label{eq:IntegratedG}
\end{equation}

\noindent where $V$ and $C_{\alpha}$ are as defined above, and $M_{\rm cl}$ is the total cloud mass. Using Equation (\ref{eq:IntegratedG}), we define the average filament/cloud extension, $\langle~X_{{\rm j},\alpha}~\rangle$; its rms along each axis, $\langle~X^2_{{\rm j},\alpha}~\rangle^{\frac{1}{2}}$; the average velocity, $\langle~v_{{\rm j},\alpha}~\rangle$; and its rms along each axis, $\langle~v^2_{{\rm j},\alpha}~\rangle^{\frac{1}{2}}$. In order to retain the scalability of our results, these quantities are normalised with respect to their initial values. Velocity measurements are the exemption to this as they are normalised with respect to the wind speed, $v_{\rm w}$.\par

A)\footnote{Note that the notation used for the list of diagnostics introduced in this section has been chosen so that it matches the notation used for the panels of the Figures presented in Section \ref{sec:Results}. This facilitates the identification of different parameters, their mathematical definitions, and their respective plots (see Table \ref{Table2}).} Using the above definitions, we estimate the length-to-width and width-to-width aspect ratio of filaments along $\rm j=2,3$, respectively, as follows:

\begin{equation}
\xi_{{\rm j},\alpha}=\frac{\iota_{{\rm j},\alpha}}{\iota_{1,\alpha}},
\label{eq:AspectRatio}
\end{equation}

\noindent where $\iota_{{\rm j},\alpha}$ are the effective radii (see \citealt{1994ApJ...420..213K}) along each axis ($\rm j=1,2,3$):

\begin{equation}
\iota_{{\rm j},\alpha}=\left[5\left(\langle~X^2_{{\rm j},\alpha}~\rangle-\langle~X_{{\rm j},\alpha}~\rangle^2\right)\right]^{\frac{1}{2}}.
\label{eq:Moments}
\end{equation}

B, H) From Equation (\ref{eq:Moments}), we define the lateral width/expansion (along $X_1$) and the displacement of the centre of mass of filaments (along the streaming axis, $X_2$) as $\iota_{1,\alpha}$ and $\langle~X_{{\rm 2},\alpha}~\rangle$, respectively.\par

C) In a similar way, we define the total (for $\rm j=1,2,3$) and transverse (for $\rm j=1,3$) velocity dispersion as follows:

\begin{equation}
\delta_{{\rm v}_{\alpha}}\equiv|\bm{\delta_{{\rm v}_{\alpha}}}|=\sqrt{\sum_{\rm j}\delta_{{\rm v}_{{\rm j},\alpha}}^2},
\label{eq:rmsVelocity}
\end{equation}

\noindent where the corresponding dispersion of the $\rm j$-component of the velocity (see \citealt{1994ApJ...433..757M}), $\delta_{{\rm v}_{{\rm j},\alpha}}$, reads

\begin{equation}
\delta_{{\rm v}_{{\rm j},\alpha}}=\left(\langle~v^2_{{\rm j},\alpha}~\rangle-\langle~v_{{\rm j},\alpha}~\rangle^2\right)^{\frac{1}{2}}.
\label{eq:rmsVelocityComponent}
\end{equation}

I) From Equation (\ref{eq:rmsVelocityComponent}), we define the bulk velocity of filaments as $\langle~v_{2,\alpha}~\rangle$. The temporal behaviour of this parameter is used to study the acceleration of the cloud.\par

D) Using Equation (\ref{eq:AveragedF}), we also measure the mean vorticity $[~\omega_{\alpha}~]$ of the gas in the filaments, where $\omega_{\alpha}=|~{\bm \omega_{\alpha}}~|=|~{\bm\nabla}\bm{\times v_{\alpha}}~|$, and the averaged value is normalised with respect to its initial value (i.e., $[~\omega_{\alpha,0}~]$). 

In order to quantify the kinetic energy densities in filament material, we decompose the total velocity field into mean, $\overline{v_{{\rm j},\alpha}}\equiv\langle~v_{{\rm j},\alpha}~\rangle$; and turbulent, $v'_{{\rm j},\alpha}$, components, i.e., $\bm{v_{\alpha}}=\bm{\overline{v_{\alpha}}+v'_{\alpha}}$ (see \citealt{2004ApJ...616..669K,2004tise.book.....D,2014MNRAS.438.2513P} for thorough discussions on statistical averaging in problems involving MHD turbulence). Thus, the corresponding turbulent kinetic energy density reads:

\begin{equation}
E'_{{\rm k},\alpha}=\frac{1}{2}\rho|\bm{v'_{\alpha}}|^2.
\label{eq:TurbKinEner}
\end{equation}

E) Using Equation (\ref{eq:TurbKinEner}), we define the averaged turbulent kinetic energy density of filaments as $[~E'_{{\rm k},\alpha}~]$.\par

Similarly, to study the magnetic energy densities in filament material we decompose the total magnetic field into mean, $\overline{B_{{\rm j},\alpha}}\equiv[~B_{{\rm j},\alpha}~]$; and turbulent, $B'_{{\rm j},\alpha}$ components, i.e., $\bm{B_{\alpha}}=\bm{\overline{B_{\alpha}}+B'_{\alpha}}$. Thus, we define the turbulent magnetic energy density,

\begin{equation}
E'_{{\rm m},\alpha}=\frac{1}{2}|\bm{B'_{\alpha}}|^2,
\label{eq:TurbMagEner}
\end{equation}

\noindent in filament material. Note that we normalise the above energy densities with respect to the wind kinetic energy density, $E_{\rm k,w}=\frac{1}{2}\rho_{\rm w}v_{\rm w}^2$.\par

F, G) Using Equation (\ref{eq:TurbMagEner}), we calculate two parameters: the averaged turbulent magnetic energy density as $[~E'_{{\rm m},\alpha}~]$, and the ratio between turbulent magnetic and turbulent kinetic energy densities, i.e., $[~E'_{{\rm m},\alpha}~]/[~E'_{{\rm k},\alpha}~]$.\par

Note that a summary of all the diagnostics described above is presented in Table \ref{Table2}.

\begin{table}\centering
\caption{Summary of the diagnostics described in Section \ref{subsec:Diagnostics}, which we employ to investigate the morphology, energetics, and dynamics of filaments produced in wind-cloud interactions. In column $1$, we provide the diagnostic identifier (ID) using the letters A through I, which we also use to label the plots presented in Section \ref{sec:Results}. In columns $2$ and $3$, we list the symbol and definition of each diagnostic.}
\begin{adjustbox}{max width=0.475\textwidth}
\begin{tabular}{c c c}
\hline
\textbf{(1)} & \textbf{(2)} & \textbf{(3)}\Tstrut\\
\textbf{ID} & \textbf{Diagnostic} &\textbf{Description}\Tstrut\Bstrut \\ \hline
A & $\xi_{2,\alpha}$ & Length-to-width aspect ratio \Tstrut \\
B & $\iota_{1,\alpha}$ & Lateral width\\
C & $\delta_{{\rm v}_{\alpha}}$ & Transverse velocity dispersion\\
D & $[~\omega_{\alpha}~]$ & Mean vorticity\\
E & $[~E'_{{\rm k},\alpha}~]$ & Averaged turbulent kinetic energy density\\
F & $[~E'_{{\rm m},\alpha}~]$ & Averaged turbulent magnetic energy density\\
G & $[~E'_{{\rm m},\alpha}~]/[~E'_{{\rm k},\alpha}~]$ & Ratio between turbulent energy densities\\
H & $\langle~X_{{\rm 2},\alpha}~\rangle$ & Displacement of the centre of mass\\
I & $\langle~v_{2,\alpha}~\rangle$ & Bulk speed in the direction of streaming\Bstrut \\ \hline
\end{tabular}
\end{adjustbox}
\label{Table2}
\end{table} 

\subsection{Reference time-scales}
\label{subsec:DynamicalTime-Scales}
The relevant dynamical time-scales in our simulations are:

\noindent a) The cloud-crushing time (see \citealt{1994ApJ...432..194J,1996ApJ...473..365J}),

\begin{equation}
t_{\rm cc}=\frac{2r_{\rm c}}{v_{\rm s}}=\left(\frac{\rho_{\rm c}}{\rho_{\rm w}}\right)^{\frac{1}{2}}\frac{2r_{\rm c}}{{\cal M_{\rm w}} c_{\rm w}}=\chi^{\frac{1}{2}}\frac{2r_{\rm c}}{{\cal M_{\rm w}} c_{\rm w}},
\label{eq:CloudCrushing}
\end{equation}

\noindent where $v_{\rm s}={\cal M_{\rm w}} c_{\rm w}\chi^{-\frac{1}{2}}$ is the approximate speed of the internal shock travelling through the cloud after the initial collision with the wind. In order to maintain scalability, all the time-scales reported in this paper are normalised with respect to the cloud-crushing time.

\noindent b) The simulation time, which in our case is:

\begin{equation}
t_{\rm sim}=1.25\,t_{\rm cc}.
\label{eq:SimulationTime}
\end{equation}

\noindent c) The wind-passage time:

\begin{equation}
t_{\rm wp}=\frac{2r_{\rm c}}{v_{\rm w}}=\frac{1}{\chi^{\frac{1}{2}}}\,t_{\rm cc}=0.032\,t_{\rm cc}.
\label{eq:WindPassage}
\end{equation}

\noindent d) The turbulence-crossing time:

\begin{equation}
t_{\rm tu}=\frac{2r_{\rm c}}{\delta v_{\rm cloud}}=\frac{2r_{\rm c}}{{\cal M_{\rm tu}}c_{\rm cloud}}
\label{eq:TurbulenceCrossing}
\end{equation}

\noindent e) The KH instability growth time (see \citealt{1961hhs..book.....C}):

\begin{equation}
\frac{t_{\rm KH}}{t_{\rm cc}}\simeq\left[\frac{\rho'_{\rm c}\rho'_{\rm w}k_{\rm KH}^2}{(\rho'_{\rm c}+\rho'_{\rm w})^2}(v'_{\rm w}-v'_{\rm c})^2-\frac{2B^{'2}k_{\rm KH}^2}{(\rho'_{\rm c}+\rho'_{\rm w})}\right]^{-\frac{1}{2}}\frac{{\cal M_{\rm w}}c_{\rm w}}{2r_{\rm c}\chi^{\frac{1}{2}}},
\label{KHtime}
\end{equation}

\noindent where $k_{\rm KH}=\frac{2\pi}{\lambda_{\rm KH}}$ is the wavenumber of the KH perturbations and the primed quantities of the physical variables correspond to their values at the location of shear layers.

\noindent f) The RT instability growth time (see \citealt{1961hhs..book.....C}):

\begin{equation}
\frac{t_{\rm RT}}{t_{\rm cc}}\simeq\left[\left(\frac{\rho'_{\rm c}-\rho'_{\rm w}}{\rho'_{\rm c}+\rho'_{\rm w}}\right)ak_{\rm RT}-\frac{2B^{'2}k_{\rm RT}^2}{(\rho'_{\rm c}+\rho'_{\rm w})}\right]^{-\frac{1}{2}}\frac{{\cal M_{\rm w}}c_{\rm w}}{2r_{\rm c}\chi^{\frac{1}{2}}},
\label{RTtime}
\end{equation}

\noindent where $k_{\rm RT}=\frac{2\pi}{\lambda_{\rm RT}}$ is the wavenumber of the RT perturbations, $a$ is the local, effective acceleration of dense gas, and the primed quantities of the physical variables correspond to their values at the leading edge of the cloud. Both the KT and RT time-scales in Equations (\ref{KHtime}) and (\ref{RTtime}), respectively, correspond to the incompressible regime, so they should only be considered as indicative for the highly compressible models considered in this series of papers.\par

To ensure that sequential snapshots adequately capture details of the evolution of filamentary tails, simulation outputs are written at intervals of $\Delta t=8.2\times 10^{-3}\,t_{\rm cc}$.\par

\section{Results}
\label{sec:Results}
We split this section into two parts. In the first part, Section \ref{subsec:FilamentFormation}, we contrast the overall process of filament formation in uniform and turbulent environments by examining the morphological properties of these structures in three models with the M domain configuration, i.e., with $R_{64}$, (see Table \ref{Table1}). In this section we also present a summary of the properties and kinematics of filaments and wind-swept clouds in different models and discuss the entrainment of these structures in supersonic winds. In the second part, Section \ref{subsec:Turbulence}, we analyse the effects of turning on and off different profiles for the turbulent density, velocity, and magnetic field distributions of six different models with the S domain configuration, i.e., with $R_{128}$, (see Table \ref{Table1}). In this section we systematically investigate models with turbulent clouds and compare the morphological, kinematic, and magnetic properties of the resulting filaments with their non-turbulent counterpart.

\subsection{Filament formation and structure in clouds with turbulent density, velocity and magnetic fields}
\label{subsec:FilamentFormation}
\begin{figure*}
\begin{center}
  \begin{tabular}{c c c c c c}
\multicolumn{6}{l}{\hspace{-0.3cm}M1) UNI-0-0}\\ 
\hspace{-0.3cm}\resizebox{31.5mm}{!}{\includegraphics{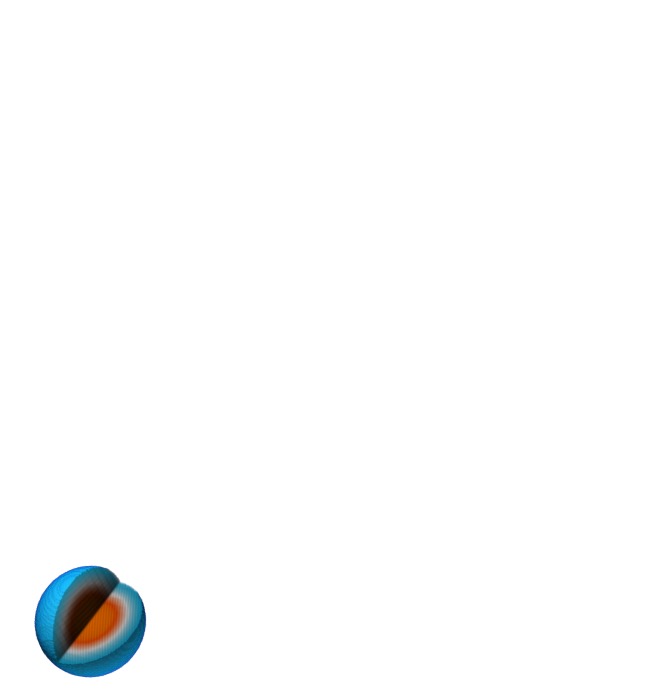}} & \hspace{-1cm}\resizebox{31.5mm}{!}{\includegraphics{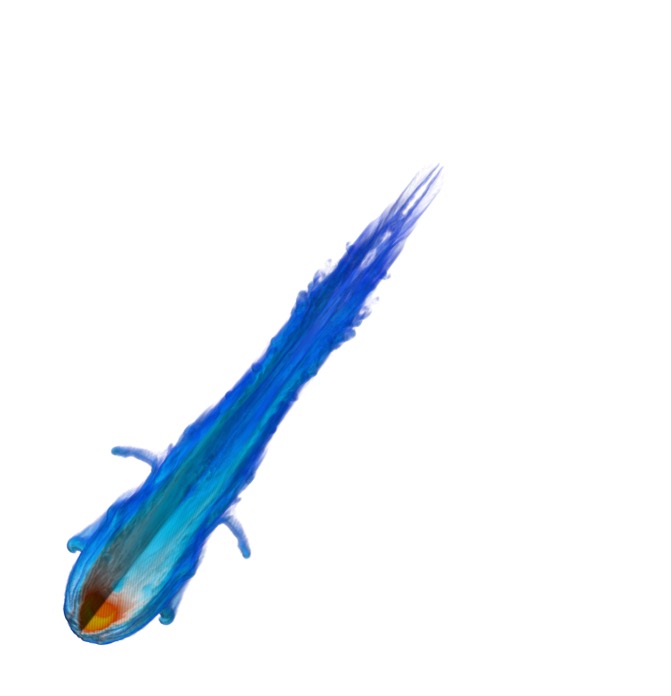}} & \hspace{-0.5cm}\resizebox{31.5mm}{!}{\includegraphics{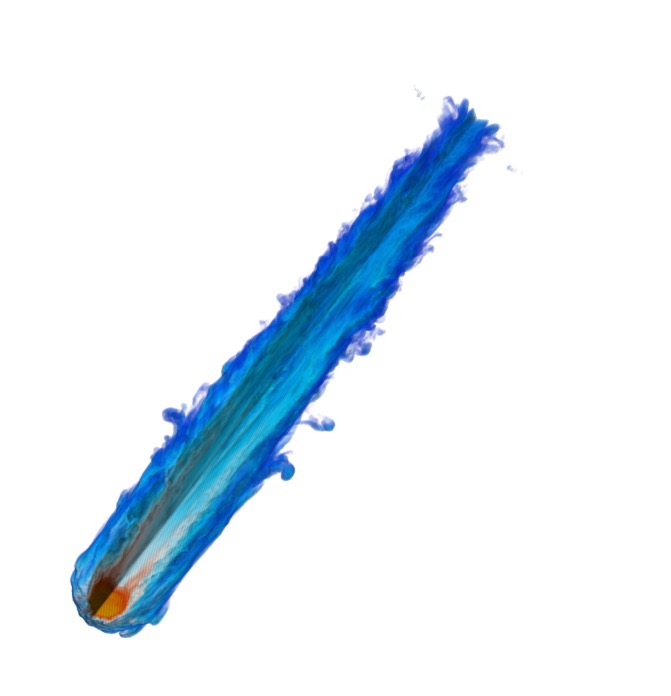}} & \hspace{-0.5cm}\resizebox{31.5mm}{!}{\includegraphics{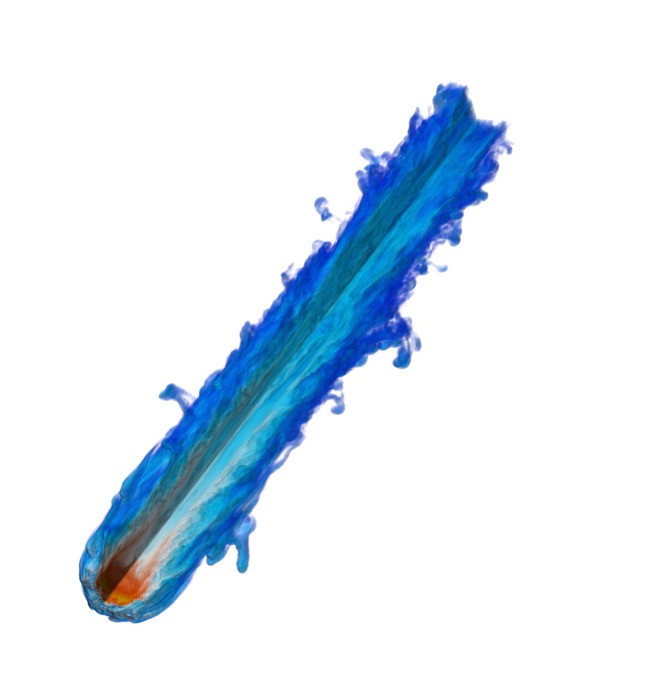}} & \hspace{-0.5cm}\resizebox{31.5mm}{!}{\includegraphics{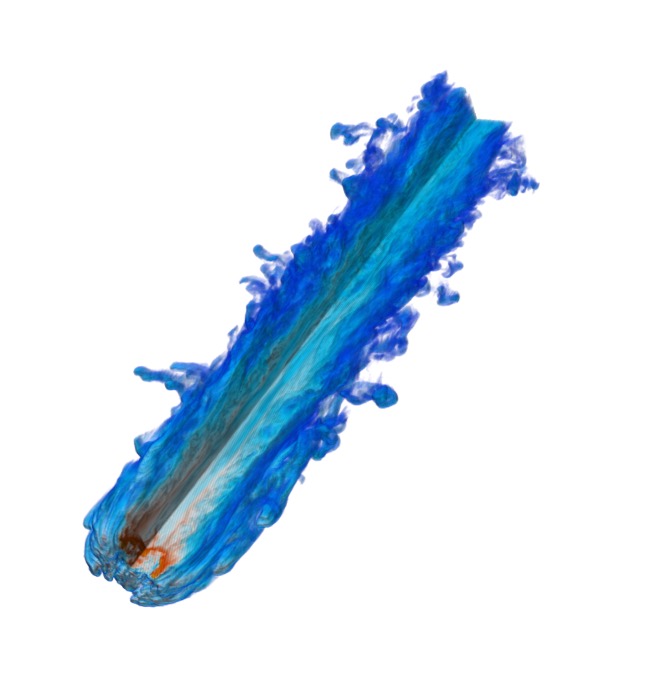}} & \hspace{-0.5cm}\resizebox{31.5mm}{!}{\includegraphics{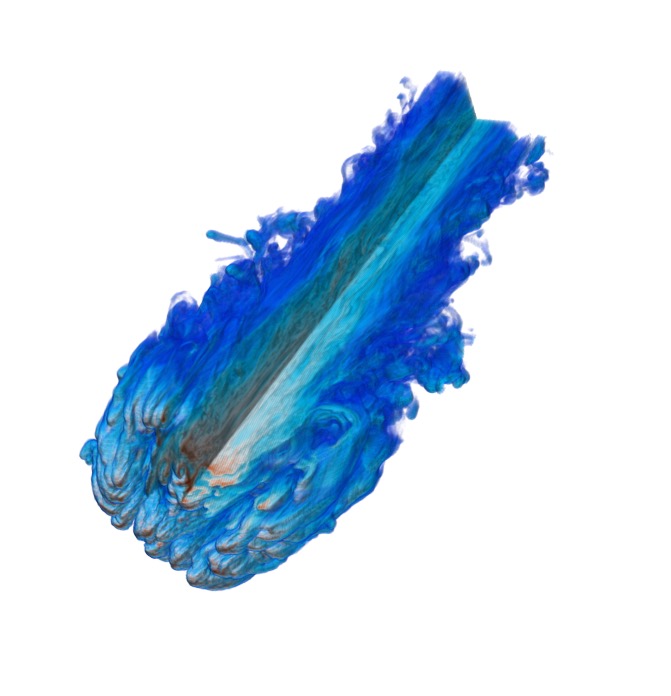}}\\
\hspace{-2cm}$t/t_{\rm cc}=0$ & \hspace{-1.5cm}$t/t_{\rm cc}=0.25$ & \hspace{-0.5cm}$t/t_{\rm cc}=0.50$ & \hspace{-0.5cm}$t/t_{\rm cc}=0.75$ & \hspace{-0.5cm}$t/t_{\rm cc}=1.00$ & \hspace{-0.5cm}$t/t_{\rm cc}=1.25$\Dstrut\\
\multicolumn{6}{l}{\hspace{-0.3cm}M2) TUR-SUP-BST}\\ 
\hspace{-0.3cm}\resizebox{31.5mm}{!}{\includegraphics{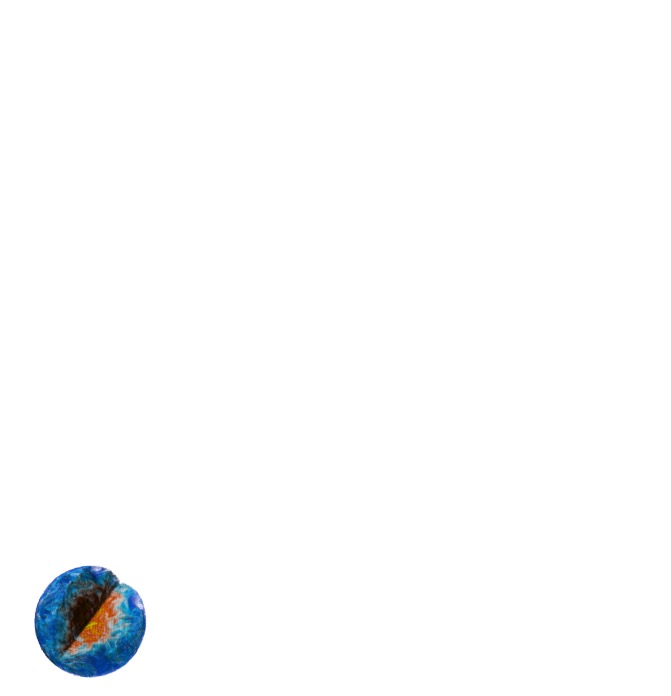}} & \hspace{-1cm}\resizebox{31.5mm}{!}{\includegraphics{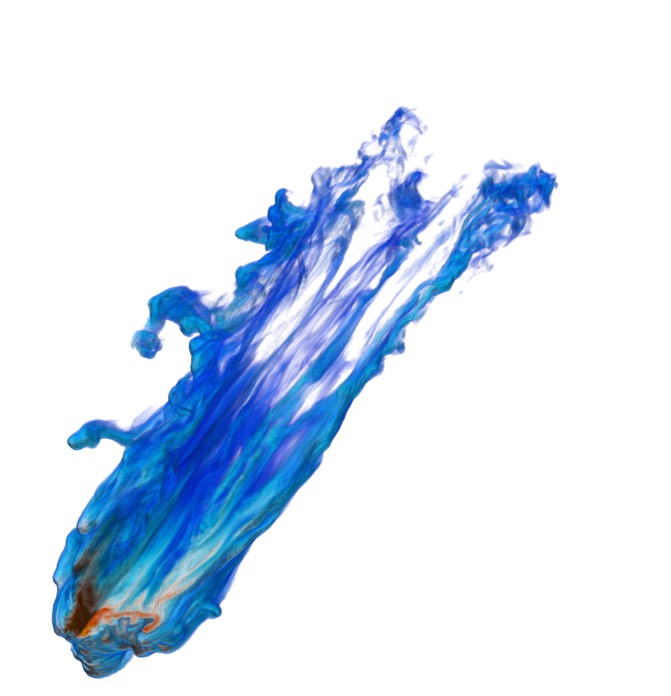}} & \hspace{-0.5cm}\resizebox{31.5mm}{!}{\includegraphics{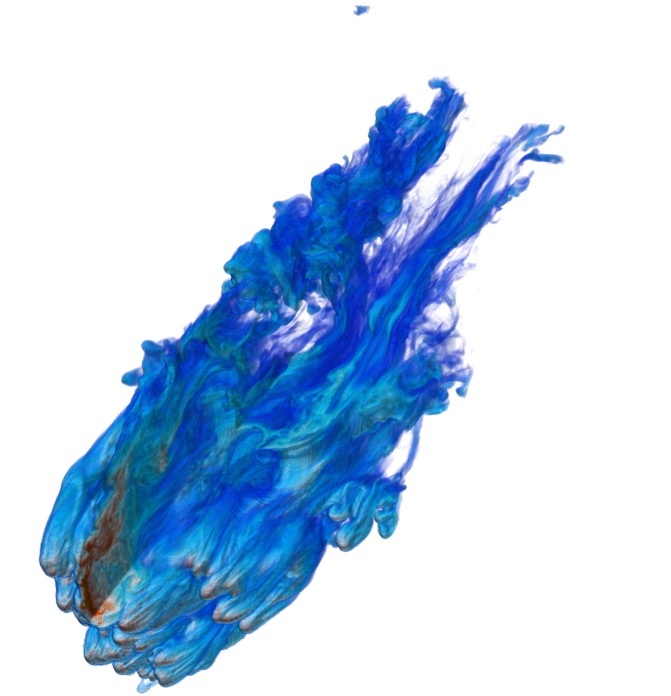}} & \hspace{-0.5cm}\resizebox{31.5mm}{!}{\includegraphics{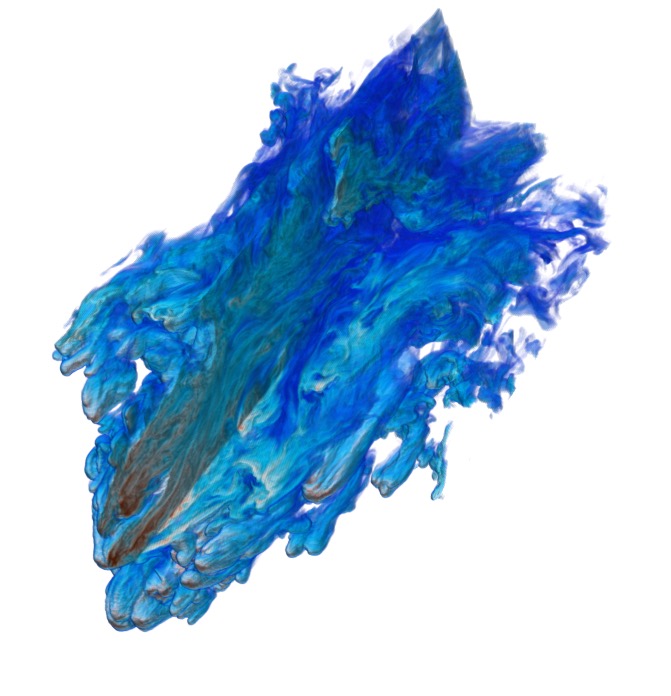}} & \hspace{-0.5cm}\resizebox{31.5mm}{!}{\includegraphics{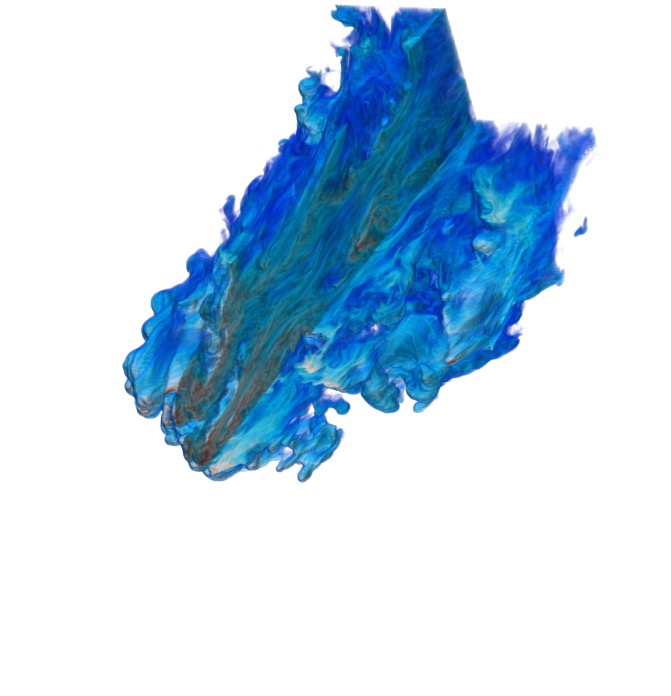}} & \hspace{-0.5cm}\resizebox{31.5mm}{!}{\includegraphics{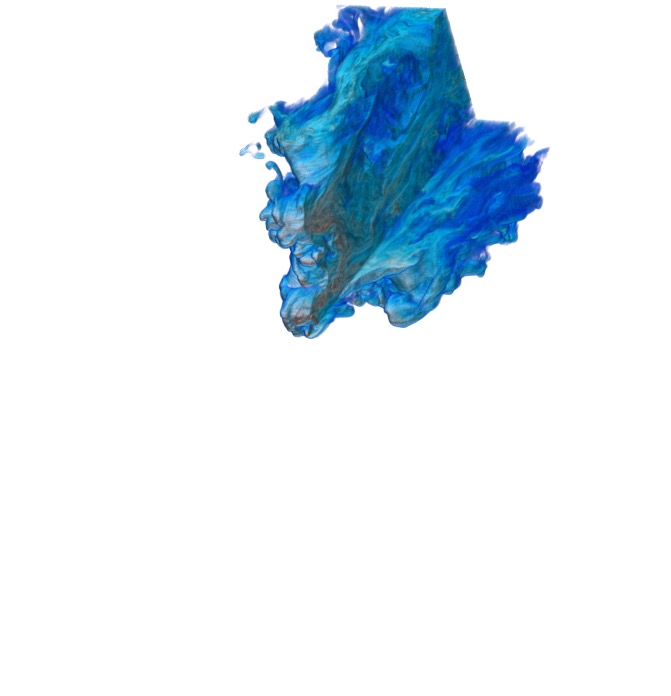}}\\
\hspace{-2cm}$t/t_{\rm cc}=0$ & \hspace{-1.5cm}$t/t_{\rm cc}=0.25$ & \hspace{-0.5cm}$t/t_{\rm cc}=0.50$ & \hspace{-0.5cm}$t/t_{\rm cc}=0.75$ & \hspace{-0.5cm}$t/t_{\rm cc}=1.00$ & \hspace{-0.5cm}$t/t_{\rm cc}=1.25$\Dstrut\\
\multicolumn{6}{l}{\hspace{-0.3cm}M3) TUR-SUP-BST-ISO}\\ 
\hspace{-0.3cm}\resizebox{31.5mm}{!}{\includegraphics{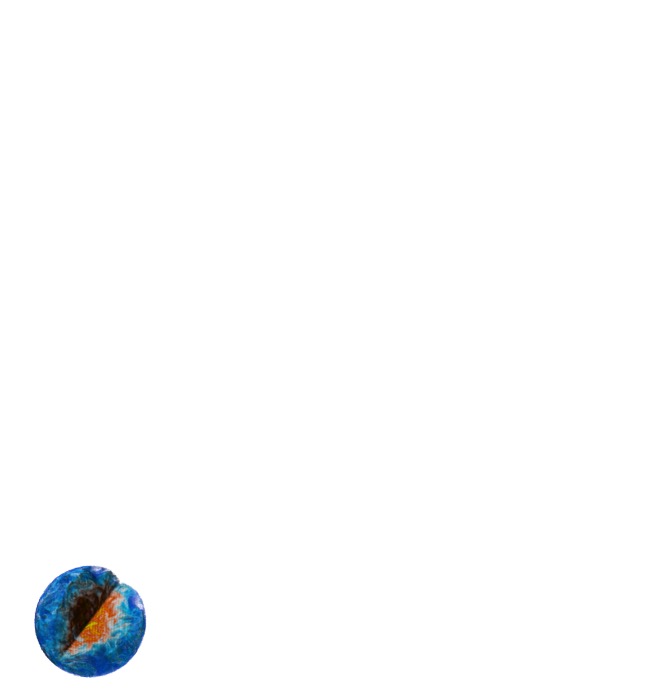}} & \hspace{-1cm}\resizebox{31.5mm}{!}{\includegraphics{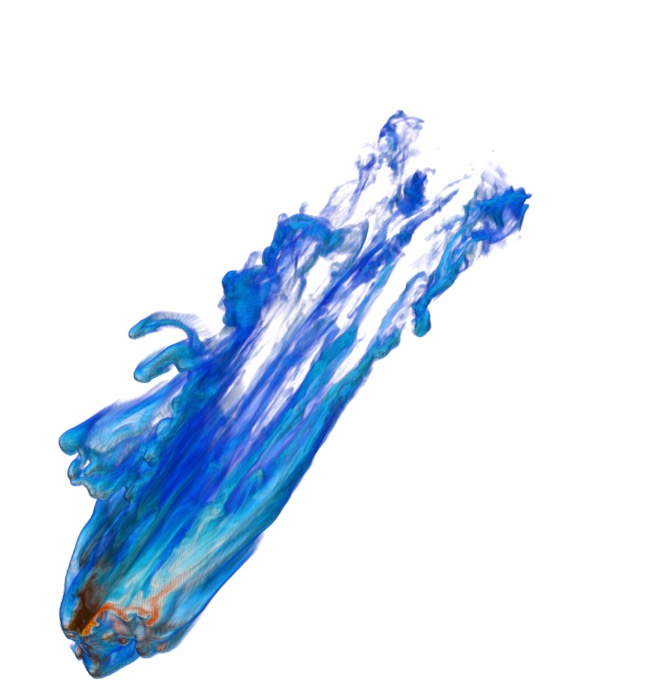}} & \hspace{-0.5cm}\resizebox{31.5mm}{!}{\includegraphics{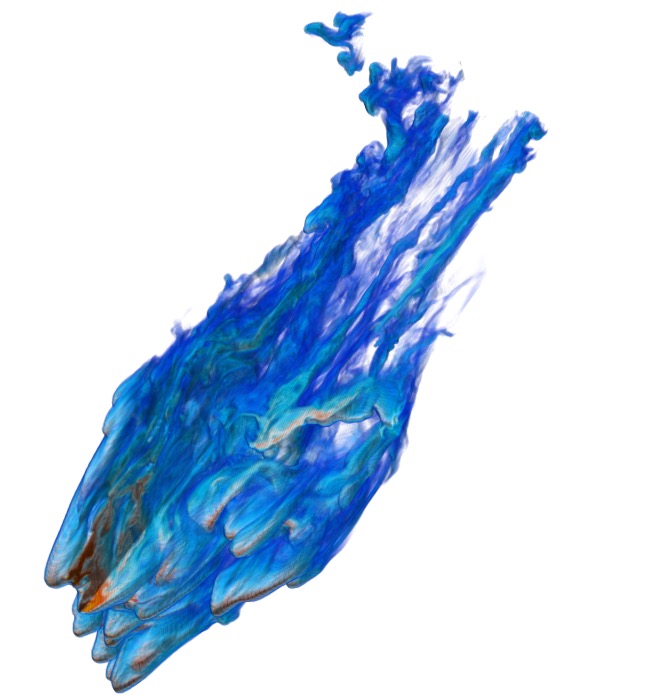}} & \hspace{-0.5cm}\resizebox{31.5mm}{!}{\includegraphics{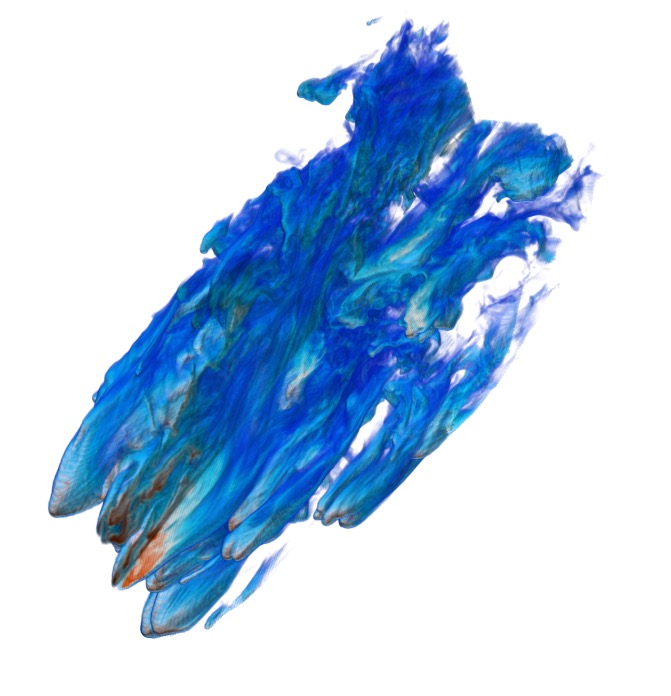}} & \hspace{-0.5cm}\resizebox{31.5mm}{!}{\includegraphics{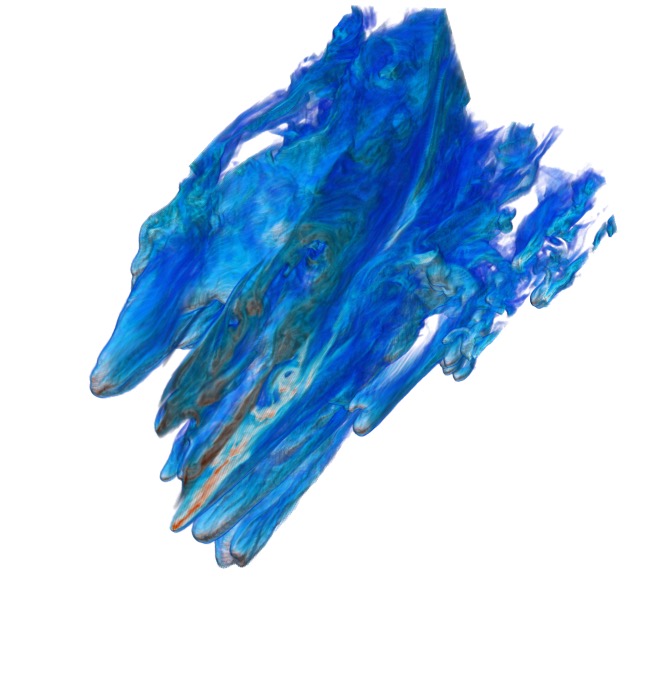}} & \hspace{-0.5cm}\resizebox{31.5mm}{!}{\includegraphics{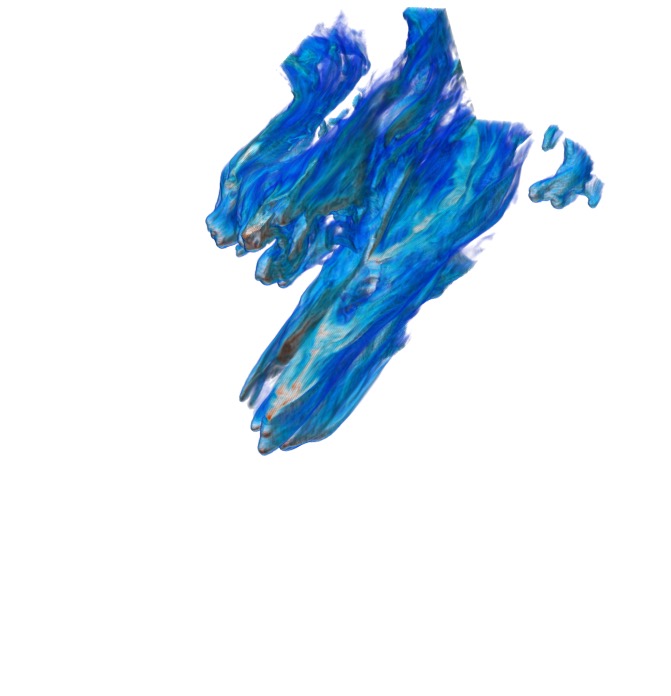}}\\
\hspace{-2cm}$t/t_{\rm cc}=0$ & \hspace{-1.5cm}$t/t_{\rm cc}=0.25$ & \hspace{-0.5cm}$t/t_{\rm cc}=0.50$ & \hspace{-0.5cm}$t/t_{\rm cc}=0.75$ & \hspace{-0.5cm}$t/t_{\rm cc}=1.00$ & \hspace{-0.5cm}$t/t_{\rm cc}=1.25$\Dstrut\\
\resizebox{!}{6mm}{\includegraphics{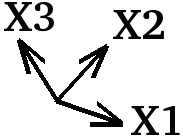}}   & \multicolumn{5}{c}{\resizebox{!}{6mm}{\includegraphics{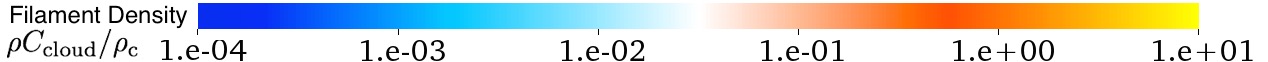}}}\\
  \end{tabular}
  \caption{3D volume renderings of the logarithm of the mass density in filaments ($\rho C_{\rm cloud}$) normalised with respect to the initial cloud density, $\rho_{\rm c}$, for $0\leq t/t_{\rm cc}\leq1.25$. Panel M1 shows the evolution of an adiabatic, uniform cloud immersed in an oblique magnetic field (UNI-0-0). Panels M2 and M3 show the evolution of turbulent clouds with log-normal density distributions, Gaussian velocity fields, and turbulent magnetic fields with two equations of state: adiabatic (TUR-SUP-BST) and quasi-isothermal (TUR-SUP-BST-ISO), respectively. Note how the inclusion of turbulent clouds leads to the formation of filaments with larger cross sectional areas, more complex density substructures, and higher displacements in the direction of streaming (i.e., higher accelerations) than the one arising from a uniform cloud. Using a quasi-isothermal (radiative) polytropic index produces less vortices at wind-filament interfaces, leading to the formation of a more laminar filamentary tail. Movies showing the full-time evolution of the models presented here are available online at \url{https://goo.gl/iXgJYk}.}
  \label{Figure2}
\end{center}
\end{figure*}

\begin{figure*}
\begin{center}
  \begin{tabular}{c c c c c c}
\multicolumn{6}{l}{\hspace{-0.3cm}M1) UNI-0-0}\\ 
\hspace{-0.3cm}\resizebox{31.5mm}{!}{\includegraphics{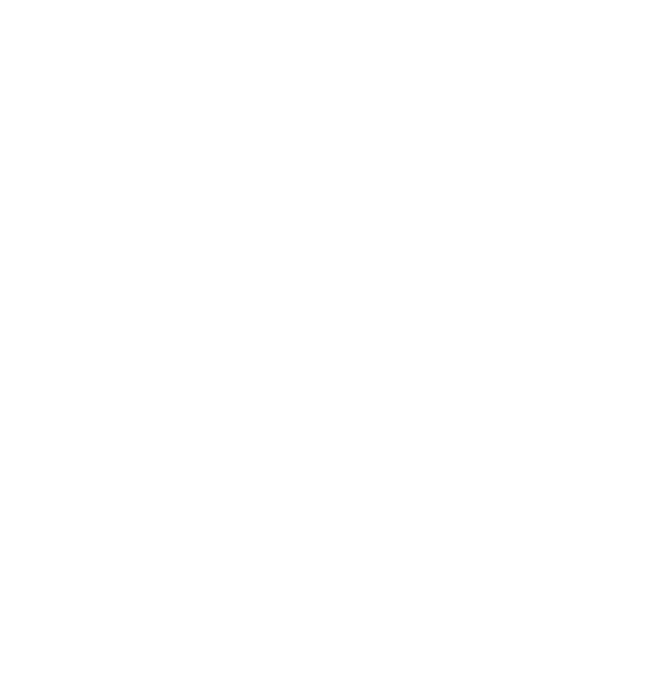}} & \hspace{-1cm}\resizebox{31.5mm}{!}{\includegraphics{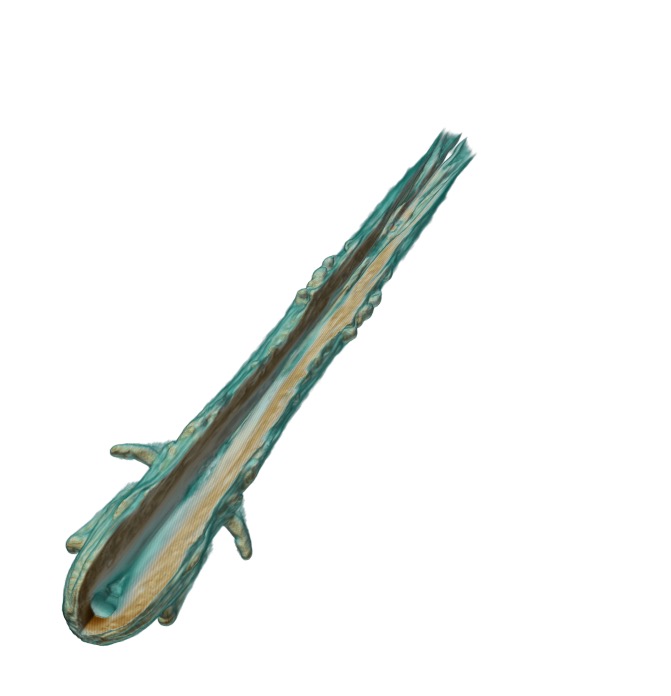}} & \hspace{-0.5cm}\resizebox{31.5mm}{!}{\includegraphics{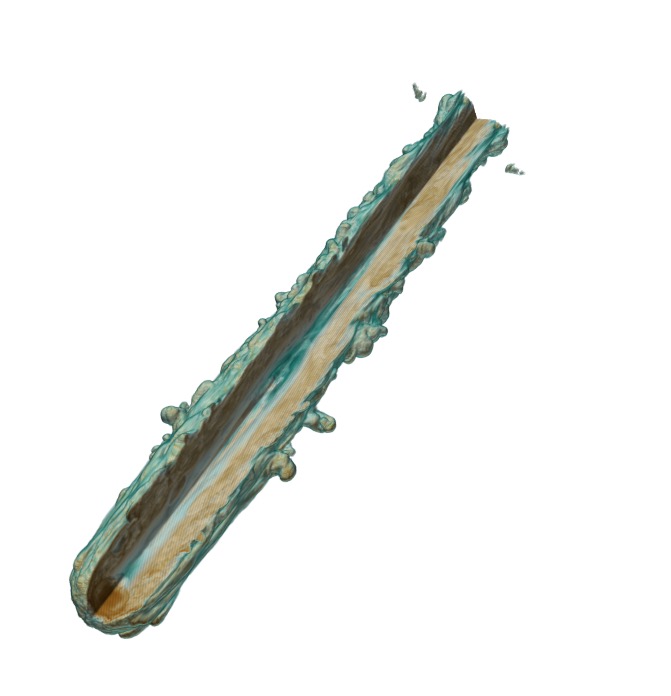}} & \hspace{-0.5cm}\resizebox{31.5mm}{!}{\includegraphics{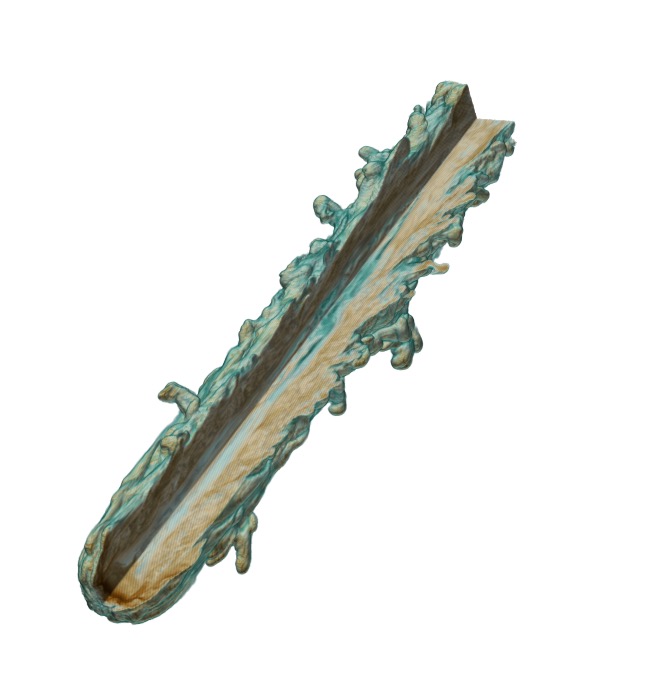}} & \hspace{-0.5cm}\resizebox{31.5mm}{!}{\includegraphics{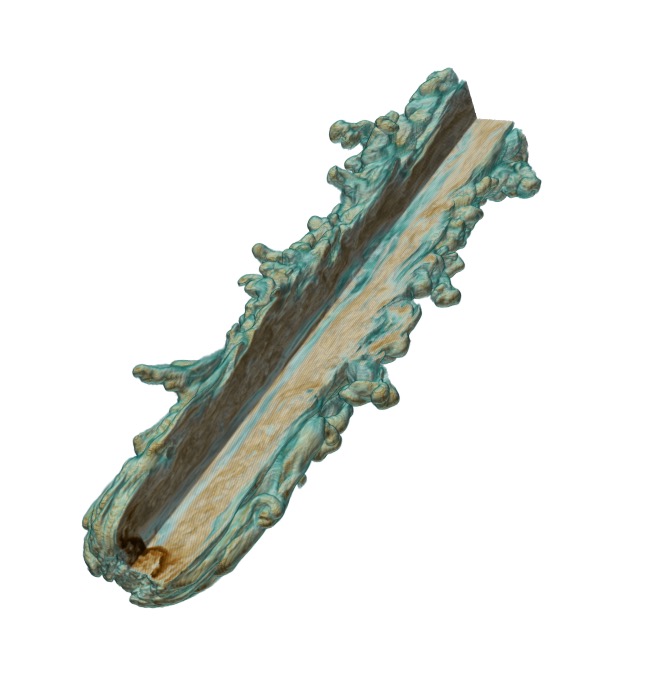}} & \hspace{-0.5cm}\resizebox{31.5mm}{!}{\includegraphics{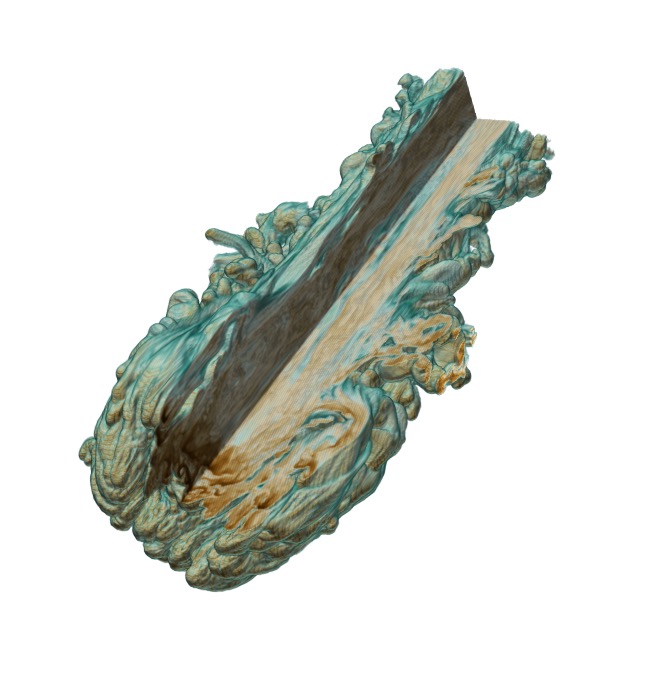}}\\
\hspace{-2cm}$t/t_{\rm cc}=0$ & \hspace{-1.5cm}$t/t_{\rm cc}=0.25$ & \hspace{-0.5cm}$t/t_{\rm cc}=0.50$ & \hspace{-0.5cm}$t/t_{\rm cc}=0.75$ & \hspace{-0.5cm}$t/t_{\rm cc}=1.00$ & \hspace{-0.5cm}$t/t_{\rm cc}=1.25$\Dstrut\\
\multicolumn{6}{l}{\hspace{-0.3cm}M2) TUR-SUP-BST}\\ 
\hspace{-0.3cm}\resizebox{31.5mm}{!}{\includegraphics{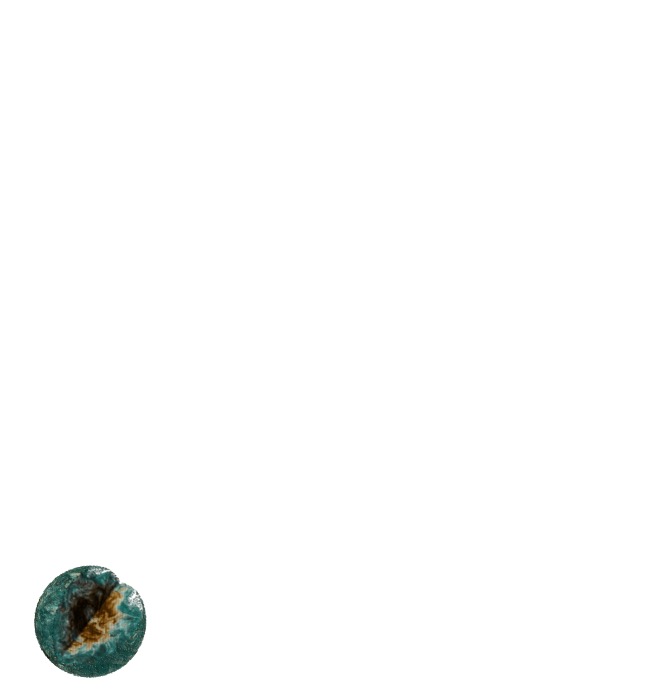}} & \hspace{-1cm}\resizebox{31.5mm}{!}{\includegraphics{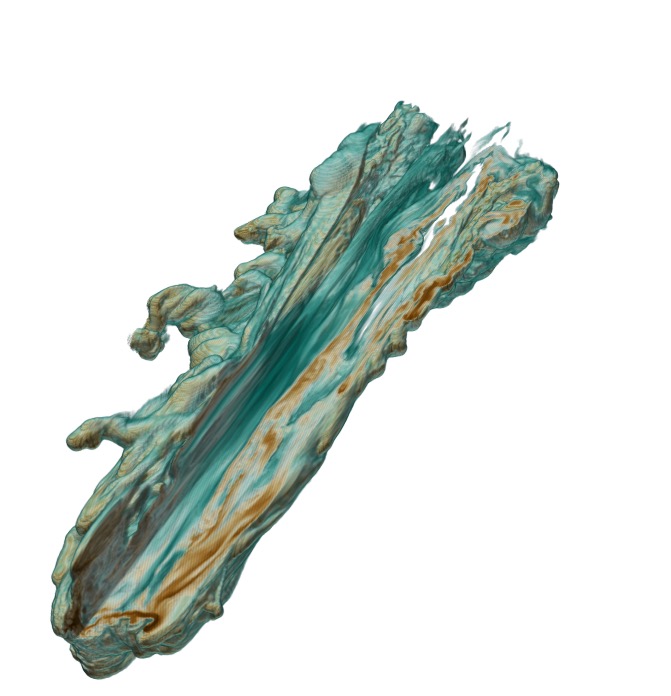}} & \hspace{-0.5cm}\resizebox{31.5mm}{!}{\includegraphics{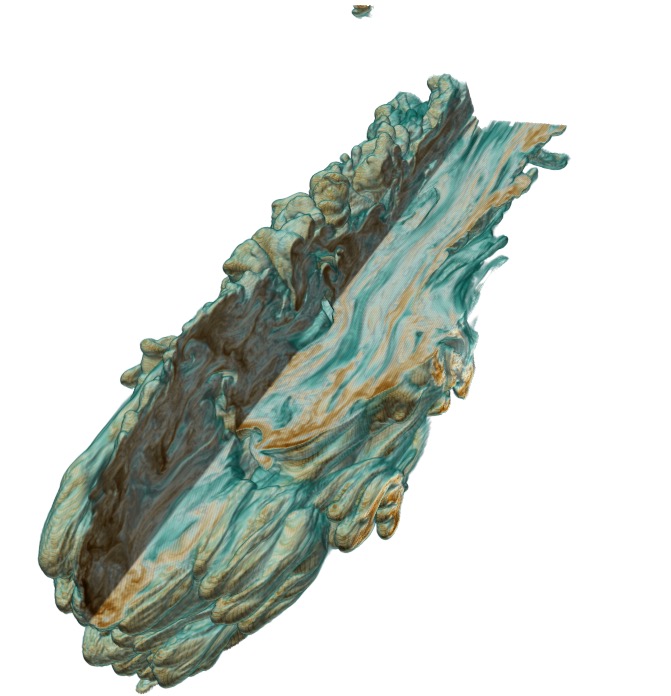}} & \hspace{-0.5cm}\resizebox{31.5mm}{!}{\includegraphics{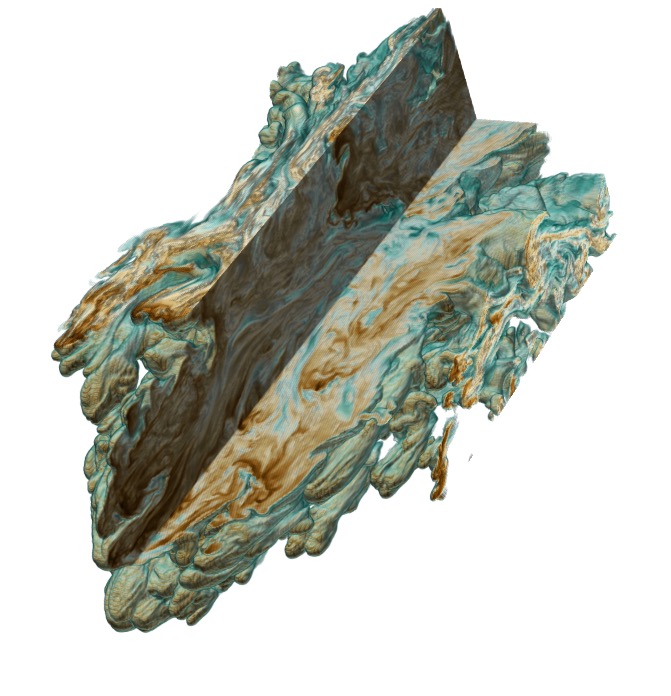}} & \hspace{-0.5cm}\resizebox{31.5mm}{!}{\includegraphics{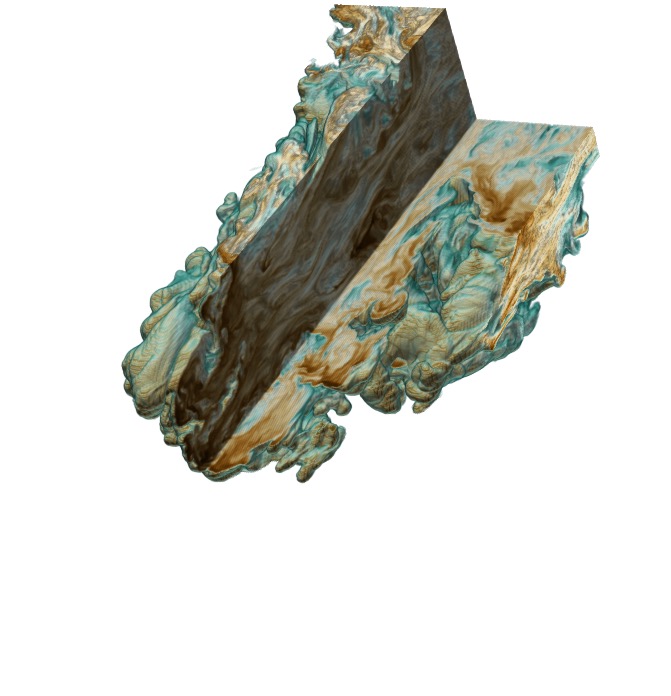}} & \hspace{-0.5cm}\resizebox{31.5mm}{!}{\includegraphics{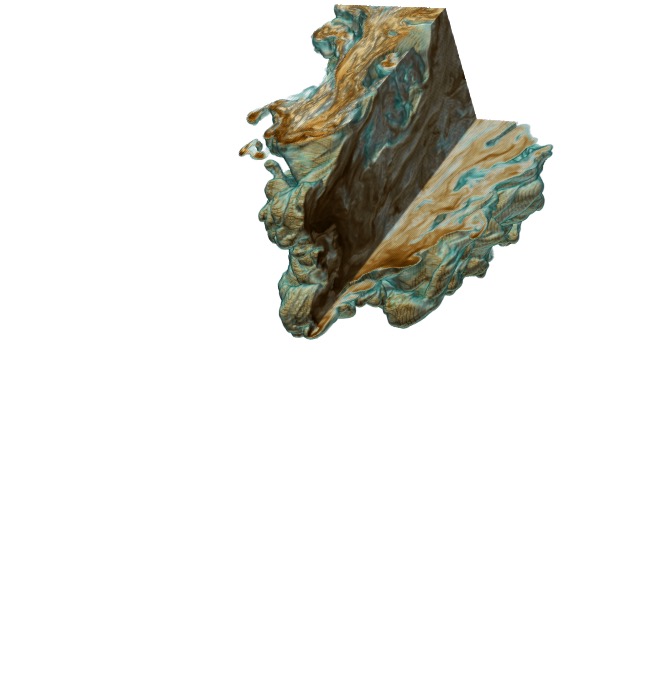}}\\
\hspace{-2cm}$t/t_{\rm cc}=0$ & \hspace{-1.5cm}$t/t_{\rm cc}=0.25$ & \hspace{-0.5cm}$t/t_{\rm cc}=0.50$ & \hspace{-0.5cm}$t/t_{\rm cc}=0.75$ & \hspace{-0.5cm}$t/t_{\rm cc}=1.00$ & \hspace{-0.5cm}$t/t_{\rm cc}=1.25$\Dstrut\\
\multicolumn{6}{l}{\hspace{-0.3cm}M3) TUR-SUP-BST-ISO}\\ 
\hspace{-0.3cm}\resizebox{31.5mm}{!}{\includegraphics{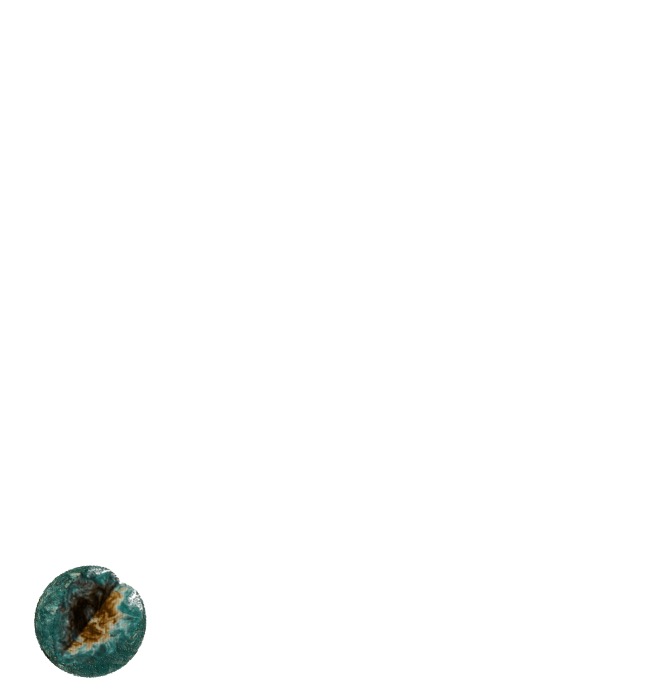}} & \hspace{-1cm}\resizebox{31.5mm}{!}{\includegraphics{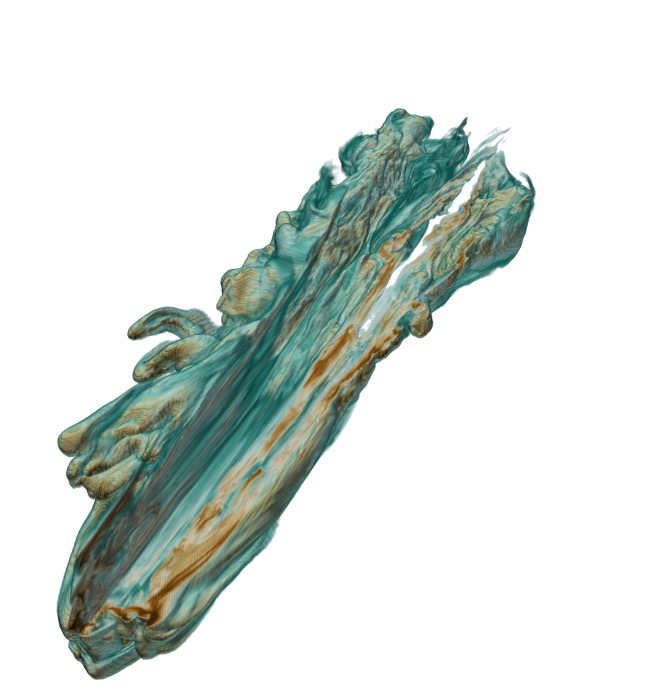}} & \hspace{-0.5cm}\resizebox{31.5mm}{!}{\includegraphics{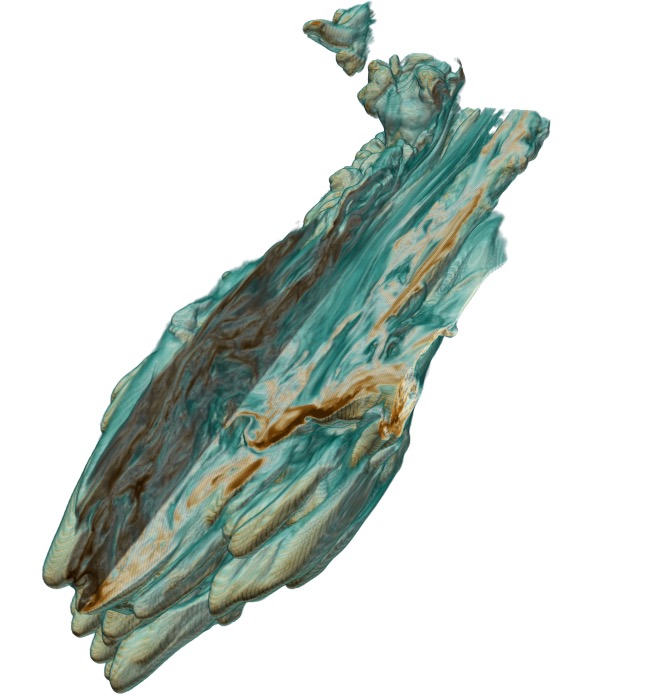}} & \hspace{-0.5cm}\resizebox{31.5mm}{!}{\includegraphics{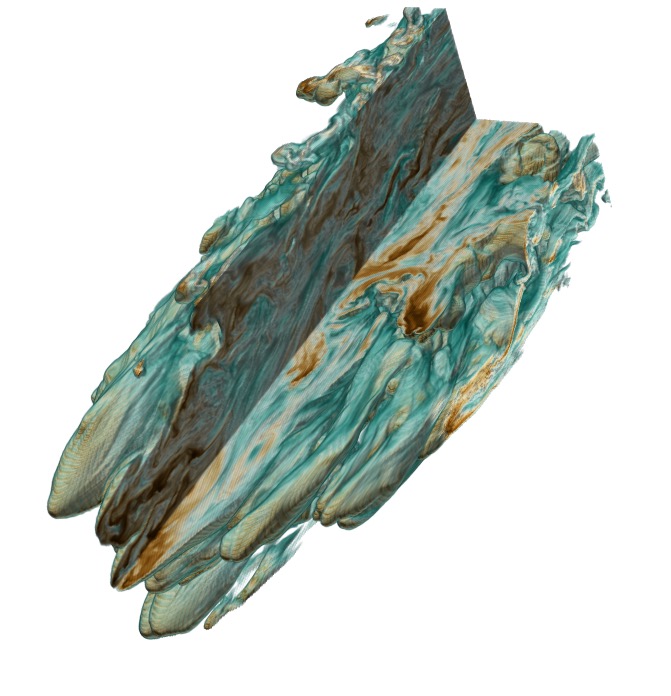}} & \hspace{-0.5cm}\resizebox{31.5mm}{!}{\includegraphics{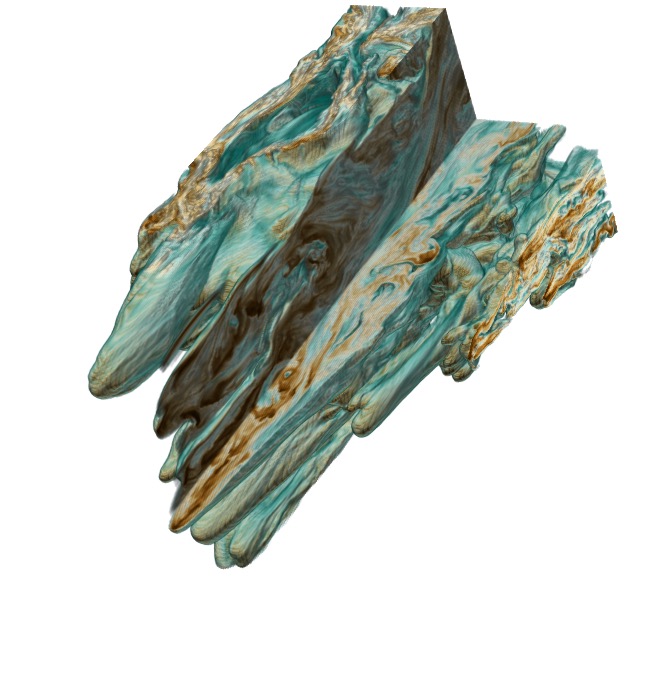}} & \hspace{-0.5cm}\resizebox{31.5mm}{!}{\includegraphics{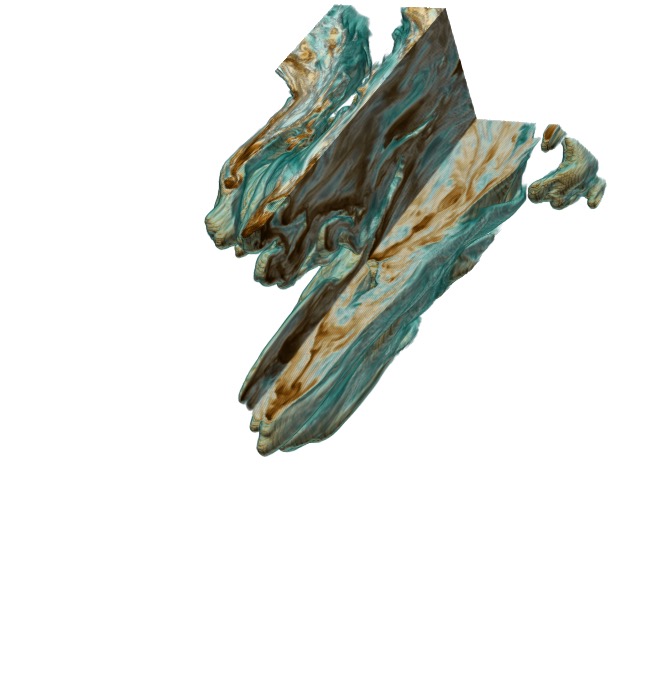}}\\
\hspace{-2cm}$t/t_{\rm cc}=0$ & \hspace{-1.5cm}$t/t_{\rm cc}=0.25$ & \hspace{-0.5cm}$t/t_{\rm cc}=0.50$ & \hspace{-0.5cm}$t/t_{\rm cc}=0.75$ & \hspace{-0.5cm}$t/t_{\rm cc}=1.00$ & \hspace{-0.5cm}$t/t_{\rm cc}=1.25$\Dstrut\\
\resizebox{!}{6mm}{\includegraphics{Axes3D.jpg}}   & \multicolumn{5}{c}{\resizebox{!}{6mm}{\includegraphics{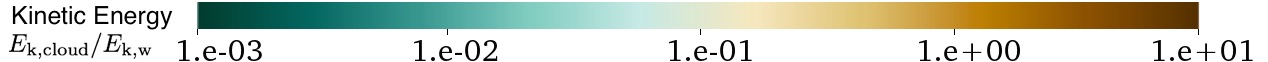}}}\\
  \end{tabular}
  \caption{Same as Figure \ref{Figure2}, but here we present 3D volume renderings of the logarithm of the kinetic energy density in filaments ($E_{\rm k,cloud}$) normalised with respect to the wind kinetic energy density, $E_{\rm k,w}$, for $0\leq t/t_{\rm cc}\leq1.25$. Panel M1 shows the evolution of a uniform cloud immersed in an oblique magnetic field (UNI-0-0) with an initially-null, internal velocity field. Panels M2 and M3 show the evolution of turbulent clouds with log-normal density distributions, Gaussian velocity fields, and turbulent magnetic fields in adiabatic (TUR-SUP-BST) and quasi-isothermal (TUR-SUP-BST-ISO) cases, respectively. The inclusion of turbulence aids cloud expansion and results in the formation of high-kinetic-energy knots and sub-filaments in the interior of the main filamentary tails. The expansion caused by the initial Gaussian velocity field increases the cross sectional area of the clouds, accelerating them to the point where some parcels of high-density gas have the same kinetic energy of the external wind (see regions in light brown colour). Both turbulent models have higher kinetic energies than the uniform model, but the inclusion of a softer equation of state (see model M3) leads to a more confined, slower turbulent filament than in model M2. Movies showing the full-time evolution of the models presented here are available online at \url{https://goo.gl/iXgJYk}.} 
  \label{Figure3}
\end{center}
\end{figure*}

\begin{figure*}
\begin{center}
  \begin{tabular}{c c c c c c}
\multicolumn{6}{l}{\hspace{-0.3cm}M1) UNI-0-0}\\ 
\hspace{-0.3cm}\resizebox{31.5mm}{!}{\includegraphics{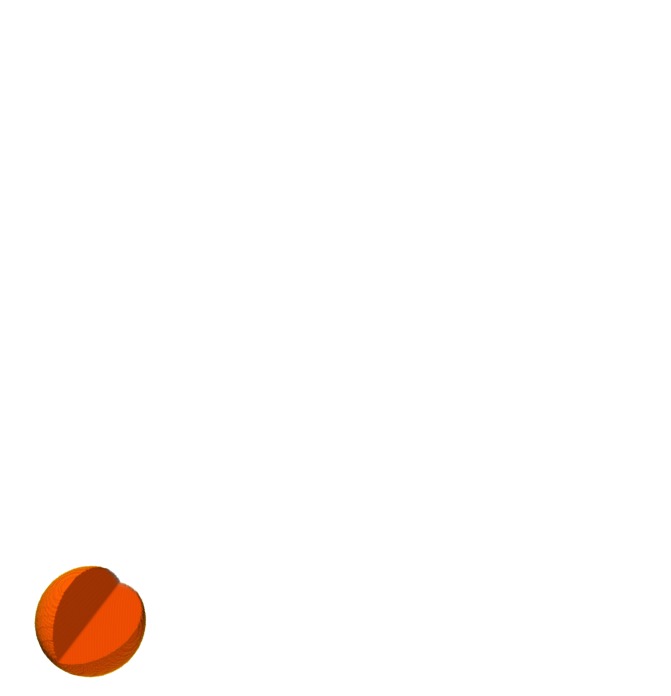}} & \hspace{-1cm}\resizebox{31.5mm}{!}{\includegraphics{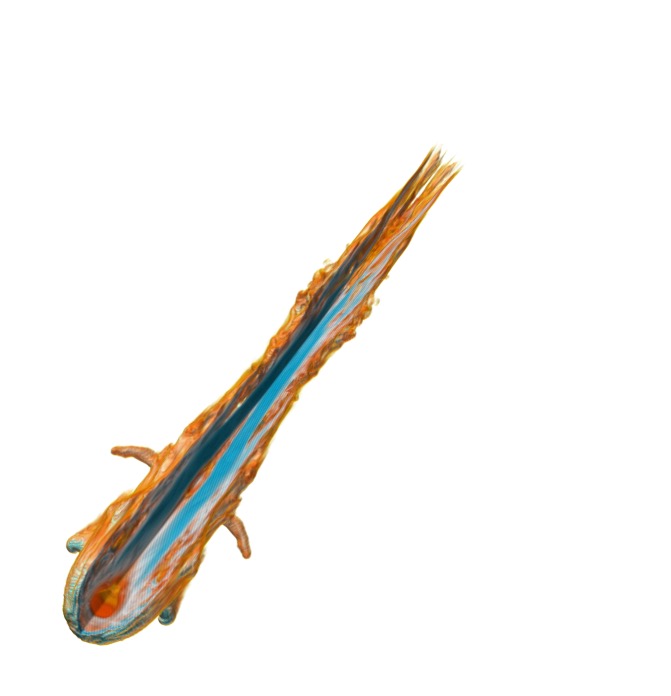}} & \hspace{-0.5cm}\resizebox{31.5mm}{!}{\includegraphics{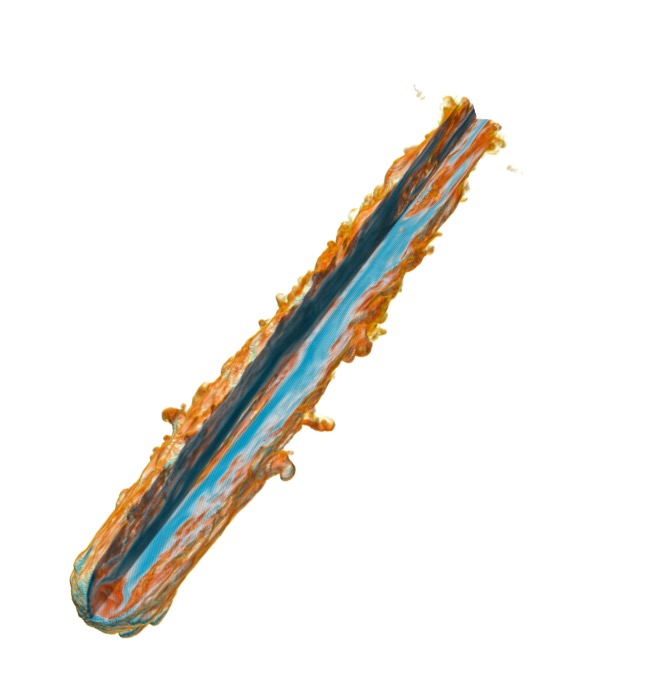}} & \hspace{-0.5cm}\resizebox{31.5mm}{!}{\includegraphics{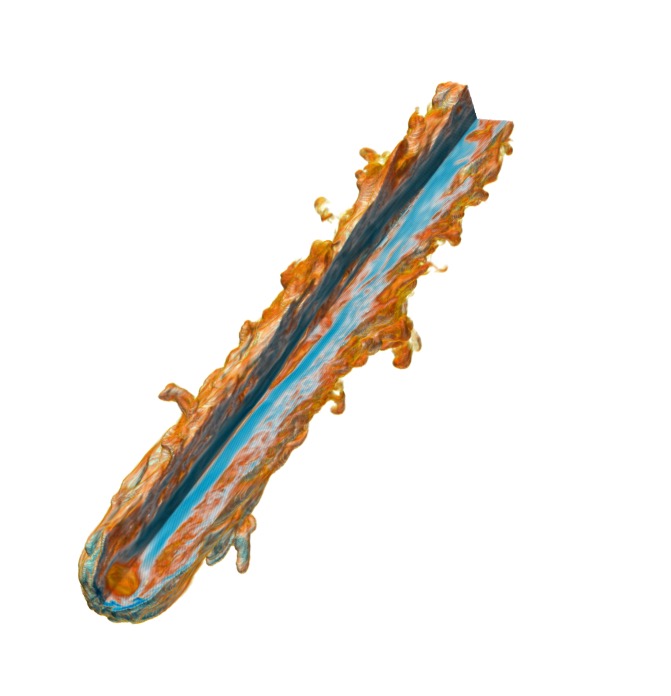}} & \hspace{-0.5cm}\resizebox{31.5mm}{!}{\includegraphics{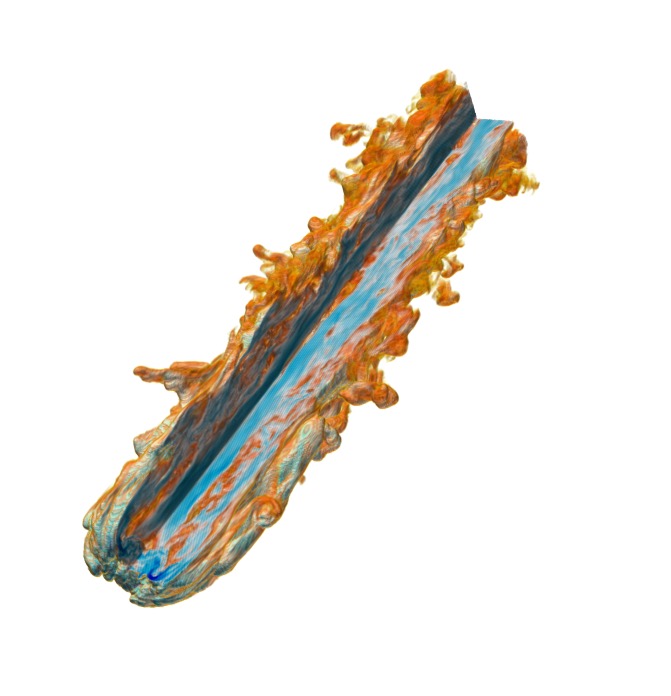}} & \hspace{-0.5cm}\resizebox{31.5mm}{!}{\includegraphics{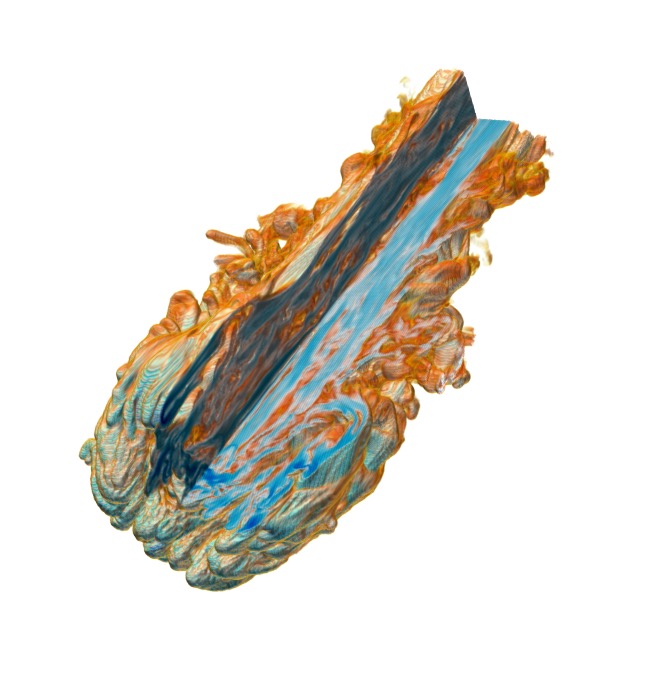}}\\
\hspace{-2cm}$t/t_{\rm cc}=0$ & \hspace{-1.5cm}$t/t_{\rm cc}=0.25$ & \hspace{-0.5cm}$t/t_{\rm cc}=0.50$ & \hspace{-0.5cm}$t/t_{\rm cc}=0.75$ & \hspace{-0.5cm}$t/t_{\rm cc}=1.00$ & \hspace{-0.5cm}$t/t_{\rm cc}=1.25$\Dstrut\\
\multicolumn{6}{l}{\hspace{-0.3cm}M2) TUR-SUP-BST}\\ 
\hspace{-0.3cm}\resizebox{31.5mm}{!}{\includegraphics{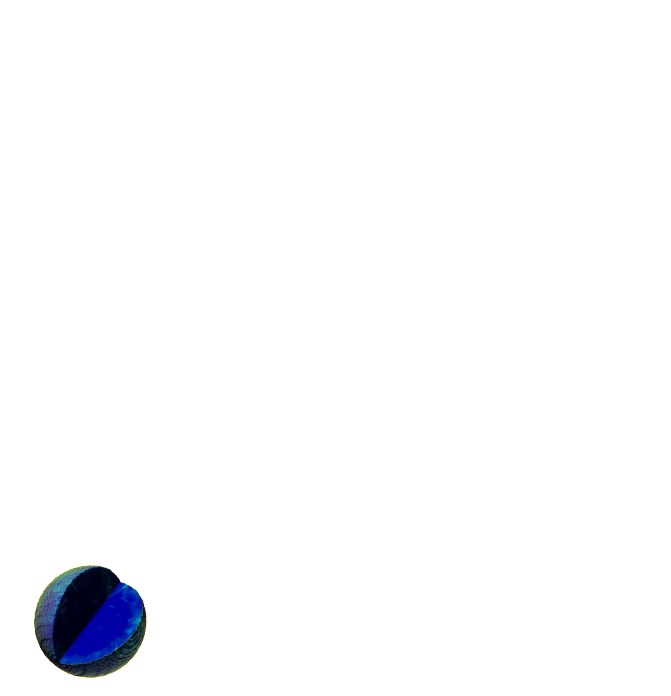}} & \hspace{-1cm}\resizebox{31.5mm}{!}{\includegraphics{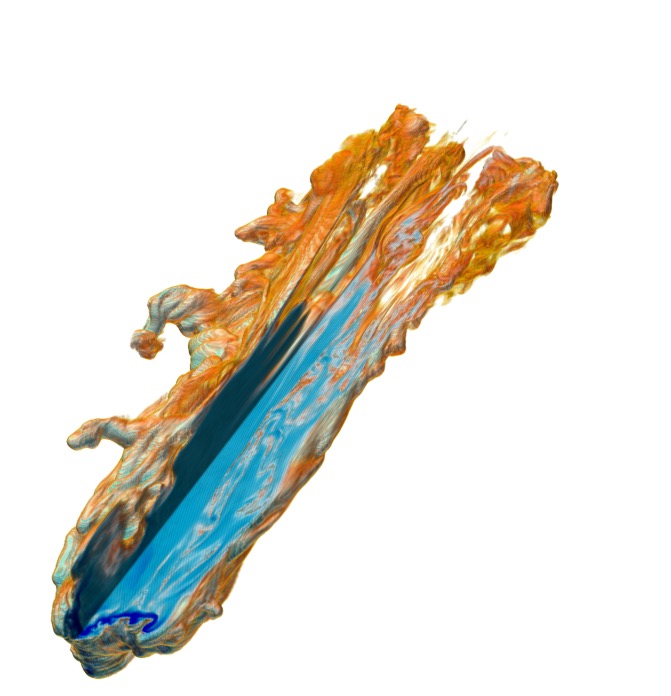}} & \hspace{-0.5cm}\resizebox{31.5mm}{!}{\includegraphics{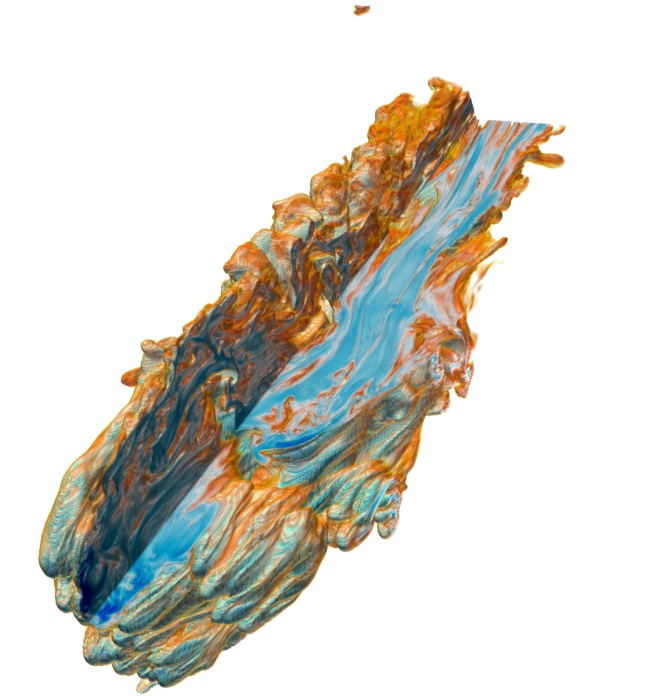}} & \hspace{-0.5cm}\resizebox{31.5mm}{!}{\includegraphics{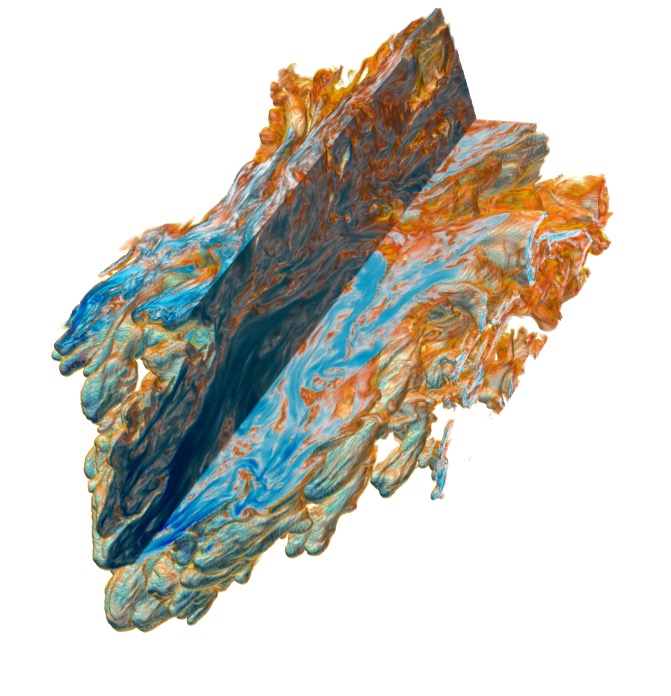}} & \hspace{-0.5cm}\resizebox{31.5mm}{!}{\includegraphics{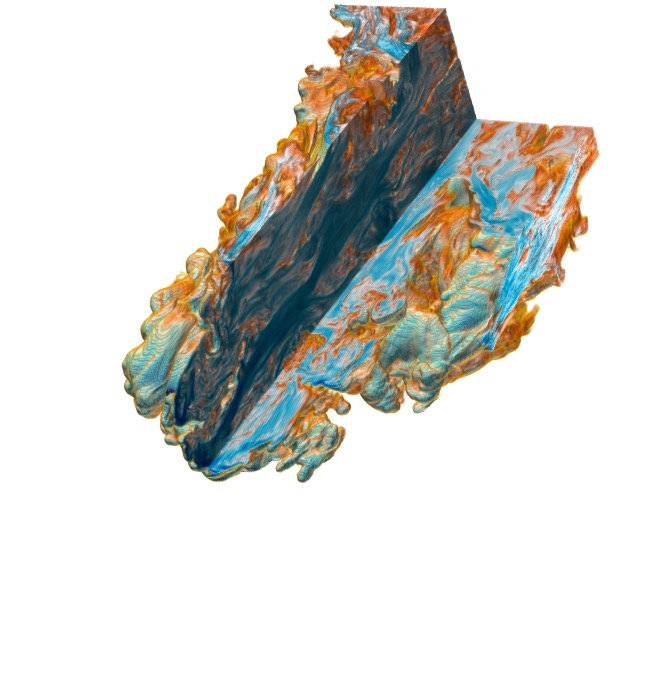}} & \hspace{-0.5cm}\resizebox{31.5mm}{!}{\includegraphics{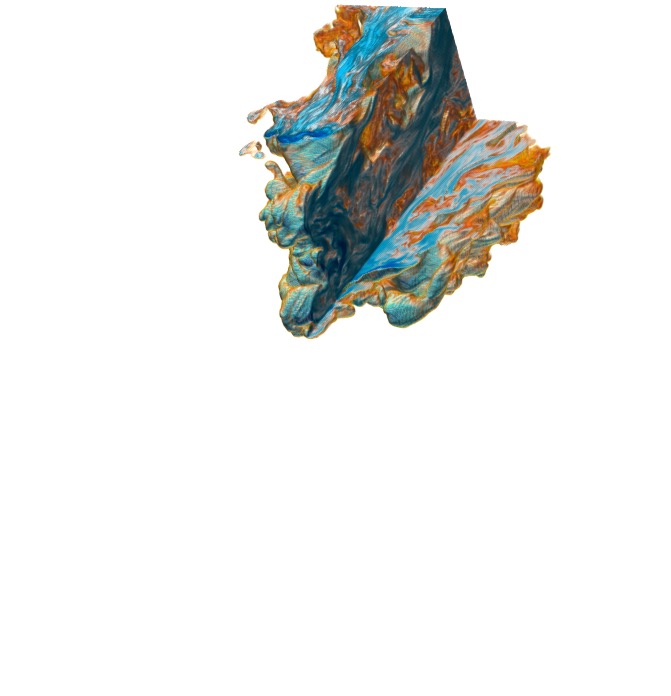}}\\
\hspace{-2cm}$t/t_{\rm cc}=0$ & \hspace{-1.5cm}$t/t_{\rm cc}=0.25$ & \hspace{-0.5cm}$t/t_{\rm cc}=0.50$ & \hspace{-0.5cm}$t/t_{\rm cc}=0.75$ & \hspace{-0.5cm}$t/t_{\rm cc}=1.00$ & \hspace{-0.5cm}$t/t_{\rm cc}=1.25$\Dstrut\\
\multicolumn{6}{l}{\hspace{-0.3cm}M3) TUR-SUP-BST-ISO}\\ 
\hspace{-0.3cm}\resizebox{31.5mm}{!}{\includegraphics{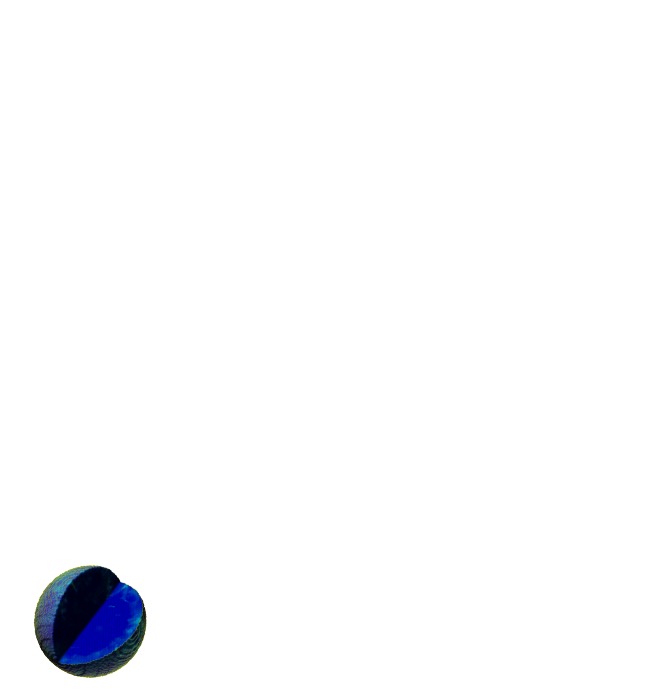}} & \hspace{-1cm}\resizebox{31.5mm}{!}{\includegraphics{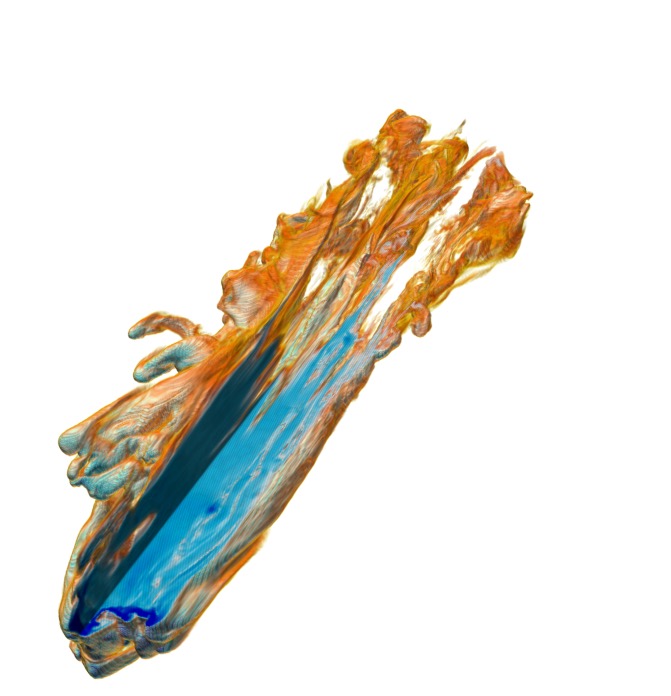}} & \hspace{-0.5cm}\resizebox{31.5mm}{!}{\includegraphics{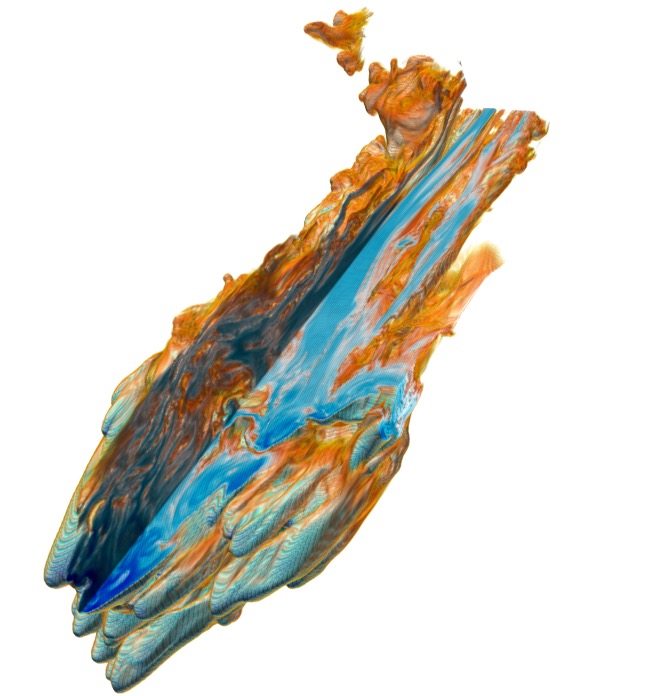}} & \hspace{-0.5cm}\resizebox{31.5mm}{!}{\includegraphics{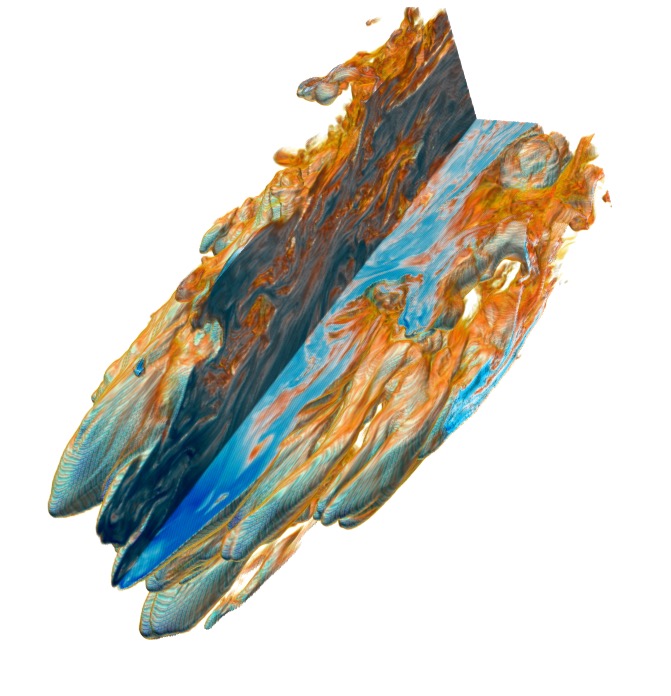}} & \hspace{-0.5cm}\resizebox{31.5mm}{!}{\includegraphics{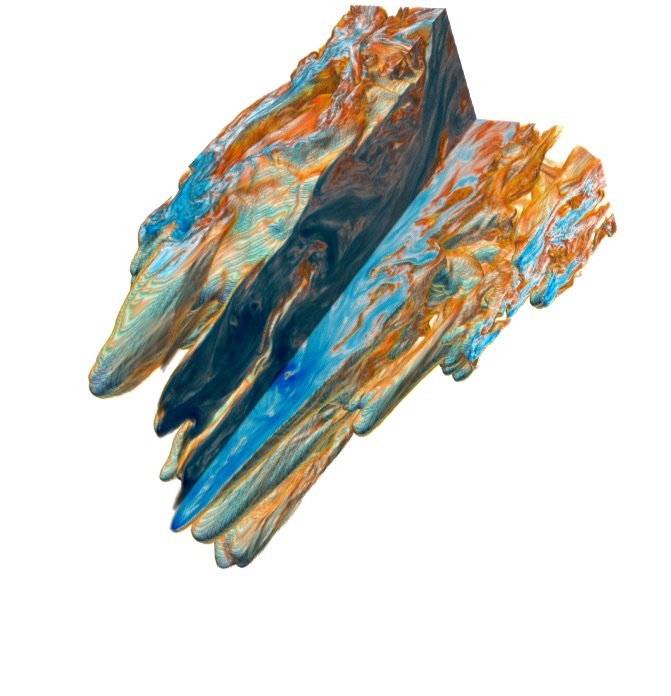}} & \hspace{-0.5cm}\resizebox{31.5mm}{!}{\includegraphics{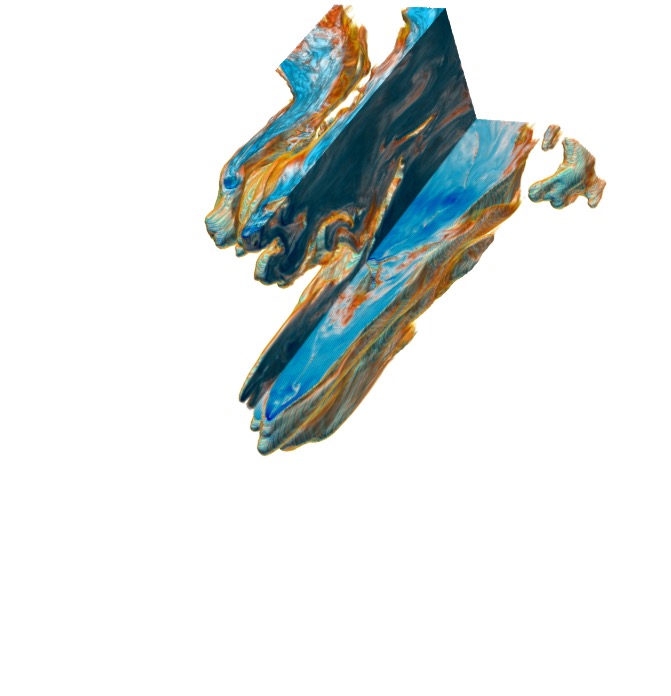}}\\
\hspace{-2cm}$t/t_{\rm cc}=0$ & \hspace{-1.5cm}$t/t_{\rm cc}=0.25$ & \hspace{-0.5cm}$t/t_{\rm cc}=0.50$ & \hspace{-0.5cm}$t/t_{\rm cc}=0.75$ & \hspace{-0.5cm}$t/t_{\rm cc}=1.00$ & \hspace{-0.5cm}$t/t_{\rm cc}=1.25$\Dstrut\\
\resizebox{!}{6mm}{\includegraphics{Axes3D.jpg}}   & \multicolumn{5}{c}{\resizebox{!}{6mm}{\includegraphics{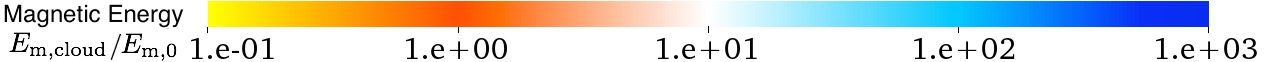}}}\\
  \end{tabular}
  \caption{Same as Figures \ref{Figure2} and \ref{Figure3}, but here we present 3D volume renderings of the logarithm of the magnetic energy density in filaments ($E_{\rm m,cloud}$) normalised with respect to the initial magnetic energy density in the wind, $E_{\rm m,0}$, for $0\leq t/t_{\rm cc}\leq1.25$. Panel M1 shows the evolution of a uniform cloud immersed in a purely oblique magnetic field (UNI-0-0). Panels M2 and M3 show the evolution of turbulent clouds with log-normal density distributions, Gaussian velocity fields, and turbulent magnetic fields in adiabatic (TUR-SUP-BST) and quasi-isothermal (TUR-SUP-BST-ISO) cases, respectively. We note that a fraction of the initially-strong magnetic energy in the turbulent magnetic field included in models TUR-SUP-BST and TUR-SUP-BST-ISO dissipates in a few tenths of $t_{\rm cc}$. After this transient effect, i.e., for times $t/t_{\rm cc}\geq0.25$, knots and sub-filaments remain strongly magnetised with magnetic energy densities as high as those in their progenitor clouds (see Appendix \ref{sec:Appendix5} for a quantitative comparison of the magnetic energy enhancement in different models), thus aiding cloud survival by reducing the disruptive effects of dynamical instabilities. In the radiative case, model TUR-SUP-BST-ISO, this effect is more significant than in the adiabatic model, owing to the extra compression caused by the softer polytropic index used in this model. Movies showing the full-time evolution of the models presented here are available online at \url{https://goo.gl/iXgJYk}.}
  \label{Figure4}
\end{center}
\end{figure*}

In this section we compare the global evolution of wind-cloud systems with uniform and turbulent clouds as they are swept up by a low-density, supersonic wind to form filaments. Figures \ref{Figure2}, \ref{Figure3}, and \ref{Figure4} show the time evolution of the mass density, kinetic energy density, and magnetic energy density, respectively, of filament gas in three models, UNI-0-0, TUR-SUP-BST, and TUR-SUP-BST-ISO, at six different times, namely $t/t_{\rm cc}= 0$, $t/t_{\rm cc}=0.25$, $t/t_{\rm cc}=0.5$, $t/t_{\rm cc}=0.75$, $t/t_{\rm cc}=1.0$, and $t/t_{\rm cc}=1.25$. All these parameters have been multiplied by the tracer $C_{\rm cloud}$, so that only filament gas is displayed in the snapshots of these figures. In addition, a quarter of the volume in the rendered images has been clipped to show the internal structure of the clouds and filaments in detail.\par

The aforementioned figures confirm our previous results presented in \citetalias{2016MNRAS.455.1309B}. They show that: i) A filament can be seen as constituted by two main substructures, namely a tail and a footpoint, and ii) The formation of filaments is a universal process characterised by four evolutionary phases: 1) A tail formation phase, in which material, mainly removed from the envelope of the cloud (in time-scales of the order of $t/t_{\rm wp}\sim2-5$; i.e., $t/t_{\rm cc}\sim 0.06-0.16$), is transported downstream to form an elongated tail; 2) A tail erosion phase in which the wind shapes the newly-formed tail on time-scales that depend on how fast KH instabilities (e.g., see \citealt{1993ApJ...407..588M}; \citealt*{1994ApJ...432..194J}; \citealt{1996ApJ...460..777F}; \citealt*{2000ApJ...545..475R}; \citealt{2016MNRAS.455.4274L}) grow at the wind-filament interface; 3) A footpoint dispersion phase in which dense nuclei in the footpoint of the turbulent cloud are disrupted by RT instabilities (e.g., see \citealt*{1982MNRAS.201..833N,1995ApJ...453..332J}; \citealt{2007ApJ...671.1726S}), producing sub-filamentation; and 4) A free floating phase in which the filamentary structure loses some coherence and becomes entrained in the wind (see Section 4 in \citetalias{2016MNRAS.455.1309B} for a full description of the dynamics and time-scales involved in the formation of filaments).\par

Despite the universality of the global process, Figures \ref{Figure2}, \ref{Figure3}, and \ref{Figure4} also show that the disruption process of non-turbulent and turbulent clouds results in filaments with different morphologies. We discuss those differences qualitatively in Sections \ref{subsubsec:UniformvsTurbulent} and \ref{subsubsec:RadiativeLosses} and quantitatively in Sections \ref{subsubsec:Properties} and \ref{subsubsec:Entrainment} below.

\subsubsection{Uniform vs. turbulent cloud models}
\label{subsubsec:UniformvsTurbulent}
Here we compare the filamentary tails in models with uniform clouds versus those in models with turbulent clouds from a qualitative perspective. Contrasting Panel M1 of Figure \ref{Figure2} with the other two panels (Panels M2 and M3 of Figure \ref{Figure2}) reveals that turbulent clouds produce filaments that are less confined and have more chaotic and sinuous tails than the one arising from a uniform cloud. While the single nucleus in the core of the uniform cloud prevents the wind from rapidly flowing through core material at early stages, the presence of multiple high-density nuclei, surrounded by a low-density inter-nucleus medium, permits a faster percolation of the wind through the footpoints in the turbulent cloud models (in agreement with \citealt{2009ApJ...703..330C,2017ApJ...834..144S}). Thus, the wind in turbulent models removes inter-nucleus material from the clouds and produces a collection of low- and high-density knots and sub-filaments along the tail.\par

Figures \ref{Figure3} and \ref{Figure4} also show the effect mentioned above. For example, Panel M1 of Figure \ref{Figure3} shows that the gas with high kinetic energy density is confined to the interior of the filament in model UNI-0-0, while Panels M2 and M3 show that the kinetic profile of the gas in the filament is much more anisotropic in models TUR-SUP-BST, and TUR-SUP-BST-ISO than in the uniform case, with varios high-kinetic-energy-density knots and sub-filaments threading the tail. The presence of parcels of gas with kinetic energy densities similar to or higher than the kinetic energy density of the wind in the snapshots of Panels M2 and M3 Figure \ref{Figure3} for times $t/t_{\rm cc}\geq 0.75$ also implies that filaments emerging from turbulent clouds become more easily entrained in a supersonic wind than their uniform counterpart.\par

In a similar fashion the snapshots of the magnetic energy density in Panels M2 and M3 of Figure \ref{Figure4} indicate the presence of several wind-entrained, magnetised sub-filaments and cloudlets with more distorted morphologies in the filamentary tails arising from turbulent clouds than in the one emerging from the uniform cloud. The observed difference in the magnetic morphology of filaments can be explained as follows. In the uniform model, UNI-0-0, the regions of high magnetic energy density are confined to the leading edge of the cloud (where the field lines pile up and are stretched by the passage of the wind) and to the rear side of the cloud (where converging shocks advect field lines and fold them to form a current sheet behind the cloud; see Section 5.5 in \citetalias{2016MNRAS.455.1309B} for a thorough discussion). In this model a turbulent magnetised tail only appears after the cloud breaks up at $t/t_{\rm cc}=1.0$. On the other hand, in the turbulent scenarios, TUR-SUP-BST and TUR-SUP-BST-ISO, the presence of the internal, turbulent magnetic field plays two crucial roles in the dynamics of the wind-swept clouds from the very beginning of the simulation: 1) it helps stabilise the cloud against turbulence- and wind-driven expansion and stripping (after an initial, transient phase of turbulence dissipation), and 2) it keeps the filament gas strongly magnetised at all times, thus preventing KH instabilities (arising at wind-cloud interfaces) from quickly disrupting the cloud/filament (in agreement to what was found by \citealt{2015MNRAS.449....2M} for tangled, internal magnetic fields). In fact, the reference time-scales for the growth of KH instabilities (see Equation \ref{KHtime}) in models UNI-0-0, TUR-SUP-BST, and TUR-SUP-BST-ISO are in the ratio $t_{\rm KH,\,M1}:t_{\rm KH,\,M2}:t_{\rm KH,\,M3}=1:1.5:1.8$, indicating that suppression of long-wavelength ($\lambda_{\rm KH}\sim 1\,r_{\rm c}$) modes of this instability occurs in models with turbulent clouds. Thus, unlike \cite{2008ApJ...674..157C,2009ApJ...703..330C,2017ApJ...834..144S}, who found that fractal/turbulent clouds are fragmented or disrupted faster than uniform clouds, we find here that the cloud-crushing time of Equation (\ref{eq:CloudCrushing}) continues to be a good estimate for the overall "break-up" time of turbulent filaments as their main structures remain coherent for the entire simulation time-scale ($1\,t_{\rm sim}$; see Equation \ref{eq:SimulationTime}), owing to the protective effects of internal, turbulent magnetic fields.\par

Note also that Panels M2 and M3 of Figure \ref{Figure4} indicate that the knots and sub-filaments formed in turbulent models are as strongly magnetised as their progenitor clouds, thus implying that the magnetic field strength in the filament is similar to the initial magnetic field strength in the cloud in models where self-consistent, strong, turbulent magnetic fields are added to the clouds. Indeed, a quantitative analysis of the magnetic energy enhancement presented in Appendix \ref{sec:Appendix5} reveals that the ratio of magnetic energy (and magnetic field strength) in the filament to that in the initial cloud remains nearly constant $\sim1$ over the entire evolution of these simulations. The constancy of this ratio in these models has important implications for astrophysical filaments as it indicates that ISM filaments have the same magnetic field strength as their progenitor clouds. This result is potentially important for the understanding of the formation and evolution of the radio filaments observed in the Galactic centre as they have been suggested to have magnetic field strengths of the order of the strengths estimated in molecular clouds (e.g., see \citealt{1999ASPC..186..483R,2001PASA...18..431B}).\par

Another difference between uniform and turbulent cloud models is that the dispersion of the filament footpoints is anisotropic in turbulent scenarios and occurs at the locations of the densest nuclei in the mass distribution of the cloud. Each of these dense regions inside the cloud undergoes a break-up phase of its own, and this occurs faster for more diffuse regions than for denser regions. Consequently, dense regions in the cloud survive longer than diffuse regions and the break-up phase of turbulent clouds is not a drastic, abrupt event in which the structure of a single nucleus is disrupted as in the uniform case (see e.g., the snapshot of model UNI-0-0 at $t/t_{\rm cc}=1.0$). It is rather a slow, steady process in which several nuclei inside the turbulent cloud are eroded by RT instabilities at distinct locations in the cloud and on different time-scales (see e.g., the snapshot of model UNI-0-0 at $t/t_{\rm cc}=1.25$). The reference time-scales for the growth of RT instabilities (see Equation \ref{RTtime}) in models UNI-0-0, TUR-SUP-BST, and TUR-SUP-BST-ISO are in the ratio $t_{\rm RT,\,M1}:t_{\rm RT,\,M2}:t_{\rm RT,\,M3}=1:0.7:0.9$, indicating RT-driven sub-filamentation occurs faster in models with turbulent clouds and confirming that strong magnetic fields hasten the growth of long-wavelength ($\lambda_{\rm RT}\sim 1\,r_{\rm c}$) modes of this instability (in agreement with the results presented in \citealt{1999ApJ...527L.113G,2000ApJ...543..775G}; and in Section 5.5.1 in \citetalias{2016MNRAS.455.1309B}).\par

Overall, we find in this section that the intrinsic porosity of turbulent clouds facilitates their lateral expansion (increasing their cross-sectional area) and makes them susceptible to a greater ram-pressure force than the one acting upon the uniform cloud. This enhanced drag force pushes these turbulent clouds (regardless of whether or not they are radiative) farther away from their original position than in the uniform scenario during the same time-scale (see e.g., the snapshots at $t/t_{\rm cc}\geq 1.0$). Even though a higher drag force and a more expanded cross section would mean that turbulent clouds are more prone to the disruptive effects of ablation and dynamical instabilities, we do not find evidence of turbulent clouds being disrupted faster than the uniform cloud as reported in previous hydrodynamical studies. The reason is that the strong, turbulent magnetic fields, that we self-consistently included in our models, stabilises the clouds against the wind ram pressure and turbulence and keeps dense gas clumped together, thus preventing cloud material from rapidly mixing with wind material via small-scale KH instabilities. This signifies that self-consistently including turbulence in the initial conditions of wind-swept clouds has the effect of increasing cloud acceleration, without affecting cloud shredding and the overall coherence of the resulting filaments. This result is crucial as it shows that the process of entrainment of cold, dense gas into hot, low-density winds is viable (see Section \ref{subsubsec:Entrainment} for further details on the dynamics of filaments).\par

\subsubsection{Effects of radiative losses}
\label{subsubsec:RadiativeLosses}
Panels M2 and M3 of Figures \ref{Figure2}, \ref{Figure3}, and \ref{Figure4} can also be compared with one another. The snapshots in these panels show that softening the gas equation of the state to $\gamma=1.1$ in model TUR-SUP-BST-ISO (in order to mimic the effects of radiative cooling) has three effects on the resulting filaments: 1) it suppresses small-scale KH instability modes at the sides of the cloud, leading to the emergence of a more laminar filament (also seen in model MHD-Ob-I in \citetalias{2016MNRAS.455.1309B}); 2) it produces a tail with a collection of linear sub-filaments anchored to denser and slower nuclei; and 3) it aids the survival of the filament by preserving its core gas denser and more strongly magnetised than in the adiabatic ($\gamma=\frac{5}{3}$), turbulent model, TUR-SUP-BST. These effects are caused by the increased compression to which the cloud gas is subjected in model TUR-SUP-BST-ISO (see also Section 5.3 in \citealt{1994ApJ...420..213K} and Section 4.5 in \citealt{2006ApJS..164..477N}). The higher density contrast at wind-cloud interfaces delays the emergence of small-scale KH instabilities and protects dense regions in the cloud from disruption, thus slowing it down and prolonging its lifetime. The reference time-scales for the growth of KH instabilities (see Equation \ref{KHtime}) in models TUR-SUP-BST and TUR-SUP-BST-ISO are in the ratio $t_{\rm KH,\,M2}:t_{\rm KH,\,M3}=1:1.2$, indicating that the emergence of long-wavelength ($\lambda_{\rm KH}\sim 1\,r_{\rm c}$) modes of this instability is delayed in the quasi-isothermal model.\par

Based on the above results, we find that the ability of a cool, dense cloud to radiate thermal energy away is another crucial element to its survival as entrained structures in a hot, supersonic wind. In adiabatic scenarios KH instabilities have a pronounced impact on the morphology and lifetime of filaments as the thermally-driven expansion accelerates gas mixing and increases the degree of mass stripping and turbulence in the downstream flow. In the radiative scenario, on the other hand, the cloud can cool via thermal radiation, which keeps its gas dense and cold and inhibits KH instabilities at fluid interfaces. As opposed to models of quasi-isothermal, uniform clouds, which produce a single, laminar filament (see Figure 10 in \citetalias{2016MNRAS.455.1309B}), in the case of turbulent cloud models, the presence of multiple high-density nuclei in their cores result in the formation of tails with several sub-filaments along them (each of these sub-filaments is supported by one of these nuclei). Sub-filamentation of wind-swept clouds is an RT instability-driven process in both uniform and turbulent clouds. We showed in the previous section that this process is more efficient in turbulent cloud models, owing to the porous nature of turbulent density structures, but the extra compression of the gas in model TUR-SUP-BST-ISO slows the cloud down and delays the emergence of RT instability modes. Indeed, the reference time-scales for the growth of long-wavelength ($\lambda_{\rm RT}\sim 1\,r_{\rm c}$) RT instabilities (see Equation \ref{RTtime}) in models TUR-SUP-BST and TUR-SUP-BST-ISO are in the ratio $t_{\rm RT,\,M2}:t_{\rm RT,\,M3}=1:1.4$, which is similar to the $1:1.6$ ratio obtained for the adiabatic and quasi-isothermal models of wind-swept uniform clouds discussed in Section 5.5.2 in \citetalias{2016MNRAS.455.1309B}.\par

Overall, we find that the ability of the cloud to radiate in model TUR-SUP-BST-ISO effectively suppresses KH instabilities at wind-filament interfaces, thus aiding cloud survival and making its entrainment into the hot, supersonic wind even more feasible than in the turbulent, adiabatic model, TUR-SUP-BST. These findings are in agreement with the conclusions presented in previous studies of radiative wind/shock-swept clouds for different cooling regimes (e.g., see \citealt{2002AA...395L..13M,2004ApJ...604...74F,2005AA...443..495M,2009ApJ...703..330C,2015ApJ...805..158S}).

\subsubsection{Filament morphology and energetics}
\label{subsubsec:Properties}

\begin{figure*}
\begin{center}
  \begin{tabular}{c c}
  \resizebox{80mm}{!}{\includegraphics{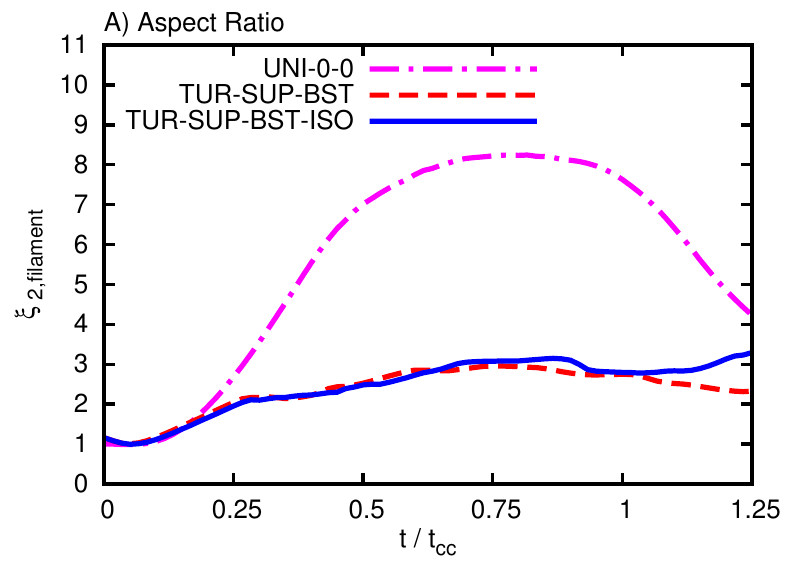}} & \resizebox{80mm}{!}{\includegraphics{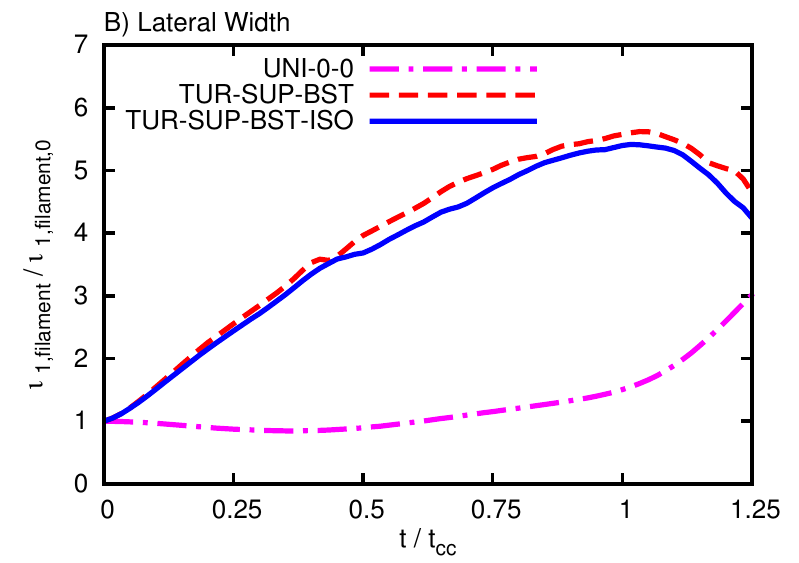}}\\
  \resizebox{80mm}{!}{\includegraphics{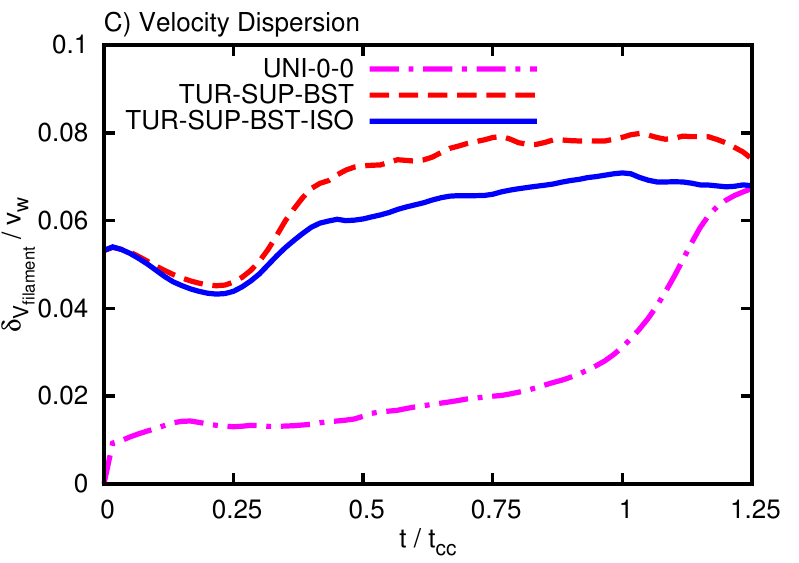}} & \resizebox{80mm}{!}{\includegraphics{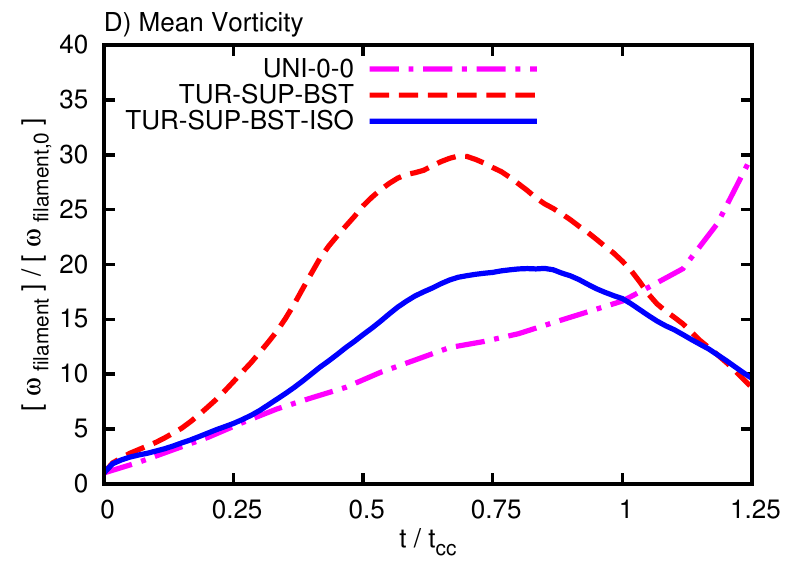}}\\
  \end{tabular}
  \caption{Time evolution of four diagnostics: the aspect ratio (Panel A), the lateral width (Panel B), the transverse velocity dispersion (Panel C), and the mean vorticity enhancement (Panel D), in models UNI-0-0 (dash-dotted line), TUR-SUP-BST (dashed line), and TUR-SUP-BST-ISO (solid line). We find that: i) turbulence leads to the formation of filaments with greater lateral widths, higher velocity dispersions, and more pronounced vorticity enhancements than the idealised uniform cloud; and ii) the inclusion of a softer polytropic index to mimic radiative losses in model TUR-SUP-BST-ISO suppresses KH instabilities, thus reducing the lateral elongation, velocity dispersion, and vorticity of the resulting filament.}
  \label{Figure5}
\end{center}
\end{figure*}

Here we discuss the role of turbulence on the morphology and energetics of wind-swept clouds (filaments) from a quantitative point of view. Figure \ref{Figure5} presents the time evolution of four parameters calculated for cloud/filament material (i.e., using the scalar $C_{\rm cloud}$). Panels A and B show the evolution of two geometrical quantities, namely the aspect ratios (see Equation \ref{eq:AspectRatio}) and lateral widths of filaments (see Equation \ref{eq:Moments}), respectively. These panels indicate that both turbulent clouds (in models TUR-SUP-BST and TUR-SUP-BST-ISO) have lower aspect ratios than the uniform cloud (in model UNI-0-0) as a result of them being more laterally elongated by a combination of shock- and turbulence-triggered expansion. This behaviour bears out the qualitative analysis presented in the preceding sections: the wind is able to travel across cloud material more easily when its gas is turbulent than when it is uniform, causing the clouds to quickly expand. Note that: i) the change in the slope of the lateral width curve in model UNI-0-0 is due to the cloud break-up via RT instabilities, while ii) the decline seen in the lateral width of the filaments in models TUR-SUP-BST and TUR-SUP-BST-ISO for times $t/t_{\rm cc}>1.1$ responds to cloud material starting to leave the computational domain through the sides of it (see Appendix \ref{sec:Appendix4} for a discussion on the effects of the computational domain size on our diagnostics).\par

Panels C and D of Figure \ref{Figure5} show the evolution of the transverse velocity dispersions (see Equation \ref{eq:rmsVelocity}) and mean vorticity enhancements (from Equation \ref{eq:AveragedF}) of filament gas. These panels show two effects: i) turbulent clouds (in models TUR-SUP-BST and TUR-SUP-BST-ISO) generate filaments with higher velocity dispersions and higher mean vorticities than the uniform cloud (in model UNI-0-0), and ii) the inclusion of a softer equation of state (with $\gamma=1.1$) in model TUR-SUP-BST-ISO results in a reduction of the velocity dispersion and small-scale vorticity in filament gas, when compared to its adiabatic counterpart in model TUR-SUP-BST. Regarding the former effect, Panel C shows that clouds with an initially turbulent velocity field remain turbulent throughout the simulation with $\delta_{{\rm v}_{{\rm filament}}}/v_{\rm w}=0.05-0.08$, while the uniform cloud has $\delta_{{\rm v}_{{\rm filament}}}/v_{\rm w}\sim0.02$ and only develops a similarly turbulent tail after its core has been disrupted by RT instabilities for $t/t_{\rm cc}\geq1.0$. The decline seen in the velocity dispersions of the turbulent models for $t/t_{\rm cc}<0.3$ is due to the initial, transient dissipation of the supersonic turbulence in shocks inside these clouds. Regarding the latter effect, Panels C and D show that the ability of a cloud to radiate energy away is crucial for its survival in a supersonic wind as it suppresses the disruptive effects of KH instabilities by inhibiting the deposit of small-scale vortices at wind-filament interfaces, thus reducing both the velocity dispersions and mean vorticities of filament gas. This is in agreement with our qualitative analysis presented in Section \ref{subsubsec:RadiativeLosses} and also with previous work of radiative clouds interacting with winds/shocks by e.g., \cite{2005ApJ...619..327F,2005A&A...444..505O,2007ApJ...668..310R,2009ApJ...703..330C}.\par

\begin{figure}
\begin{center}
  \begin{tabular}{c}
  \resizebox{80mm}{!}{\includegraphics{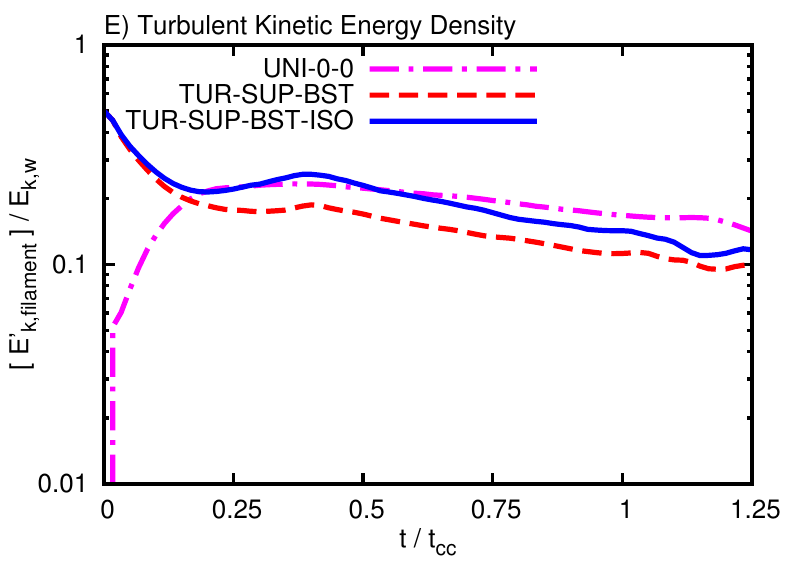}}\\
  \resizebox{80mm}{!}{\includegraphics{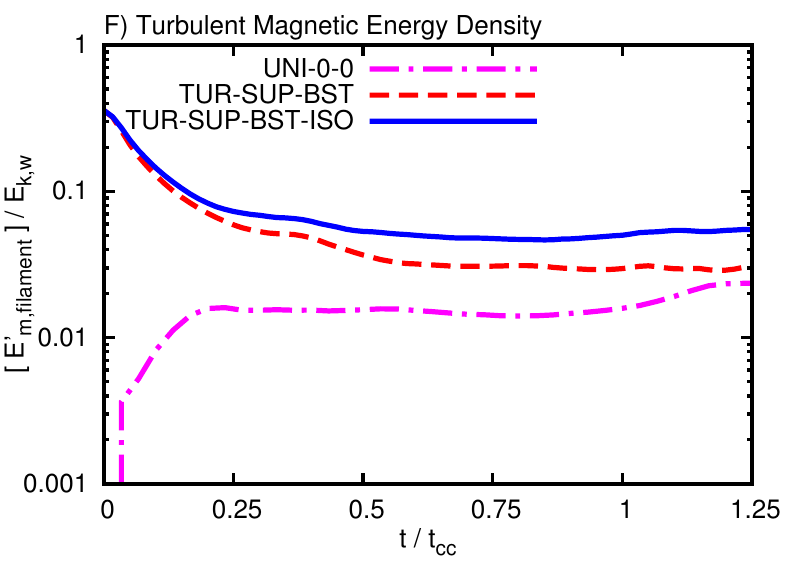}}\\
  \resizebox{80mm}{!}{\includegraphics{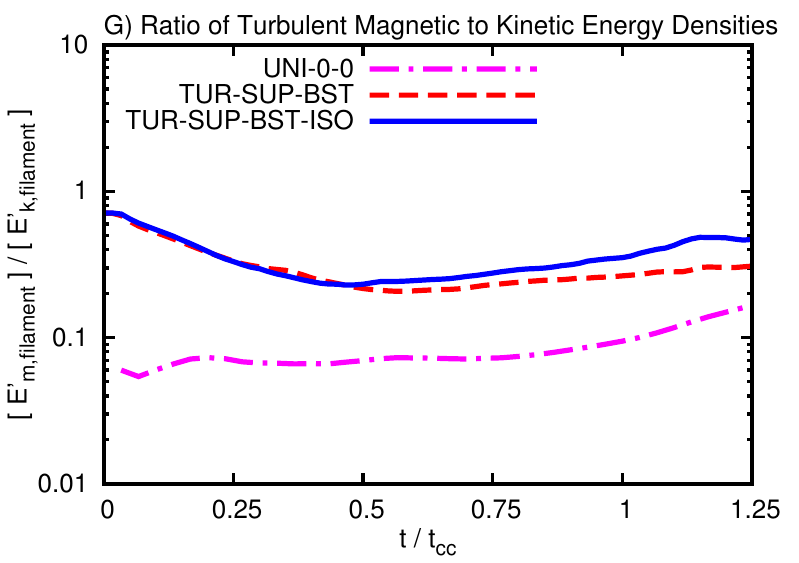}}\\
  \end{tabular}
  \caption{Same as Figure \ref{Figure5}, but here Panels E and F show the average turbulent kinetic and magnetic energy densities, respectively, and Panel G shows the ratio of the two energy densities. We find that: i) while turbulent clouds undergo a transient period of rapid dissipation of their supersonic turbulence (in shocks), their uniform counterpart becomes turbulent at the beginning of the interaction, and ii) the ratios of turbulent magnetic to turbulent kinetic energy densities indicate sub-equipartition and remain nearly constant with values of $[~E'_{\rm m, filament}~]/[~E'_{\rm k, filament}~]=0.1-0.4$ in all models.}
  \label{Figure6}
\end{center}
\end{figure}

The turbulence energetics of our models of wind-swept clouds is another important aspect to be analysed in this section. Panels E, F, and G of Figure \ref{Figure6} show the evolution of three parameters, namely the average turbulent kinetic energy density (see Equation \ref{eq:TurbKinEner}), the average turbulent magnetic energy density (see Equation \ref{eq:TurbMagEner}), and the ratio of these two energy densities, respectively. These panels reveal four effects: i) Turbulent clouds in models TUR-SUP-BST and TUR-SUP-BST-ISO experience a short (transient) period ($0\leq t/t_{\rm cc}\leq 0.15$) of rapid dissipation of the initially supersonic turbulence prescribed for them, ii) This dissipation is slightly less significant when the radiative, turbulent cloud in model TUR-SUP-BST-ISO is considered, iii) The cloud gas in the uniform model UNI-0-0 experiences the opposite effect, becoming turbulent very quickly (also over short time-scales: $0\leq t/t_{\rm cc}\leq 0.15$), and iv) After the curves become flat (for times $t/t_{\rm cc}\geq 0.15$), the ratios of turbulent magnetic to turbulent kinetic energy densities remain nearly constant in all models.\par

In order to explain the aforementioned effects, let us discuss the curves presented in Panels E and F of Figure \ref{Figure6}. In both cases we find a similar time evolution. We observe a decline of the energy densities of turbulence in the filaments of models TUR-SUP-BST and TUR-SUP-BST-ISO. This energy dissipation continues until $t/t_{\rm cc}=0.15$, when the ram pressure of the external wind equates the pressures (thermal plus magnetic) inside the cloud, and induces the initial expansion of the cloud (see Panel B of Figure \ref{Figure5}). After this time, both the turbulent kinetic and turbulent magnetic energy densities remain nearly constant around $[~E'_{\rm k, filament}~]/E_{\rm k, w}\sim0.1-0.2$ and $[~E'_{\rm m, filament}~]/E_{\rm k, w}\sim0.03-0.05$, respectively, until the end of the simulations. The inclusion of a softer polytropic index (i.e., of radiative cooling) in model TUR-SUP-BST-ISO prevents cloud gas from being overheated and keeps the gas dense and strongly magnetised, thus reducing its turbulent dissipation and quenching its expansion (see Panel B of Figure \ref{Figure5}). As a result, the radiative filament in this model exhibits turbulent kinetic and turbulent magnetic energy densities $\sim 30\,\%$ and $\sim 60\,\%$ higher, respectively, than the adiabatic filament in model TUR-SUP-BST (despite being more collimated than it) at all times. In the case of the filament arising from the non-turbulent cloud model, UNI-0-0, Panels E and F of Figure \ref{Figure6} show that the energy densities increase very rapidly at the beginning of the interaction (until $t/t_{\rm cc}\sim0.2$) as a result of the cloud gas being shock-heated and exposed to small-scale instabilities. After this time, both turbulent energy densities also remain nearly constant around $[~E'_{\rm k, filament}~]/E_{\rm k, w}\sim0.15$ and $[~E'_{\rm m, filament}~]/E_{\rm k, w}\sim0.02$, respectively, until $t/t_{\rm cc}=1.0$, when the RT instability-driven break-up of the spherical cloud tangles the magnetic field (note e.g., how the turbulent magnetic energy densities in models UNI-0-0 and TUR-SUP-BST approach $[~E'_{\rm m, filament}~]/E_{\rm k, w}\sim0.03$ towards the end of the evolution).\par

Overall, we find that the ratio of turbulent magnetic to turbulent kinetic energy densities in the filaments considered in this section remain nearly constant throughout the evolution of both uniform and turbulent cloud models, with values in the range  $[~E'_{\rm m, filament}~]/[~E'_{\rm k, filament}~]=0.1-0.4$. Filaments arising from uniform clouds favour the lower limit of this range, while turbulent filaments favour the upper limit (owing to their enhanced magnetic energy density). This result has important implications for observations as it indicates that the magnetic field in wind-swept clouds and filaments is in sub-equipartition with respect to the turbulent kinetic energy density, suggesting that this property can be used to constrain the magnetic field strength of wind-swept ISM clouds and filaments from their observed kinetic properties.


\subsubsection{Dynamics and entrainment of wind-swept clouds}
\label{subsubsec:Entrainment}
\begin{figure*}
\begin{center}
\begin{tabular}{c c}
\resizebox{80mm}{!}{\includegraphics{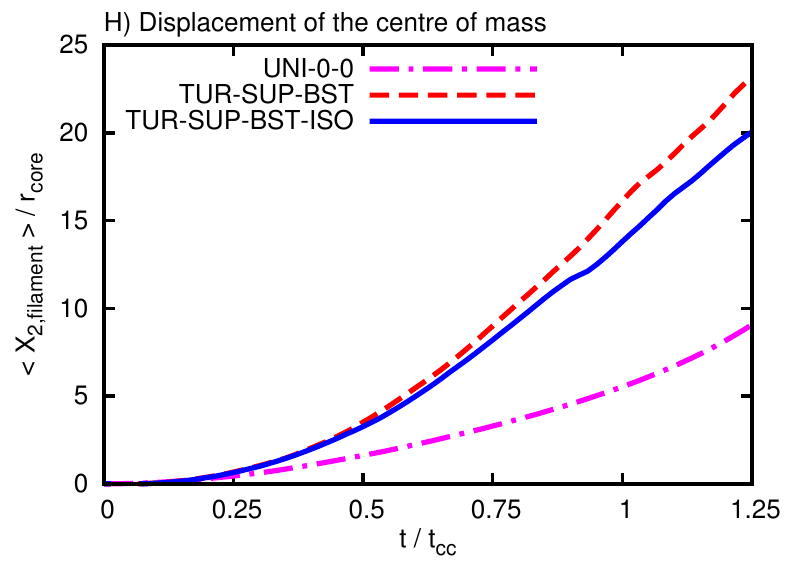}} & \resizebox{80mm}{!}{\includegraphics{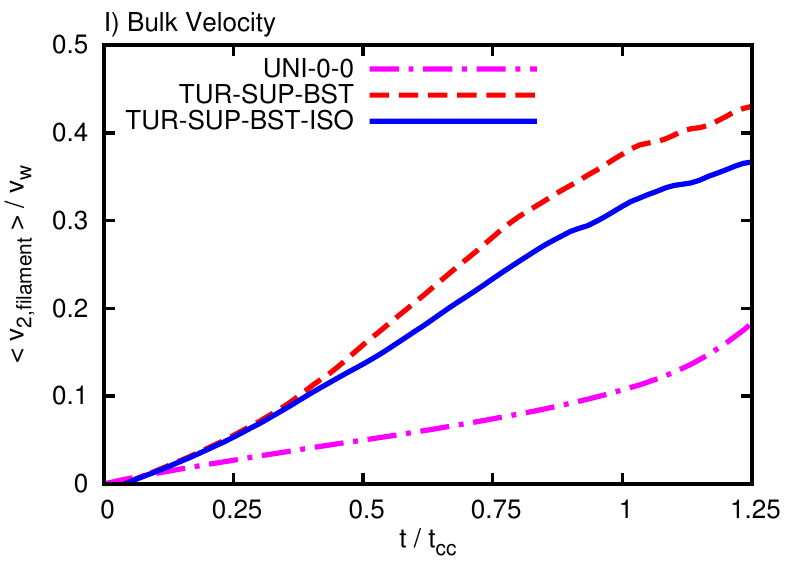}}\\
\end{tabular}
\caption{Same as Figures \ref{Figure5} and \ref{Figure6}, but here we show the displacement of the centre of mass (Panel H) and the bulk velocity (in the direction of streaming) of wind-swept clouds (filaments) entrained in the wind (Panel I). We find that the cloud with an initially-uniform density is $3-4$ times slower and travel distances $\sim 3$ times shorter than its turbulent counterparts (at $t/t_{\rm cc}=1.0$), implying that the self-consistent inclusion of both turbulence and magnetic fields is crucial for the full understanding of the dynamics and entrainment of cold, wind-swept clouds and filaments into hot, supersonic winds.}
\label{Figure7}
\end{center}
\end{figure*}

As mentioned above, turbulent clouds are more easily expanded by shocks and turbulence than uniform clouds, but does this affect the bulk dynamics of the resulting filaments? Do turbulent clouds move faster than uniform clouds to reach larger distances when immersed in a supersonic wind? Recent studies of fractal or turbulent clouds show that turbulent clouds are more easily disrupted by dynamical instabilities than uniform clouds (e.g., see \citealt{2009ApJ...703..330C,2017ApJ...834..144S}), implying that interstellar clouds (either spherical or turbulent) are unlikely to be ram-pressure accelerated for longer times before being fully disrupted (e.g., see \citealt{2015arXiv150701951Z,2016MNRAS.455.1830T}). However, these models did not consider the effects of tangled or turbulent magnetic fields threading the clouds, which have been demonstrated to provide support to spherical, wind-swept clouds by reducing the mixing of cloud material with ambient gas and thus prolonging their lifetime (e.g., see \citealt{2015MNRAS.449....2M}; \citetalias{2016MNRAS.455.1309B}). Here we use Figure \ref{Figure7} to discuss a broader view of the dynamics of clouds than previous models by investigating the motion of clouds and filaments that self-consistently incorporate turbulence and magnetic fields. We show that both turbulence and magnetic fields play significant roles in accelerating clouds and prolonging their lifetimes.\par

Figure \ref{Figure7} shows the displacement of the centre of mass (Panel H) and the bulk velocity (Panel I) of filaments in models UNI-0-0, TUR-SUP-BST, and TUR-SUP-BST-ISO as a function of time. Panel H indicates the distances travelled by each filament as measured by $\langle~X_{{\rm 2},{\rm filament}}~\rangle$, normalised with respect to the initial radius of the cloud core, $r_{\rm core}$ (from Equation \ref{eq:IntegratedG}). We see that the wind transports turbulent clouds/filaments over distances equivalent to $\langle~X_{\rm 2,filament}~\rangle/r_{\rm core}\sim14-16$, in the direction of streaming (measured at $t/t_{\rm cc}=1.0$). These distances are $2-3$ times as large as the distances to which uniform clouds are transported over the same time-scale (i.e., $\langle~X_{\rm 2,filament}~\rangle/r_{\rm core}\sim6$ at $t/t_{\rm cc}=1.0$), implying that the inclusion of turbulence aids cloud acceleration. The main driver of the cloud dynamics is the supersonic wind in all models, but self-consistent, turbulent clouds undergo higher accelerations than their uniform counterpart as a result of their larger cross sectional areas (see Section \ref{subsubsec:Properties}). As mentioned earlier, the presence of internal, turbulent magnetic fields in models TUR-SUP-BST and TUR-SUP-BST-ISO is crucial as it forms an effective magnetic shield (in agreement with \citealt{1996ApJ...473..365J,1999ApJ...517..242M}; \citetalias{2016MNRAS.455.1309B}) that prevents their enhanced accelerations from disrupting the cloud by suppressing KH instabilities at wind-cloud interfaces.\par

To confirm the above result we also investigate the range of velocities that are characteristic of the wind-embedded filaments at $t/t_{\rm cc}=1.0$. We use our definition of the mass-weighted bulk velocity, i.e., $\langle~v_{\rm 2,filament}~\rangle$, normalised with respect to the wind speed, $v_{\rm w}$ (from Equation \ref{eq:IntegratedG}), to study the bulk motion of filaments in the direction of streaming. Panel I of Figure \ref{Figure7} provides these measurements and shows that the bulk speed in turbulent models has values $\langle~v_{\rm 2,filament}~\rangle/v_{\rm w}\sim0.32-0.37$ at $t/t_{\rm cc}=1.0$, which are $3-4$ times larger than the bulk speed acquired by the uniform cloud, $\langle~v_{\rm 2,filament}~\rangle/v_{\rm w}\sim0.1$, over the same time-scale. The bulk speed in the uniform scenario, UNI-0-0, only increases after the RT-instability-driven break-up of its footpoint at $t/t_{\rm cc}=1.0$ as a result of the rapid growth of its cross sectional area. Despite this, the cloud/filament in the uniform scenario is always slower than its turbulent counterparts, confirming that the inclusion of turbulence and self-consistent magnetic fields favours the entrainment of cold, dense, high-speed gas into hot, diffuse, supersonic winds. Note that our simulations show that dense clouds and their associated filamentary tails can be effectively advected by a global, supersonic wind to reach larger distances, provided that realistic levels of turbulence and magnetic fields are included self-consistently.\par

As mentioned in Section \ref{subsubsec:UniformvsTurbulent}, turbulent clouds favour the formation of smaller sub-filaments and cloudlets along their filamentary tails. These substructures are not destroyed in one cloud-crushing time, but they also become entrained in the wind and quickly accelerate after $t/t_{\rm cc}=1.0$. Panels H and I of Figure \ref{Figure7} show that these wind-entrained structures reach distances $\langle~X_{\rm 2,filament}~\rangle/r_{\rm core}\sim 20-24$ and attain bulk speeds $\langle~v_{\rm 2,filament}~\rangle/v_{\rm w}\sim0.36-0.42$, respectively, in turbulent models. By contrast, the uniform (non-turbulent) cloud only reaches distances of $\langle~X_{\rm 2,filament}~\rangle/r_{\rm core}\sim 8$ and bulk speeds of $\langle~v_{\rm 2,filament}~\rangle/v_{\rm w}\sim0.18$ at $t/t_{\rm cc}=1.25$. These results are crucial for our understanding of the transport of dense material from low to high latitudes in galactic winds and outflows (e.g., see \citealt{1997A&A...320..378S,1999ApJ...523..575L,1984Natur.310..568S,2003ApJ...582..246B,2010ApJ...724.1044S}), but our current setups do not allow us to follow the full evolution of these substructures for times longer than $1\,t_{\rm sim}$. Thus, future numerical work, including larger simulation domains is warranted to investigate the fate of these substructures.

\subsection{Disentangling the relative contributions of turbulent density, velocity, and magnetic fields}
\label{subsec:Turbulence}
In the previous sections we compared the evolution of filaments emerging from a uniform cloud and from self-consistent, turbulent cloud models with the aim of understanding the effects of turbulence and magnetic fields on the morphology, energetics, and dynamics of filaments. However, self-consistent models do not allow us to differentiate between the roles of the different components of turbulence. Thus, the relative effects of turbulent density profiles, turbulent velocity fields, and turbulent magnetic fields need to be disentangled by exploring their effects on clouds separately. In this section we discuss the qualitative and quantitative effects of systematically adding turbulent density, velocity, and magnetic field distributions to the initial clouds on the evolution of several diagnostics. We use wind-cloud models with higher resolutions, i.e., $R_{128}$, (in smaller computational domains) to perform this comparison. In Section \ref{subsubsec:Morphology} we discuss qualitative aspects, whilst in Sections \ref{subsubsec:AspectRatio}, \ref{subsubsec:EnergyDensities}, and \ref{subsubsec:BulkSpeedandDistance} we discuss the implications of adding turbulence to cloud models on the formation and evolution of filaments in a quantitative manner.

\subsubsection{On the morphology of filaments}
\label{subsubsec:Morphology}

\begin{figure*}
\begin{center}
  \begin{tabular}{c c c c c c}
\multicolumn{6}{l}{\hspace{-0.3cm}S4) Uni-0-0}\\ 
\resizebox{24mm}{!}{\includegraphics{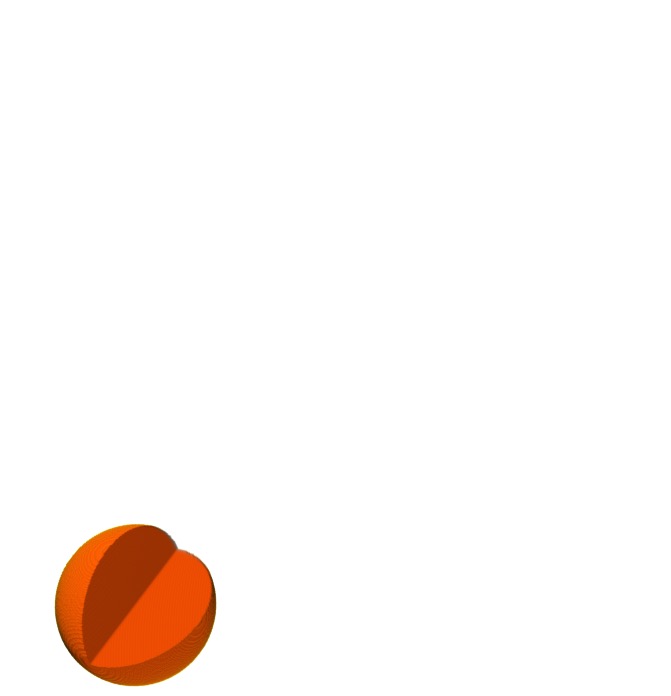}} & \resizebox{24mm}{!}{\includegraphics{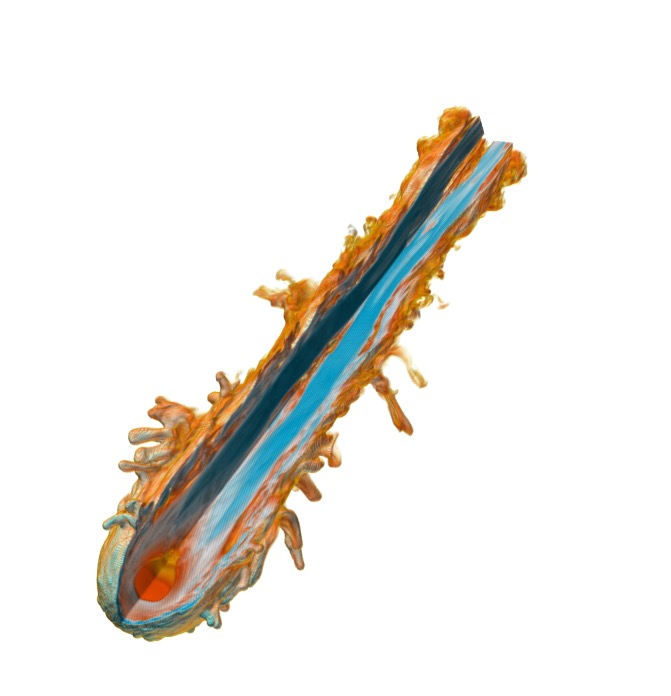}} & \resizebox{24mm}{!}{\includegraphics{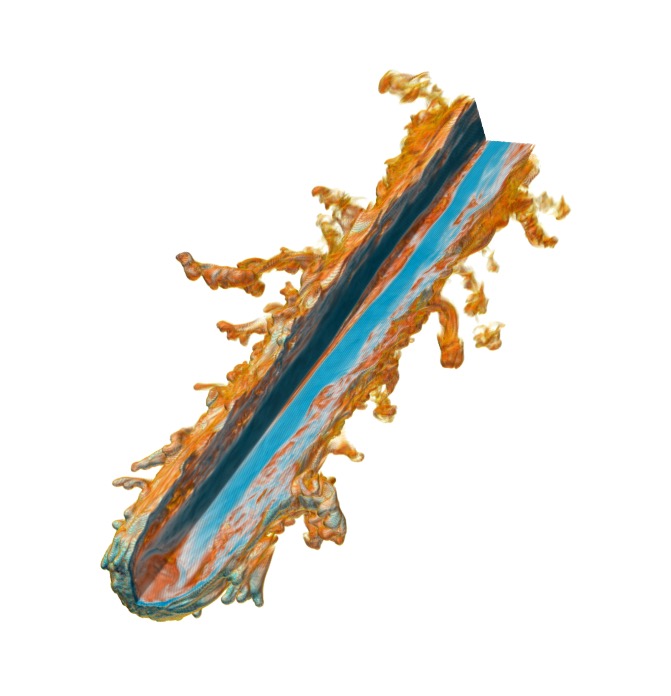}} & \resizebox{24mm}{!}{\includegraphics{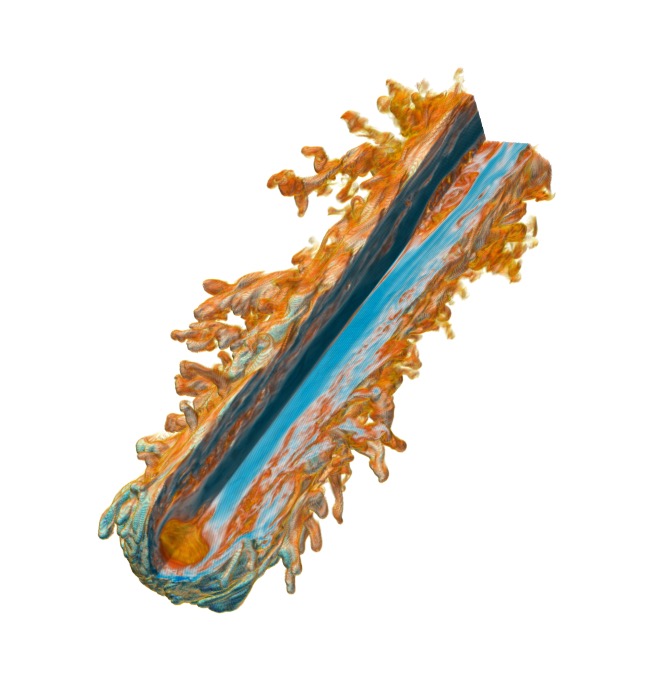}} & \resizebox{24mm}{!}{\includegraphics{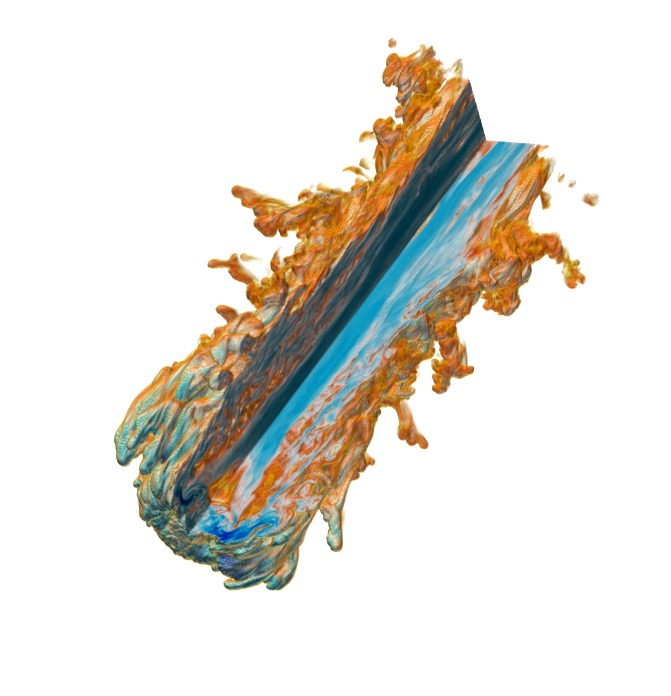}} & \resizebox{24mm}{!}{\includegraphics{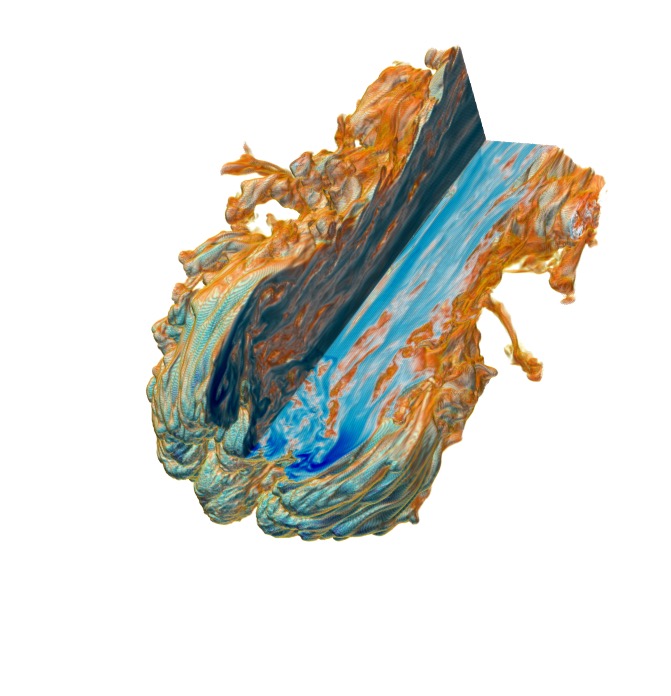}}\\
\hspace{-2cm}$t/t_{\rm cc}=0$ & \hspace{-1.5cm}$t/t_{\rm cc}=0.25$ & \hspace{-0.5cm}$t/t_{\rm cc}=0.50$ & \hspace{-0.5cm}$t/t_{\rm cc}=0.75$ & \hspace{-0.5cm}$t/t_{\rm cc}=1.00$ & \hspace{-0.5cm}$t/t_{\rm cc}=1.25$\Dstrut\\
\multicolumn{6}{l}{\hspace{-0.3cm}S5) Tur-0-0}\\ 
\resizebox{24mm}{!}{\includegraphics{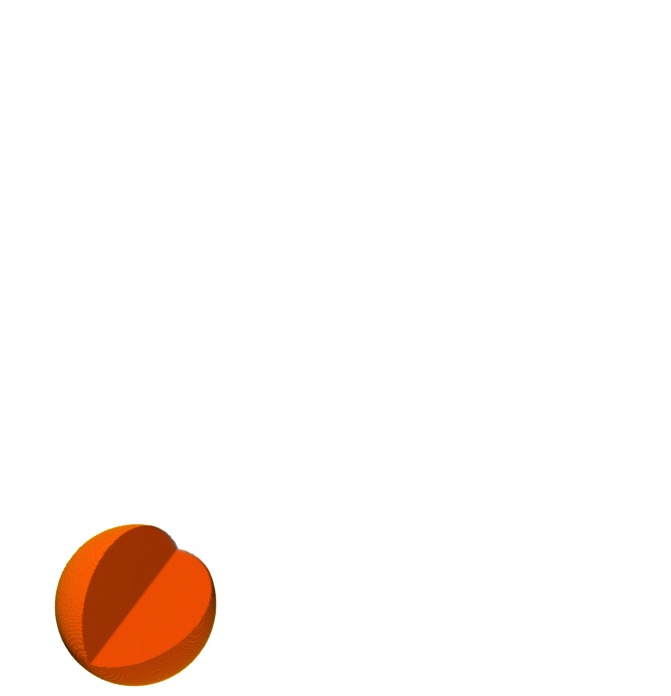}} & \resizebox{24mm}{!}{\includegraphics{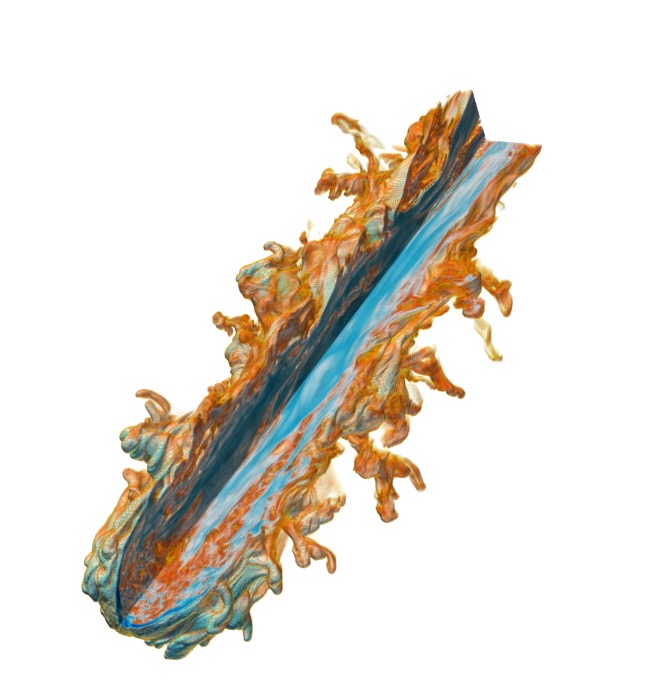}} & \resizebox{24mm}{!}{\includegraphics{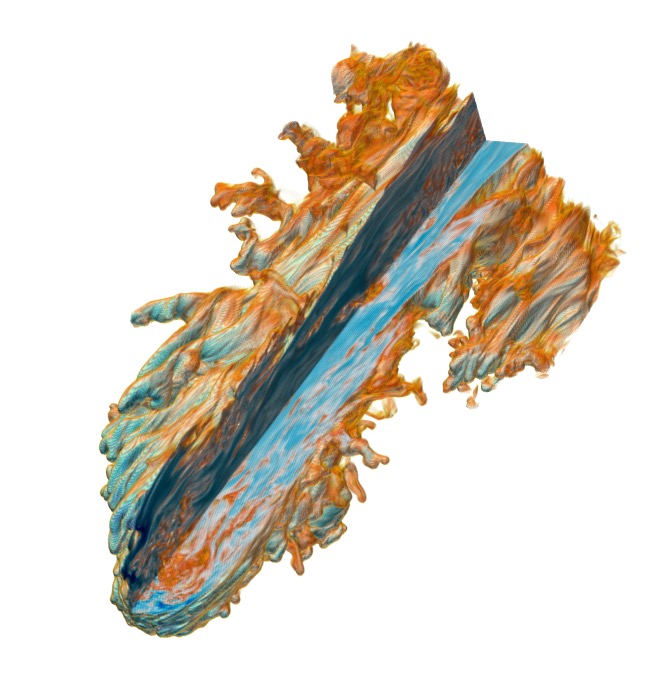}} & \resizebox{24mm}{!}{\includegraphics{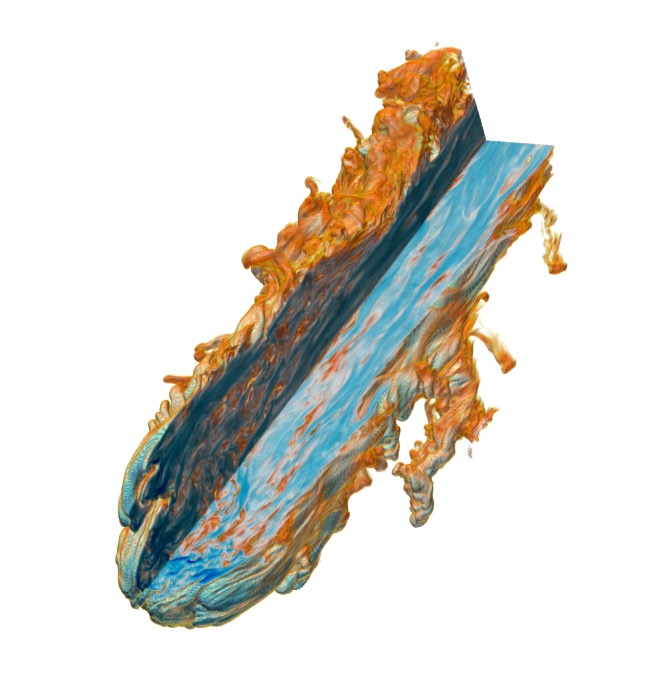}} & \resizebox{24mm}{!}{\includegraphics{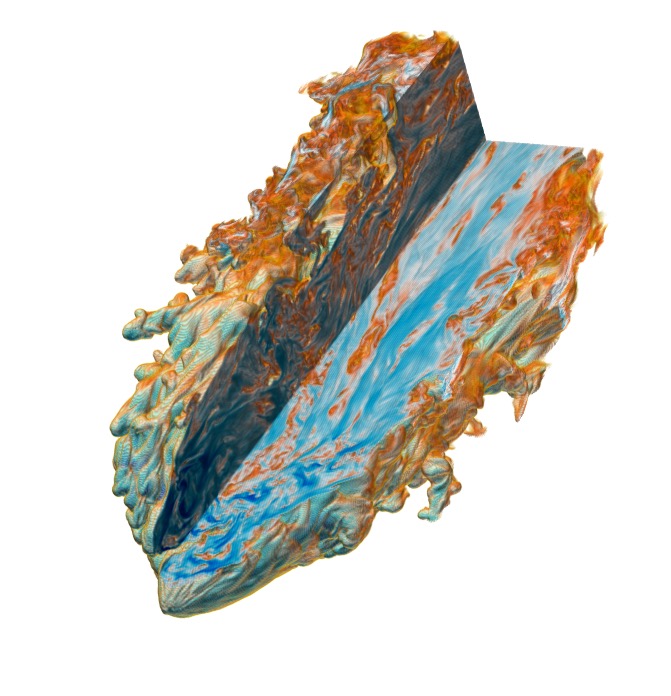}} & \resizebox{24mm}{!}{\includegraphics{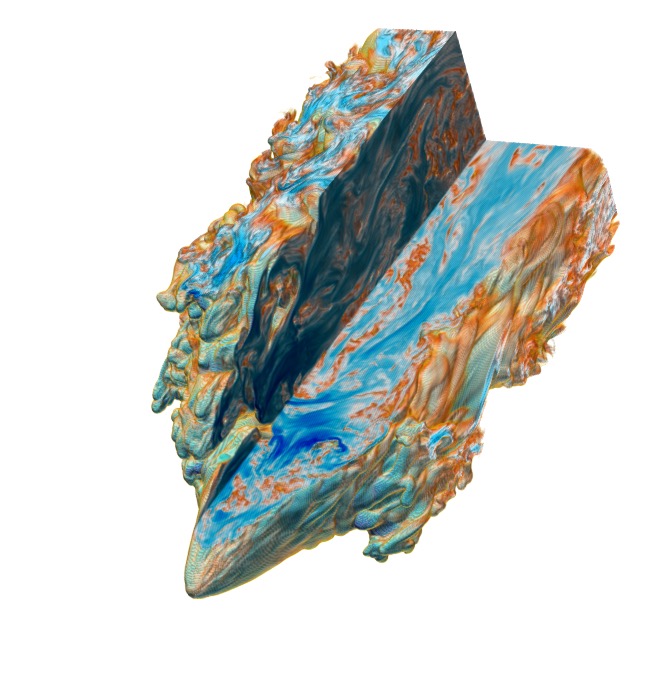}}\\
\hspace{-2cm}$t/t_{\rm cc}=0$ & \hspace{-1.5cm}$t/t_{\rm cc}=0.25$ & \hspace{-0.5cm}$t/t_{\rm cc}=0.50$ & \hspace{-0.5cm}$t/t_{\rm cc}=0.75$ & \hspace{-0.5cm}$t/t_{\rm cc}=1.00$ & \hspace{-0.5cm}$t/t_{\rm cc}=1.25$\Dstrut\\
\multicolumn{6}{l}{\hspace{-0.3cm}S6) Tur-Sub-0}\\ 
\resizebox{24mm}{!}{\includegraphics{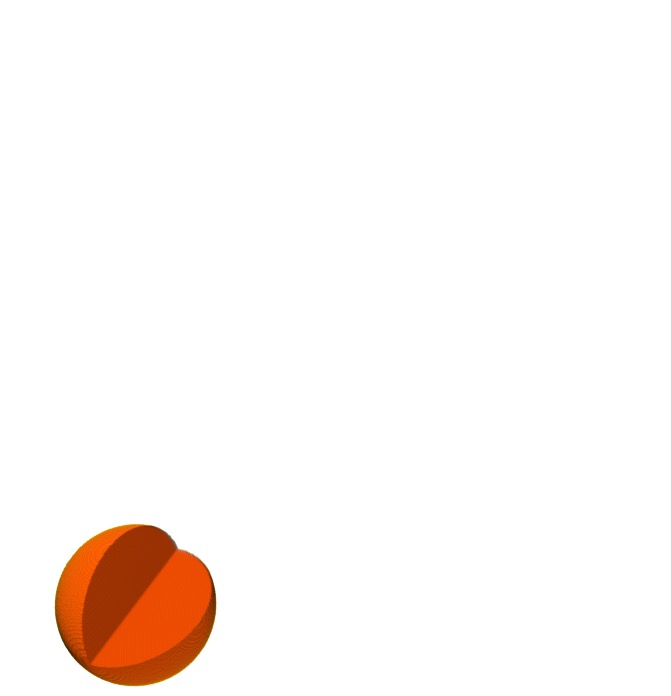}} & \resizebox{24mm}{!}{\includegraphics{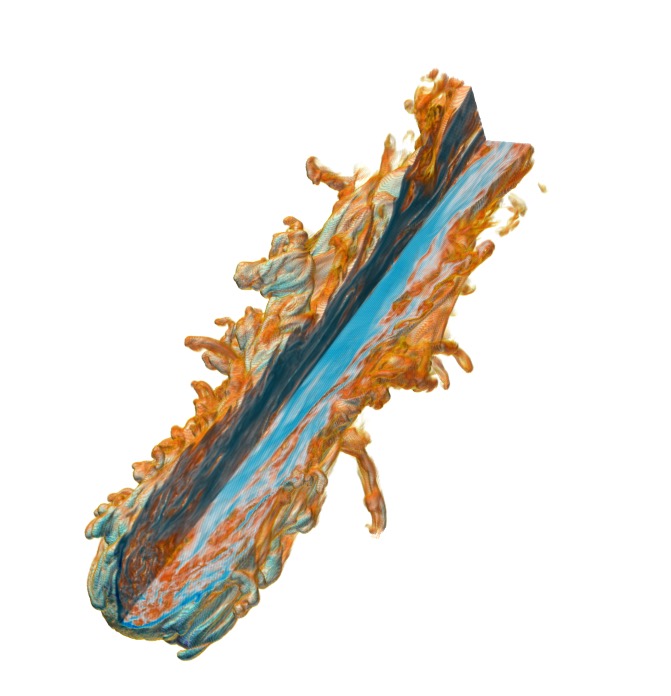}} & \resizebox{24mm}{!}{\includegraphics{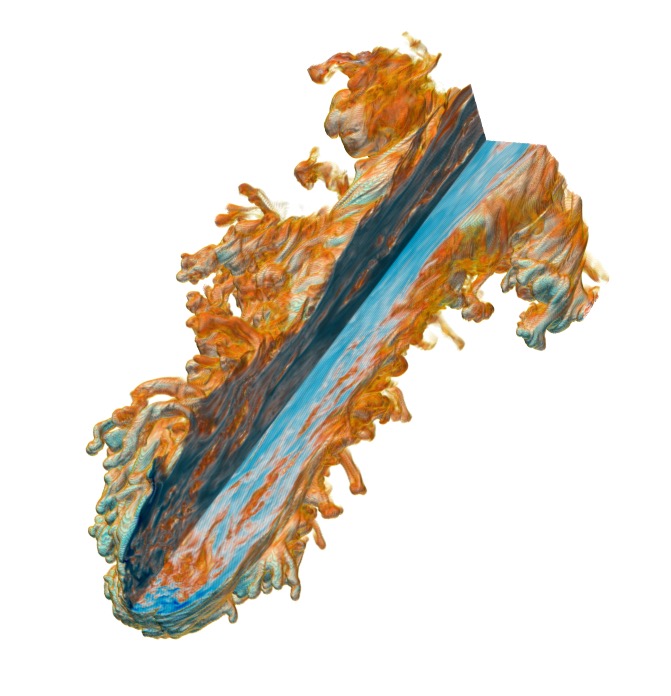}} & \resizebox{24mm}{!}{\includegraphics{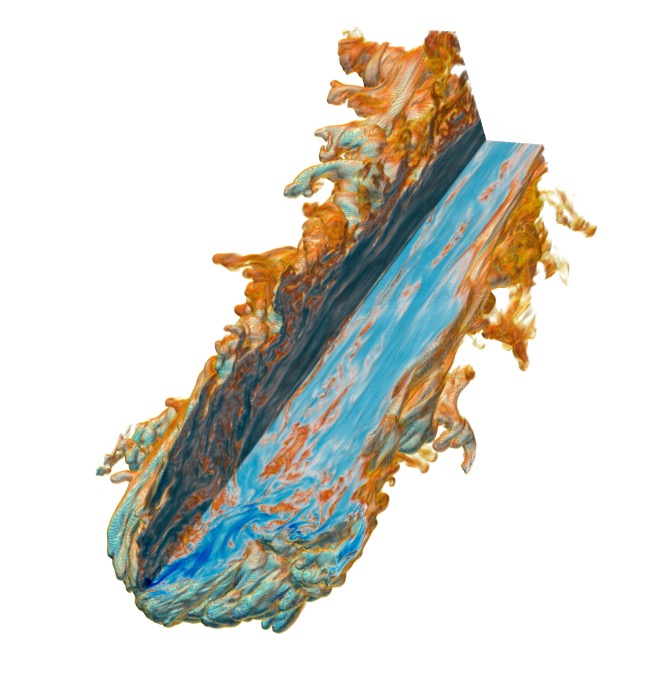}} & \resizebox{24mm}{!}{\includegraphics{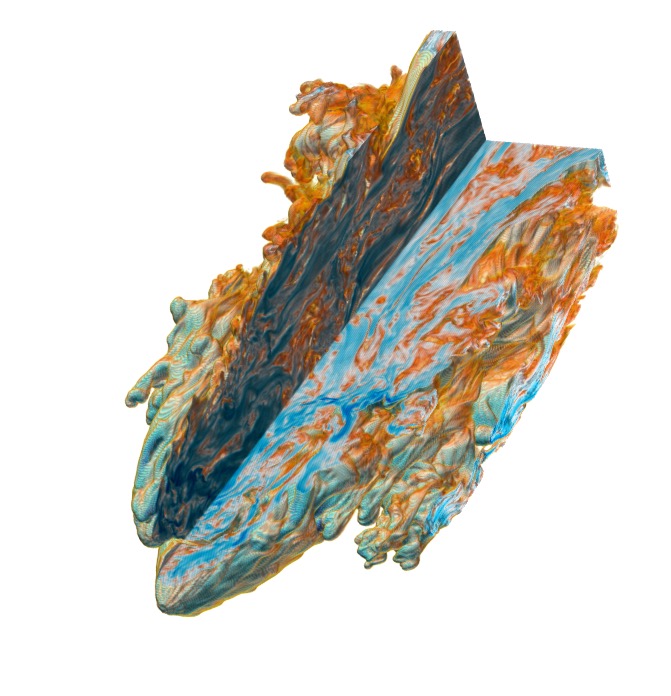}} & \resizebox{24mm}{!}{\includegraphics{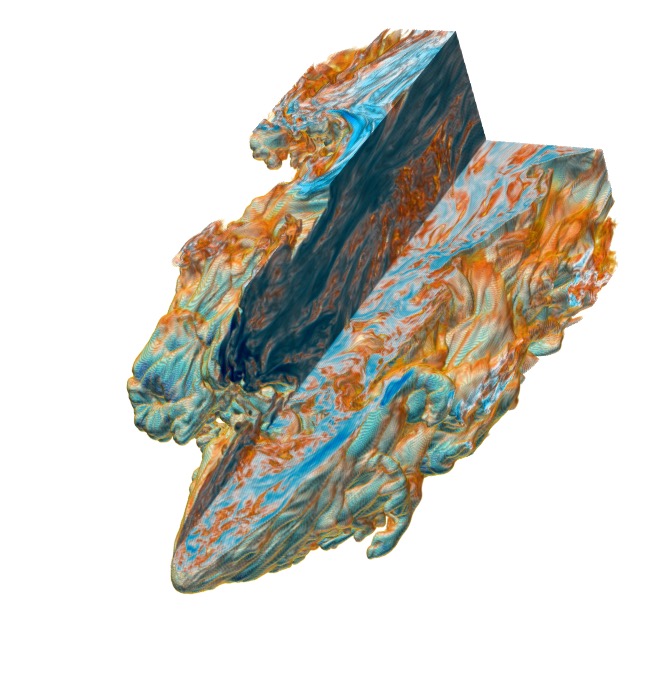}}\\
\hspace{-2cm}$t/t_{\rm cc}=0$ & \hspace{-1.5cm}$t/t_{\rm cc}=0.25$ & \hspace{-0.5cm}$t/t_{\rm cc}=0.50$ & \hspace{-0.5cm}$t/t_{\rm cc}=0.75$ & \hspace{-0.5cm}$t/t_{\rm cc}=1.00$ & \hspace{-0.5cm}$t/t_{\rm cc}=1.25$\Dstrut\\
\multicolumn{6}{l}{\hspace{-0.3cm}S7) Tur-Sup-0}\\ 
\resizebox{24mm}{!}{\includegraphics{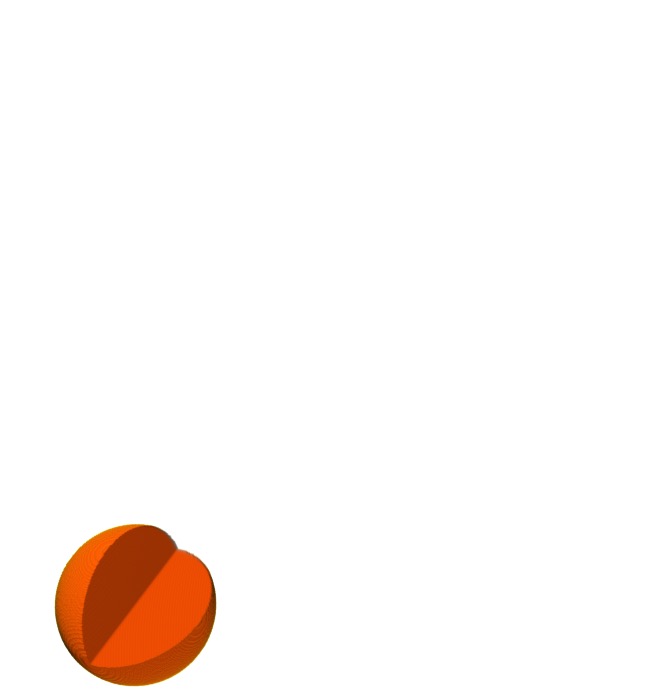}} & \resizebox{24mm}{!}{\includegraphics{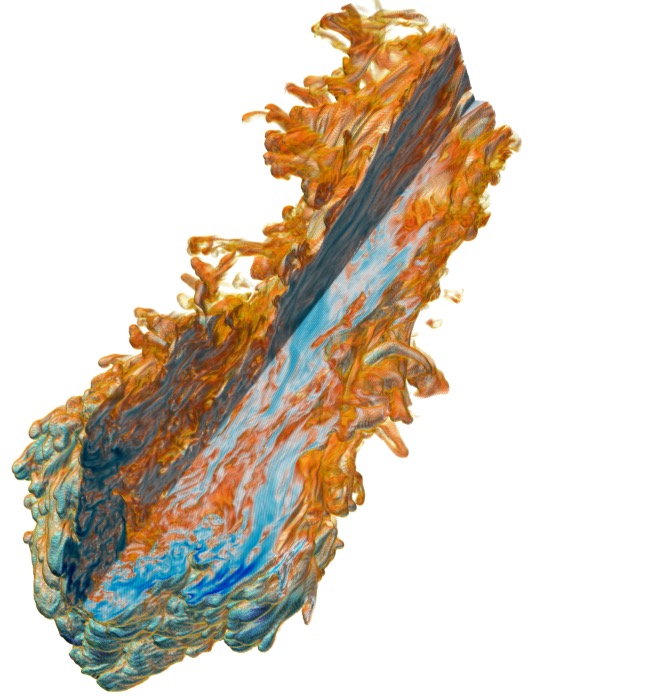}} & \resizebox{24mm}{!}{\includegraphics{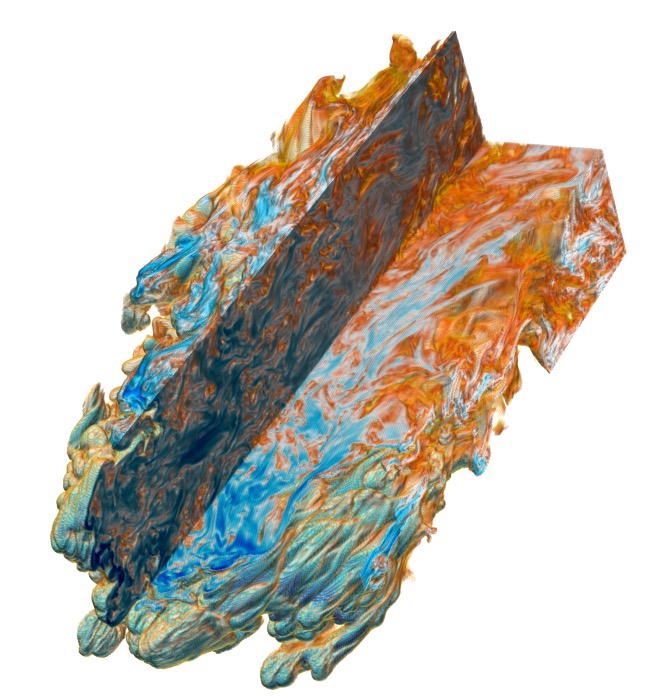}} & \resizebox{24mm}{!}{\includegraphics{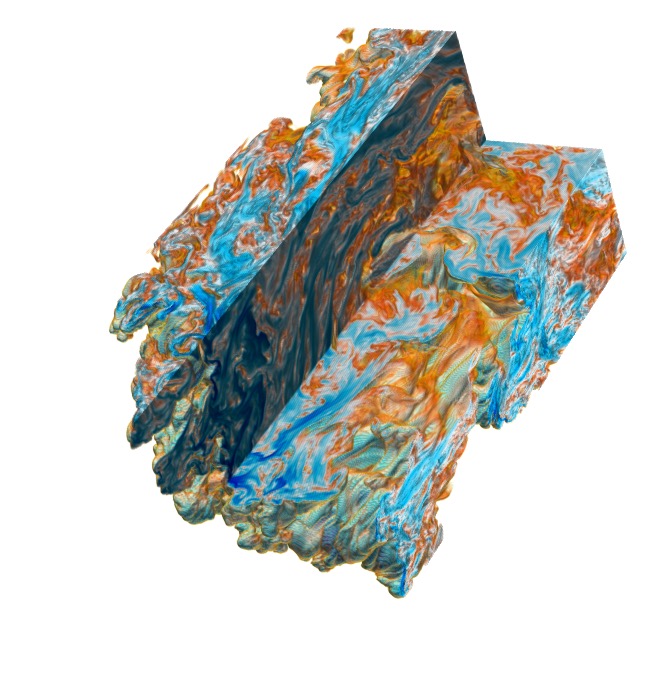}} & \resizebox{24mm}{!}{\includegraphics{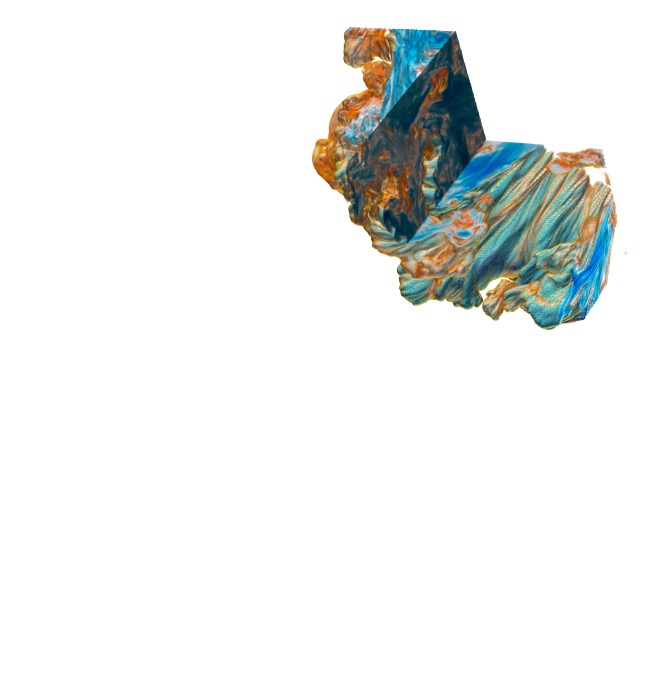}} & \resizebox{24mm}{!}{\includegraphics{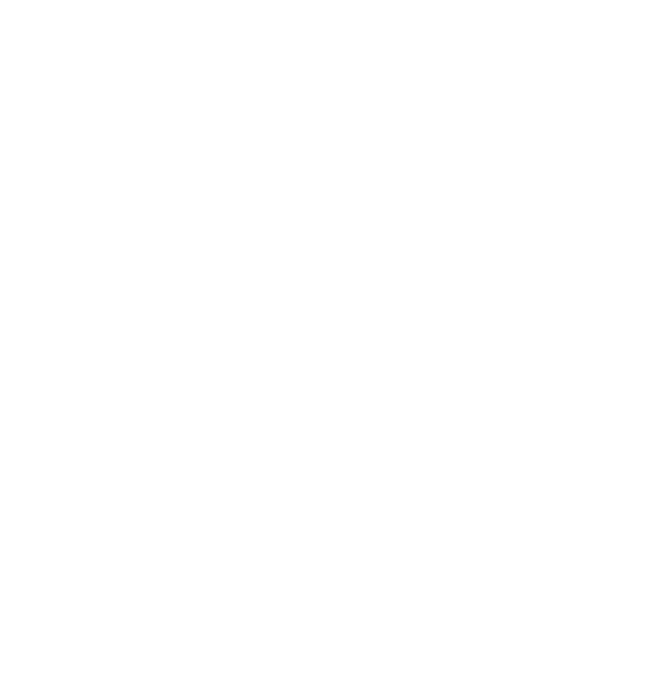}}\\
\hspace{-2cm}$t/t_{\rm cc}=0$ & \hspace{-1.5cm}$t/t_{\rm cc}=0.25$ & \hspace{-0.5cm}$t/t_{\rm cc}=0.50$ & \hspace{-0.5cm}$t/t_{\rm cc}=0.75$ & \hspace{-0.5cm}$t/t_{\rm cc}=1.00$ & \hspace{-0.5cm}$t/t_{\rm cc}=1.25$\Dstrut\\
\multicolumn{6}{l}{\hspace{-0.3cm}S8) Tur-Sub-Bwk}\\ 
\resizebox{24mm}{!}{\includegraphics{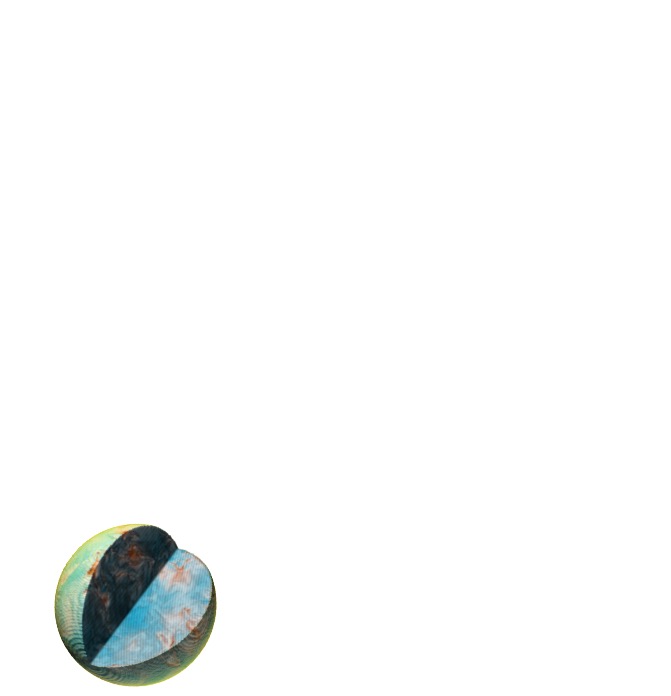}} & \resizebox{24mm}{!}{\includegraphics{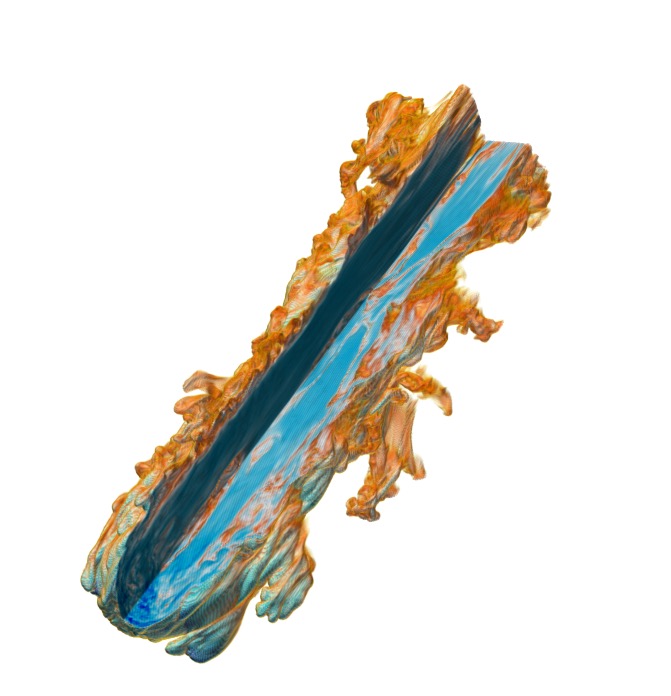}} & \resizebox{24mm}{!}{\includegraphics{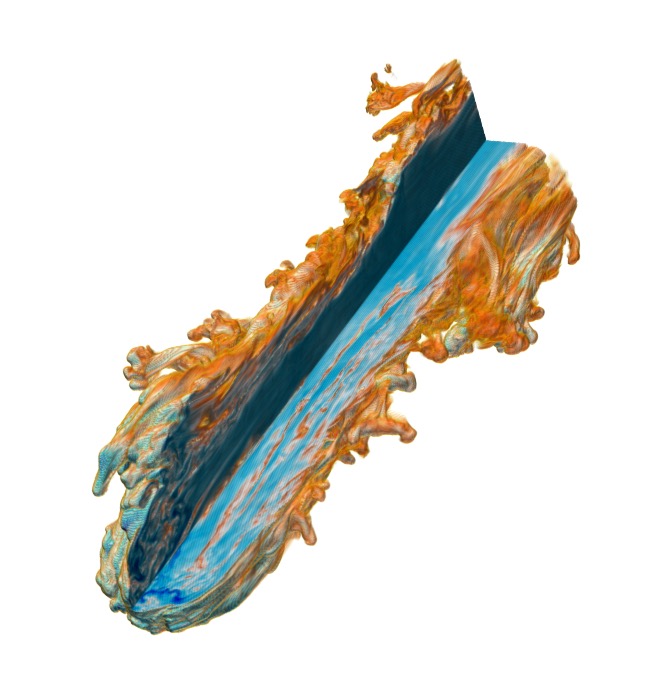}} & \resizebox{24mm}{!}{\includegraphics{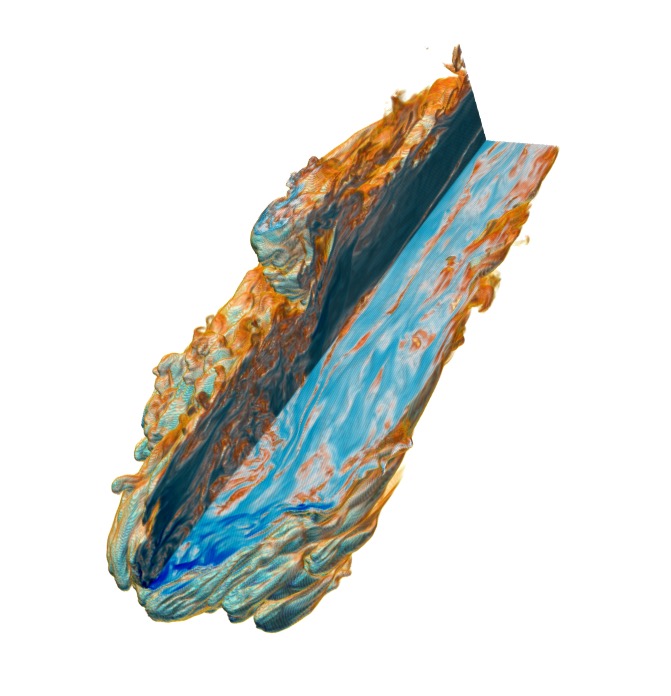}} & \resizebox{24mm}{!}{\includegraphics{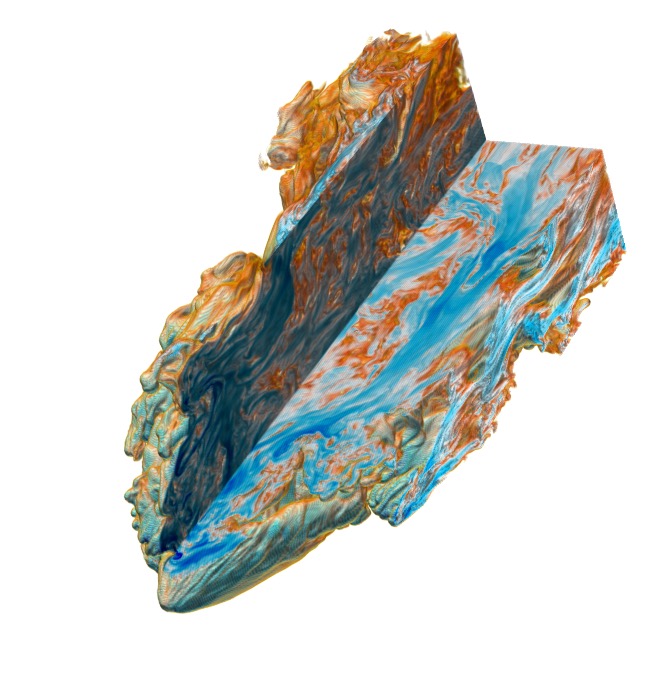}} & \resizebox{24mm}{!}{\includegraphics{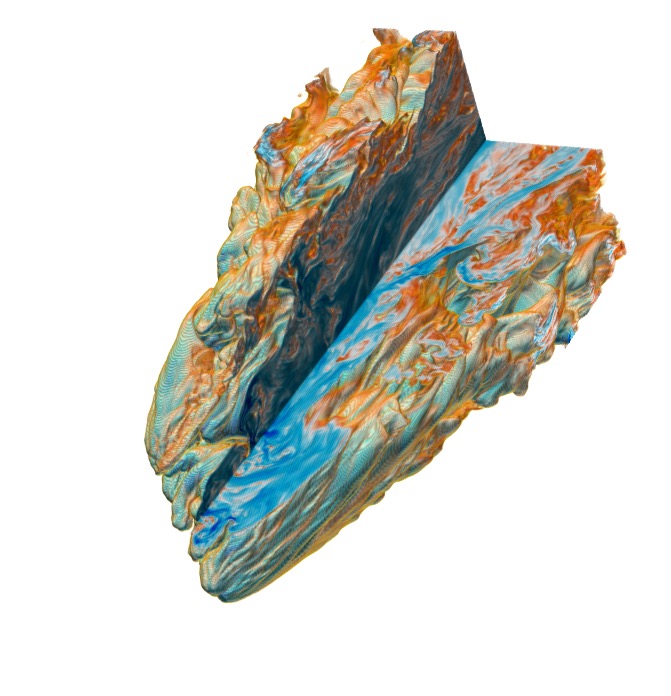}}\\
\hspace{-2cm}$t/t_{\rm cc}=0$ & \hspace{-1.5cm}$t/t_{\rm cc}=0.25$ & \hspace{-0.5cm}$t/t_{\rm cc}=0.50$ & \hspace{-0.5cm}$t/t_{\rm cc}=0.75$ & \hspace{-0.5cm}$t/t_{\rm cc}=1.00$ & \hspace{-0.5cm}$t/t_{\rm cc}=1.25$\Dstrut\\
\multicolumn{6}{l}{\hspace{-0.3cm}S9) Tur-Sub-Bst}\\ 
\resizebox{24mm}{!}{\includegraphics{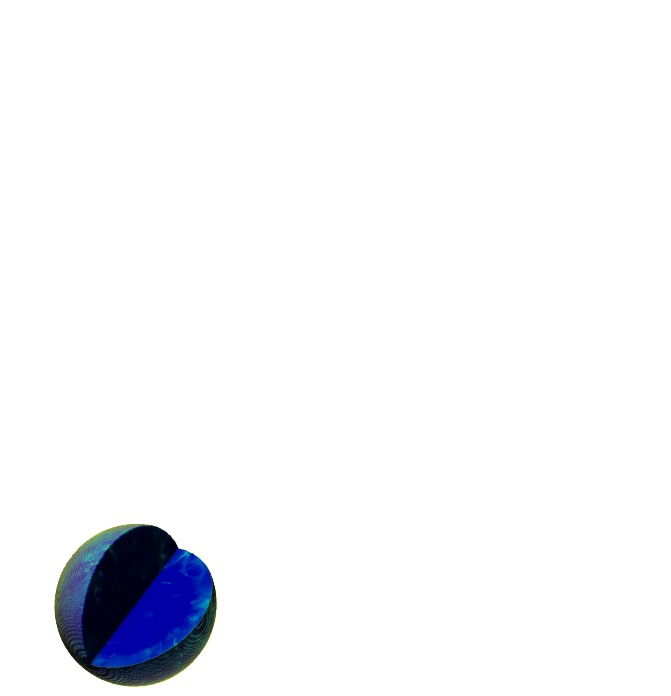}} & \resizebox{24mm}{!}{\includegraphics{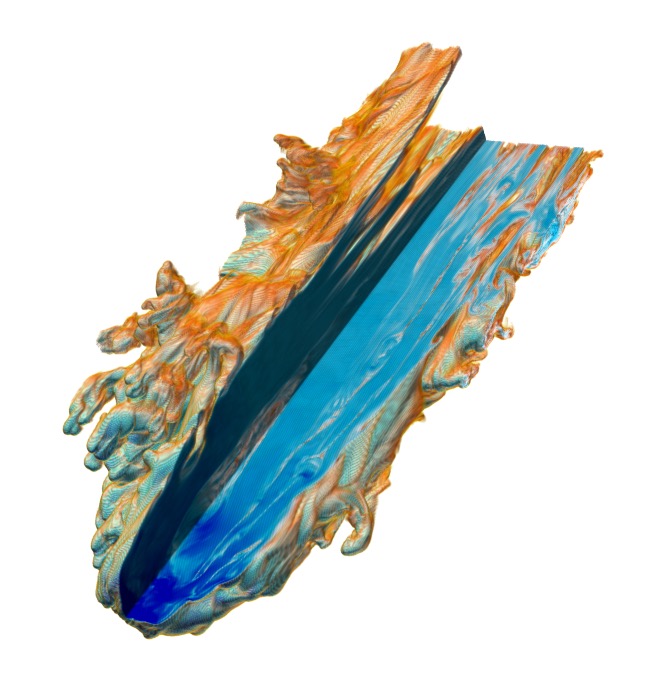}} & \resizebox{24mm}{!}{\includegraphics{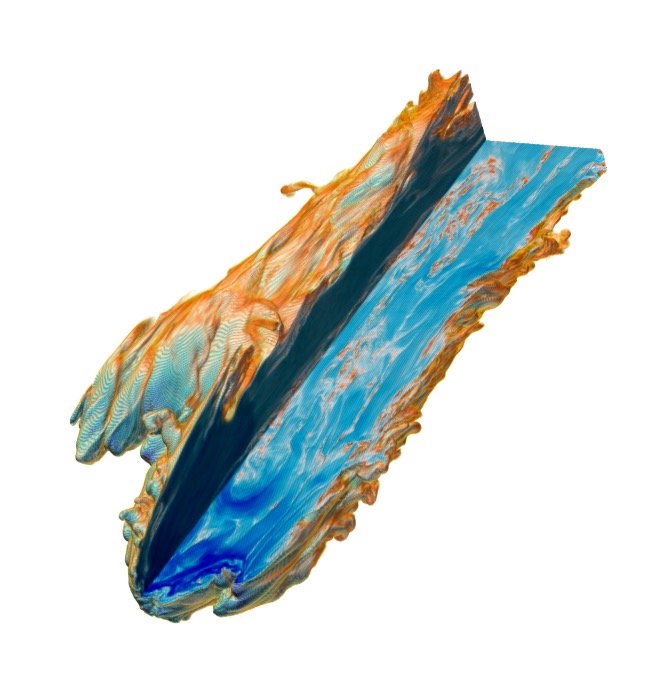}} & \resizebox{24mm}{!}{\includegraphics{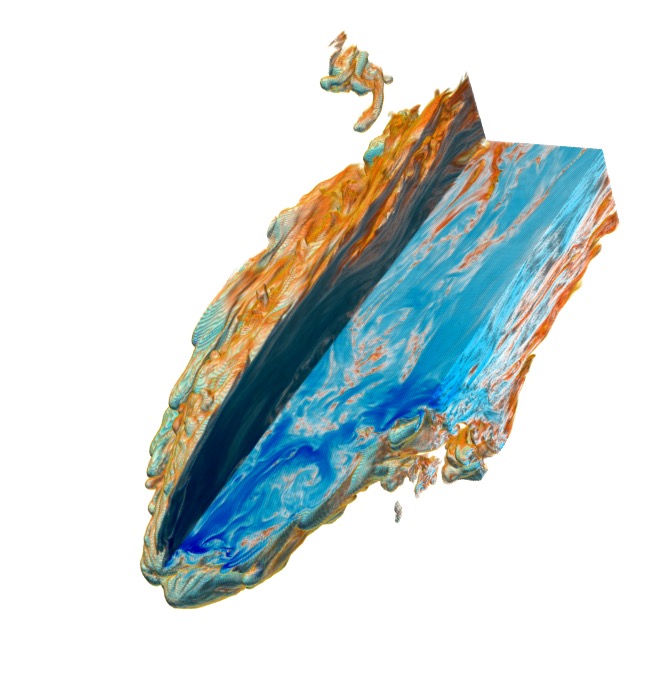}} & \resizebox{24mm}{!}{\includegraphics{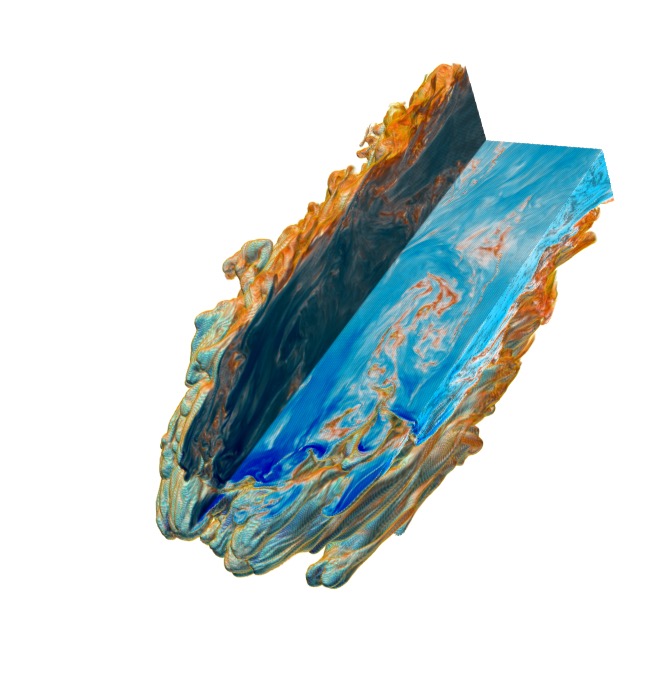}} & \resizebox{24mm}{!}{\includegraphics{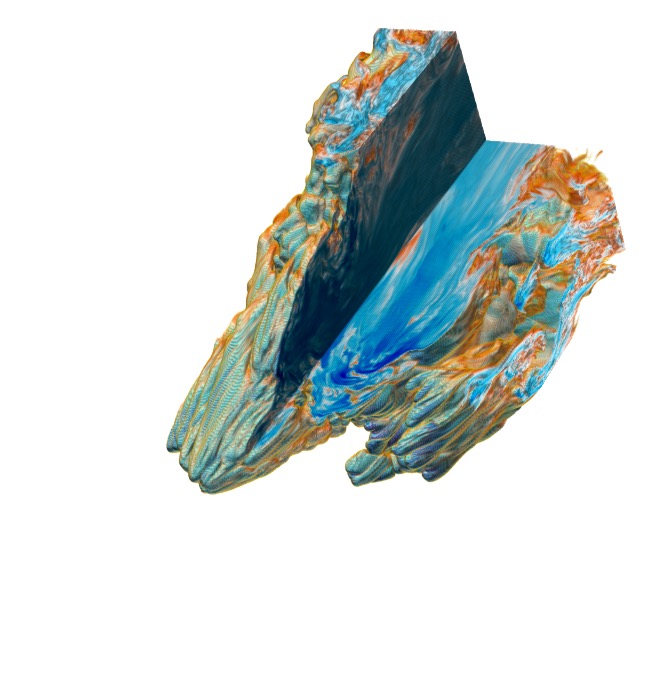}}\\
\hspace{-2cm}$t/t_{\rm cc}=0$ & \hspace{-1.5cm}$t/t_{\rm cc}=0.25$ & \hspace{-0.5cm}$t/t_{\rm cc}=0.50$ & \hspace{-0.5cm}$t/t_{\rm cc}=0.75$ & \hspace{-0.5cm}$t/t_{\rm cc}=1.00$ & \hspace{-0.5cm}$t/t_{\rm cc}=1.25$\Dstrut\\
\resizebox{!}{6mm}{\includegraphics{Axes3D.jpg}}   & \multicolumn{5}{c}{\resizebox{!}{6mm}{\includegraphics{bar_magenergy.jpg}}}\\
  \end{tabular}
  \caption{Same as Figure \ref{Figure4}, but here we present the evolution of the normalised magnetic energy density in six models with the S domain configuration for $0\leq t/t_{\rm cc}\leq1.25$. Panels S4 and S5 show clouds with uniform (Uni-0-0) and turbulent (Tur-0-0) densities (but without turbulent velocity or magnetic fields), respectively. Panels S6 and S7 show turbulent clouds with turbulent, subsonic (Tur-Sub-0) and supersonic (Tur-Sup-0) velocity fields, respectively. In models S4-S7 the clouds are immersed in a uniform, oblique magnetic field. Panels S8 and S9 present turbulent clouds with weak (Tur-Sub-Bwk) and strong (Tur-Sub-Bst) turbulent magnetic fields, in addition to the turbulent density and velocity fields of the previous models.}
  \label{Figure8}
\end{center}
\end{figure*}


In this section we describe the magnetic structure of filaments in different models. The 3D volume renderings of Figure \ref{Figure8} show the time evolution, for $0\leq t/t_{\rm cc}\leq1.25$, of the normalised magnetic energy density of wind-swept clouds in six different models with the S domain configuration (see Table \ref{Table1}). The reference time-scales for the growth of KH instabilities (see Equation \ref{KHtime}) and RT instabilities (see Equation \ref{RTtime}) at wind-filament interfaces in models Uni-0-0, Tur-0-0, Tur-Sub-0, Tur-Sup-0, Tur-Sub-Bwk, and Tur-Sub-Bst are in the ratios $t_{\rm KH,\,S4}:t_{\rm KH,\,S5}:t_{\rm KH,\,S6}:t_{\rm KH,\,S7}:t_{\rm KH,\,S8}:t_{\rm KH,\,S9}=1:0.9:0.9:0.6:1.1:1.5$ and $t_{\rm RT,\,S4}:t_{\rm RT,\,S5}:t_{\rm RT,\,S6}:t_{\rm RT,\,S7}:t_{\rm RT,\,S8}:t_{\rm RT,\,S9}=1:1:1:0.4:0.9:0.8$, respectively, for $\lambda_{\rm KH}=\lambda_{\rm RT}\sim 1\,r_{\rm c}$. These calculations indicate that the inclusion of different turbulence profiles in the clouds leads to varying levels of suppression or enhancement of dynamical instabilities and thus to morphological differences in the resulting filaments. We discuss these differences below.\par

Panel S4 of Figure \ref{Figure8} shows the magnetic structure of the filament in model Uni-0-0. In this model a uniform, spherical cloud is uniformly magnetised at the beginning of the simulation (same as model UNI-0-0, but in a smaller computational domain). As a result of compression, and folding and stretching of magnetic field lines, respectively, two regions of high magnetic energy are identified at $t/t_{\rm cc}=0.25$: the first one is located at the front end of the filament footpoint, while the second one extends along the tail embedding an obliquely-oriented current sheet (same as in model UNI-0-0). The magnetic energy in the core of the cloud remains unchanged at this time, but as the late expansion (for $t/t_{\rm cc}\geq0.5$) of the core takes place, its magnetic energy is progressively amplified $\sim10^2-10^3$ times (see Section 5.5 of \citetalias{2016MNRAS.455.1309B} for further details). At $t/t_{\rm cc}=1.0$, the footpoint is dispersed and the magnetic field at the leading edge of the cloud becomes turbulent with small-scale vortical motions dominating at later times.\par 

Panel S5 of Figure \ref{Figure8} shows the evolution of the magnetic energy density in model Tur-0-0. In this model the wind strikes a static cloud with a turbulent density distribution, initially immersed in a uniformly magnetised medium. After the filamentary tail forms downstream, both the footpoint and the tail of the filament are affected by shock-triggered turbulence and vortical motions. As a result, the structure of this filament is more chaotic than the one in model Uni-0-0. In model Tur-0-0 the magnetic field lines fold and stretch around the most massive nuclei in the cloud's core. Since the distribution of these nuclei is anisotropic, the magnetotail becomes inhomogeneous. This results in the magnetic field being locally enhanced in regions sheltered by or in between dense nuclei in the core (see the snapshots for $0.25\leq t/t_{\rm cc}\leq1.00$), while remaining unchanged at other locations. Thus, models Uni-0-0 and Tur-0-0 produce filaments that are structurally different: a uniform cloud favours the formation of a filament with a single current sheet while a turbulent cloud produces a filamentary tail filled with several highly-magnetised knots and sub-filaments (at which $E_{\rm m, cloud}/E_{{\rm m}_{0}}\sim 10^2-10^3$). Similar structures have been found in purely HD and MHD simulations of shocks interacting with inhomogeneous media, i.e., systems that have more than one cloud (e.g., see \citealt{2002ApJ...576..832P,2005MNRAS.361.1077P,2009MNRAS.392..964R,2012MNRAS.425.2212A,2014MNRAS.444..971A}; \citealt*{2016rscd.conf..146R}).\par

Panel S6 of Figure \ref{Figure8} shows the magnetic structure of the filament in model Tur-Sub-0. In this model the cloud is initialised with the same density distribution used for model Tur-0-0, plus a subsonic, Gaussian velocity field (with a Mach number of ${\cal M_{\rm tu}}=0.33$). We find no qualitative difference in the magnetic structure of filaments in models Tur-0-0 and Tur-Sub-0 throughout the entire evolution. Both models, Tur-0-0 and Tur-Sub-0, produce filaments with non-uniform structures characterised by the presence of strongly-magnetised knots and sub-filaments along their tails. This implies that a subsonically-turbulent velocity field does not provide a sufficiently-high (extra) kinetic energy density to the cloud to have significant effects on the morphology and dynamics of the resulting filament. This result is expected from analytical estimates of the ratio of the turbulence-crossing and the cloud-crushing time-scales, i.e., $t_{\rm tu}/t_{\rm cc}\sim12$, which indicates that a subsonically-turbulent velocity field with ${\cal M_{\rm tu}}=0.33$ would need $\sim12$ cloud-crushing times to have a dynamical impact on the cloud (see Equation \ref{eq:TurbulenceCrossing}).\par


Panel S7 of Figure \ref{Figure8} shows the evolution of the magnetic energy in model Tur-Sup-0. This model is started with the same density PDF and magnetic field configuration previously assigned to the above models, plus a supersonic, turbulent velocity field with Mach number of ${\cal M_{\rm tu}}=8.9$ (consistent with the original Mach number of the cloud extracted from \citealt{2012ApJ...761..156F}; see Section \ref{subsec:Initial and Boundary Conditions}). The turbulence-crossing time (see Equation \ref{eq:TurbulenceCrossing}) for this model is of the order of $t_{\rm tu}/t_{\rm cc}\sim0.4$, i.e., turbulence is dynamically important for this system. The cloud in this model expands quickly from the very beginning of the interaction as a result of internal turbulent motions. This increases the effective cross section upon which the ram-pressure force acts and the cloud becomes prone to longer-wavelength, highly-disruptive unstable modes. Both the KH and RT instabilities grow faster in this model than in any of the other turbulent cloud models, e.g., the growth time-scales of KH and RT modes with $\lambda_{\rm KH}=\lambda_{\rm RT}\sim 1\,r_{\rm c}$ are in the ratios $t_{\rm KH,\,S5}:t_{\rm KH,\,S7}=1:0.7$ and $t_{\rm RT,\,S5}:t_{\rm RT,\,S7}=1:0.4$, respectively, in models Tur-0-0 and Tur-Sup-0. This signifies that the supersonic cloud is dispersed, mixed with the wind, and disrupted faster than in the turbulent models discussed above. In fact, we find that the cloud break-up occurs on time-scales of the order of the turbulence-crossing time-scale rather than in the typical $t/t_{\rm cc}\sim 1.0$. After the break-up, the cloud expands beyond the boundaries of the computational domain, making the bow shock at its leading edge vanish and biasing the qualitative results to low-velocity-dispersion gas. Since the cloud moves very quickly out of the computational domain, we stop this simulation shortly after $t/t_{\rm cc}=1.0$.\par

Panel S8 of Figure \ref{Figure8} shows the evolution of the magnetic energy density in model Tur-Sub-Bwk, in which a weak, turbulent magnetic field (with $[~\beta_{\rm tu}~]=4$) is added to the initial cloud, alongside the turbulent density and velocity fields of the previous models. In this simulation we observe a filament with a similar structure to the ones emerging from models Tur-0-0 and Tur-Sub-0, but its interior harbours a larger number of strongly-magnetised sub-filaments and a more laminar magnetotail (see the snapshots for $0.25\leq t/t_{\rm cc}\leq0.75$). The magnetic field strength of the knots and sub-filaments in this model is also higher than in the cases without turbulent magnetic fields, owing to the stretching of magnetic field lines anchored to regions of high density (dense nuclei) in the footpoint. By the end of the evolution, vortical motions dominate and the filament in model Tur-Sub-Bst resembles the others in models Tur-0-0 and Tur-Sub-0, displaying knots and sub-filaments with similar magnetic field strengths ($E_{\rm m, cloud}/E_{{\rm m}_{0}}\sim 10^2-10^3$). The principal effect of the weak, turbulent magnetic field in this model is to mildly protect the cloud/filament from KH instabilities emerging at wind-cloud interfaces as revealed by the enhanced laminarity of its filament with respect to those in models Tur-0-0 and Tur-Sub-0. Indeed, the reference growth time-scales of KH instabilities with $\lambda_{\rm KH}\sim 1\,r_{\rm c}$ are in the ratio $t_{\rm KH,\,S6}:t_{\rm KH,\,S8}=1:1.2$ in models Tur-Sub-0 and Tur-Sub-Bwk.\par

Panel S9 of Figure \ref{Figure8} shows the morphology of the filament in model Tur-Sub-Bst. The cloud in this model is initialised with a strong, turbulent magnetic field (with $[~\beta_{\rm tu}~]=0.04$) on top of the turbulent density and velocity fields used in the above models. The previously-mentioned effects of a turbulent magnetic field on the morphology of the cloud are also seen in this model. The tail of this filament is inhabited by magnetised knots and sub-filaments with higher magnetic energies ($E_{\rm m, cloud}/E_{{\rm m}_{0}}\sim 10^3-10^4$) and it is more laminar than in the weak-field case (model Tur-Sub-Bwk). The higher magnetic pressure produces two effects: a) it further shields the magnetotail, suppressing KH instabilities at wind-filament boundaries (the reference time-scales for the growth of KH instabilities with $\lambda_{\rm KH}\sim 1\,r_{\rm c}$ are in the ratio $t_{\rm KH,\,S6}:t_{\rm KH,\,S9}=1:1.6$ in models Tur-Sub-0 and Tur-Sub-Bst); and b) it enhances the growth of RT instabilities at the leading edge of the cloud (the reference time-scales for the growth of RT instabilities with $\lambda_{\rm RT}\sim 1\,r_{\rm c}$ are in the ratio $t_{\rm RT,\,S6}:t_{\rm RT,\,S9}=1:0.9$ in models Tur-Sub-0 and Tur-Sub-Bst), in agreement with previous MHD studies, e.g., \citealt{1996ApJ...473..365J,1999ApJ...510..726M,1999ApJ...527L.113G,2000ApJ...543..775G}; \citetalias{2016MNRAS.455.1309B}. In fact, after $t/t_{\rm cc}=0.75$, small-scale RT bubbles penetrate the front end of the cloud and push material laterally, thus forming a series of sub-filaments along the tail.\par

Overall, the panels of Figure \ref{Figure8} reveal an important property of interstellar filaments that are produced by wind-cloud interactions. In agreement with our result in Section \ref{subsubsec:UniformvsTurbulent} for the turbulent models TUR-SUP-BST and TUR-SUP-BST-ISO, the 3D renderings in Panel S9 of Figure \ref{Figure8} (for model Tur-Sub-Bst) show that several knots and sub-filaments have magnetic energy densities similar to that in the initial cloud, confirming that the inclusion of realistic, strong magnetic fields into the initial cloud results in a filament with similarly-strong magnetic fields (see Appendix \ref{sec:Appendix5} for further details). In addition, the above results also highlight the importance of including self-consistent turbulent magnetic fields when considering supersonic clouds as they prevent the cloud from being rapidly shredded while it expands. The turbulent destruction of clouds has been studied in both inviscid, turbulent models (e.g., see \citealt{2009ApJ...703..330C,2017ApJ...834..144S}) and sub-grid turbulent viscosity models (e.g., see \citealt{2009MNRAS.394.1351P,2010MNRAS.405..821P,2011Ap&SS.336..239P,2016MNRAS.457.4470P,2017MNRAS.468.3184G}) of wind/shock cloud systems. In agreement with the conclusions drawn from these studies, our results here show that turbulence by itself may potentially have the ability to disrupt a cloud by increasing the mixing of cloud and wind material via hastened dynamical instabilities. Thus, magnetic fields should be a crucial ingredient of any realistic wind-cloud system as they help maintain the stability of wind-cloud interfaces and delay cloud/filament disruption (as we also pointed out in Section \ref{subsec:FilamentFormation}).\par

\subsubsection{On the lateral width, velocity dispersion, and vorticity}
\label{subsubsec:AspectRatio}

\begin{figure*}
\begin{center}
  \begin{tabular}{c c}
  \textbf{Filament Tail (Cloud Envelope)} & \textbf{Filament Footpoint (Cloud Core)}\\
  \resizebox{80mm}{!}{\includegraphics{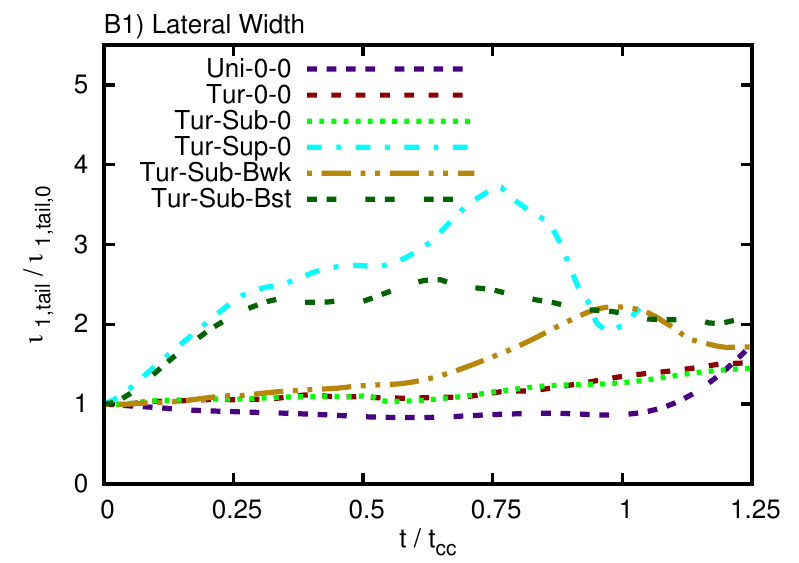}} & \resizebox{80mm}{!}{\includegraphics{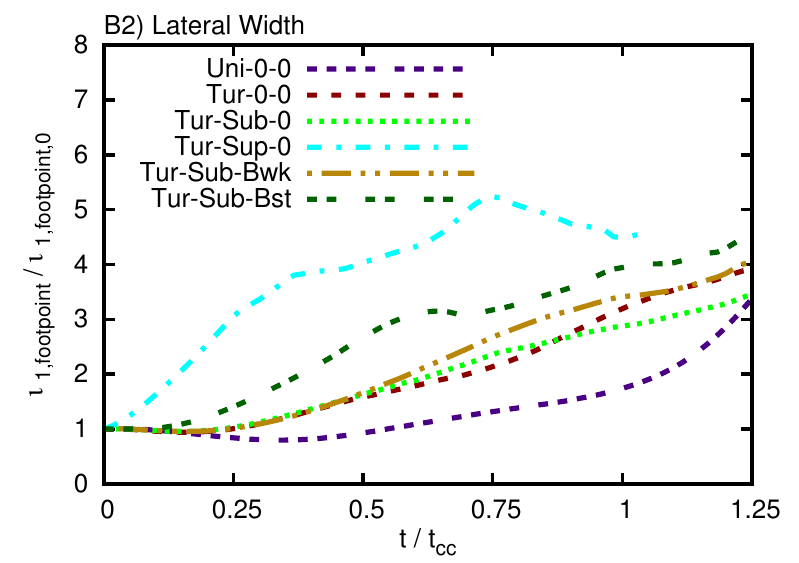}}\\
  \resizebox{80mm}{!}{\includegraphics{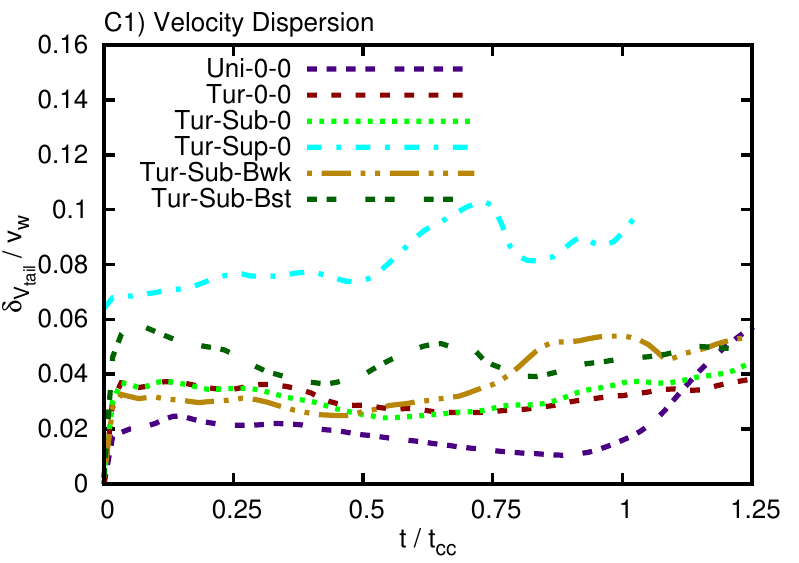}} & \resizebox{80mm}{!}{\includegraphics{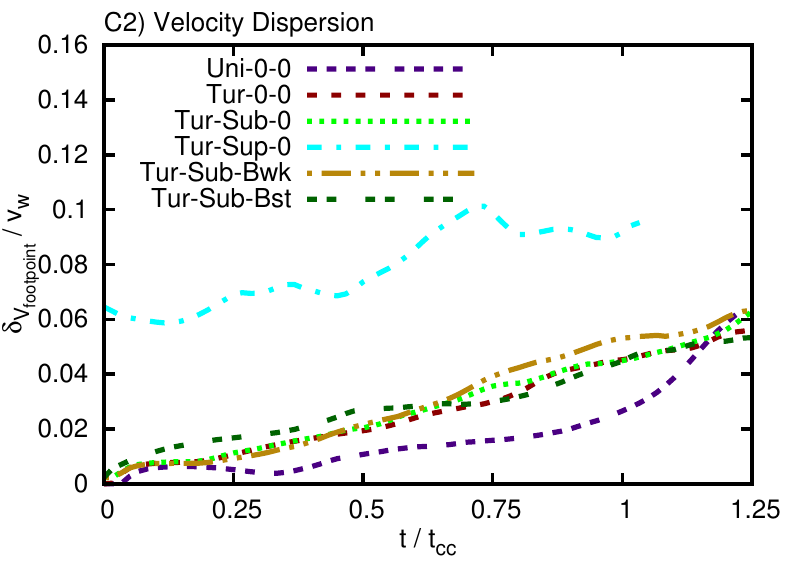}}\\  
  \resizebox{80mm}{!}{\includegraphics{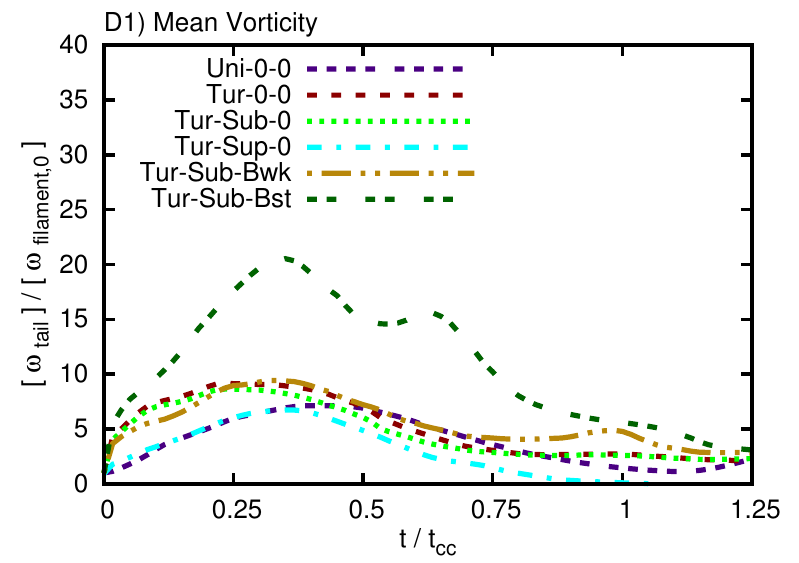}} & \resizebox{80mm}{!}{\includegraphics{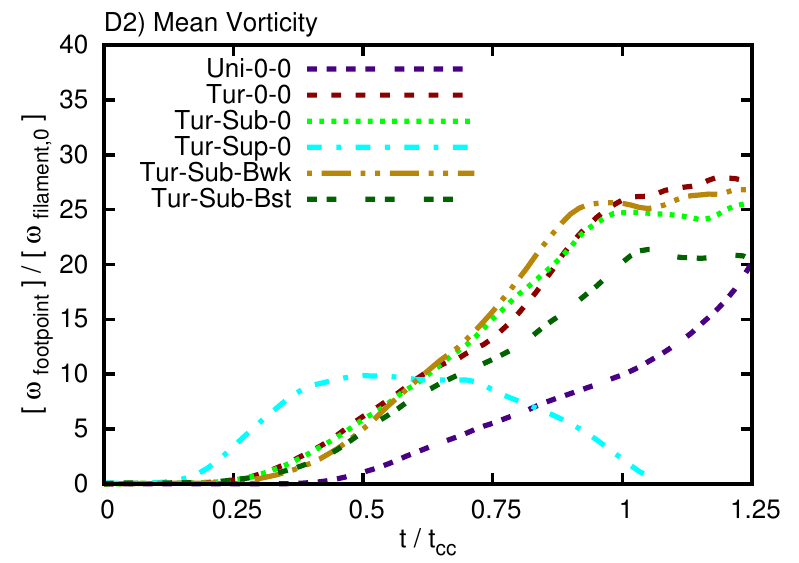}}\\ 
  \end{tabular}
  \caption{Time evolution of three diagnostics in the tails (left-hand side column) and footpoints (right-hand side column) of filaments in six models with the S configuration (see Table \ref{Table1}). Panels B1 and B2 show the lateral widths, Panels C1 and C2 show the transverse velocity dispersions, and Panels D1 and D2 show the mean vorticity enhancements, in models Uni-0-0 (four-dashed line), Tur-0-0 (long-dashed line), Tur-Sub-0 (dotted line), Tur-Sup-0 (short dash-dotted line), Tur-Sub-Bwk (long-dash-two-dotted line), and Tur-Sub-Bst (two-dashed line). We find that turbulent cloud models produce filaments with greater lateral elongations, higher velocity dispersions, and higher vorticity enhancements than the uniform cloud model. The largest effect on these diagnostics is produced by supersonic velocity fields, followed by strong magnetic fields, and then by turbulent density distributions.}
  \label{Figure9}
\end{center}
\end{figure*}

In this and subsequent sections we describe the above models by examining the evolution of different diagnostics in the tails (using the scalar $C_{\rm envelope}$) and footpoints (using the scalar $C_{\rm core}$) of filaments, separately. Figure \ref{Figure9} shows the time evolution of three parameters: the lateral width (Panels B1 and B2), velocity dispersion (Panels C1 and C2), and mean vorticity enhancement (Panels D1 and D2), for tail (left-hand side column) and footpoint (right-hand side column) material, respectively.\par

These diagnostics reveal that turbulent models produce filaments with greater lateral elongations, higher velocity dispersions, and higher relative vorticities than the uniform model. In order of significance, we observe that supersonic velocity fields (see model Tur-Sup-0) cause the most pronounced effect, owing to the fast expansion of cloud gas triggered by the thermal energy injected into the cloud via turbulence dissipation (which mainly occurs in internal shocks). Lateral widths are $3-4$ times larger than in models Uni-0-0 and Tur-0-0, with values of $\iota_{1,{\rm tail}}/\iota_{1,{\rm tail,0}}\sim3$ and $\iota_{1,{\rm footpoint}}/\iota_{1,{\rm footpoint,0}}\sim4-5$ being characteristic of the tail (Panel B1) and footpoint (Panel B2), respectively, of the filament in model Tur-Sup-0. Similar ratios between the velocity dispersions in models Tur-Sup-0 and Tur-0-0 are observed in Panels C1 and C2, in which the supersonically-turbulent velocity field produces transverse velocity dispersions $\delta_{v_{\rm tail}}/v_{\rm w}\sim\delta_{v_{\rm footpoint}}/v_{\rm w}\sim 0.06-0.08$ in both the tail and footpoint of the filament. Since the cloud in this model is already supersonically-turbulent at the beginning of the simulation, little additional vorticity is deposited in both the tail and footpoint of its filament, as indicated by the evolution of the mean vorticity enhancement in Panels D1 and D2. Note that the unbound, turbulence-triggered cloud expansion causes a large fraction of cloud material in model Tur-Sup-0 to leave the computational domain through its sides after $t/t_{\rm cc}=0.4$. By $t/t_{\rm cc}=1.0$, $\sim80\,\%$ of the cloud has left the simulation domain, so we stop the simulation at that point. In addition, we find that the inclusion of a subsonically-turbulent velocity field (see model Tur-Sub-0) does not produce any significant effect on the lateral width, velocity dispersion, and mean vorticity enhancement of the filament, in agreement with our qualitative analysis in the previous section. The subsonic turbulence in model Tur-Sub-0 also decays, but dissipation in this case occurs in small-scale eddies on the viscous scale.\par

The second largest effect on the above diagnostics is caused by turbulent magnetic fields. The stronger the magnetic field (i.e., the larger the magnetic pressure) in the cloud, the greater the magnetically-driven expansion of filament gas. This is revealed by the curves of models Tur-Sub-0, Tur-Sub-Bwk, and Tur-Sub-Bst of Panels B1 and B2 of Figure \ref{Figure9}, which show that lateral widths become systematically larger as we increase the strength of the initial magnetic field threading the cloud. In the weak-field model, Tur-Sub-Bwk, magnetically-driven expansion becomes important for $t/t_{\rm cc}\geq0.5$, producing lateral elongations up to $\sim80\,\%$ (in the tail) and $\sim25\,\%$ (in the footpoint) larger than in model Tur-Sub-0. In the strong-field model, Tur-Sub-Bst, the effect is more significant and occurs from the beginning of the interaction, producing lateral elongations $\sim160\,\%$ (in the tail) and $\sim50\,\%$ (in the footpoint) larger than in model Tur-Sub-0. Panel B2 also shows that the footpoint disruption of turbulent filaments occurs in a steadier manner than in their uniform counterpart (where an abrupt break-up takes place at $t/t_{\rm cc}=1.0$). In the case of velocity dispersions (Panels C1 and C2) and mean vorticity enhancements (Panels D1 and D2), we find that increasing the strength of the initially-turbulent magnetic field produces different effects on the tails and footpoints of filaments. In the tails turbulent magnetic fields lead to velocity dispersions up to $\sim 60\,\%$ (in the weak-field case) and $\sim 80\,\%$ (in the strong-field case) higher than in model Tur-Sub-0; as well as to mean vorticity enhancements up to $\sim 100\,\%$ (in the weak-field case) and $\sim 180\,\%$ (in the strong-field case) greater than in model Tur-Sub-0. In the footpoints, on the other hand, there is no clear trend with increasing magnetic field strength. In the weak-field case both the velocity dispersions and vorticity enhancements are higher than in model Tur-Sub-0, whilst in the strong-field case both diagnostics decrease. We attribute this behaviour to the shielding effects that strong magnetic fields have on the footpoint of this filament, via suppression of gas mixing (see a discussion on mixing fractions in Appendix \ref{sec:Appendix6}).\par

The third largest effect on the aforementioned diagnostics is produced by turbulent density distributions as they allow a faster percolation of wind-driven shocks through the porous medium of filaments. By comparing model Uni-0-0 with any of the turbulent models in Figure \ref{Figure9}, we find that the inclusion of turbulent density profiles in the initial conditions of clouds produces a rapid development of vortical motions inside the clouds, which result in the formation of highly turbulent, less confined filaments downstream (e.g., compare the curves of models Uni-0-0 and Tur-0-0 on the left-hand side panels for the tails and on the right-hand side panels for the footpoints of filaments in Figure \ref{Figure9}). Panels B1 and B2 show that the turbulent cloud of model Tur-0-0 produces a filament with tail elongations $\sim 20\,\%$ and footpoint elongations $\sim 70\,\%$ greater than in model Uni-0-0, whose cloud only expands after its break-up at $t/t_{\rm cc}=1.0$. In addition, Panels C1 and C2 show that the turbulent cloud of model Tur-0-0 exhibits velocity dispersions $\sim100\,\%$ larger than their uniform counterpart (model Uni-0-0) in both the tail and footpoint of the filament throughout the simulation. The values of all models only approach each other at the end of the simulation, when the abrupt break-up of the uniform cloud leads to a more turbulent velocity distribution in this model. Panels D1 and D2 show a similar behaviour. Mean vorticity enhancements in the turbulent cloud of model Tur-0-0 are twice as high as in the uniform case, Uni-0-0, during most of the evolution in both tail and footpoint material.\par

\subsubsection{On the energy densities of the filaments}
\label{subsubsec:EnergyDensities}
\begin{figure*}
\begin{center}
\begin{tabular}{c c}
  	\textbf{Filament Tail (Cloud Envelope)} & \textbf{Filament Footpoint (Cloud Core)}\\
	\resizebox{80mm}{!}{\includegraphics{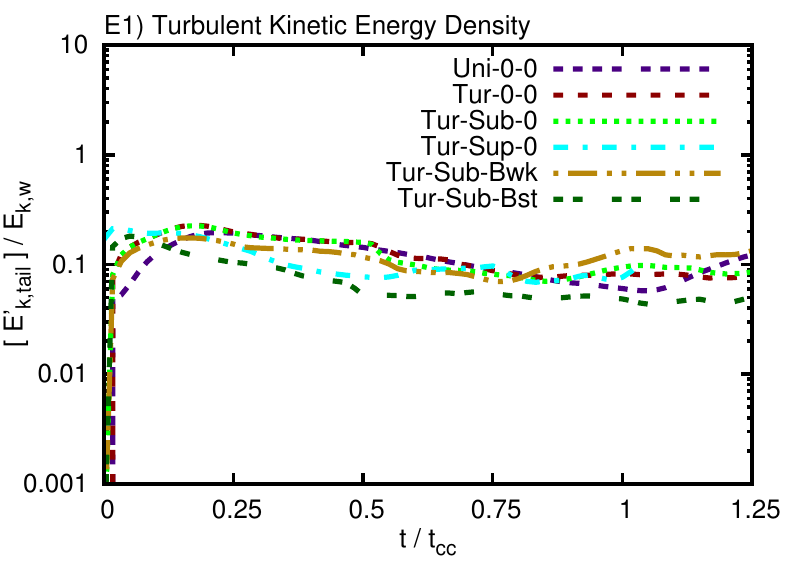}} & \resizebox{80mm}{!}{\includegraphics{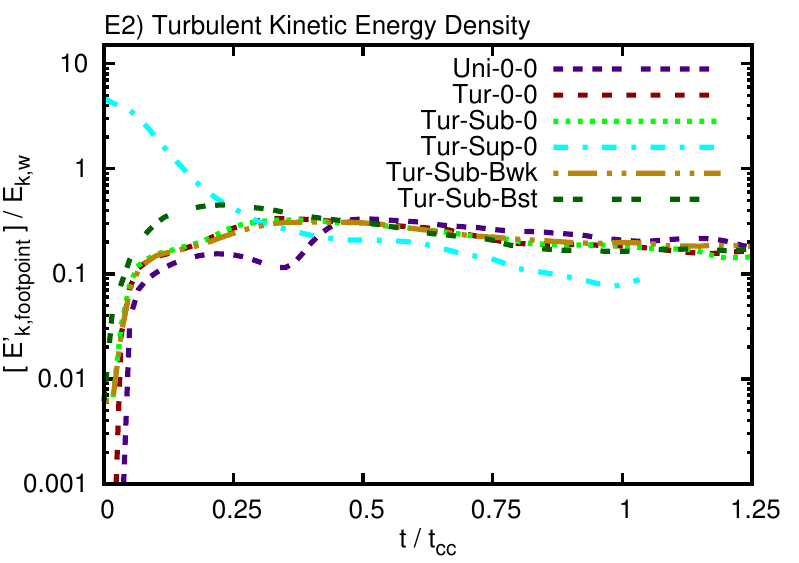}}\\
	\resizebox{80mm}{!}{\includegraphics{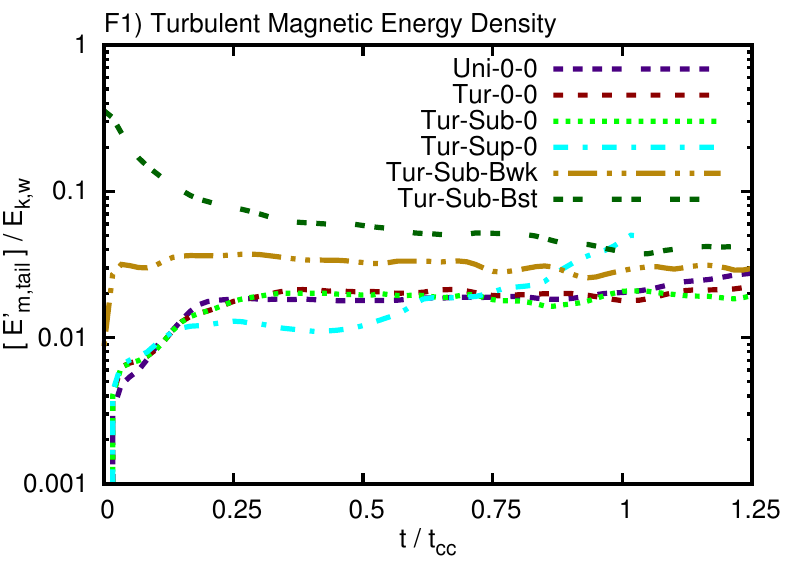}} & \resizebox{80mm}{!}{\includegraphics{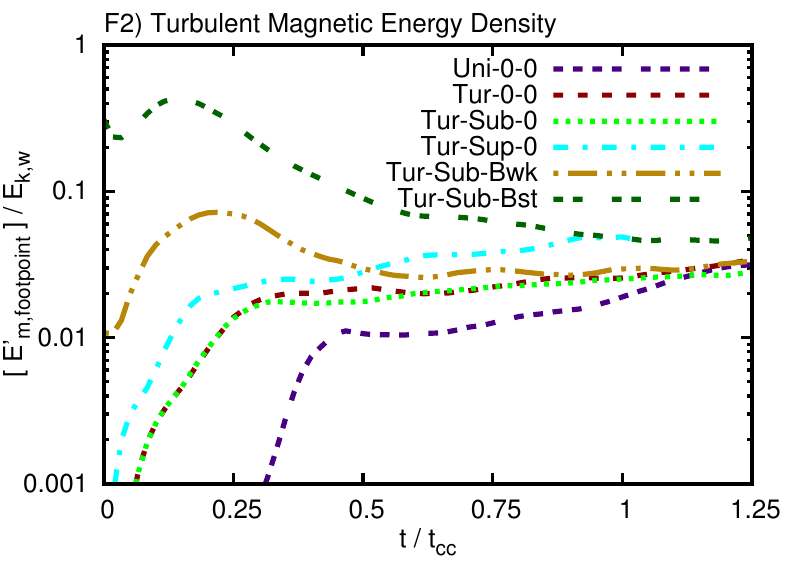}}\\
	\resizebox{80mm}{!}{\includegraphics{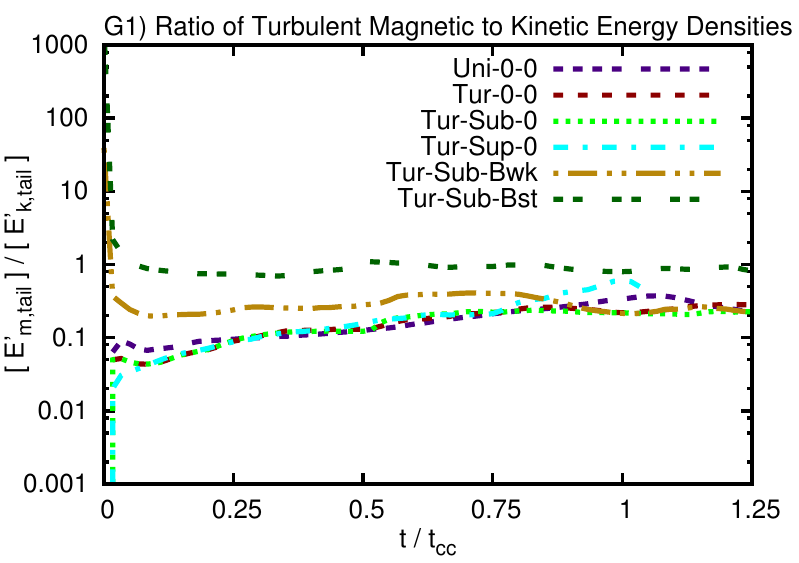}} & \resizebox{80mm}{!}{\includegraphics{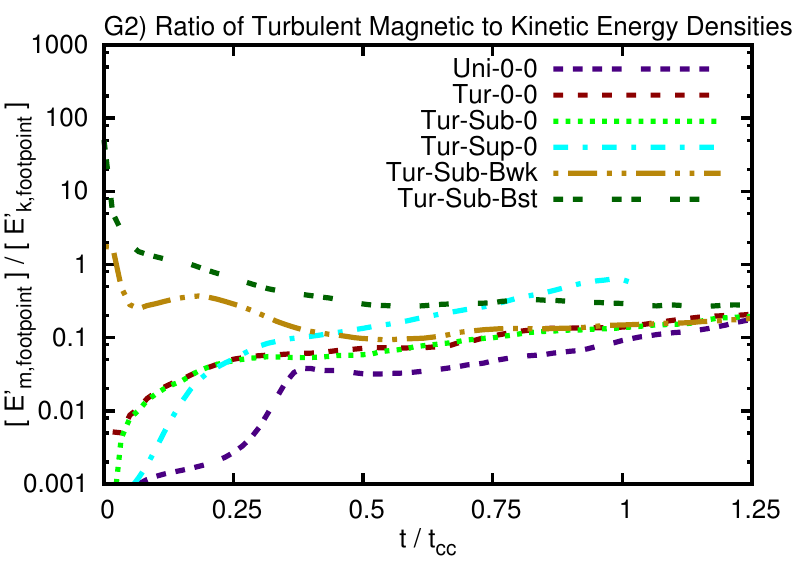}}\\
\end{tabular}
\caption{Same as Figure \ref{Figure9}, but here Panels E1 and E2 show the turbulent kinetic energy density, Panels F1 and F2 show the turbulent magnetic energy density, and Panels G1 and G2 show the evolution of the ratio of turbulent magnetic to turbulent kinetic energy densities, in the tails (left-hand side column) and footpoints (right-hand side column) of filaments. We find that: a) regardless of the model, as the simulations progress, the ratio of turbulent magnetic to turbulent kinetic energy densities becomes constant in all models with $[~E'_{{\rm m},\alpha}~]/[~E'_{{\rm k},\alpha}~]\sim 0.25-0.75$, confirming sub-equipartition, and b) strong magnetic fields produce magnetically-confined filaments.}
\label{Figure10}
\end{center}
\end{figure*}

Here we examine how the different contributors to the total energy density in filaments evolve in different models. Figure \ref{Figure10} shows the average energy densities, normalised with respect to the kinetic energy density of the wind, $E_{\rm k,w}$, in tail (left-hand side column) and footpoint (right-hand side column) material separately.\par

Panels E1 and E2 of Figure \ref{Figure10} indicate that the turbulent kinetic energy density in the tails and footpoints of turbulent filaments rises more rapidly than that of the uniform model. After the rapid increase observed in all models, the curves of the turbulent kinetic energy density flatten for $t/t_{\rm cc}\geq0.2$ and steadily decrease to reach $[~E'_{{\rm k,tail}}~]/E_{\rm k,w}\sim0.1$ and $[~E'_{{\rm k,footpoint}}~]/E_{\rm k,w}\sim0.2$ of the initial wind kinetic energy density in tail and footpoint material, respectively. Even when a supersonically-turbulent velocity field is incorporated (see model Tur-Sup-0), the early-stage dissipation of turbulence into thermal energy that this cloud experiences leads to kinetic energy densities $[~E'_{{\rm k,tail}}~]/E_{\rm k,w}\sim[~E'_{{\rm k,footpoint}}~]/E_{\rm k,w}\sim0.1$ in both the tail and footpoint of the filament. Panels E1 and E2 of Figure \ref{Figure10} also show that the inclusion of a strong, turbulent magnetic field in the initial conditions (see model Tur-Sub-Bst) leads to the formation of a more laminar tail with lower kinetic energy densities, owing to the higher densities produced by magnetic confinement of cloud/filament gas. This is in agreement with our qualitative analysis reported in Sections \ref{subsubsec:UniformvsTurbulent} and \ref{subsubsec:Morphology}, in which we showed that strong magnetic fields produce a denser, more strongly-magnetised, and less mixed filamentary tail than models with null or weak turbulent magnetic fields.\par

Panels F1 and F2 of Figure \ref{Figure10} show that the turbulent magnetic energy density in both tail and footpoint material converges as time progresses in all models, regardless of the initial conditions. Models that are started without turbulent magnetic fields (i.e., models Uni-0-0, Tur-0-0, Tur-Sub-0, and Tur-Sup-0) develop turbulent magnetic field components on short time-scales ($t/t_{\rm cc}<0.1$) as a result of vortical motions rapidly arising in the interior of the cloud via wind-driven shock heating and dynamical instabilities. Except for the delay observed in model Uni-0-0 in Panel F2, the curves of models Uni-0-0, Tur-0-0, Tur-Sub-0, and Tur-Sup-0 evolve similarly to converge to $[~E'_{{\rm m,tail}}~]/E_{\rm k,w}\sim [~E'_{{\rm m,footpoint}}~]/E_{\rm k,w}\sim0.02-0.03$ of the initial wind kinetic energy density in both tail and footpoint material. On the other hand, models that are started with turbulent magnetic fields (Tur-Sub-Bwk and Tur-Sub-Bst) show evidence of turbulence dissipation. In the case of the tails (see Panel F1), dissipation of turbulent magnetic energy into thermal energy occurs from the very beginning of the simulations, while in the case of the footpoints (see Panel F2), dissipation is delayed by the initial compression of field lines in the cores of these clouds, leading to a transient enhancement of their turbulent magnetic energy density. At the end of the simulation, however, the magnetic energy densities of the footpoints in models Tur-Sup-Bwk and Tur-Sup-Bst also approach $[~E'_{{\rm m,tail}}~]/E_{\rm k,w}\sim [~E'_{{\rm m,footpoint}}~]/E_{\rm k,w}\sim0.03-0.04$. Overall, Panels F1 and F2 of Figure \ref{Figure10} show that weakly-magnetised filaments reach magnetic saturation as a result of the turbulent twisting, stretching, and folding of the magnetic field lines being stopped by large Lorentz forces, while in strongly-magnetised filaments these forces are already large at the beginning of the interaction, so they lead to dissipation of magnetic energy into thermal energy and consequently to cloud expansion.\par

We also find that the effects of turbulent magnetic fields on the formation and evolution of filaments are more dependent on the strength than on the topology of the initial, turbulent magnetic fields. In \cite{2016PhDT.......154B} we showed that the presence of a very weak, turbulent magnetic field (with $[~\beta_{\rm tu}~]=100$) inside the cloud has little impact on the evolution of the turbulent energy densities in the resulting filament, and here we show that systematically increasing the strength of the turbulent magnetic field to $[~\beta_{\rm tu}~]=4$ first (in model Tur-Sub-Bwk) and then to $[~\beta_{\rm tu}~]=0.04$ (in model Tur-Sub-Bst) with the same topology leads to increasing shielding effects on the cloud and higher turbulent energy densities. The stronger the initially-turbulent magnetic field, the longer its dissipation time-scale and the higher its impact on the dynamics and energetics of the clouds and filaments. For example, in core material (Panel F2) we find that the turbulent magnetic energy densities of models Tur-0-0, Tur-Sub-0, and Tur-Sub-Bwk converge to $[~E'_{{\rm m,footpoint}}~]/E_{\rm k,w}\sim0.03$ at $t/t_{\rm cc}\sim0.4$. This is much earlier than the time it takes for the turbulent magnetic energy to dissipate in model Tur-Sub-Bst, which only approaches the magnetic energy densities of models Tur-0-0 and Tur-Sub-0 at the end of the interaction (i.e. for $t/t_{\rm cc}\geq1$). Thus, our models indicate that the tangling of the initially-turbulent magnetic field lines in the cloud is insufficient to modify the kinetic and magnetic properties of the filament by itself. The fast growth of vortical motions in models with solely the oblique, uniform magnetic field (e.g., models Tur-0-0 and Tur-Sub-0), rapidly leads to the formation of a turbulent field with similar energy densities to the magnetic field prescribed for model Tur-Sub-Bwk after $t/t_{\rm cc}>0.4$. The evolution of model Tur-Sub-Bst shows, on the other hand, that if the initial turbulent magnetic field is also strong, the additional magnetic pressure provided to the cloud produces a filament with less turbulent motions and higher magnetic fluctuations (strongly magnetised sub-filaments and knots) than its counterparts.\par

\begin{figure*}
\begin{center}
\begin{tabular}{c c}
  	\textbf{Filament Tail (Cloud Envelope)} & \textbf{Filament Footpoint (Cloud Core)}\\
	\resizebox{80mm}{!}{\includegraphics{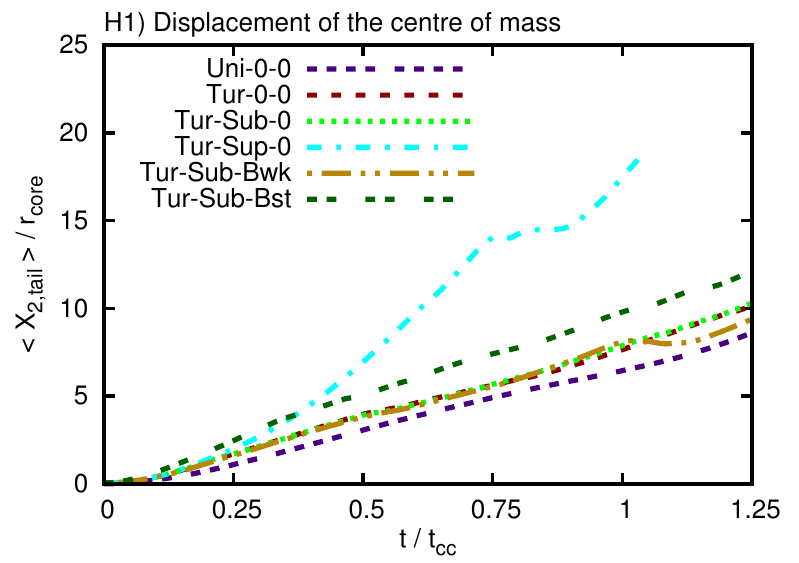}} & \resizebox{80mm}{!}{\includegraphics{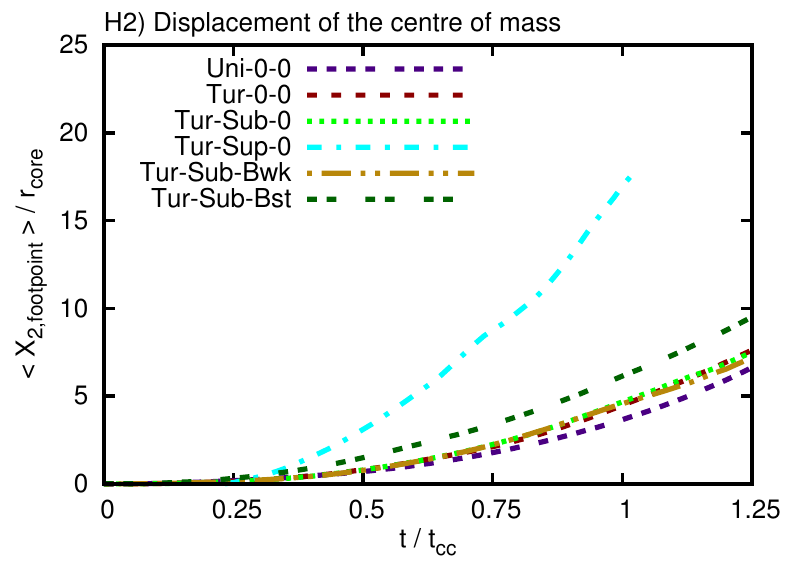}}\\
	\resizebox{80mm}{!}{\includegraphics{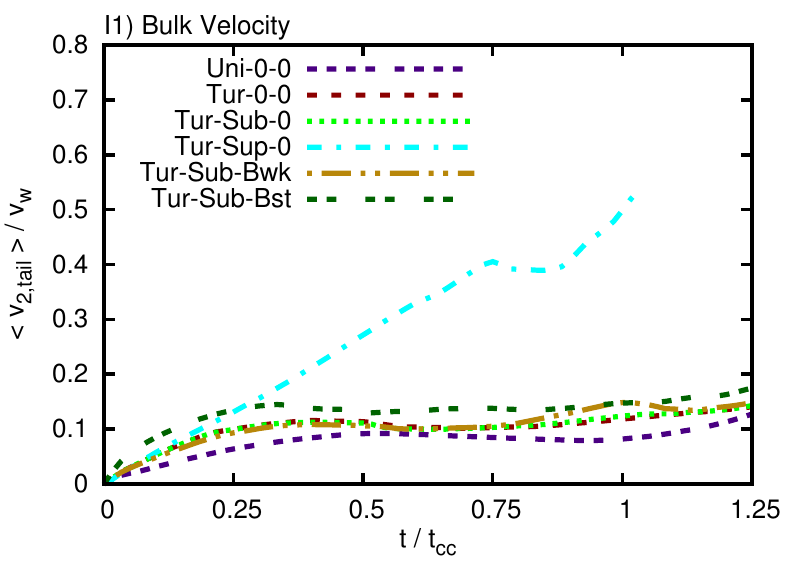}} & \resizebox{80mm}{!}{\includegraphics{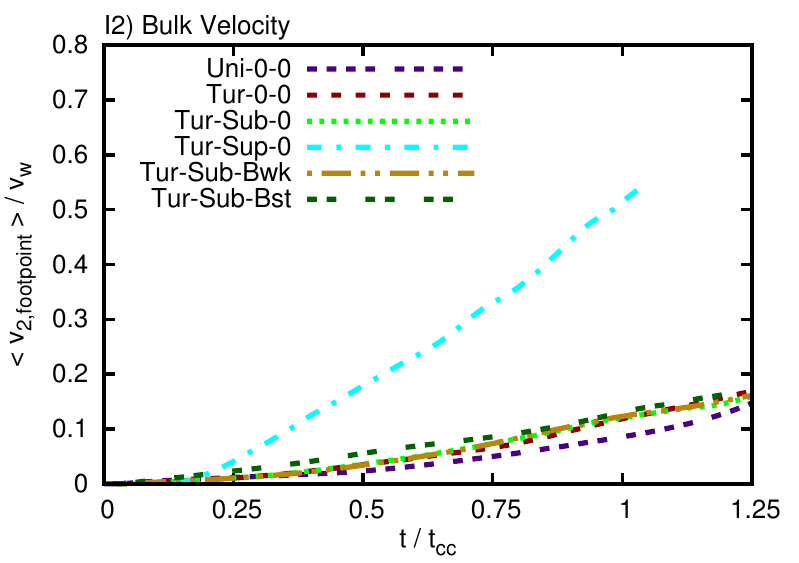}}\\
\end{tabular}
\caption{Same as Figures \ref{Figure9} and \ref{Figure10}, but here Panels H1 and H2 show the displacement of the centre of mass and Panels I1 and I2 show the bulk velocity (in the direction of streaming) of the tails (left-hand side column) and footpoints (right-hand side column) of wind-swept clouds (filaments) entrained in the wind. We find that turbulence and magnetic fields alter the dynamics of filaments. Supersonic velocity fields produce the largest effect, followed by strong magnetic fields, and then by turbulent density distributions, owing to enhanced cross sectional areas in all cases.}
\label{Figure11}
\end{center}
\end{figure*}

Based on the above results, we investigate now the evolution of the ratio of turbulent magnetic to turbulent kinetic energy densities in both the tails and footpoints of filaments (see Panels G1 and G2 of Figure \ref{Figure10}, respectively). This is an important quantity in observations of magnetic clouds and filaments in the ISM, where the kinetic properties of these structures are measured from e.g., emission/absorption line profiles, but the strength of the total magnetic field is unknown or poorly constrained. Overall, Panels G1 and G2 show similar trends for both tails and filaments, with ratios of turbulent magnetic to turbulent kinetic energy densities converging to values $[~E'_{{\rm m},\alpha}~]/[~E'_{{\rm k},\alpha}~]\sim 0.25-0.75$ in all models after $t/t_{\rm cc}=0.4$. This is similar to the range found in our models with the M configuration, i.e., $[~E'_{{\rm m},\alpha}~]/[~E'_{{\rm k},\alpha}~]\sim 0.1-0.4$ (see Section \ref{subsubsec:Properties}). Models with supersonically-turbulent velocity and/or strong, turbulent magnetic fields favour the upper limits of these ranges, while models with uniform clouds favour the lower limits. The overall result, however, indicates that wind-swept clouds always evolve into stages at which the turbulent magnetic energy density is in sub-equipartition with respect to the turbulent kinetic energy density, regardless of the initial conditions.\par

\subsubsection{On the cloud/filament dynamics}
\label{subsubsec:BulkSpeedandDistance}

As mentioned above, turbulent clouds are more easily expanded by turbulence dissipation and shock heating than uniform clouds. In Section \ref{subsubsec:Entrainment} we showed that this causes turbulent clouds to accelerate more rapidly than uniform clouds, owing to their larger cross sectional areas. Here we complement our previous conclusions by investigating the role of different turbulent densities, velocities, and magnetic fields on the dynamics of clouds, separately. Figure \ref{Figure11} shows the displacement of the centre of mass (Panels H1 and H2) and the bulk velocity (Panel I1 and I2) of tails (left-hand side column) and footpoints (right-hand side column) in different models as a function of time. These panels indicate that the inclusion of turbulence radically changes our expectations on the dynamics of wind-swept clouds. In agreement with our results presented in Section \ref{subsubsec:AspectRatio}, we find here that supersonically-turbulent velocity fields produce the largest effect on the displacement of the centre of mass and the bulk speed of clouds/filaments. The fast expansion experienced by this cloud leads to an increased cross sectional area and consequently to an enhanced drag force, which accelerates the cloud to higher velocities and allows it reach farther distances than in other models. Both the tail and footpoint of the filament in model Tur-Sup-0 travel distances three times larger than in model Uni-0-0 and $50\,\%$ larger than in model Tur-Sub-Bst (see Panels H1 and H2), and acquire bulk speeds $5$ times greater than in any other model.\par

The second largest effect on the parameters of Figure \ref{Figure11} is produced by strong, turbulent magnetic fields as they also have the ability to expand the cross sectional area of the cloud (see Section \ref{subsubsec:AspectRatio}). In model Tur-Sub-Bst both the tail and footpoint of the filament travel distances $30\,\%$ times larger than in model Uni-0-0 and $20\,\%$ larger than in model Tur-0-0 (see Panels H1 and H2), and acquire bulk speeds $50-70\,\%$ times greater than in model Uni-0-0 and $10-20\,\%$ greater than in model Tur-0-0 (see Panels I1 and I2). The third largest effect on the dynamics of filaments is produced by turbulent density distributions, which allow wind-driven shocks to move more easily through cloud material, thus expanding it more than in the uniform cloud model. In model Tur-0-0 both the tail and footpoint of the filament travel distances $30\,\%$ times larger than in model Uni-0-0 (see Panels H1 and H2), and acquire bulk speeds $60\,\%$ times greater than in model Uni-0-0 (see Panels I1 and I2). Note also that the inclusion of a subsonically-turbulent velocity field (in model Tur-Sub-0) and of a weak, turbulent magnetic field (in model Tur-Sub-Bwk) does not produce significant effects on the filament dynamics.\par

Overall, the above results and the ones presented in previous sections show that self-consistently adding turbulence to wind-cloud systems enhances cloud acceleration without hastening its disruption. Since the main driver of the cloud dynamics is the supersonic wind, the turbulence-driven increase of the cross sectional area in turbulent cloud models results in a higher effective drag force on these clouds. Thus, in turbulent models the wind is able to transport cold, dense clouds/filaments over distances at least twice as large as the distance travelled by uniform clouds. These entrained structures also travel at least $>50\,\%$ faster than uniform clouds.\par

\section{Caveats and Limitations}
\label{sec:Caveats}
An important note regarding the self-consistency of the turbulent models presented in this paper is that in all models (but one, i.e., model TUR-SUP-BST-ISO, which is fully self-consistent) we take the original distributions of density, velocity, and magnetic fields from \cite{2012ApJ...761..156F}, but 1) we use an adiabatic index of $\gamma=\frac{5}{3}$ (as opposed to an isothermal index; see model TUR-SUP-BST), and 2) we re-scale the mean values of the distributions to pre-defined target values (see models Tur-0-0, Tur-Sub-0, Tur-Sup-0, Tur-Sub-Bwk, and Tur-Sub-Bst), specifically chosen so that we can compare them with the results presented in \citetalias{2016MNRAS.455.1309B} and conduct a systematic study over different turbulence parameters. Hence, the adiabatic turbulent models described here do preserve the original distribution function of these fields, but they are only partially self-consistent (e.g., scaling the turbulent velocity field to a different target Mach number would result in changes in the density and magnetic field distributions as well, and this is not accounted for in the adiabatic models presented here). Despite this, the turbulent simulations presented in this paper: 1) are more realistic than any previous model used to describe wind-cloud systems in the ISM, 2) are designed so that they can be used to study a wide parameter space and analyse how different turbulence parameters for the cloud affect the dynamics and morphology of filaments, and 3) form the basis for more sophisticated models (currently in preparation) of fully-radiative, self-consistent turbulent clouds being swept up by supersonic winds.\par

\section{Summary and concluding remarks}
\label{sec:Summary}
We have presented a detailed numerical study of the formation of filamentary structures arising from the interplay between supersonic winds and turbulent clouds in the ISM. We have expanded our previous work (see \citetalias{2016MNRAS.455.1309B}) by incorporating clouds with turbulent density, velocity, and magnetic fields. The aim of this work is to investigate how the inclusion of turbulence affects the formation, morphology, and dynamics of filaments, and in particular how the strength and topology of the magnetic field in and around wind-swept clouds changes when the magnetic field in the cloud is self-contained and turbulent. We summarise the main conclusions of our study below:

\begin{enumerate}
\item Our results show that the mechanism by which turbulent clouds are disrupted to form filaments is a universal process. Filaments are composed of two substructures, namely tails and footpoints, which evolve in a similar fashion in both uniform and turbulent cloud models. The evolution of wind-swept clouds involves four phases: 1) A tail formation phase in which material, stripped from the sides and the interior of the cloud, is transported to the rear side of the cloud to form an elongated tail, 2) A tail erosion phase in which KH instabilities at the wind-filament interface continuously re-shape the morphology of the tail, 3) A footpoint dispersion phase characterised by dense regions in the cloud being disrupted by RT unstable modes; and 4) A free-floating phase in which sub-filaments and cloudlets become entrained in the wind. Movies showing the full-time evolution of the models presented in Figures \ref{Figure2}, \ref{Figure3}, and \ref{Figure4} are available online\footnote{\url{https://goo.gl/iXgJYk}}.
\item The inclusion of turbulence in the initial conditions for the clouds produces several effects on the morphology, energetics, and dynamics of the resulting filaments. Turbulent clouds result in the formation of filaments with lower aspect (length-to-width) ratios, larger lateral widths ($\iota_{1,\rm{filament}}/\iota_{1,\rm{filament},0}\sim 5$), higher transverse velocity dispersions ($\delta_{{\rm v}_{\rm filament}}/v_{\rm w}\sim0.06$), and higher vorticity enhancements than those arising from uniform clouds. The evolution of the turbulent energy densities also differs in turbulent and uniform cloud models. While turbulent clouds undergo a short period ($t/t_{\rm cc}<0.2$) of dissipation of their turbulent energy into thermal energy, a uniform cloud experiences the opposite effect, rapidly becoming turbulent over similar time-scales (i.e., for $t/t_{\rm cc}<0.2$). After this time, both the kinetic and magnetic energy densities in all models saturate until the end of the evolution and maintain a fixed ratio of $[~E'_{\rm m,filament}~]/[~E'_{\rm k,filament}~]\sim 0.1-0.4$. This indicates that the magnetic energy density in wind-swept clouds is in sub-equipartition with respect to the turbulent kinetic energy density. The near universality of $[~E'_{\rm m}~]/[~E'_{\rm k}~]$ found here may be used to infer magnetic field strengths from measuring the line widths of wind-swept clouds and filaments in observations.
\item Regarding the dynamics of wind-swept clouds, we have shown that the self-consistent inclusion of turbulent density, velocity, and magnetic fields in the initial conditions for clouds produces significant effects on the displacement of the centre of mass and the bulk velocity (in the streaming direction) of the filaments. Turbulence aids cloud expansion and effectively increases the cross sectional area upon which the wind ram-pressure force acts. The enhanced drag force accelerates turbulent clouds more than in uniform cloud models, allowing them to travel distances of $\langle~X_{\rm 2,footpoint}~\rangle/r_{\rm core}\sim15-20$ and reach bulk speeds of $\langle~v_{\rm 2,footpoint}~\rangle/v_{\rm w}\sim0.3-0.4$ by $t/t_{\rm cc}=1.0$, which are both three times larger than the distances and bulk speeds acquired by uniform clouds over the same time-scales.
\item Considering the systematic inclusion of turbulent density, velocity, and magnetic fields, we find that a turbulent density profile allows the wind-driven shocks to propagate faster through low-density regions in the cloud, thus producing a rapid development of vortical motions. This results in the formation of more turbulent and less confined filaments than in uniform modes. Filamentary tails in turbulent models consist of a collection of knots and sub-filaments, rather than of a single structure as in uniform models. The inclusion of turbulent velocity fields in the clouds has varying effects on the global evolution of filaments. When a subsonically-turbulent velocity field is considered, we find no significant effect on the evolution compared to models with null velocity fields. On the other hand, when a supersonically-turbulent velocity field is considered, the cloud undergoes a fast expansion phase, as a result of the turbulence-crossing time-scale being lower than the dynamical, cloud-crushing time-scale of the system.
\item The role of a turbulent magnetic field on the morphology and dynamics of wind-swept clouds highly depends on its initial strength. The stronger the initial magnetic field, the greater the suppression of KH instabilities at wind-cloud interfaces and the enhancement of RT instabilities at the front of the cloud. The presence of turbulent magnetic fields also triggers cloud expansion (thus aiding cloud acceleration) and enhances the internal vorticity of the filamentary tails (thus aiding sub-filamentation), but it also keeps the cloud protected from disruption. Turbulent magnetic fields shield the cloud from the disruptive effects of dynamical instabilities, which prevents the fast stripping of cloud material and reduces the mixing of wind and cloud gas. The ability of cloud gas to radiate its energy away and keep itself cool also produces shielding effects on the resulting filament, as our fully self-consistent model with a softer (nearly isothermal) equation of state shows. In this case the cloud remains dense and keeps its turbulent magnetic energy density high at all times. The resulting high density contrast and the strong magnetic tension at wind-cloud interfaces suppress KH instabilities, enhance flow laminarity, and aid cloud survival in the supersonic wind.
\item Our simulations reveal that filaments produced in ISM wind-cloud interactions are expected to have similar magnetic field strengths as their progenitor clouds. We have shown that the ratio of magnetic energy density in the filament to the initial magnetic energy density in the cloud remains constant $\sim 1$ over the entire evolution of models that include self-consistently strong, turbulent magnetic fields. The regions with the highest magnetic energy densities in turbulent models are the knots and sub-filaments (i.e., the anisotropies) along their tails. The fact that filaments arising from turbulent clouds with strong magnetic fields remain highly magnetised as the simulations progress provides a numerical basis for observations of ISM filaments with similar strengths as the clouds that potentially originated them.
\item Overall, we conclude that introducing turbulence in a self-consistent manner is crucial to understanding entrainment of high-density, cold gas in low-density, hot winds and outflows. Our simulations show that both the porosity of the turbulent density field and its corresponding turbulent velocity field enhance cloud acceleration via dissipation of supersonic turbulence, while at the same time the strong, turbulent magnetic field prevents cloud ablation by shielding the cloud from dynamical shredding. In other words, turbulent clouds produce filaments with anisotropic substructures that travel faster and reach larger distances than uniform clouds, without being fully disrupted in the process. This is particularly important for explaining the presence of high-latitude dense gas embedded in hot galactic outflows as our simulations show that wind-swept clouds are able to survive disruption aided by turbulence, magnetic fields, and radiative cooling much longer than suggested by previous models.
\item Finally, we determine the required numerical resolution and domain size needed to obtain converged results (see Appendices \ref{sec:Appendix3} and \ref{sec:Appendix4}). We find that our chosen resolutions of $R_{64}$ and $R_{128}$ for our computational domain configurations M and S, respectively, adequately capture the evolution of wind-swept uniform and turbulent clouds for the parameter space explored in this series of papers.
\end{enumerate}

\subsection*{Acknowledgements}
We thank Ross Parkin, Cornelia Lang, Naomi McClure-Griffiths, and Helga Den\'{e}s for insightful discussions on the numerical setups and on the potential applications of the simulations presented here, and the anonymous referee for their comments which helped improve this work. WBB thanks the National Secretariat of Higher Education, Science, Technology, and Innovation of Ecuador, SENESCYT, for funding this project through a doctoral scholarship (CI:1711298438).  WBB is the recipient of the Olin Eggen Scholarship at Mount Stromlo Observatory. CF acknowledges funding provided by the Australian Research Council's Discovery Projects (grants DP130102078 and DP150104329). RMC is the recipient of an Australian Research Council Future Fellowship (FT110100108). This research was supported by the National Computational Infrastructure at ANU and the Pawsey Supercomputing Centre, with funding from the Australian Government and the Government of Western Australia, through grants n72 and ek9. CF thanks for high-performance computing resources provided by the Leibniz Rechenzentrum and the Gauss Centre for Supercomputing (grants pr32lo, pr48pi and GCS Large-scale project 10391), and the Partnership for Advanced Computing in Europe (PRACE grant pr89mu). The 3D plots and movies reported in this series of papers were generated using the VisIt visualisation software \citep{HPV:VisIt}.





\bibliography{paper2.bib}

\begin{thebibliography}{}
\makeatletter
\relax
\def\mn@urlcharsother{\let\do\@makeother \do\$\do\&\do\#\do\^\do\_\do\%\do\~}
\def\mn@doi{\begingroup\mn@urlcharsother \@ifnextchar [ {\mn@doi@}
  {\mn@doi@[]}}
\def\mn@doi@[#1]#2{\def\@tempa{#1}\ifx\@tempa\@empty \href
  {http://dx.doi.org/#2} {doi:#2}\else \href {http://dx.doi.org/#2} {#1}\fi
  \endgroup}
\def\mn@eprint#1#2{\mn@eprint@#1:#2::\@nil}
\def\mn@eprint@arXiv#1{\href {http://arxiv.org/abs/#1} {{\tt arXiv:#1}}}
\def\mn@eprint@dblp#1{\href {http://dblp.uni-trier.de/rec/bibtex/#1.xml}
  {dblp:#1}}
\def\mn@eprint@#1:#2:#3:#4\@nil{\def\@tempa {#1}\def\@tempb {#2}\def\@tempc
  {#3}\ifx \@tempc \@empty \let \@tempc \@tempb \let \@tempb \@tempa \fi \ifx
  \@tempb \@empty \def\@tempb {arXiv}\fi \@ifundefined
  {mn@eprint@\@tempb}{\@tempb:\@tempc}{\expandafter \expandafter \csname
  mn@eprint@\@tempb\endcsname \expandafter{\@tempc}}}

\bibitem[\protect\citeauthoryear{{Abramson} \& {Kenney}}{{Abramson} \&
  {Kenney}}{2014}]{2014AJ....147...63A}
{Abramson} A.,  {Kenney} J.~D.~P.,  2014, \mn@doi [\aj]
  {10.1088/0004-6256/147/3/63}, \href
  {http://adsabs.harvard.edu/abs/2014AJ....147...63A} {147, 63}

\bibitem[\protect\citeauthoryear{{Al{\= u}zas}, {Pittard}, {Hartquist}, {Falle}
   \& {Langton}}{{Al{\= u}zas} et~al.}{2012}]{2012MNRAS.425.2212A}
{Al{\= u}zas} R.,  {Pittard} J.~M.,  {Hartquist} T.~W.,  {Falle} S.~A.~E.~G.,
  {Langton} R.,  2012, \mn@doi [\mnras] {10.1111/j.1365-2966.2012.21598.x},
  \href {http://adsabs.harvard.edu/abs/2012MNRAS.425.2212A} {425, 2212}

\bibitem[\protect\citeauthoryear{{Al{\= u}zas}, {Pittard}, {Falle}  \&
  {Hartquist}}{{Al{\= u}zas} et~al.}{2014}]{2014MNRAS.444..971A}
{Al{\= u}zas} R.,  {Pittard} J.~M.,  {Falle} S.~A.~E.~G.,   {Hartquist} T.~W.,
  2014, \mn@doi [\mnras] {10.1093/mnras/stu1501}, \href
  {http://adsabs.harvard.edu/abs/2014MNRAS.444..971A} {444, 971}

\bibitem[\protect\citeauthoryear{{Ballone} et~al.,}{{Ballone}
  et~al.}{2013}]{2013ApJ...776...13B}
{Ballone} A.,  et~al., 2013, \mn@doi [\apj] {10.1088/0004-637X/776/1/13}, \href
  {http://adsabs.harvard.edu/abs/2013ApJ...776...13B} {776, 13}

\bibitem[\protect\citeauthoryear{{Ballone} et~al.,}{{Ballone}
  et~al.}{2016}]{2016ApJ...819L..28B}
{Ballone} A.,  et~al., 2016, \mn@doi [\apjl] {10.3847/2041-8205/819/2/L28},
  \href {http://adsabs.harvard.edu/abs/2016ApJ...819L..28B} {819, L28}

\bibitem[\protect\citeauthoryear{{Banda-Barrag{\'a}n}}{{Banda-Barrag{\'a}n}}{2016}]{2016PhDT.......154B}
{Banda-Barrag{\'a}n} W.~E.,  2016, PhD thesis, Research School of Astronomy and
  Astrophysics, Australian National University
  (\url{http://adsabs.harvard.edu/abs/2016PhDT.......154B})

\bibitem[\protect\citeauthoryear{{Banda-Barrag{\'a}n}, {Parkin}, {Federrath},
  {Crocker}  \& {Bicknell}}{{Banda-Barrag{\'a}n}
  et~al.}{2016}]{2016MNRAS.455.1309B}
{Banda-Barrag{\'a}n} W.~E.,  {Parkin} E.~R.,  {Federrath} C.,  {Crocker} R.~M.,
    {Bicknell} G.~V.,  2016, \mn@doi [\mnras] {10.1093/mnras/stv2405}, \href
  {http://adsabs.harvard.edu/abs/2016MNRAS.455.1309B} {455, 1309}

\bibitem[\protect\citeauthoryear{{Batchelor}}{{Batchelor}}{2000}]{2000ifd..book.....B}
{Batchelor} G.~K.,  2000, {An Introduction to Fluid Dynamics}

\bibitem[\protect\citeauthoryear{{Benedettini} et~al.,}{{Benedettini}
  et~al.}{2015}]{2015MNRAS.453.2036B}
{Benedettini} M.,  et~al., 2015, \mn@doi [\mnras] {10.1093/mnras/stv1750},
  \href {http://adsabs.harvard.edu/abs/2015MNRAS.453.2036B} {453, 2036}

\bibitem[\protect\citeauthoryear{{Bhat}, {Subramanian}  \&
  {Brandenburg}}{{Bhat} et~al.}{2016}]{2016MNRAS.461..240B}
{Bhat} P.,  {Subramanian} K.,   {Brandenburg} A.,  2016, \mn@doi [\mnras]
  {10.1093/mnras/stw1257}, \href
  {http://adsabs.harvard.edu/abs/2016MNRAS.461..240B} {461, 240}

\bibitem[\protect\citeauthoryear{{Bicknell} \& {Li}}{{Bicknell} \&
  {Li}}{2001}]{2001PASA...18..431B}
{Bicknell} G.~V.,  {Li} J.,  2001, \mn@doi [\pasa] {10.1071/AS01058}, \href
  {http://adsabs.harvard.edu/abs/2001PASA...18..431B} {18, 431}

\bibitem[\protect\citeauthoryear{{Bland} \& {Tully}}{{Bland} \&
  {Tully}}{1988}]{1988Natur.334...43B}
{Bland} J.,  {Tully} B.,  1988, \mn@doi [\nat] {10.1038/334043a0}, \href
  {http://adsabs.harvard.edu/abs/1988Natur.334...43B} {334, 43}

\bibitem[\protect\citeauthoryear{{Bland-Hawthorn} \& {Cohen}}{{Bland-Hawthorn}
  \& {Cohen}}{2003}]{2003ApJ...582..246B}
{Bland-Hawthorn} J.,  {Cohen} M.,  2003, \mn@doi [\apj] {10.1086/344573}, \href
  {http://adsabs.harvard.edu/abs/2003ApJ...582..246B} {582, 246}

\bibitem[\protect\citeauthoryear{{Brandenburg} \& {Subramanian}}{{Brandenburg}
  \& {Subramanian}}{2005}]{2005PhR...417....1B}
{Brandenburg} A.,  {Subramanian} K.,  2005, \mn@doi [\physrep]
  {10.1016/j.physrep.2005.06.005}, \href
  {http://adsabs.harvard.edu/abs/2005PhR...417....1B} {417, 1}

\bibitem[\protect\citeauthoryear{{Br{\"u}ggen} \& {Scannapieco}}{{Br{\"u}ggen}
  \& {Scannapieco}}{2016}]{2016ApJ...822...31B}
{Br{\"u}ggen} M.,  {Scannapieco} E.,  2016, \mn@doi [\apj]
  {10.3847/0004-637X/822/1/31}, \href
  {http://adsabs.harvard.edu/abs/2016ApJ...822...31B} {822, 31}

\bibitem[\protect\citeauthoryear{{Brunt}, {Heyer}  \& {Mac Low}}{{Brunt}
  et~al.}{2009}]{2009A&A...504..883B}
{Brunt} C.~M.,  {Heyer} M.~H.,   {Mac Low} M.-M.,  2009, \mn@doi [\aap]
  {10.1051/0004-6361/200911797}, \href
  {http://adsabs.harvard.edu/abs/2009A%26A...504..883B} {504, 883}

\bibitem[\protect\citeauthoryear{{Burkert}, {Schartmann}, {Alig}, {Gillessen},
  {Genzel}, {Fritz}  \& {Eisenhauer}}{{Burkert}
  et~al.}{2012}]{2012ApJ...750...58B}
{Burkert} A.,  {Schartmann} M.,  {Alig} C.,  {Gillessen} S.,  {Genzel} R.,
  {Fritz} T.~K.,   {Eisenhauer} F.,  2012, \mn@doi [\apj]
  {10.1088/0004-637X/750/1/58}, \href
  {http://adsabs.harvard.edu/abs/2012ApJ...750...58B} {750, 58}

\bibitem[\protect\citeauthoryear{{Calder{\'o}n}, {Ballone}, {Cuadra},
  {Schartmann}, {Burkert}  \& {Gillessen}}{{Calder{\'o}n}
  et~al.}{2016}]{2015arXiv150707012C}
{Calder{\'o}n} D.,  {Ballone} A.,  {Cuadra} J.,  {Schartmann} M.,  {Burkert}
  A.,   {Gillessen} S.,  2016, \mn@doi [\mnras] {10.1093/mnras/stv2644}, \href
  {http://adsabs.harvard.edu/abs/2016MNRAS.455.4388C} {455, 4388}

\bibitem[\protect\citeauthoryear{{Carlqvist}, {Gahm}  \& {Kristen}}{{Carlqvist}
  et~al.}{2003}]{2003A&A...403..399C}
{Carlqvist} P.,  {Gahm} G.~F.,   {Kristen} H.,  2003, \mn@doi [\aap]
  {10.1051/0004-6361:20030365}, \href
  {http://adsabs.harvard.edu/abs/2003A%26A...403..399C} {403, 399}

\bibitem[\protect\citeauthoryear{{Carral}, {Hollenbach}, {Lord}, {Colgan},
  {Haas}, {Rubin}  \& {Erickson}}{{Carral} et~al.}{1994}]{1994ApJ...423..223C}
{Carral} P.,  {Hollenbach} D.~J.,  {Lord} S.~D.,  {Colgan} S.~W.~J.,  {Haas}
  M.~R.,  {Rubin} R.~H.,   {Erickson} E.~F.,  1994, \mn@doi [\apj]
  {10.1086/173801}, \href {http://adsabs.harvard.edu/abs/1994ApJ...423..223C}
  {423, 223}

\bibitem[\protect\citeauthoryear{{Cecil}, {Bland-Hawthorn}  \&
  {Veilleux}}{{Cecil} et~al.}{2002}]{2002ApJ...576..745C}
{Cecil} G.,  {Bland-Hawthorn} J.,   {Veilleux} S.,  2002, \mn@doi [\apj]
  {10.1086/341861}, \href {http://adsabs.harvard.edu/abs/2002ApJ...576..745C}
  {576, 745}

\bibitem[\protect\citeauthoryear{{Chandrasekhar}}{{Chandrasekhar}}{1961}]{1961hhs..book.....C}
{Chandrasekhar} S.,  1961, {Hydrodynamic and hydromagnetic stability}

\bibitem[\protect\citeauthoryear{{Chevalier} \& {Clegg}}{{Chevalier} \&
  {Clegg}}{1985}]{1985Natur.317...44C}
{Chevalier} R.~A.,  {Clegg} A.~W.,  1985, \mn@doi [\nat] {10.1038/317044a0},
  \href {http://adsabs.harvard.edu/abs/1985Natur.317...44C} {317, 44}

\bibitem[\protect\citeauthoryear{Childs et~al.,}{Childs
  et~al.}{2012}]{HPV:VisIt}
Childs H.,  et~al., 2012, in , {High Performance Visualization--Enabling
  Extreme-Scale Scientific Insight}.
pp 357--372

\bibitem[\protect\citeauthoryear{{Chynoweth}, {Langston}, {Yun}, {Lockman},
  {Rubin}  \& {Scoles}}{{Chynoweth} et~al.}{2008}]{2008AJ....135.1983C}
{Chynoweth} K.~M.,  {Langston} G.~I.,  {Yun} M.~S.,  {Lockman} F.~J.,  {Rubin}
  K.~H.~R.,   {Scoles} S.~A.,  2008, \mn@doi [\aj]
  {10.1088/0004-6256/135/6/1983}, \href
  {http://adsabs.harvard.edu/abs/2008AJ....135.1983C} {135, 1983}

\bibitem[\protect\citeauthoryear{{Cooper}, {Bicknell}, {Sutherland}  \&
  {Bland-Hawthorn}}{{Cooper} et~al.}{2008}]{2008ApJ...674..157C}
{Cooper} J.~L.,  {Bicknell} G.~V.,  {Sutherland} R.~S.,   {Bland-Hawthorn} J.,
  2008, \mn@doi [\apj] {10.1086/524918}, \href
  {http://adsabs.harvard.edu/abs/2008ApJ...674..157C} {674, 157}

\bibitem[\protect\citeauthoryear{{Cooper}, {Bicknell}, {Sutherland}  \&
  {Bland-Hawthorn}}{{Cooper} et~al.}{2009}]{2009ApJ...703..330C}
{Cooper} J.~L.,  {Bicknell} G.~V.,  {Sutherland} R.~S.,   {Bland-Hawthorn} J.,
  2009, \mn@doi [\apj] {10.1088/0004-637X/703/1/330}, \href
  {http://adsabs.harvard.edu/abs/2009ApJ...703..330C} {703, 330}

\bibitem[\protect\citeauthoryear{{Cowling}}{{Cowling}}{1976}]{1976magn.book.....C}
{Cowling} T.~G.,  1976, {Magnetohydrodynamics}

\bibitem[\protect\citeauthoryear{{Crawford}, {Hatch}, {Fabian}  \&
  {Sanders}}{{Crawford} et~al.}{2005}]{2005MNRAS.363..216C}
{Crawford} C.~S.,  {Hatch} N.~A.,  {Fabian} A.~C.,   {Sanders} J.~S.,  2005,
  \mn@doi [\mnras] {10.1111/j.1365-2966.2005.09463.x}, \href
  {http://adsabs.harvard.edu/abs/2005MNRAS.363..216C} {363, 216}

\bibitem[\protect\citeauthoryear{{Crutcher}, {Wandelt}, {Heiles}, {Falgarone}
  \& {Troland}}{{Crutcher} et~al.}{2010}]{2010ApJ...725..466C}
{Crutcher} R.~M.,  {Wandelt} B.,  {Heiles} C.,  {Falgarone} E.,   {Troland}
  T.~H.,  2010, \mn@doi [\apj] {10.1088/0004-637X/725/1/466}, \href
  {http://adsabs.harvard.edu/abs/2010ApJ...725..466C} {725, 466}

\bibitem[\protect\citeauthoryear{{Dahlburg}, {Einaudi}, {LaRosa}  \&
  {Shore}}{{Dahlburg} et~al.}{2002}]{2002ApJ...568..220D}
{Dahlburg} R.~B.,  {Einaudi} G.,  {LaRosa} T.~N.,   {Shore} S.~N.,  2002,
  \mn@doi [\apj] {10.1086/338842}, \href
  {http://adsabs.harvard.edu/abs/2002ApJ...568..220D} {568, 220}

\bibitem[\protect\citeauthoryear{{Davidson}}{{Davidson}}{2004}]{2004tise.book.....D}
{Davidson} P.~A.,  2004, {Turbulence : an introduction for scientists and
  engineers}

\bibitem[\protect\citeauthoryear{{Dedner}, {Kemm}, {Kr{\"o}ner}, {Munz},
  {Schnitzer}  \& {Wesenberg}}{{Dedner} et~al.}{2002}]{2002JCoPh.175..645D}
{Dedner} A.,  {Kemm} F.,  {Kr{\"o}ner} D.,  {Munz} C.-D.,  {Schnitzer} T.,
  {Wesenberg} M.,  2002, \mn@doi [Journal of Computational Physics]
  {10.1006/jcph.2001.6961}, \href
  {http://adsabs.harvard.edu/abs/2002JCoPh.175..645D} {175, 645}

\bibitem[\protect\citeauthoryear{{Dgani}, {Walder}  \& {Nussbaumer}}{{Dgani}
  et~al.}{1993}]{1993A&A...267..155D}
{Dgani} R.,  {Walder} R.,   {Nussbaumer} H.,  1993, \aap, \href
  {http://adsabs.harvard.edu/abs/1993A%26A...267..155D} {267, 155}

\bibitem[\protect\citeauthoryear{Drazin}{Drazin}{1970}]{FLM:383106}
Drazin P.~G.,  1970, \mn@doi [Journal of Fluid Mechanics]
  {10.1017/S0022112070001295}, 42, 321

\bibitem[\protect\citeauthoryear{{Drazin}}{{Drazin}}{2002}]{2002ihs..book.....D}
{Drazin} P.~G.,  2002, {Introduction to Hydrodynamic Stability}

\bibitem[\protect\citeauthoryear{{Drazin} \& {Reid}}{{Drazin} \&
  {Reid}}{2004}]{2004hyst.book.....D}
{Drazin} P.~G.,  {Reid} W.~H.,  2004, {Hydrodynamic Stability}

\bibitem[\protect\citeauthoryear{{Elmegreen} \& {Scalo}}{{Elmegreen} \&
  {Scalo}}{2004}]{2004ARA&A..42..211E}
{Elmegreen} B.~G.,  {Scalo} J.,  2004, \mn@doi [\araa]
  {10.1146/annurev.astro.41.011802.094859}, \href
  {http://adsabs.harvard.edu/abs/2004ARA%26A..42..211E} {42, 211}

\bibitem[\protect\citeauthoryear{{Enokiya} et~al.,}{{Enokiya}
  et~al.}{2014}]{2014ApJ...780...72E}
{Enokiya} R.,  et~al., 2014, \mn@doi [\apj] {10.1088/0004-637X/780/1/72}, \href
  {http://adsabs.harvard.edu/abs/2014ApJ...780...72E} {780, 72}

\bibitem[\protect\citeauthoryear{{Federrath}}{{Federrath}}{2013}]{2013MNRAS.436.1245F}
{Federrath} C.,  2013, \mn@doi [\mnras] {10.1093/mnras/stt1644}, \href
  {http://adsabs.harvard.edu/abs/2013MNRAS.436.1245F} {436, 1245}

\bibitem[\protect\citeauthoryear{{Federrath} \& {Banerjee}}{{Federrath} \&
  {Banerjee}}{2015}]{2015MNRAS.448.3297F}
{Federrath} C.,  {Banerjee} S.,  2015, \mn@doi [\mnras] {10.1093/mnras/stv180},
  \href {http://adsabs.harvard.edu/abs/2015MNRAS.448.3297F} {448, 3297}

\bibitem[\protect\citeauthoryear{{Federrath} \& {Klessen}}{{Federrath} \&
  {Klessen}}{2012}]{2012ApJ...761..156F}
{Federrath} C.,  {Klessen} R.~S.,  2012, \mn@doi [\apj]
  {10.1088/0004-637X/761/2/156}, \href
  {http://adsabs.harvard.edu/abs/2012ApJ...761..156F} {761, 156}

\bibitem[\protect\citeauthoryear{{Federrath}, {Klessen}  \&
  {Schmidt}}{{Federrath} et~al.}{2008}]{2008ApJ...688L..79F}
{Federrath} C.,  {Klessen} R.~S.,   {Schmidt} W.,  2008, \mn@doi [\apjl]
  {10.1086/595280}, \href {http://adsabs.harvard.edu/abs/2008ApJ...688L..79F}
  {688, L79}

\bibitem[\protect\citeauthoryear{{Federrath}, {Roman-Duval}, {Klessen},
  {Schmidt}  \& {Mac Low}}{{Federrath} et~al.}{2010}]{2010A&A...512A..81F}
{Federrath} C.,  {Roman-Duval} J.,  {Klessen} R.~S.,  {Schmidt} W.,   {Mac Low}
  M.-M.,  2010, \mn@doi [\aap] {10.1051/0004-6361/200912437}, \href
  {http://adsabs.harvard.edu/abs/2010A%26A...512A..81F} {512, A81}

\bibitem[\protect\citeauthoryear{{Federrath}, {Schober}, {Bovino}  \&
  {Schleicher}}{{Federrath} et~al.}{2014}]{2014ApJ...797L..19F}
{Federrath} C.,  {Schober} J.,  {Bovino} S.,   {Schleicher} D.~R.~G.,  2014,
  \mn@doi [\apjl] {10.1088/2041-8205/797/2/L19}, \href
  {http://adsabs.harvard.edu/abs/2014ApJ...797L..19F} {797, L19}

\bibitem[\protect\citeauthoryear{{Ferri{\`e}re}}{{Ferri{\`e}re}}{2001}]{2001RvMP...73.1031F}
{Ferri{\`e}re} K.~M.,  2001, \mn@doi [Reviews of Modern Physics]
  {10.1103/RevModPhys.73.1031}, \href
  {http://adsabs.harvard.edu/abs/2001RvMP...73.1031F} {73, 1031}

\bibitem[\protect\citeauthoryear{{Ferri{\`e}re}}{{Ferri{\`e}re}}{2007}]{2007EAS....23....3F}
{Ferri{\`e}re} K.,  2007, in {Miville-Desch{\^e}nes} M.-A.,  {Boulanger} F.,
  eds,  EAS Publications Series Vol. 23, EAS Publications Series. pp 3--17,
  \mn@doi{10.1051/eas:2007002}

\bibitem[\protect\citeauthoryear{{Ferri{\`e}re}}{{Ferri{\`e}re}}{2011}]{2011MmSAI..82..824F}
{Ferri{\`e}re} K.,  2011, \memsai, \href
  {http://adsabs.harvard.edu/abs/2011MmSAI..82..824F} {82, 824}

\bibitem[\protect\citeauthoryear{{Field}}{{Field}}{1965}]{1965ApJ...142..531F}
{Field} G.~B.,  1965, \mn@doi [\apj] {10.1086/148317}, \href
  {http://adsabs.harvard.edu/abs/1965ApJ...142..531F} {142, 531}

\bibitem[\protect\citeauthoryear{{Fragile}, {Murray}, {Anninos}  \& {van
  Breugel}}{{Fragile} et~al.}{2004}]{2004ApJ...604...74F}
{Fragile} P.~C.,  {Murray} S.~D.,  {Anninos} P.,   {van Breugel} W.,  2004,
  \mn@doi [\apj] {10.1086/381726}, \href
  {http://adsabs.harvard.edu/abs/2004ApJ...604...74F} {604, 74}

\bibitem[\protect\citeauthoryear{{Fragile}, {Anninos}, {Gustafson}  \&
  {Murray}}{{Fragile} et~al.}{2005}]{2005ApJ...619..327F}
{Fragile} P.~C.,  {Anninos} P.,  {Gustafson} K.,   {Murray} S.~D.,  2005,
  \mn@doi [\apj] {10.1086/426313}, \href
  {http://adsabs.harvard.edu/abs/2005ApJ...619..327F} {619, 327}

\bibitem[\protect\citeauthoryear{{Frank}, {Jones}, {Ryu}  \& {Gaalaas}}{{Frank}
  et~al.}{1996}]{1996ApJ...460..777F}
{Frank} A.,  {Jones} T.~W.,  {Ryu} D.,   {Gaalaas} J.~B.,  1996, \mn@doi [\apj]
  {10.1086/177009}, \href {http://adsabs.harvard.edu/abs/1996ApJ...460..777F}
  {460, 777}

\bibitem[\protect\citeauthoryear{{Gaensler} et~al.,}{{Gaensler}
  et~al.}{2011}]{2011Natur.478..214G}
{Gaensler} B.~M.,  et~al., 2011, \mn@doi [\nat] {10.1038/nature10446}, \href
  {http://adsabs.harvard.edu/abs/2011Natur.478..214G} {478, 214}

\bibitem[\protect\citeauthoryear{{Gardiner} \& {Stone}}{{Gardiner} \&
  {Stone}}{2005}]{2005JCoPh.205..509G}
{Gardiner} T.~A.,  {Stone} J.~M.,  2005, \mn@doi [Journal of Computational
  Physics] {10.1016/j.jcp.2004.11.016}, \href
  {http://adsabs.harvard.edu/abs/2005JCoPh.205..509G} {205, 509}

\bibitem[\protect\citeauthoryear{{Gardiner} \& {Stone}}{{Gardiner} \&
  {Stone}}{2008}]{2008JCoPh.227.4123G}
{Gardiner} T.~A.,  {Stone} J.~M.,  2008, \mn@doi [Journal of Computational
  Physics] {10.1016/j.jcp.2007.12.017}, \href
  {http://adsabs.harvard.edu/abs/2008JCoPh.227.4123G} {227, 4123}

\bibitem[\protect\citeauthoryear{{Gillessen} et~al.,}{{Gillessen}
  et~al.}{2012}]{2012Natur.481...51G}
{Gillessen} S.,  et~al., 2012, \mn@doi [\nat] {10.1038/nature10652}, \href
  {http://adsabs.harvard.edu/abs/2012Natur.481...51G} {481, 51}

\bibitem[\protect\citeauthoryear{{Gillessen} et~al.,}{{Gillessen}
  et~al.}{2013}]{2013ApJ...763...78G}
{Gillessen} S.,  et~al., 2013, \mn@doi [\apj] {10.1088/0004-637X/763/2/78},
  \href {http://adsabs.harvard.edu/abs/2013ApJ...763...78G} {763, 78}

\bibitem[\protect\citeauthoryear{{Goldsmith} \& {Pittard}}{{Goldsmith} \&
  {Pittard}}{2016}]{2016MNRAS.461..578G}
{Goldsmith} K.~J.~A.,  {Pittard} J.~M.,  2016, \mn@doi [\mnras]
  {10.1093/mnras/stw1365}, \href
  {http://adsabs.harvard.edu/abs/2016MNRAS.461..578G} {461, 578}

\bibitem[\protect\citeauthoryear{{Goodson}, {Heitsch}, {Eklund}  \&
  {Williams}}{{Goodson} et~al.}{2017}]{2017MNRAS.468.3184G}
{Goodson} M.~D.,  {Heitsch} F.,  {Eklund} K.,   {Williams} V.~A.,  2017,
  \mn@doi [\mnras] {10.1093/mnras/stx720}, \href
  {http://adsabs.harvard.edu/abs/2017MNRAS.468.3184G} {468, 3184}

\bibitem[\protect\citeauthoryear{{Gregori}, {Miniati}, {Ryu}  \&
  {Jones}}{{Gregori} et~al.}{1999}]{1999ApJ...527L.113G}
{Gregori} G.,  {Miniati} F.,  {Ryu} D.,   {Jones} T.~W.,  1999, \mn@doi [\apjl]
  {10.1086/312402}, \href {http://adsabs.harvard.edu/abs/1999ApJ...527L.113G}
  {527, L113}

\bibitem[\protect\citeauthoryear{{Gregori}, {Miniati}, {Ryu}  \&
  {Jones}}{{Gregori} et~al.}{2000}]{2000ApJ...543..775G}
{Gregori} G.,  {Miniati} F.,  {Ryu} D.,   {Jones} T.~W.,  2000, \mn@doi [\apj]
  {10.1086/317130}, \href {http://adsabs.harvard.edu/abs/2000ApJ...543..775G}
  {543, 775}

\bibitem[\protect\citeauthoryear{{Greve}}{{Greve}}{2004}]{2004A&A...416...67G}
{Greve} A.,  2004, \mn@doi [\aap] {10.1051/0004-6361:20031709}, \href
  {http://adsabs.harvard.edu/abs/2004A%26A...416...67G} {416, 67}

\bibitem[\protect\citeauthoryear{{Heckman} \& {Thompson}}{{Heckman} \&
  {Thompson}}{2017}]{2017arXiv170109062H}
{Heckman} T.~M.,  {Thompson} T.~A.,  2017, preprint, \href
  {http://adsabs.harvard.edu/abs/2017arXiv170109062H} {} (\mn@eprint {arXiv}
  {1701.09062})

\bibitem[\protect\citeauthoryear{{Heiles}}{{Heiles}}{2004}]{2004ASPC..323...79H}
{Heiles} C.,  2004, in {Johnstone} D.,  {Adams} F.~C.,  {Lin} D.~N.~C.,
  {Neufeeld} D.~A.,   {Ostriker} E.~C.,  eds,  Astronomical Society of the
  Pacific Conference Series Vol. 323, Star Formation in the Interstellar
  Medium: In Honor of David Hollenbach. p.~79

\bibitem[\protect\citeauthoryear{{Heiles} \& {Troland}}{{Heiles} \&
  {Troland}}{2003}]{2003ApJ...586.1067H}
{Heiles} C.,  {Troland} T.~H.,  2003, \mn@doi [\apj] {10.1086/367828}, \href
  {http://adsabs.harvard.edu/abs/2003ApJ...586.1067H} {586, 1067}

\bibitem[\protect\citeauthoryear{{Hennebelle} \& {Falgarone}}{{Hennebelle} \&
  {Falgarone}}{2012}]{2012A&ARv..20...55H}
{Hennebelle} P.,  {Falgarone} E.,  2012, \mn@doi [\aapr]
  {10.1007/s00159-012-0055-y}, \href
  {http://adsabs.harvard.edu/abs/2012A%26ARv..20...55H} {20, 55}

\bibitem[\protect\citeauthoryear{{Hester} et~al.,}{{Hester}
  et~al.}{1996}]{1996ApJ...456..225H}
{Hester} J.~J.,  et~al., 1996, \mn@doi [\apj] {10.1086/176643}, \href
  {http://adsabs.harvard.edu/abs/1996ApJ...456..225H} {456, 225}

\bibitem[\protect\citeauthoryear{{Higdon}, {Lingenfelter}  \&
  {Rothschild}}{{Higdon} et~al.}{2009}]{2009ApJ...698..350H}
{Higdon} J.~C.,  {Lingenfelter} R.~E.,   {Rothschild} R.~E.,  2009, \mn@doi
  [\apj] {10.1088/0004-637X/698/1/350}, \href
  {http://adsabs.harvard.edu/abs/2009ApJ...698..350H} {698, 350}

\bibitem[\protect\citeauthoryear{{Hopkins}}{{Hopkins}}{2013}]{2013MNRAS.430.1880H}
{Hopkins} P.~F.,  2013, \mn@doi [\mnras] {10.1093/mnras/stt010}, \href
  {http://adsabs.harvard.edu/abs/2013MNRAS.430.1880H} {430, 1880}

\bibitem[\protect\citeauthoryear{{Inoue} \& {Inutsuka}}{{Inoue} \&
  {Inutsuka}}{2012}]{2012ApJ...759...35I}
{Inoue} T.,  {Inutsuka} S.-i.,  2012, \mn@doi [\apj]
  {10.1088/0004-637X/759/1/35}, \href
  {http://adsabs.harvard.edu/abs/2012ApJ...759...35I} {759, 35}

\bibitem[\protect\citeauthoryear{{Johansson} \& {Ziegler}}{{Johansson} \&
  {Ziegler}}{2013}]{2013ApJ...766...45J}
{Johansson} E.~P.~G.,  {Ziegler} U.,  2013, \mn@doi [\apj]
  {10.1088/0004-637X/766/1/45}, \href
  {http://adsabs.harvard.edu/abs/2013ApJ...766...45J} {766, 45}

\bibitem[\protect\citeauthoryear{{Jones}, {Kang}  \& {Tregillis}}{{Jones}
  et~al.}{1994}]{1994ApJ...432..194J}
{Jones} T.~W.,  {Kang} H.,   {Tregillis} I.~L.,  1994, \mn@doi [\apj]
  {10.1086/174560}, \href {http://adsabs.harvard.edu/abs/1994ApJ...432..194J}
  {432, 194}

\bibitem[\protect\citeauthoryear{{Jones}, {Ryu}  \& {Tregillis}}{{Jones}
  et~al.}{1996}]{1996ApJ...473..365J}
{Jones} T.~W.,  {Ryu} D.,   {Tregillis} I.~L.,  1996, \mn@doi [\apj]
  {10.1086/178151}, \href {http://adsabs.harvard.edu/abs/1996ApJ...473..365J}
  {473, 365}

\bibitem[\protect\citeauthoryear{{Jun}, {Norman}  \& {Stone}}{{Jun}
  et~al.}{1995}]{1995ApJ...453..332J}
{Jun} B.-I.,  {Norman} M.~L.,   {Stone} J.~M.,  1995, \mn@doi [\apj]
  {10.1086/176393}, \href {http://adsabs.harvard.edu/abs/1995ApJ...453..332J}
  {453, 332}

\bibitem[\protect\citeauthoryear{{Kainulainen}, {Beuther}, {Henning}  \&
  {Plume}}{{Kainulainen} et~al.}{2009}]{2009A&A...508L..35K}
{Kainulainen} J.,  {Beuther} H.,  {Henning} T.,   {Plume} R.,  2009, \mn@doi
  [\aap] {10.1051/0004-6361/200913605}, \href
  {http://adsabs.harvard.edu/abs/2009A%26A...508L..35K} {508, L35}

\bibitem[\protect\citeauthoryear{{Kainulainen}, {Federrath}  \&
  {Henning}}{{Kainulainen} et~al.}{2013}]{2013A&A...553L...8K}
{Kainulainen} J.,  {Federrath} C.,   {Henning} T.,  2013, \mn@doi [\aap]
  {10.1051/0004-6361/201321431}, \href
  {http://adsabs.harvard.edu/abs/2013A%26A...553L...8K} {553, L8}

\bibitem[\protect\citeauthoryear{{Kainulainen}, {Federrath}  \&
  {Henning}}{{Kainulainen} et~al.}{2014}]{2014Sci...344..183K}
{Kainulainen} J.,  {Federrath} C.,   {Henning} T.,  2014, \mn@doi [Science]
  {10.1126/science.1248724}, \href
  {http://adsabs.harvard.edu/abs/2014Sci...344..183K} {344, 183}

\bibitem[\protect\citeauthoryear{{Kenney}, {Abramson}  \&
  {Bravo-Alfaro}}{{Kenney} et~al.}{2015}]{2015AJ....150...59K}
{Kenney} J.~D.~P.,  {Abramson} A.,   {Bravo-Alfaro} H.,  2015, \mn@doi [\aj]
  {10.1088/0004-6256/150/2/59}, \href
  {http://adsabs.harvard.edu/abs/2015AJ....150...59K} {150, 59}

\bibitem[\protect\citeauthoryear{{Klein}, {McKee}  \& {Colella}}{{Klein}
  et~al.}{1994}]{1994ApJ...420..213K}
{Klein} R.~I.,  {McKee} C.~F.,   {Colella} P.,  1994, \mn@doi [\apj]
  {10.1086/173554}, \href {http://adsabs.harvard.edu/abs/1994ApJ...420..213K}
  {420, 213}

\bibitem[\protect\citeauthoryear{{Konstandin}, {Girichidis}, {Federrath}  \&
  {Klessen}}{{Konstandin} et~al.}{2012}]{2012ApJ...761..149K}
{Konstandin} L.,  {Girichidis} P.,  {Federrath} C.,   {Klessen} R.~S.,  2012,
  \mn@doi [\apj] {10.1088/0004-637X/761/2/149}, \href
  {http://adsabs.harvard.edu/abs/2012ApJ...761..149K} {761, 149}

\bibitem[\protect\citeauthoryear{{Koo}, {Moon}, {Lee}, {Lee}  \&
  {Matthews}}{{Koo} et~al.}{2007}]{2007ApJ...657..308K}
{Koo} B.-C.,  {Moon} D.-S.,  {Lee} H.-G.,  {Lee} J.-J.,   {Matthews} K.,  2007,
  \mn@doi [\apj] {10.1086/510550}, \href
  {http://adsabs.harvard.edu/abs/2007ApJ...657..308K} {657, 308}

\bibitem[\protect\citeauthoryear{{Kornreich} \& {Scalo}}{{Kornreich} \&
  {Scalo}}{2000}]{2000ApJ...531..366K}
{Kornreich} P.,  {Scalo} J.,  2000, \mn@doi [\apj] {10.1086/308425}, \href
  {http://adsabs.harvard.edu/abs/2000ApJ...531..366K} {531, 366}

\bibitem[\protect\citeauthoryear{{Kronberger}, {Kapferer}, {Ferrari},
  {Unterguggenberger}  \& {Schindler}}{{Kronberger}
  et~al.}{2008}]{2008A&A...481..337K}
{Kronberger} T.,  {Kapferer} W.,  {Ferrari} C.,  {Unterguggenberger} S.,
  {Schindler} S.,  2008, \mn@doi [\aap] {10.1051/0004-6361:20078904}, \href
  {http://adsabs.harvard.edu/abs/2008A%26A...481..337K} {481, 337}

\bibitem[\protect\citeauthoryear{{Krumholz} \& {Kruijssen}}{{Krumholz} \&
  {Kruijssen}}{2015}]{2015MNRAS.453..739K}
{Krumholz} M.~R.,  {Kruijssen} J.~M.~D.,  2015, \mn@doi [\mnras]
  {10.1093/mnras/stv1670}, \href
  {http://adsabs.harvard.edu/abs/2015MNRAS.453..739K} {453, 739}

\bibitem[\protect\citeauthoryear{{Krumholz} \& {Thompson}}{{Krumholz} \&
  {Thompson}}{2012}]{2012ApJ...760..155K}
{Krumholz} M.~R.,  {Thompson} T.~A.,  2012, \mn@doi [\apj]
  {10.1088/0004-637X/760/2/155}, \href
  {http://adsabs.harvard.edu/abs/2012ApJ...760..155K} {760, 155}

\bibitem[\protect\citeauthoryear{{Krumholz} \& {Thompson}}{{Krumholz} \&
  {Thompson}}{2013}]{2013MNRAS.434.2329K}
{Krumholz} M.~R.,  {Thompson} T.~A.,  2013, \mn@doi [\mnras]
  {10.1093/mnras/stt1174}, \href
  {http://adsabs.harvard.edu/abs/2013MNRAS.434.2329K} {434, 2329}

\bibitem[\protect\citeauthoryear{{Krumholz}, {Kruijssen}  \&
  {Crocker}}{{Krumholz} et~al.}{2017}]{2017MNRAS.466.1213K}
{Krumholz} M.~R.,  {Kruijssen} J.~M.~D.,   {Crocker} R.~M.,  2017, \mn@doi
  [\mnras] {10.1093/mnras/stw3195}, \href
  {http://adsabs.harvard.edu/abs/2017MNRAS.466.1213K} {466, 1213}

\bibitem[\protect\citeauthoryear{{Kuncic} \& {Bicknell}}{{Kuncic} \&
  {Bicknell}}{2004}]{2004ApJ...616..669K}
{Kuncic} Z.,  {Bicknell} G.~V.,  2004, \mn@doi [\apj] {10.1086/425032}, \href
  {http://adsabs.harvard.edu/abs/2004ApJ...616..669K} {616, 669}

\bibitem[\protect\citeauthoryear{{LaRosa}, {Kassim}, {Lazio}  \&
  {Hyman}}{{LaRosa} et~al.}{2000}]{2000AJ....119..207L}
{LaRosa} T.~N.,  {Kassim} N.~E.,  {Lazio} T.~J.~W.,   {Hyman} S.~D.,  2000,
  \mn@doi [\aj] {10.1086/301168}, \href
  {http://adsabs.harvard.edu/abs/2000AJ....119..207L} {119, 207}

\bibitem[\protect\citeauthoryear{{LaRosa}, {Nord}, {Lazio}  \&
  {Kassim}}{{LaRosa} et~al.}{2004}]{2004ApJ...607..302L}
{LaRosa} T.~N.,  {Nord} M.~E.,  {Lazio} T.~J.~W.,   {Kassim} N.~E.,  2004,
  \mn@doi [\apj] {10.1086/383233}, \href
  {http://adsabs.harvard.edu/abs/2004ApJ...607..302L} {607, 302}

\bibitem[\protect\citeauthoryear{{Lang}, {Anantharamaiah}, {Kassim}  \&
  {Lazio}}{{Lang} et~al.}{1999}]{1999ApJ...521L..41L}
{Lang} C.~C.,  {Anantharamaiah} K.~R.,  {Kassim} N.~E.,   {Lazio} T.~J.~W.,
  1999, \mn@doi [\apjl] {10.1086/312180}, \href
  {http://adsabs.harvard.edu/abs/1999ApJ...521L..41L} {521, L41}

\bibitem[\protect\citeauthoryear{{Larson}}{{Larson}}{1981}]{1981MNRAS.194..809L}
{Larson} R.~B.,  1981, \mn@doi [\mnras] {10.1093/mnras/194.4.809}, \href
  {http://adsabs.harvard.edu/abs/1981MNRAS.194..809L} {194, 809}

\bibitem[\protect\citeauthoryear{{Lazarian}}{{Lazarian}}{2014}]{2014SSRv..181....1L}
{Lazarian} A.,  2014, \mn@doi [\ssr] {10.1007/s11214-013-0031-5}, \href
  {http://adsabs.harvard.edu/abs/2014SSRv..181....1L} {181, 1}

\bibitem[\protect\citeauthoryear{{Lazarian} \& {Vishniac}}{{Lazarian} \&
  {Vishniac}}{1999}]{1999ApJ...517..700L}
{Lazarian} A.,  {Vishniac} E.~T.,  1999, \mn@doi [\apj] {10.1086/307233}, \href
  {http://adsabs.harvard.edu/abs/1999ApJ...517..700L} {517, 700}

\bibitem[\protect\citeauthoryear{{Le{\~a}o}, {de Gouveia Dal Pino},
  {Falceta-Gon{\c c}alves}, {Melioli}  \& {Geraissate}}{{Le{\~a}o}
  et~al.}{2009}]{2009MNRAS.394..157L}
{Le{\~a}o} M.~R.~M.,  {de Gouveia Dal Pino} E.~M.,  {Falceta-Gon{\c c}alves}
  D.,  {Melioli} C.,   {Geraissate} F.~G.,  2009, \mn@doi [\mnras]
  {10.1111/j.1365-2966.2008.14337.x}, \href
  {http://adsabs.harvard.edu/abs/2009MNRAS.394..157L} {394, 157}

\bibitem[\protect\citeauthoryear{{Le{\~a}o}, {de Gouveia Dal Pino},
  {Santos-Lima}  \& {Lazarian}}{{Le{\~a}o} et~al.}{2013}]{2013ApJ...777...46L}
{Le{\~a}o} M.~R.~M.,  {de Gouveia Dal Pino} E.~M.,  {Santos-Lima} R.,
  {Lazarian} A.,  2013, \mn@doi [\apj] {10.1088/0004-637X/777/1/46}, \href
  {http://adsabs.harvard.edu/abs/2013ApJ...777...46L} {777, 46}

\bibitem[\protect\citeauthoryear{{Lecoanet} et~al.,}{{Lecoanet}
  et~al.}{2016}]{2016MNRAS.455.4274L}
{Lecoanet} D.,  et~al., 2016, \mn@doi [\mnras] {10.1093/mnras/stv2564}, \href
  {http://adsabs.harvard.edu/abs/2016MNRAS.455.4274L} {455, 4274}

\bibitem[\protect\citeauthoryear{{Lehnert}, {Heckman}  \& {Weaver}}{{Lehnert}
  et~al.}{1999}]{1999ApJ...523..575L}
{Lehnert} M.~D.,  {Heckman} T.~M.,   {Weaver} K.~A.,  1999, \mn@doi [\apj]
  {10.1086/307762}, \href {http://adsabs.harvard.edu/abs/1999ApJ...523..575L}
  {523, 575}

\bibitem[\protect\citeauthoryear{{Li}, {Frank}  \& {Blackman}}{{Li}
  et~al.}{2013}]{2013ApJ...774..133L}
{Li} S.,  {Frank} A.,   {Blackman} E.~G.,  2013, \mn@doi [\apj]
  {10.1088/0004-637X/774/2/133}, \href
  {http://adsabs.harvard.edu/abs/2013ApJ...774..133L} {774, 133}

\bibitem[\protect\citeauthoryear{{Lockman} \& {McClure-Griffiths}}{{Lockman} \&
  {McClure-Griffiths}}{2016}]{2016ApJ...826..215L}
{Lockman} F.~J.,  {McClure-Griffiths} N.~M.,  2016, \mn@doi [\apj]
  {10.3847/0004-637X/826/2/215}, \href
  {http://adsabs.harvard.edu/abs/2016ApJ...826..215L} {826, 215}

\bibitem[\protect\citeauthoryear{{Mac Low} \& {Klessen}}{{Mac Low} \&
  {Klessen}}{2004}]{2004RvMP...76..125M}
{Mac Low} M.-M.,  {Klessen} R.~S.,  2004, \mn@doi [Reviews of Modern Physics]
  {10.1103/RevModPhys.76.125}, \href
  {http://adsabs.harvard.edu/abs/2004RvMP...76..125M} {76, 125}

\bibitem[\protect\citeauthoryear{{Mac Low}, {McKee}, {Klein}, {Stone}  \&
  {Norman}}{{Mac Low} et~al.}{1994}]{1994ApJ...433..757M}
{Mac Low} M.-M.,  {McKee} C.~F.,  {Klein} R.~I.,  {Stone} J.~M.,   {Norman}
  M.~L.,  1994, \mn@doi [\apj] {10.1086/174685}, \href
  {http://adsabs.harvard.edu/abs/1994ApJ...433..757M} {433, 757}

\bibitem[\protect\citeauthoryear{{Mac Low}, {Toraskar}, {Oishi}  \&
  {Abel}}{{Mac Low} et~al.}{2007}]{2007ApJ...668..980M}
{Mac Low} M.-M.,  {Toraskar} J.,  {Oishi} J.~S.,   {Abel} T.,  2007, \mn@doi
  [\apj] {10.1086/521292}, \href
  {http://adsabs.harvard.edu/abs/2007ApJ...668..980M} {668, 980}

\bibitem[\protect\citeauthoryear{{Mackey} \& {Lim}}{{Mackey} \&
  {Lim}}{2010}]{2010MNRAS.403..714M}
{Mackey} J.,  {Lim} A.~J.,  2010, \mn@doi [\mnras]
  {10.1111/j.1365-2966.2009.16181.x}, \href
  {http://adsabs.harvard.edu/abs/2010MNRAS.403..714M} {403, 714}

\bibitem[\protect\citeauthoryear{{Marcolini}, {Brighenti}  \&
  {D'Ercole}}{{Marcolini} et~al.}{2003}]{2003MNRAS.345.1329M}
{Marcolini} A.,  {Brighenti} F.,   {D'Ercole} A.,  2003, \mn@doi [\mnras]
  {10.1046/j.1365-2966.2003.07054.x}, \href
  {http://adsabs.harvard.edu/abs/2003MNRAS.345.1329M} {345, 1329}

\bibitem[\protect\citeauthoryear{{Martin}}{{Martin}}{2005}]{2005ApJ...621..227M}
{Martin} C.~L.,  2005, \mn@doi [\apj] {10.1086/427277}, \href
  {http://adsabs.harvard.edu/abs/2005ApJ...621..227M} {621, 227}

\bibitem[\protect\citeauthoryear{{Matsubayashi}, {Sugai}, {Hattori}, {Kawai},
  {Ozaki}, {Kosugi}, {Ishigaki}  \& {Shimono}}{{Matsubayashi}
  et~al.}{2009}]{2009ApJ...701.1636M}
{Matsubayashi} K.,  {Sugai} H.,  {Hattori} T.,  {Kawai} A.,  {Ozaki} S.,
  {Kosugi} G.,  {Ishigaki} T.,   {Shimono} A.,  2009, \mn@doi [\apj]
  {10.1088/0004-637X/701/2/1636}, \href
  {http://adsabs.harvard.edu/abs/2009ApJ...701.1636M} {701, 1636}

\bibitem[\protect\citeauthoryear{{McClure-Griffiths}, {Dickey}, {Gaensler},
  {Green}, {Green}  \& {Haverkorn}}{{McClure-Griffiths}
  et~al.}{2012}]{2012ApJS..199...12M}
{McClure-Griffiths} N.~M.,  {Dickey} J.~M.,  {Gaensler} B.~M.,  {Green} A.~J.,
  {Green} J.~A.,   {Haverkorn} M.,  2012, \mn@doi [\apjs]
  {10.1088/0067-0049/199/1/12}, \href
  {http://adsabs.harvard.edu/abs/2012ApJS..199...12M} {199, 12}

\bibitem[\protect\citeauthoryear{{McClure-Griffiths}, {Green}, {Hill},
  {Lockman}, {Dickey}, {Gaensler}  \& {Green}}{{McClure-Griffiths}
  et~al.}{2013}]{2013ApJ...770L...4M}
{McClure-Griffiths} N.~M.,  {Green} J.~A.,  {Hill} A.~S.,  {Lockman} F.~J.,
  {Dickey} J.~M.,  {Gaensler} B.~M.,   {Green} A.~J.,  2013, \mn@doi [\apjl]
  {10.1088/2041-8205/770/1/L4}, \href
  {http://adsabs.harvard.edu/abs/2013ApJ...770L...4M} {770, L4}

\bibitem[\protect\citeauthoryear{{McCourt}, {O'Leary}, {Madigan}  \&
  {Quataert}}{{McCourt} et~al.}{2015}]{2015MNRAS.449....2M}
{McCourt} M.,  {O'Leary} R.~M.,  {Madigan} A.-M.,   {Quataert} E.,  2015,
  \mn@doi [\mnras] {10.1093/mnras/stv355}, \href
  {http://adsabs.harvard.edu/abs/2015MNRAS.449....2M} {449, 2}

\bibitem[\protect\citeauthoryear{{McCourt}, {Oh}, {O'Leary}  \&
  {Madigan}}{{McCourt} et~al.}{2016}]{2016arXiv161001164M}
{McCourt} M.,  {Oh} S.~P.,  {O'Leary} R.~M.,   {Madigan} A.-M.,  2016,
  preprint, \href {http://adsabs.harvard.edu/abs/2016arXiv161001164M} {}
  (\mn@eprint {arXiv} {1610.01164})

\bibitem[\protect\citeauthoryear{{McEntaffer}, {Grieves}, {DeRoo}  \&
  {Brantseg}}{{McEntaffer} et~al.}{2013}]{2013ApJ...774..120M}
{McEntaffer} R.~L.,  {Grieves} N.,  {DeRoo} C.,   {Brantseg} T.,  2013, \mn@doi
  [\apj] {10.1088/0004-637X/774/2/120}, \href
  {http://adsabs.harvard.edu/abs/2013ApJ...774..120M} {774, 120}

\bibitem[\protect\citeauthoryear{{McKee} \& {Ostriker}}{{McKee} \&
  {Ostriker}}{2007}]{2007ARA&A..45..565M}
{McKee} C.~F.,  {Ostriker} E.~C.,  2007, \mn@doi [\araa]
  {10.1146/annurev.astro.45.051806.110602}, \href
  {http://adsabs.harvard.edu/abs/2007ARA%26A..45..565M} {45, 565}

\bibitem[\protect\citeauthoryear{{Melioli}, {de Gouveia dal Pino}  \&
  {Raga}}{{Melioli} et~al.}{2005}]{2005AA...443..495M}
{Melioli} C.,  {de Gouveia dal Pino} E.~M.,   {Raga} A.,  2005, \mn@doi [\aap]
  {10.1051/0004-6361:20052679}, \href
  {http://adsabs.harvard.edu/abs/2005A%26A...443..495M} {443, 495}

\bibitem[\protect\citeauthoryear{{Melioli}, {de Gouveia Dal Pino}, {de La Reza}
   \& {Raga}}{{Melioli} et~al.}{2006}]{2006MNRAS.373..811M}
{Melioli} C.,  {de Gouveia Dal Pino} E.~M.,  {de La Reza} R.,   {Raga} A.,
  2006, \mn@doi [\mnras] {10.1111/j.1365-2966.2006.11076.x}, \href
  {http://adsabs.harvard.edu/abs/2006MNRAS.373..811M} {373, 811}

\bibitem[\protect\citeauthoryear{{Melioli}, {Brighenti}, {D'Ercole}  \& {de
  Gouveia Dal Pino}}{{Melioli} et~al.}{2008}]{2008MNRAS.388..573M}
{Melioli} C.,  {Brighenti} F.,  {D'Ercole} A.,   {de Gouveia Dal Pino} E.~M.,
  2008, \mn@doi [\mnras] {10.1111/j.1365-2966.2008.13446.x}, \href
  {http://adsabs.harvard.edu/abs/2008MNRAS.388..573M} {388, 573}

\bibitem[\protect\citeauthoryear{{Melioli}, {Brighenti}, {D'Ercole}  \& {de
  Gouveia Dal Pino}}{{Melioli} et~al.}{2009}]{2009MNRAS.399.1089M}
{Melioli} C.,  {Brighenti} F.,  {D'Ercole} A.,   {de Gouveia Dal Pino} E.~M.,
  2009, \mn@doi [\mnras] {10.1111/j.1365-2966.2009.14725.x}, \href
  {http://adsabs.harvard.edu/abs/2009MNRAS.399.1089M} {399, 1089}

\bibitem[\protect\citeauthoryear{{Melioli}, {de Gouveia Dal Pino}  \&
  {Geraissate}}{{Melioli} et~al.}{2013}]{2013MNRAS.430.3235M}
{Melioli} C.,  {de Gouveia Dal Pino} E.~M.,   {Geraissate} F.~G.,  2013,
  \mn@doi [\mnras] {10.1093/mnras/stt126}, \href
  {http://adsabs.harvard.edu/abs/2013MNRAS.430.3235M} {430, 3235}

\bibitem[\protect\citeauthoryear{{Mellema}, {Kurk}  \&
  {R{\"o}ttgering}}{{Mellema} et~al.}{2002}]{2002AA...395L..13M}
{Mellema} G.,  {Kurk} J.~D.,   {R{\"o}ttgering} H.~J.~A.,  2002, \mn@doi [\aap]
  {10.1051/0004-6361:20021408}, \href
  {http://adsabs.harvard.edu/abs/2002A%26A...395L..13M} {395, L13}

\bibitem[\protect\citeauthoryear{{Mellema}, {Arthur}, {Henney}, {Iliev}  \&
  {Shapiro}}{{Mellema} et~al.}{2006}]{2006ApJ...647..397M}
{Mellema} G.,  {Arthur} S.~J.,  {Henney} W.~J.,  {Iliev} I.~T.,   {Shapiro}
  P.~R.,  2006, \mn@doi [\apj] {10.1086/505294}, \href
  {http://adsabs.harvard.edu/abs/2006ApJ...647..397M} {647, 397}

\bibitem[\protect\citeauthoryear{{Mignone} \& {Tzeferacos}}{{Mignone} \&
  {Tzeferacos}}{2010}]{2010JCoPh.229.2117M}
{Mignone} A.,  {Tzeferacos} P.,  2010, \mn@doi [Journal of Computational
  Physics] {10.1016/j.jcp.2009.11.026}, \href
  {http://adsabs.harvard.edu/abs/2010JCoPh.229.2117M} {229, 2117}

\bibitem[\protect\citeauthoryear{{Mignone}, {Bodo}, {Massaglia}, {Matsakos},
  {Tesileanu}, {Zanni}  \& {Ferrari}}{{Mignone}
  et~al.}{2007}]{2007ApJS..170..228M}
{Mignone} A.,  {Bodo} G.,  {Massaglia} S.,  {Matsakos} T.,  {Tesileanu} O.,
  {Zanni} C.,   {Ferrari} A.,  2007, \mn@doi [\apjs] {10.1086/513316}, \href
  {http://adsabs.harvard.edu/abs/2007ApJS..170..228M} {170, 228}

\bibitem[\protect\citeauthoryear{{Mignone}, {Tzeferacos}  \& {Bodo}}{{Mignone}
  et~al.}{2010}]{2010JCoPh.229.5896M}
{Mignone} A.,  {Tzeferacos} P.,   {Bodo} G.,  2010, \mn@doi [Journal of
  Computational Physics] {10.1016/j.jcp.2010.04.013}, \href
  {http://adsabs.harvard.edu/abs/2010JCoPh.229.5896M} {229, 5896}

\bibitem[\protect\citeauthoryear{{Mignone}, {Zanni}, {Tzeferacos}, {van
  Straalen}, {Colella}  \& {Bodo}}{{Mignone}
  et~al.}{2012}]{2012ApJS..198....7M}
{Mignone} A.,  {Zanni} C.,  {Tzeferacos} P.,  {van Straalen} B.,  {Colella} P.,
    {Bodo} G.,  2012, \mn@doi [\apjs] {10.1088/0067-0049/198/1/7}, \href
  {http://adsabs.harvard.edu/abs/2012ApJS..198....7M} {198, 7}

\bibitem[\protect\citeauthoryear{{Miniati}, {Ryu}, {Ferrara}  \&
  {Jones}}{{Miniati} et~al.}{1999a}]{1999ApJ...510..726M}
{Miniati} F.,  {Ryu} D.,  {Ferrara} A.,   {Jones} T.~W.,  1999a, \mn@doi [\apj]
  {10.1086/306599}, \href {http://adsabs.harvard.edu/abs/1999ApJ...510..726M}
  {510, 726}

\bibitem[\protect\citeauthoryear{{Miniati}, {Jones}  \& {Ryu}}{{Miniati}
  et~al.}{1999b}]{1999ApJ...517..242M}
{Miniati} F.,  {Jones} T.~W.,   {Ryu} D.,  1999b, \mn@doi [\apj]
  {10.1086/307162}, \href {http://adsabs.harvard.edu/abs/1999ApJ...517..242M}
  {517, 242}

\bibitem[\protect\citeauthoryear{{Miyoshi} \& {Kusano}}{{Miyoshi} \&
  {Kusano}}{2005}]{Miyoshi:2005}
{Miyoshi} T.,  {Kusano} K.,  2005, \mn@doi [\jcp] {10.1016/j.jcp.2005.02.017},
  \href {http://adsabs.harvard.edu/abs/2005JCoPh.208..315M} {208, 315}

\bibitem[\protect\citeauthoryear{{Molina}, {Glover}, {Federrath}  \&
  {Klessen}}{{Molina} et~al.}{2012}]{2012MNRAS.423.2680M}
{Molina} F.~Z.,  {Glover} S.~C.~O.,  {Federrath} C.,   {Klessen} R.~S.,  2012,
  \mn@doi [\mnras] {10.1111/j.1365-2966.2012.21075.x}, \href
  {http://adsabs.harvard.edu/abs/2012MNRAS.423.2680M} {423, 2680}

\bibitem[\protect\citeauthoryear{{Monceau-Baroux} \&
  {Keppens}}{{Monceau-Baroux} \& {Keppens}}{2017}]{2017A&A...600A.134M}
{Monceau-Baroux} R.,  {Keppens} R.,  2017, \mn@doi [\aap]
  {10.1051/0004-6361/201629796}, \href
  {http://adsabs.harvard.edu/abs/2017A%26A...600A.134M} {600, A134}

\bibitem[\protect\citeauthoryear{{Morris} \& {Yusef-Zadeh}}{{Morris} \&
  {Yusef-Zadeh}}{1985}]{1985AJ.....90.2511M}
{Morris} M.,  {Yusef-Zadeh} F.,  1985, \mn@doi [\aj] {10.1086/113955}, \href
  {http://adsabs.harvard.edu/abs/1985AJ.....90.2511M} {90, 2511}

\bibitem[\protect\citeauthoryear{{Morris}, {Zhao}  \& {Goss}}{{Morris}
  et~al.}{2014}]{2014IAUS..303..369M}
{Morris} M.~R.,  {Zhao} J.-H.,   {Goss} W.~M.,  2014, in {Sjouwerman} L.~O.,
  {Lang} C.~C.,   {Ott} J.,  eds,  IAU Symposium Vol. 303, IAU Symposium. pp
  369--373 (\mn@eprint {arXiv} {1312.2238}), \mn@doi{10.1017/S1743921314000933}

\bibitem[\protect\citeauthoryear{{Mouri}, {Takaoka}, {Hori}  \&
  {Kawashima}}{{Mouri} et~al.}{2002}]{2002PhRvE..65e6304M}
{Mouri} H.,  {Takaoka} M.,  {Hori} A.,   {Kawashima} Y.,  2002, \mn@doi [\pre]
  {10.1103/PhysRevE.65.056304}, \href
  {http://adsabs.harvard.edu/abs/2002PhRvE..65e6304M} {65, 056304}

\bibitem[\protect\citeauthoryear{{Murray}, {White}, {Blondin}  \&
  {Lin}}{{Murray} et~al.}{1993}]{1993ApJ...407..588M}
{Murray} S.~D.,  {White} S.~D.~M.,  {Blondin} J.~M.,   {Lin} D.~N.~C.,  1993,
  \mn@doi [\apj] {10.1086/172540}, \href
  {http://adsabs.harvard.edu/abs/1993ApJ...407..588M} {407, 588}

\bibitem[\protect\citeauthoryear{{Murray}, {Quataert}  \& {Thompson}}{{Murray}
  et~al.}{2005}]{2005ApJ...618..569M}
{Murray} N.,  {Quataert} E.,   {Thompson} T.~A.,  2005, \mn@doi [\apj]
  {10.1086/426067}, \href {http://adsabs.harvard.edu/abs/2005ApJ...618..569M}
  {618, 569}

\bibitem[\protect\citeauthoryear{{Nakamura}, {McKee}, {Klein}  \&
  {Fisher}}{{Nakamura} et~al.}{2006}]{2006ApJS..164..477N}
{Nakamura} F.,  {McKee} C.~F.,  {Klein} R.~I.,   {Fisher} R.~T.,  2006, \mn@doi
  [\apjs] {10.1086/501530}, \href
  {http://adsabs.harvard.edu/abs/2006ApJS..164..477N} {164, 477}

\bibitem[\protect\citeauthoryear{{Niederhaus}}{{Niederhaus}}{2007}]{2007PhDT........54N}
{Niederhaus} J.~H.~J.,  2007, PhD thesis, The University of Wisconsin - Madison

\bibitem[\protect\citeauthoryear{{Nittmann}, {Falle}  \& {Gaskell}}{{Nittmann}
  et~al.}{1982}]{1982MNRAS.201..833N}
{Nittmann} J.,  {Falle} S.~A.~E.~G.,   {Gaskell} P.~H.,  1982, \mnras, \href
  {http://adsabs.harvard.edu/abs/1982MNRAS.201..833N} {201, 833}

\bibitem[\protect\citeauthoryear{{Nolan}, {Federrath}  \& {Sutherland}}{{Nolan}
  et~al.}{2015}]{2015MNRAS.451.1380N}
{Nolan} C.~A.,  {Federrath} C.,   {Sutherland} R.~S.,  2015, \mn@doi [\mnras]
  {10.1093/mnras/stv1030}, \href
  {http://adsabs.harvard.edu/abs/2015MNRAS.451.1380N} {451, 1380}

\bibitem[\protect\citeauthoryear{{Nordlund} \& {Padoan}}{{Nordlund} \&
  {Padoan}}{1999}]{1999intu.conf..218N}
{Nordlund} {\AA}.~K.,  {Padoan} P.,  1999, in {Franco} J.,  {Carraminana} A.,
  eds, Interstellar Turbulence. p.~218 (\mn@eprint {} {astro-ph/9810074})

\bibitem[\protect\citeauthoryear{{Nynka} et~al.,}{{Nynka}
  et~al.}{2015}]{2015ApJ...800..119N}
{Nynka} M.,  et~al., 2015, \mn@doi [\apj] {10.1088/0004-637X/800/2/119}, \href
  {http://adsabs.harvard.edu/abs/2015ApJ...800..119N} {800, 119}

\bibitem[\protect\citeauthoryear{{Orlando}, {Peres}, {Reale}, {Bocchino},
  {Rosner}, {Plewa}  \& {Siegel}}{{Orlando} et~al.}{2005}]{2005A&A...444..505O}
{Orlando} S.,  {Peres} G.,  {Reale} F.,  {Bocchino} F.,  {Rosner} R.,  {Plewa}
  T.,   {Siegel} A.,  2005, \mn@doi [\aap] {10.1051/0004-6361:20052896}, \href
  {http://adsabs.harvard.edu/abs/2005A%26A...444..505O} {444, 505}

\bibitem[\protect\citeauthoryear{{Orlando}, {Bocchino}, {Peres}, {Reale},
  {Plewa}  \& {Rosner}}{{Orlando} et~al.}{2006}]{2006A&A...457..545O}
{Orlando} S.,  {Bocchino} F.,  {Peres} G.,  {Reale} F.,  {Plewa} T.,   {Rosner}
  R.,  2006, \mn@doi [\aap] {10.1051/0004-6361:20065652}, \href
  {http://adsabs.harvard.edu/abs/2006A%26A...457..545O} {457, 545}

\bibitem[\protect\citeauthoryear{{Orlando}, {Bocchino}, {Reale}, {Peres}  \&
  {Pagano}}{{Orlando} et~al.}{2008}]{2008ApJ...678..274O}
{Orlando} S.,  {Bocchino} F.,  {Reale} F.,  {Peres} G.,   {Pagano} P.,  2008,
  \mn@doi [\apj] {10.1086/529420}, \href
  {http://adsabs.harvard.edu/abs/2008ApJ...678..274O} {678, 274}

\bibitem[\protect\citeauthoryear{{Ossenkopf} \& {Mac Low}}{{Ossenkopf} \& {Mac
  Low}}{2002}]{2002A&A...390..307O}
{Ossenkopf} V.,  {Mac Low} M.-M.,  2002, \mn@doi [\aap]
  {10.1051/0004-6361:20020629}, \href
  {http://adsabs.harvard.edu/abs/2002A%26A...390..307O} {390, 307}

\bibitem[\protect\citeauthoryear{{Padoan} \& {Nordlund}}{{Padoan} \&
  {Nordlund}}{1999}]{1999ApJ...526..279P}
{Padoan} P.,  {Nordlund} {\AA}.,  1999, \mn@doi [\apj] {10.1086/307956}, \href
  {http://adsabs.harvard.edu/abs/1999ApJ...526..279P} {526, 279}

\bibitem[\protect\citeauthoryear{{Padoan}, {Nordlund}  \& {Jones}}{{Padoan}
  et~al.}{1997}]{1997MNRAS.288..145P}
{Padoan} P.,  {Nordlund} A.,   {Jones} B.~J.~T.,  1997, \mnras, \href
  {http://adsabs.harvard.edu/abs/1997MNRAS.288..145P} {288, 145}

\bibitem[\protect\citeauthoryear{{Padoan}, {Federrath}, {Chabrier}, {Evans},
  {Johnstone}, {J{\o}rgensen}, {McKee}  \& {Nordlund}}{{Padoan}
  et~al.}{2014}]{2014prpl.conf...77P}
{Padoan} P.,  {Federrath} C.,  {Chabrier} G.,  {Evans} II N.~J.,  {Johnstone}
  D.,  {J{\o}rgensen} J.~K.,  {McKee} C.~F.,   {Nordlund} {\AA}.,  2014,
  \mn@doi [Protostars and Planets VI]
  {10.2458/azu_uapress_9780816531240-ch004}, \href
  {http://adsabs.harvard.edu/abs/2014prpl.conf...77P} {pp 77--100}

\bibitem[\protect\citeauthoryear{{Parkin}}{{Parkin}}{2014}]{2014MNRAS.438.2513P}
{Parkin} E.~R.,  2014, \mn@doi [\mnras] {10.1093/mnras/stt2379}, \href
  {http://adsabs.harvard.edu/abs/2014MNRAS.438.2513P} {438, 2513}

\bibitem[\protect\citeauthoryear{{Parkin}, {Pittard}, {Corcoran}  \&
  {Hamaguchi}}{{Parkin} et~al.}{2011}]{2011ApJ...726..105P}
{Parkin} E.~R.,  {Pittard} J.~M.,  {Corcoran} M.~F.,   {Hamaguchi} K.,  2011,
  \mn@doi [\apj] {10.1088/0004-637X/726/2/105}, \href
  {http://adsabs.harvard.edu/abs/2011ApJ...726..105P} {726, 105}

\bibitem[\protect\citeauthoryear{{Passot} \& {V{\'a}zquez-Semadeni}}{{Passot}
  \& {V{\'a}zquez-Semadeni}}{1998}]{1998PhRvE..58.4501P}
{Passot} T.,  {V{\'a}zquez-Semadeni} E.,  1998, \mn@doi [\pre]
  {10.1103/PhysRevE.58.4501}, \href
  {http://adsabs.harvard.edu/abs/1998PhRvE..58.4501P} {58, 4501}

\bibitem[\protect\citeauthoryear{{Patnaude} \& {Fesen}}{{Patnaude} \&
  {Fesen}}{2009}]{2009ApJ...697..535P}
{Patnaude} D.~J.,  {Fesen} R.~A.,  2009, \mn@doi [\apj]
  {10.1088/0004-637X/697/1/535}, \href
  {http://adsabs.harvard.edu/abs/2009ApJ...697..535P} {697, 535}

\bibitem[\protect\citeauthoryear{{Pfrommer} \& {Dursi}}{{Pfrommer} \&
  {Dursi}}{2010}]{2010NatPh...6..520P}
{Pfrommer} C.,  {Dursi} J.,  2010, \mn@doi [Nature Physics]
  {10.1038/nphys1657}, \href
  {http://adsabs.harvard.edu/abs/2010NatPh...6..520P} {6, 520}

\bibitem[\protect\citeauthoryear{{Pietarila Graham}, {Danilovic}  \&
  {Sch{\"u}ssler}}{{Pietarila Graham} et~al.}{2009}]{2009ApJ...693.1728P}
{Pietarila Graham} J.,  {Danilovic} S.,   {Sch{\"u}ssler} M.,  2009, \mn@doi
  [\apj] {10.1088/0004-637X/693/2/1728}, \href
  {http://esoads.eso.org/abs/2009ApJ...693.1728P} {693, 1728}

\bibitem[\protect\citeauthoryear{{Pittard}}{{Pittard}}{2011}]{2011MNRAS.411L..41P}
{Pittard} J.~M.,  2011, \mn@doi [\mnras] {10.1111/j.1745-3933.2010.00988.x},
  \href {http://adsabs.harvard.edu/abs/2011MNRAS.411L..41P} {411, L41}

\bibitem[\protect\citeauthoryear{{Pittard} \& {Goldsmith}}{{Pittard} \&
  {Goldsmith}}{2016}]{2016MNRAS.458.1139P}
{Pittard} J.~M.,  {Goldsmith} K.~J.~A.,  2016, \mn@doi [\mnras]
  {10.1093/mnras/stw378}, \href
  {http://adsabs.harvard.edu/abs/2016MNRAS.458.1139P} {458, 1139}

\bibitem[\protect\citeauthoryear{{Pittard} \& {Parkin}}{{Pittard} \&
  {Parkin}}{2016}]{2016MNRAS.457.4470P}
{Pittard} J.~M.,  {Parkin} E.~R.,  2016, \mn@doi [\mnras]
  {10.1093/mnras/stw025}, \href
  {http://adsabs.harvard.edu/abs/2016MNRAS.457.4470P} {457, 4470}

\bibitem[\protect\citeauthoryear{{Pittard}, {Dyson}, {Falle}  \&
  {Hartquist}}{{Pittard} et~al.}{2005}]{2005MNRAS.361.1077P}
{Pittard} J.~M.,  {Dyson} J.~E.,  {Falle} S.~A.~E.~G.,   {Hartquist} T.~W.,
  2005, \mn@doi [\mnras] {10.1111/j.1365-2966.2005.09268.x}, \href
  {http://adsabs.harvard.edu/abs/2005MNRAS.361.1077P} {361, 1077}

\bibitem[\protect\citeauthoryear{{Pittard}, {Falle}, {Hartquist}  \&
  {Dyson}}{{Pittard} et~al.}{2009}]{2009MNRAS.394.1351P}
{Pittard} J.~M.,  {Falle} S.~A.~E.~G.,  {Hartquist} T.~W.,   {Dyson} J.~E.,
  2009, \mn@doi [\mnras] {10.1111/j.1365-2966.2009.13759.x}, \href
  {http://adsabs.harvard.edu/abs/2009MNRAS.394.1351P} {394, 1351}

\bibitem[\protect\citeauthoryear{{Pittard}, {Hartquist}  \& {Falle}}{{Pittard}
  et~al.}{2010}]{2010MNRAS.405..821P}
{Pittard} J.~M.,  {Hartquist} T.~W.,   {Falle} S.~A.~E.~G.,  2010, \mn@doi
  [\mnras] {10.1111/j.1365-2966.2010.16504.x}, \href
  {http://adsabs.harvard.edu/abs/2010MNRAS.405..821P} {405, 821}

\bibitem[\protect\citeauthoryear{{Pittard}, {Hartquist}, {Dyson}  \&
  {Falle}}{{Pittard} et~al.}{2011}]{2011Ap&SS.336..239P}
{Pittard} J.~M.,  {Hartquist} T.~W.,  {Dyson} J.~E.,   {Falle} S.~A.~E.~G.,
  2011, \mn@doi [\apss] {10.1007/s10509-010-0526-4}, \href
  {http://adsabs.harvard.edu/abs/2011Ap%26SS.336..239P} {336, 239}

\bibitem[\protect\citeauthoryear{{Poludnenko}, {Frank}  \&
  {Blackman}}{{Poludnenko} et~al.}{2002}]{2002ApJ...576..832P}
{Poludnenko} A.~Y.,  {Frank} A.,   {Blackman} E.~G.,  2002, \mn@doi [\apj]
  {10.1086/341886}, \href {http://adsabs.harvard.edu/abs/2002ApJ...576..832P}
  {576, 832}

\bibitem[\protect\citeauthoryear{{Price}, {Federrath}  \& {Brunt}}{{Price}
  et~al.}{2011}]{2011ApJ...727L..21P}
{Price} D.~J.,  {Federrath} C.,   {Brunt} C.~M.,  2011, \mn@doi [\apjl]
  {10.1088/2041-8205/727/1/L21}, \href
  {http://adsabs.harvard.edu/abs/2011ApJ...727L..21P} {727, L21}

\bibitem[\protect\citeauthoryear{{Proga} \& {Waters}}{{Proga} \&
  {Waters}}{2015}]{2015ApJ...804..137P}
{Proga} D.,  {Waters} T.,  2015, \mn@doi [\apj] {10.1088/0004-637X/804/2/137},
  \href {http://adsabs.harvard.edu/abs/2015ApJ...804..137P} {804, 137}

\bibitem[\protect\citeauthoryear{{Raga}, {Steffen}  \& {Gonz{\'a}lez}}{{Raga}
  et~al.}{2005}]{2005RMxAA..41...45R}
{Raga} A.,  {Steffen} W.,   {Gonz{\'a}lez} R.,  2005, Revista Mexicana, \href
  {http://adsabs.harvard.edu/abs/2005RMxAA..41...45R} {41, 45}

\bibitem[\protect\citeauthoryear{{Raga}, {Esquivel}, {Riera}  \&
  {Vel{\'a}zquez}}{{Raga} et~al.}{2007}]{2007ApJ...668..310R}
{Raga} A.~C.,  {Esquivel} A.,  {Riera} A.,   {Vel{\'a}zquez} P.~F.,  2007,
  \mn@doi [\apj] {10.1086/521143}, \href
  {http://adsabs.harvard.edu/abs/2007ApJ...668..310R} {668, 310}

\bibitem[\protect\citeauthoryear{{Raga}, {Henney}, {Vasconcelos}, {Cerqueira},
  {Esquivel}  \& {Rodr{\'{\i}}guez-Gonz{\'a}lez}}{{Raga}
  et~al.}{2009}]{2009MNRAS.392..964R}
{Raga} A.~C.,  {Henney} W.,  {Vasconcelos} J.,  {Cerqueira} A.,  {Esquivel} A.,
    {Rodr{\'{\i}}guez-Gonz{\'a}lez} A.,  2009, \mn@doi [\mnras]
  {10.1111/j.1365-2966.2008.14097.x}, \href
  {http://adsabs.harvard.edu/abs/2009MNRAS.392..964R} {392, 964}

\bibitem[\protect\citeauthoryear{{Redfield} \& {Linsky}}{{Redfield} \&
  {Linsky}}{2004}]{2004ApJ...613.1004R}
{Redfield} S.,  {Linsky} J.~L.,  2004, \mn@doi [\apj] {10.1086/423311}, \href
  {http://adsabs.harvard.edu/abs/2004ApJ...613.1004R} {613, 1004}

\bibitem[\protect\citeauthoryear{{Roberts}}{{Roberts}}{1999}]{1999ASPC..186..483R}
{Roberts} D.~A.,  1999, in {Falcke} H.,  {Cotera} A.,  {Duschl} W.~J.,  {Melia}
  F.,   {Rieke} M.~J.,  eds,  Astronomical Society of the Pacific Conference
  Series Vol. 186, The Central Parsecs of the Galaxy. p.~483

\bibitem[\protect\citeauthoryear{{Rybakin}, {Smirnov}  \&
  {Goryachev}}{{Rybakin} et~al.}{2016}]{2016rscd.conf..146R}
{Rybakin} B.,  {Smirnov} N.,   {Goryachev} V.,  2016, in Voevodin V., Sobolev
  S. (eds) Supercomputing: Second Russian Supercomputing Days. Springer
  International Publishing. Communications in Computer and Information Science,
  vol 687. pp 146--157, \mn@doi{10.1007/978-3-319-55669-7_12}

\bibitem[\protect\citeauthoryear{{Ryu}, {Jones}  \& {Frank}}{{Ryu}
  et~al.}{2000}]{2000ApJ...545..475R}
{Ryu} D.,  {Jones} T.~W.,   {Frank} A.,  2000, \mn@doi [\apj] {10.1086/317789},
  \href {http://adsabs.harvard.edu/abs/2000ApJ...545..475R} {545, 475}

\bibitem[\protect\citeauthoryear{{Sahai}, {Morris}  \& {Claussen}}{{Sahai}
  et~al.}{2012a}]{2012ApJ...751...69S}
{Sahai} R.,  {Morris} M.~R.,   {Claussen} M.~J.,  2012a, \mn@doi [\apj]
  {10.1088/0004-637X/751/1/69}, \href
  {http://adsabs.harvard.edu/abs/2012ApJ...751...69S} {751, 69}

\bibitem[\protect\citeauthoryear{{Sahai}, {G{\"u}sten}  \& {Morris}}{{Sahai}
  et~al.}{2012b}]{2012ApJ...761L..21S}
{Sahai} R.,  {G{\"u}sten} R.,   {Morris} M.~R.,  2012b, \mn@doi [\apjl]
  {10.1088/2041-8205/761/2/L21}, \href
  {http://adsabs.harvard.edu/abs/2012ApJ...761L..21S} {761, L21}

\bibitem[\protect\citeauthoryear{{Salim}, {Federrath}  \& {Kewley}}{{Salim}
  et~al.}{2015}]{2015ApJ...806L..36S}
{Salim} D.~M.,  {Federrath} C.,   {Kewley} L.~J.,  2015, \mn@doi [\apjl]
  {10.1088/2041-8205/806/2/L36}, \href
  {http://adsabs.harvard.edu/abs/2015ApJ...806L..36S} {806, L36}

\bibitem[\protect\citeauthoryear{{Santos-Lima}, {Lazarian}, {de Gouveia Dal
  Pino}  \& {Cho}}{{Santos-Lima} et~al.}{2010}]{2010ApJ...714..442S}
{Santos-Lima} R.,  {Lazarian} A.,  {de Gouveia Dal Pino} E.~M.,   {Cho} J.,
  2010, \mn@doi [\apj] {10.1088/0004-637X/714/1/442}, \href
  {http://adsabs.harvard.edu/abs/2010ApJ...714..442S} {714, 442}

\bibitem[\protect\citeauthoryear{{Santos-Lima}, {de Gouveia Dal Pino}  \&
  {Lazarian}}{{Santos-Lima} et~al.}{2012}]{2012ApJ...747...21S}
{Santos-Lima} R.,  {de Gouveia Dal Pino} E.~M.,   {Lazarian} A.,  2012, \mn@doi
  [\apj] {10.1088/0004-637X/747/1/21}, \href
  {http://adsabs.harvard.edu/abs/2012ApJ...747...21S} {747, 21}

\bibitem[\protect\citeauthoryear{{Santos-Lima}, {de Gouveia Dal Pino}, {Kowal},
  {Falceta-Gon{\c c}alves}, {Lazarian}  \& {Nakwacki}}{{Santos-Lima}
  et~al.}{2014}]{2014ApJ...781...84S}
{Santos-Lima} R.,  {de Gouveia Dal Pino} E.~M.,  {Kowal} G.,  {Falceta-Gon{\c
  c}alves} D.,  {Lazarian} A.,   {Nakwacki} M.~S.,  2014, \mn@doi [\apj]
  {10.1088/0004-637X/781/2/84}, \href
  {http://adsabs.harvard.edu/abs/2014ApJ...781...84S} {781, 84}

\bibitem[\protect\citeauthoryear{{Saury}, {Miville-Desch{\^e}nes},
  {Hennebelle}, {Audit}  \& {Schmidt}}{{Saury}
  et~al.}{2014}]{2014A&A...567A..16S}
{Saury} E.,  {Miville-Desch{\^e}nes} M.-A.,  {Hennebelle} P.,  {Audit} E.,
  {Schmidt} W.,  2014, \mn@doi [\aap] {10.1051/0004-6361/201321113}, \href
  {http://adsabs.harvard.edu/abs/2014A%26A...567A..16S} {567, A16}

\bibitem[\protect\citeauthoryear{{Scalo} \& {Elmegreen}}{{Scalo} \&
  {Elmegreen}}{2004}]{2004ARA&A..42..275S}
{Scalo} J.,  {Elmegreen} B.~G.,  2004, \mn@doi [\araa]
  {10.1146/annurev.astro.42.120403.143327}, \href
  {http://adsabs.harvard.edu/abs/2004ARA%26A..42..275S} {42, 275}

\bibitem[\protect\citeauthoryear{{Scannapieco}}{{Scannapieco}}{2017}]{2017ApJ...837...28S}
{Scannapieco} E.,  2017, \mn@doi [\apj] {10.3847/1538-4357/aa5d0d}, \href
  {http://adsabs.harvard.edu/abs/2017ApJ...837...28S} {837, 28}

\bibitem[\protect\citeauthoryear{{Scannapieco} \& {Br{\"u}ggen}}{{Scannapieco}
  \& {Br{\"u}ggen}}{2015}]{2015ApJ...805..158S}
{Scannapieco} E.,  {Br{\"u}ggen} M.,  2015, \mn@doi [\apj]
  {10.1088/0004-637X/805/2/158}, \href
  {http://adsabs.harvard.edu/abs/2015ApJ...805..158S} {805, 158}

\bibitem[\protect\citeauthoryear{{Schartmann}, {Burkert}, {Alig}, {Gillessen},
  {Genzel}, {Eisenhauer}  \& {Fritz}}{{Schartmann}
  et~al.}{2012}]{2012ApJ...755..155S}
{Schartmann} M.,  {Burkert} A.,  {Alig} C.,  {Gillessen} S.,  {Genzel} R.,
  {Eisenhauer} F.,   {Fritz} T.~K.,  2012, \mn@doi [\apj]
  {10.1088/0004-637X/755/2/155}, \href
  {http://adsabs.harvard.edu/abs/2012ApJ...755..155S} {755, 155}

\bibitem[\protect\citeauthoryear{{Schartmann} et~al.,}{{Schartmann}
  et~al.}{2015}]{2015ApJ...811..155S}
{Schartmann} M.,  et~al., 2015, \mn@doi [\apj] {10.1088/0004-637X/811/2/155},
  \href {http://adsabs.harvard.edu/abs/2015ApJ...811..155S} {811, 155}

\bibitem[\protect\citeauthoryear{{Schleicher}, {Schober}, {Federrath}, {Bovino}
   \& {Schmidt}}{{Schleicher} et~al.}{2013}]{2013NJPh...15b3017S}
{Schleicher} D.~R.~G.,  {Schober} J.,  {Federrath} C.,  {Bovino} S.,
  {Schmidt} W.,  2013, \mn@doi [New Journal of Physics]
  {10.1088/1367-2630/15/2/023017}, \href
  {http://adsabs.harvard.edu/abs/2013NJPh...15b3017S} {15, 023017}

\bibitem[\protect\citeauthoryear{{Schneider} \& {Robertson}}{{Schneider} \&
  {Robertson}}{2017}]{2017ApJ...834..144S}
{Schneider} E.~E.,  {Robertson} B.~E.,  2017, \mn@doi [\apj]
  {10.3847/1538-4357/834/2/144}, \href
  {http://adsabs.harvard.edu/abs/2017ApJ...834..144S} {834, 144}

\bibitem[\protect\citeauthoryear{{Schneider} et~al.,}{{Schneider}
  et~al.}{2012}]{2012A&A...540L..11S}
{Schneider} N.,  et~al., 2012, \mn@doi [\aap] {10.1051/0004-6361/201118566},
  \href {http://adsabs.harvard.edu/abs/2012A%26A...540L..11S} {540, L11}

\bibitem[\protect\citeauthoryear{{Schneider} et~al.,}{{Schneider}
  et~al.}{2013}]{2013ApJ...766L..17S}
{Schneider} N.,  et~al., 2013, \mn@doi [\apjl] {10.1088/2041-8205/766/2/L17},
  \href {http://adsabs.harvard.edu/abs/2013ApJ...766L..17S} {766, L17}

\bibitem[\protect\citeauthoryear{{Schneider} et~al.,}{{Schneider}
  et~al.}{2015}]{2015A&A...575A..79S}
{Schneider} N.,  et~al., 2015, \mn@doi [\aap] {10.1051/0004-6361/201423569},
  \href {http://adsabs.harvard.edu/abs/2015A%26A...575A..79S} {575, A79}

\bibitem[\protect\citeauthoryear{{Schneider} et~al.,}{{Schneider}
  et~al.}{2016}]{2016A&A...587A..74S}
{Schneider} N.,  et~al., 2016, \mn@doi [\aap] {10.1051/0004-6361/201527144},
  \href {http://adsabs.harvard.edu/abs/2016A%26A...587A..74S} {587, A74}

\bibitem[\protect\citeauthoryear{{Schober}, {Schleicher}, {Federrath}, {Bovino}
   \& {Klessen}}{{Schober} et~al.}{2015}]{2015PhRvE..92b3010S}
{Schober} J.,  {Schleicher} D.~R.~G.,  {Federrath} C.,  {Bovino} S.,
  {Klessen} R.~S.,  2015, \mn@doi [\pre] {10.1103/PhysRevE.92.023010}, \href
  {http://adsabs.harvard.edu/abs/2015PhRvE..92b3010S} {92, 023010}

\bibitem[\protect\citeauthoryear{{Sharp}}{{Sharp}}{1984}]{1984PhyD...12....3S}
{Sharp} D.~H.,  1984, \mn@doi [Physica D Nonlinear Phenomena]
  {10.1016/0167-2789(84)90510-4}, \href
  {http://adsabs.harvard.edu/abs/1984PhyD...12....3S} {12, 3}

\bibitem[\protect\citeauthoryear{{Shin}, {Stone}  \& {Snyder}}{{Shin}
  et~al.}{2008}]{2008ApJ...680..336S}
{Shin} M.-S.,  {Stone} J.~M.,   {Snyder} G.~F.,  2008, \mn@doi [\apj]
  {10.1086/587775}, \href {http://adsabs.harvard.edu/abs/2008ApJ...680..336S}
  {680, 336}

\bibitem[\protect\citeauthoryear{{Shinn}, {Koo}, {Burton}, {Lee}  \&
  {Moon}}{{Shinn} et~al.}{2009}]{2009ApJ...693.1883S}
{Shinn} J.-H.,  {Koo} B.-C.,  {Burton} M.~G.,  {Lee} H.-G.,   {Moon} D.-S.,
  2009, \mn@doi [\apj] {10.1088/0004-637X/693/2/1883}, \href
  {http://adsabs.harvard.edu/abs/2009ApJ...693.1883S} {693, 1883}

\bibitem[\protect\citeauthoryear{{Shopbell} \& {Bland-Hawthorn}}{{Shopbell} \&
  {Bland-Hawthorn}}{1998}]{1998ApJ...493..129S}
{Shopbell} P.~L.,  {Bland-Hawthorn} J.,  1998, \mn@doi [\apj] {10.1086/305108},
  \href {http://adsabs.harvard.edu/abs/1998ApJ...493..129S} {493, 129}

\bibitem[\protect\citeauthoryear{{Shore} \& {LaRosa}}{{Shore} \&
  {LaRosa}}{1999}]{1999ApJ...521..587S}
{Shore} S.~N.,  {LaRosa} T.~N.,  1999, \mn@doi [\apj] {10.1086/307601}, \href
  {http://adsabs.harvard.edu/abs/1999ApJ...521..587S} {521, 587}

\bibitem[\protect\citeauthoryear{{Sofue} \& {Handa}}{{Sofue} \&
  {Handa}}{1984}]{1984Natur.310..568S}
{Sofue} Y.,  {Handa} T.,  1984, \mn@doi [\nat] {10.1038/310568a0}, \href
  {http://adsabs.harvard.edu/abs/1984Natur.310..568S} {310, 568}

\bibitem[\protect\citeauthoryear{{Sofue}, {Kigure}  \& {Shibata}}{{Sofue}
  et~al.}{2005}]{2005PASJ...57L..39S}
{Sofue} Y.,  {Kigure} H.,   {Shibata} K.,  2005, \mn@doi [\pasj]
  {10.1093/pasj/57.5.L39}, \href
  {http://adsabs.harvard.edu/abs/2005PASJ...57L..39S} {57, L39}

\bibitem[\protect\citeauthoryear{{Stevens}, {Blondin}  \& {Pollock}}{{Stevens}
  et~al.}{1992}]{1992ApJ...386..265S}
{Stevens} I.~R.,  {Blondin} J.~M.,   {Pollock} A.~M.~T.,  1992, \mn@doi [\apj]
  {10.1086/171013}, \href {http://adsabs.harvard.edu/abs/1992ApJ...386..265S}
  {386, 265}

\bibitem[\protect\citeauthoryear{{Stone} \& {Gardiner}}{{Stone} \&
  {Gardiner}}{2007}]{2007ApJ...671.1726S}
{Stone} J.~M.,  {Gardiner} T.,  2007, \mn@doi [\apj] {10.1086/523099}, \href
  {http://adsabs.harvard.edu/abs/2007ApJ...671.1726S} {671, 1726}

\bibitem[\protect\citeauthoryear{{Stone} \& {Norman}}{{Stone} \&
  {Norman}}{1992}]{1992ApJ...390L..17S}
{Stone} J.~M.,  {Norman} M.~L.,  1992, \mn@doi [\apjl] {10.1086/186361}, \href
  {http://adsabs.harvard.edu/abs/1992ApJ...390L..17S} {390, L17}

\bibitem[\protect\citeauthoryear{{Strickland} \& {Stevens}}{{Strickland} \&
  {Stevens}}{2000}]{2000MNRAS.314..511S}
{Strickland} D.~K.,  {Stevens} I.~R.,  2000, \mn@doi [\mnras]
  {10.1046/j.1365-8711.2000.03391.x}, \href
  {http://adsabs.harvard.edu/abs/2000MNRAS.314..511S} {314, 511}

\bibitem[\protect\citeauthoryear{{Strickland}, {Ponman}  \&
  {Stevens}}{{Strickland} et~al.}{1997}]{1997A&A...320..378S}
{Strickland} D.~K.,  {Ponman} T.~J.,   {Stevens} I.~R.,  1997, \aap, \href
  {http://adsabs.harvard.edu/abs/1997A%26A...320..378S} {320, 378}

\bibitem[\protect\citeauthoryear{{Su}, {Slatyer}  \& {Finkbeiner}}{{Su}
  et~al.}{2010}]{2010ApJ...724.1044S}
{Su} M.,  {Slatyer} T.~R.,   {Finkbeiner} D.~P.,  2010, \mn@doi [\apj]
  {10.1088/0004-637X/724/2/1044}, \href
  {http://adsabs.harvard.edu/abs/2010ApJ...724.1044S} {724, 1044}

\bibitem[\protect\citeauthoryear{{Subramanian}}{{Subramanian}}{1999}]{1999PhRvL..83.2957S}
{Subramanian} K.,  1999, \mn@doi [Physical Review Letters]
  {10.1103/PhysRevLett.83.2957}, \href
  {http://adsabs.harvard.edu/abs/1999PhRvL..83.2957S} {83, 2957}

\bibitem[\protect\citeauthoryear{{Sutherland} \& {Bicknell}}{{Sutherland} \&
  {Bicknell}}{2007}]{2007ApJS..173...37S}
{Sutherland} R.~S.,  {Bicknell} G.~V.,  2007, \mn@doi [\apjs] {10.1086/520640},
  \href {http://adsabs.harvard.edu/abs/2007ApJS..173...37S} {173, 37}

\bibitem[\protect\citeauthoryear{{Thompson}, {Quataert}, {Zhang}  \&
  {Weinberg}}{{Thompson} et~al.}{2016}]{2016MNRAS.455.1830T}
{Thompson} T.~A.,  {Quataert} E.,  {Zhang} D.,   {Weinberg} D.~H.,  2016,
  \mn@doi [\mnras] {10.1093/mnras/stv2428}, \href
  {http://adsabs.harvard.edu/abs/2016MNRAS.455.1830T} {455, 1830}

\bibitem[\protect\citeauthoryear{{Tombesi}, {Mel{\'e}ndez}, {Veilleux},
  {Reeves}, {Gonz{\'a}lez-Alfonso}  \& {Reynolds}}{{Tombesi}
  et~al.}{2015}]{2015Natur.519..436T}
{Tombesi} F.,  {Mel{\'e}ndez} M.,  {Veilleux} S.,  {Reeves} J.~N.,
  {Gonz{\'a}lez-Alfonso} E.,   {Reynolds} C.~S.,  2015, \mn@doi [\nat]
  {10.1038/nature14261}, \href
  {http://adsabs.harvard.edu/abs/2015Natur.519..436T} {519, 436}

\bibitem[\protect\citeauthoryear{{Torii}, {Enokiya}, {Morris}, {Hasegawa},
  {Kudo}  \& {Fukui}}{{Torii} et~al.}{2014}]{2014ApJS..213....8T}
{Torii} K.,  {Enokiya} R.,  {Morris} M.~R.,  {Hasegawa} K.,  {Kudo} N.,
  {Fukui} Y.,  2014, \mn@doi [\apjs] {10.1088/0067-0049/213/1/8}, \href
  {http://adsabs.harvard.edu/abs/2014ApJS..213....8T} {213, 8}

\bibitem[\protect\citeauthoryear{{Tricco} \& {Price}}{{Tricco} \&
  {Price}}{2012}]{2012JCoPh.231.7214T}
{Tricco} T.~S.,  {Price} D.~J.,  2012, \mn@doi [Journal of Computational
  Physics] {10.1016/j.jcp.2012.06.039}, \href
  {http://adsabs.harvard.edu/abs/2012JCoPh.231.7214T} {231, 7214}

\bibitem[\protect\citeauthoryear{{Vazquez-Semadeni}}{{Vazquez-Semadeni}}{1994}]{1994ApJ...423..681V}
{Vazquez-Semadeni} E.,  1994, \mn@doi [\apj] {10.1086/173847}, \href
  {http://adsabs.harvard.edu/abs/1994ApJ...423..681V} {423, 681}

\bibitem[\protect\citeauthoryear{{Veilleux}, {Cecil}  \&
  {Bland-Hawthorn}}{{Veilleux} et~al.}{2005}]{2005ARA&A..43..769V}
{Veilleux} S.,  {Cecil} G.,   {Bland-Hawthorn} J.,  2005, \mn@doi [\araa]
  {10.1146/annurev.astro.43.072103.150610}, \href
  {http://adsabs.harvard.edu/abs/2005ARA%26A..43..769V} {43, 769}

\bibitem[\protect\citeauthoryear{{Veilleux}, {Bolatto}, {Tombesi}, {Melendez},
  {Sturm}, {Gonzalez-Alfonso}, {Fischer}  \& {Rupke}}{{Veilleux}
  et~al.}{2017}]{2017arXiv170600443V}
{Veilleux} S.,  {Bolatto} A.,  {Tombesi} F.,  {Melendez} M.,  {Sturm} E.,
  {Gonzalez-Alfonso} E.,  {Fischer} J.,   {Rupke} D.~S.~N.,  2017, preprint,
  \href {http://adsabs.harvard.edu/abs/2017arXiv170600443V} {} (\mn@eprint
  {arXiv} {1706.00443})

\bibitem[\protect\citeauthoryear{{Vieser} \& {Hensler}}{{Vieser} \&
  {Hensler}}{2007}]{2007AA...472..141V}
{Vieser} W.,  {Hensler} G.,  2007, \mn@doi [\aap] {10.1051/0004-6361:20042120},
  \href {http://adsabs.harvard.edu/abs/2007A%26A...472..141V} {472, 141}

\bibitem[\protect\citeauthoryear{{Vijayaraghavan} \& {Ricker}}{{Vijayaraghavan}
  \& {Ricker}}{2015}]{2015MNRAS.449.2312V}
{Vijayaraghavan} R.,  {Ricker} P.~M.,  2015, \mn@doi [\mnras]
  {10.1093/mnras/stv476}, \href
  {http://adsabs.harvard.edu/abs/2015MNRAS.449.2312V} {449, 2312}

\bibitem[\protect\citeauthoryear{{Vishniac}}{{Vishniac}}{1994}]{1994ApJ...428..186V}
{Vishniac} E.~T.,  1994, \mn@doi [\apj] {10.1086/174231}, \href
  {http://adsabs.harvard.edu/abs/1994ApJ...428..186V} {428, 186}

\bibitem[\protect\citeauthoryear{{Wang}}{{Wang}}{1995}]{1995ApJ...444..590W}
{Wang} B.,  1995, \mn@doi [\apj] {10.1086/175633}, \href
  {http://adsabs.harvard.edu/abs/1995ApJ...444..590W} {444, 590}

\bibitem[\protect\citeauthoryear{{Warhaft}}{{Warhaft}}{2000}]{2000AnRFM..32..203W}
{Warhaft} Z.,  2000, \mn@doi [Annual Review of Fluid Mechanics]
  {10.1146/annurev.fluid.32.1.203}, \href
  {http://adsabs.harvard.edu/abs/2000AnRFM..32..203W} {32, 203}

\bibitem[\protect\citeauthoryear{{Weidl}, {Jenko}, {Teaca}  \&
  {Schlickeiser}}{{Weidl} et~al.}{2015}]{2015ApJ...811....8W}
{Weidl} M.~S.,  {Jenko} F.,  {Teaca} B.,   {Schlickeiser} R.,  2015, \mn@doi
  [\apj] {10.1088/0004-637X/811/1/8}, \href
  {http://adsabs.harvard.edu/abs/2015ApJ...811....8W} {811, 8}

\bibitem[\protect\citeauthoryear{{Williams}}{{Williams}}{1999}]{1999intu.conf..190W}
{Williams} J.,  1999, in {Franco} J.,  {Carraminana} A.,  eds, Interstellar
  Turbulence. p.~190

\bibitem[\protect\citeauthoryear{{Wolfire}, {Tielens}  \&
  {Hollenbach}}{{Wolfire} et~al.}{1990}]{1990ApJ...358..116W}
{Wolfire} M.~G.,  {Tielens} A.~G.~G.~M.,   {Hollenbach} D.,  1990, \mn@doi
  [\apj] {10.1086/168966}, \href
  {http://adsabs.harvard.edu/abs/1990ApJ...358..116W} {358, 116}

\bibitem[\protect\citeauthoryear{{Wright}, {Drake}, {Drew}, {Guarcello},
  {Gutermuth}, {Hora}  \& {Kraemer}}{{Wright}
  et~al.}{2012}]{2012ApJ...746L..21W}
{Wright} N.~J.,  {Drake} J.~J.,  {Drew} J.~E.,  {Guarcello} M.~G.,  {Gutermuth}
  R.~A.,  {Hora} J.~L.,   {Kraemer} K.~E.,  2012, \mn@doi [\apjl]
  {10.1088/2041-8205/746/2/L21}, \href
  {http://adsabs.harvard.edu/abs/2012ApJ...746L..21W} {746, L21}

\bibitem[\protect\citeauthoryear{{Xu} \& {Stone}}{{Xu} \&
  {Stone}}{1995}]{1995ApJ...454..172X}
{Xu} J.,  {Stone} J.~M.,  1995, \mn@doi [\apj] {10.1086/176475}, \href
  {http://adsabs.harvard.edu/abs/1995ApJ...454..172X} {454, 172}

\bibitem[\protect\citeauthoryear{{Yamada} \& {Nishi}}{{Yamada} \&
  {Nishi}}{2001}]{2001ApJ...547...99Y}
{Yamada} M.,  {Nishi} R.,  2001, \mn@doi [\apj] {10.1086/318329}, \href
  {http://adsabs.harvard.edu/abs/2001ApJ...547...99Y} {547, 99}

\bibitem[\protect\citeauthoryear{{Yan} \& {Lazarian}}{{Yan} \&
  {Lazarian}}{2002}]{2002PhRvL..89B1102Y}
{Yan} H.,  {Lazarian} A.,  2002, \mn@doi [Physical Review Letters]
  {10.1103/PhysRevLett.89.281102}, \href
  {http://adsabs.harvard.edu/abs/2002PhRvL..89B1102Y} {89, 1102}

\bibitem[\protect\citeauthoryear{{Yirak}, {Frank}  \& {Cunningham}}{{Yirak}
  et~al.}{2010}]{2010ApJ...722..412Y}
{Yirak} K.,  {Frank} A.,   {Cunningham} A.~J.,  2010, \mn@doi [\apj]
  {10.1088/0004-637X/722/1/412}, \href
  {http://adsabs.harvard.edu/abs/2010ApJ...722..412Y} {722, 412}

\bibitem[\protect\citeauthoryear{{Yusef-Zadeh}, {Morris}  \&
  {Chance}}{{Yusef-Zadeh} et~al.}{1984}]{1984Natur.310..557Y}
{Yusef-Zadeh} F.,  {Morris} M.,   {Chance} D.,  1984, \mn@doi [\nat]
  {10.1038/310557a0}, \href {http://adsabs.harvard.edu/abs/1984Natur.310..557Y}
  {310, 557}

\bibitem[\protect\citeauthoryear{{Yusef-Zadeh}, {Hewitt}  \&
  {Cotton}}{{Yusef-Zadeh} et~al.}{2004}]{2004ApJS..155..421Y}
{Yusef-Zadeh} F.,  {Hewitt} J.~W.,   {Cotton} W.,  2004, \mn@doi [\apjs]
  {10.1086/425257}, \href {http://adsabs.harvard.edu/abs/2004ApJS..155..421Y}
  {155, 421}

\bibitem[\protect\citeauthoryear{{Zhang} \& {Thompson}}{{Zhang} \&
  {Thompson}}{2012}]{2012MNRAS.424.1170Z}
{Zhang} D.,  {Thompson} T.~A.,  2012, \mn@doi [\mnras]
  {10.1111/j.1365-2966.2012.21291.x}, \href
  {http://adsabs.harvard.edu/abs/2012MNRAS.424.1170Z} {424, 1170}

\bibitem[\protect\citeauthoryear{{Zhang}, {Thompson}, {Quataert}  \&
  {Murray}}{{Zhang} et~al.}{2015}]{2015arXiv150701951Z}
{Zhang} D.,  {Thompson} T.~A.,  {Quataert} E.,   {Murray} N.,  2015, preprint,
  \href {http://adsabs.harvard.edu/abs/2015arXiv150701951Z} {} (\mn@eprint
  {arXiv} {1507.01951})

\bibitem[\protect\citeauthoryear{{van Loo}, {Falle}, {Hartquist}  \&
  {Moore}}{{van Loo} et~al.}{2007}]{2007A&A...471..213V}
{van Loo} S.,  {Falle} S.~A.~E.~G.,  {Hartquist} T.~W.,   {Moore} T.~J.~T.,
  2007, \mn@doi [\aap] {10.1051/0004-6361:20077430}, \href
  {http://adsabs.harvard.edu/abs/2007A%26A...471..213V} {471, 213}

\bibitem[\protect\citeauthoryear{{van Loo}, {Falle}  \& {Hartquist}}{{van Loo}
  et~al.}{2010}]{2010MNRAS.406.1260V}
{van Loo} S.,  {Falle} S.~A.~E.~G.,   {Hartquist} T.~W.,  2010, \mn@doi
  [\mnras] {10.1111/j.1365-2966.2010.16761.x}, \href
  {http://adsabs.harvard.edu/abs/2010MNRAS.406.1260V} {406, 1260}

\makeatother
\end{thebibliography}



\appendix

\section{Effects of numerical resolution}
\label{sec:Appendix3}

\begin{figure*}
\begin{center}
  \begin{tabular}{c c c}
 \textbf{Filament (Cloud)} &  \textbf{Filament Tail (Cloud Envelope)} & \textbf{Filament Footpoint (Cloud Core)}\\
  \resizebox{60mm}{!}{\includegraphics{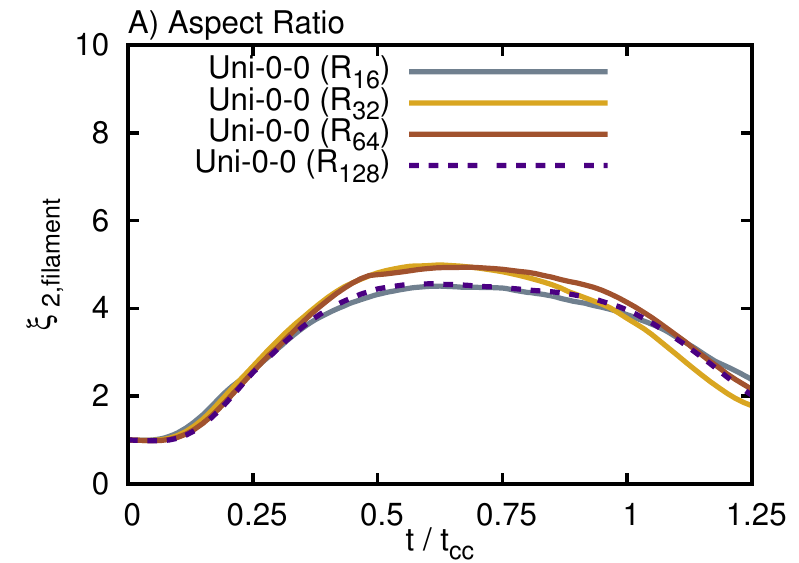}} & \hspace{-0.7cm}\resizebox{60mm}{!}{\includegraphics{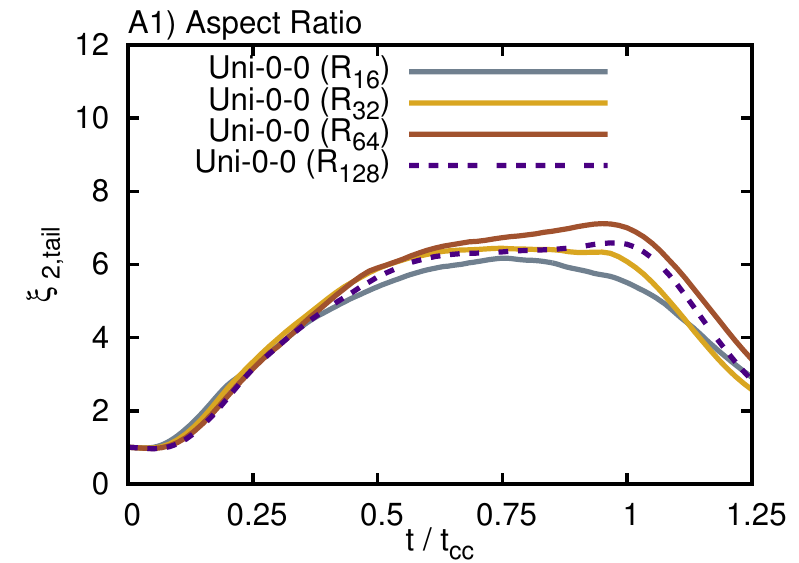}} & \hspace{-0.7cm}\resizebox{60mm}{!}{\includegraphics{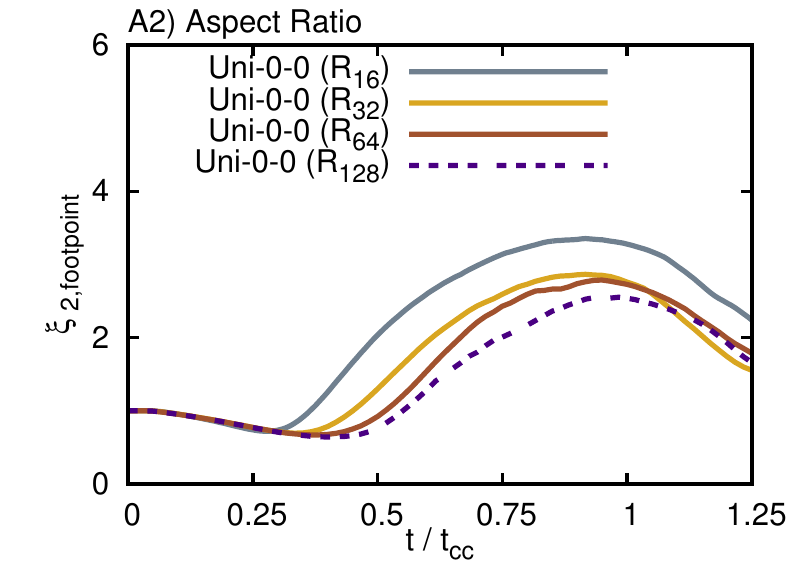}}\\
 \resizebox{60mm}{!}{\includegraphics{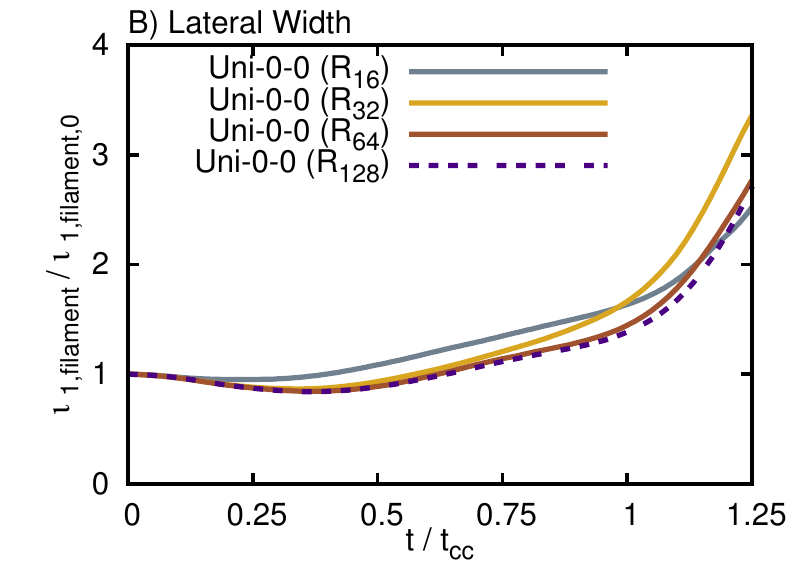}} & \hspace{-0.7cm}\resizebox{60mm}{!}{\includegraphics{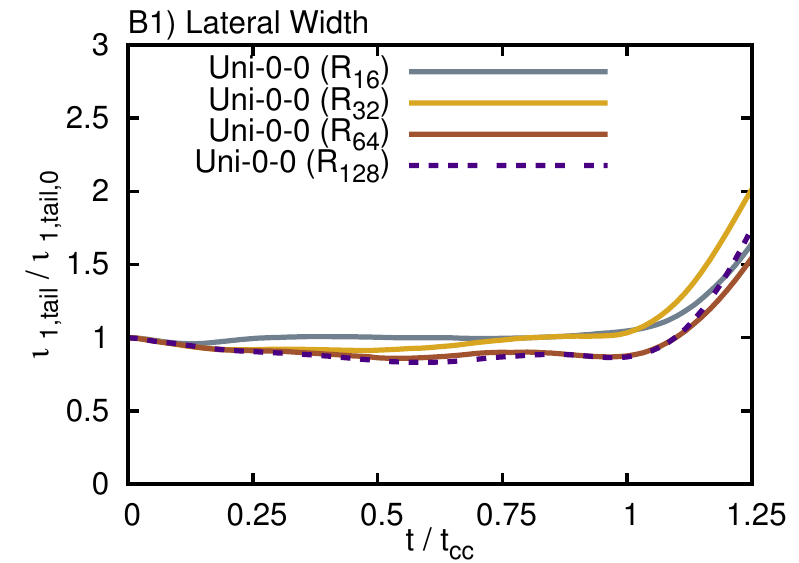}} & \hspace{-0.7cm}\resizebox{60mm}{!}{\includegraphics{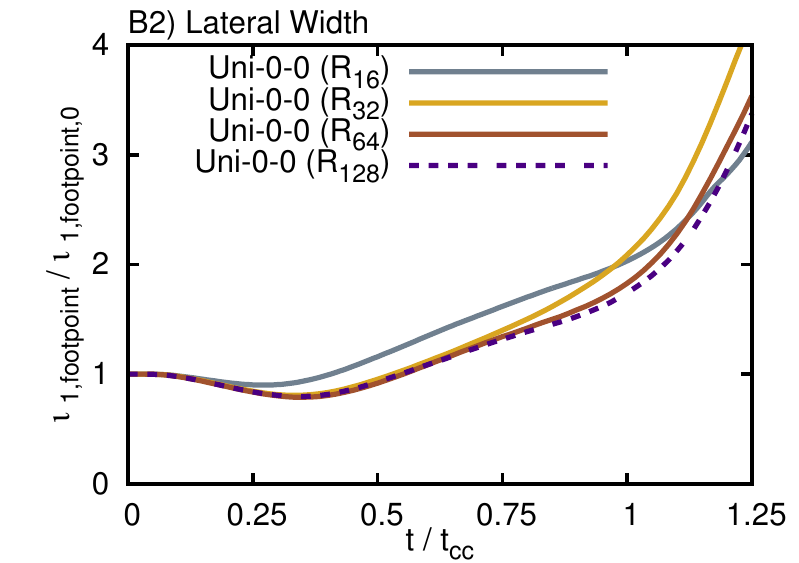}}\\
 \resizebox{60mm}{!}{\includegraphics{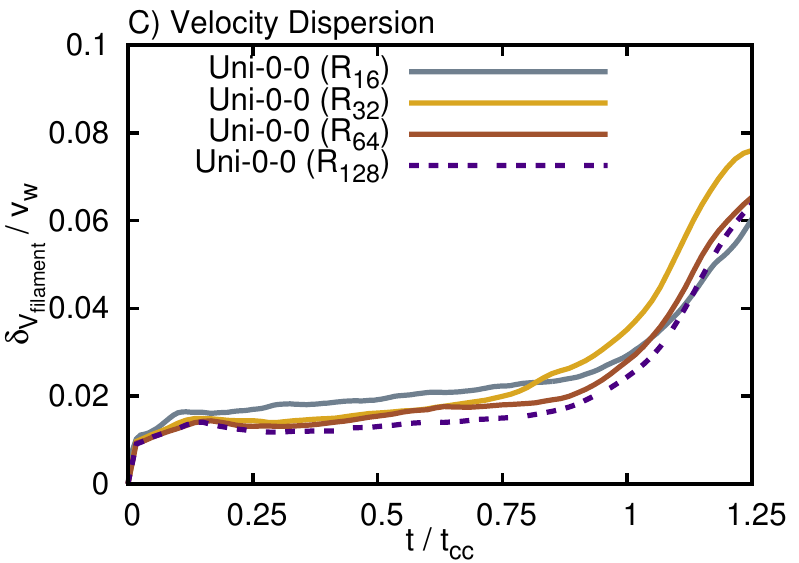}} &  \hspace{-0.7cm}\resizebox{60mm}{!}{\includegraphics{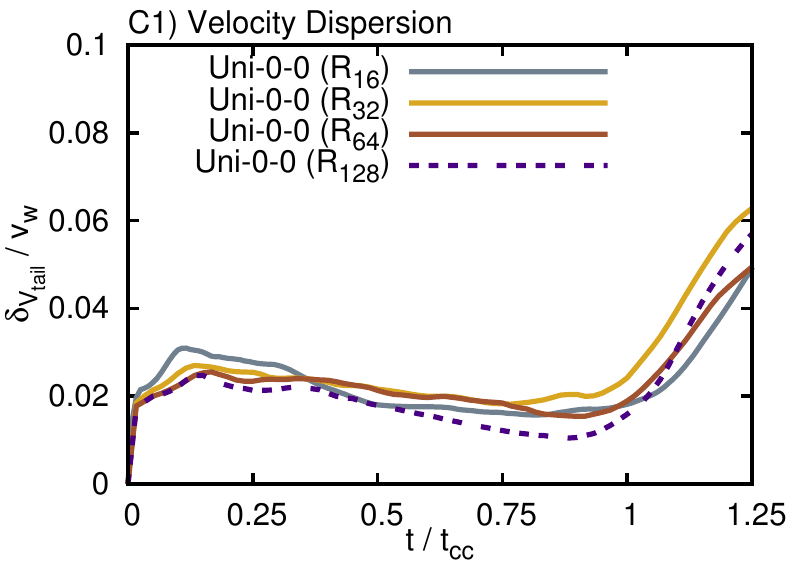}} & \hspace{-0.7cm}\resizebox{60mm}{!}{\includegraphics{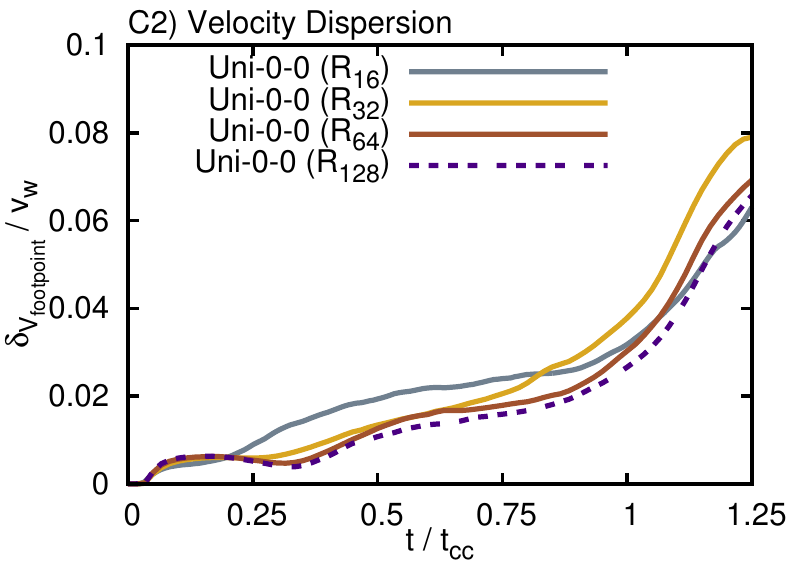}}\\
\resizebox{60mm}{!}{\includegraphics{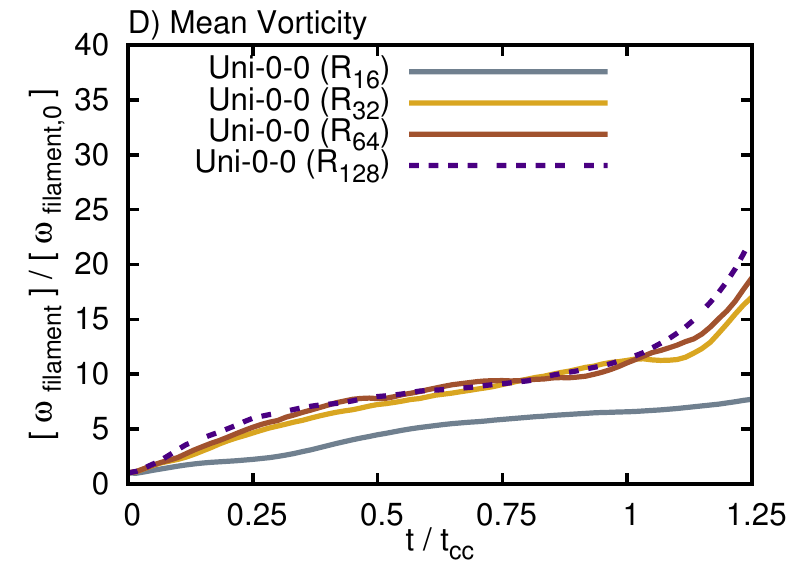}} &  \hspace{-0.7cm}\resizebox{60mm}{!}{\includegraphics{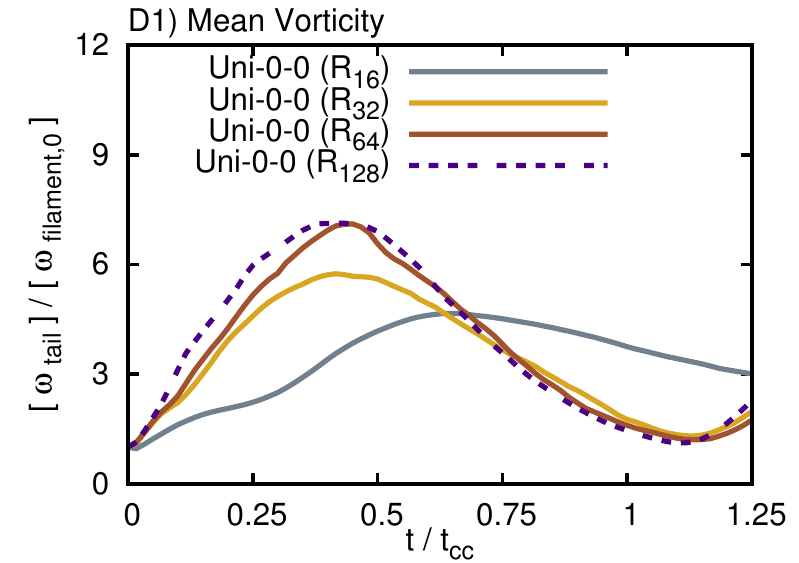}} & \hspace{-0.7cm}\resizebox{60mm}{!}{\includegraphics{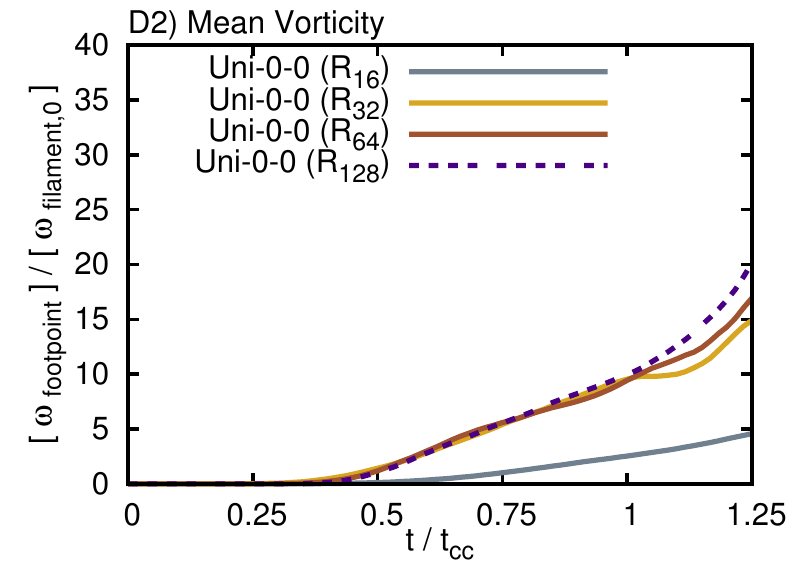}}\\
  \end{tabular}
  \caption{Time evolution of the diagnostics reported in Figures \ref{Figure5} and \ref{Figure9} in model Uni-0-0 at different resolutions ($R_{16-128}$).}
  \label{FigureC1}
\end{center}
\end{figure*}

\begin{figure*}
\begin{center}
  \begin{tabular}{c c c}
 \textbf{Filament (Cloud)} &  \textbf{Filament Tail (Cloud Envelope)} & \textbf{Filament Footpoint (Cloud Core)}\\
  \resizebox{60mm}{!}{\includegraphics{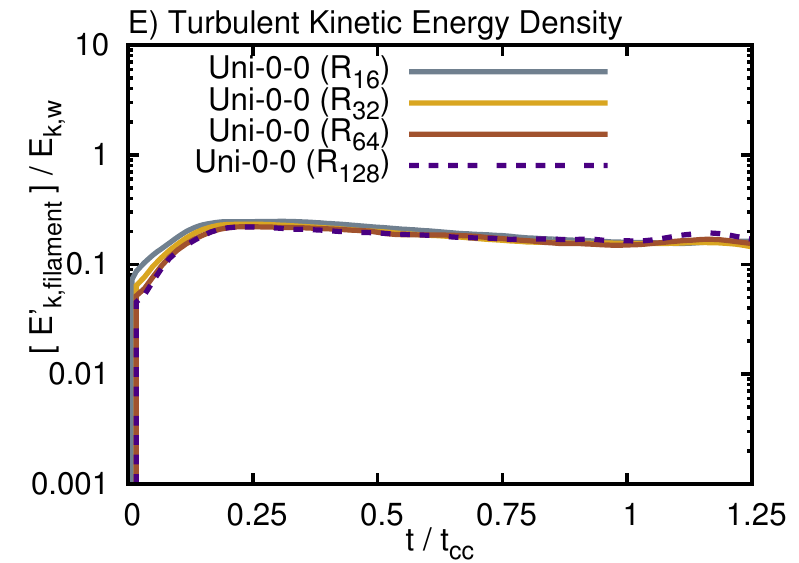}} & \hspace{-0.7cm}\resizebox{60mm}{!}{\includegraphics{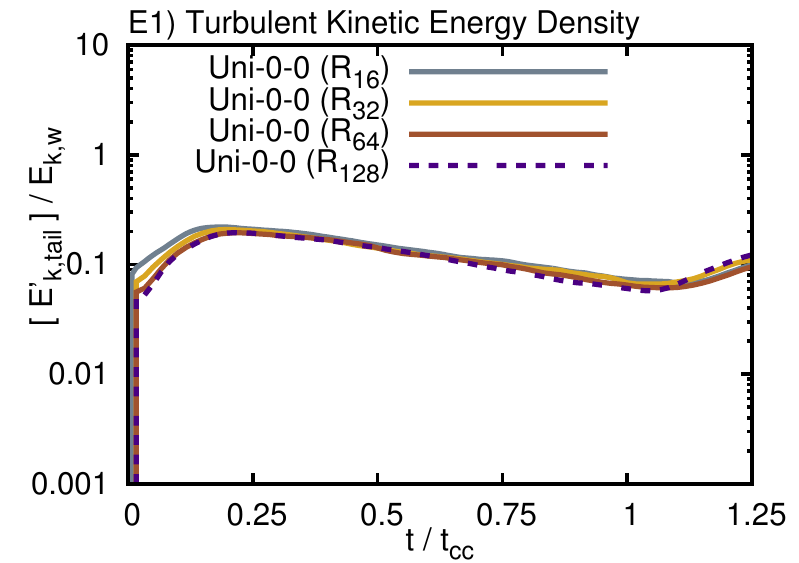}} & \hspace{-0.7cm}\resizebox{60mm}{!}{\includegraphics{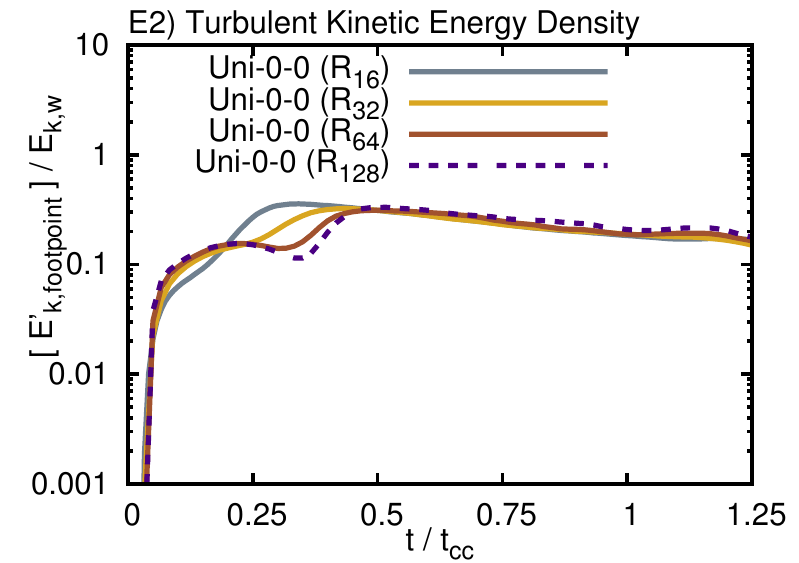}}\\
 \resizebox{60mm}{!}{\includegraphics{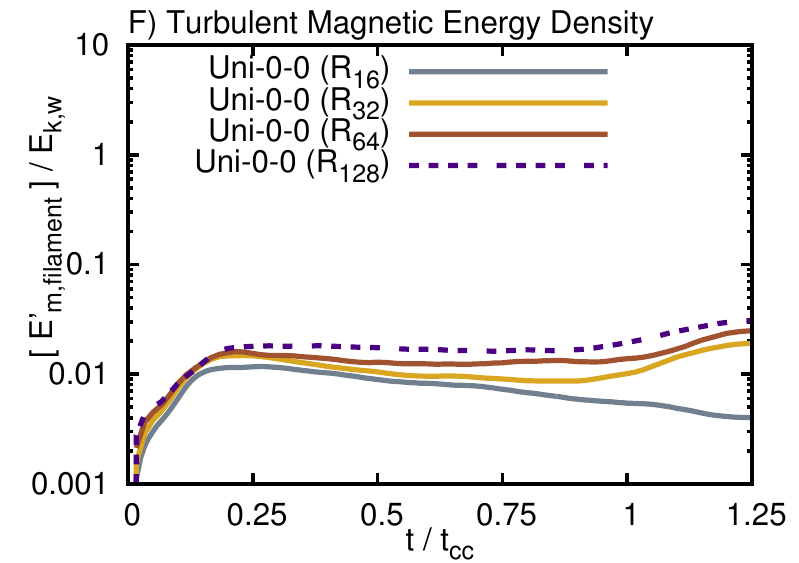}} & \hspace{-0.7cm}\resizebox{60mm}{!}{\includegraphics{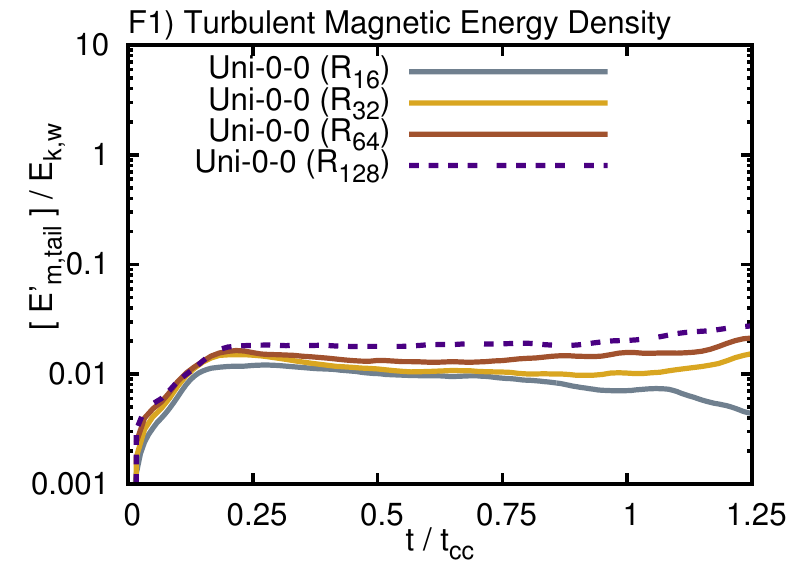}} & \hspace{-0.7cm}\resizebox{60mm}{!}{\includegraphics{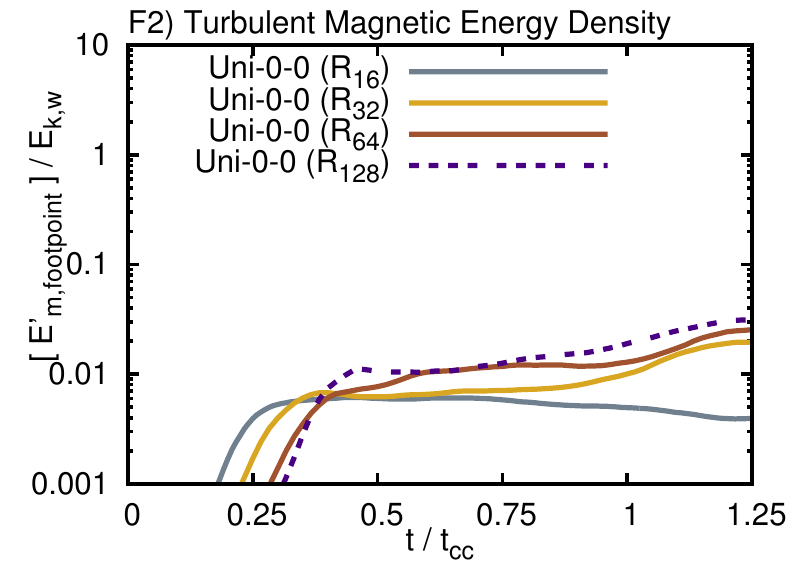}}\\
 \resizebox{60mm}{!}{\includegraphics{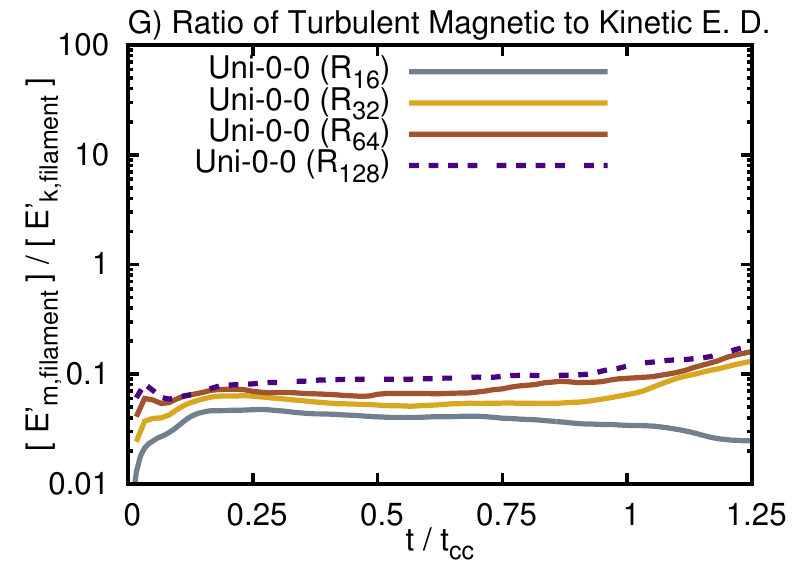}} &  \hspace{-0.7cm}\resizebox{60mm}{!}{\includegraphics{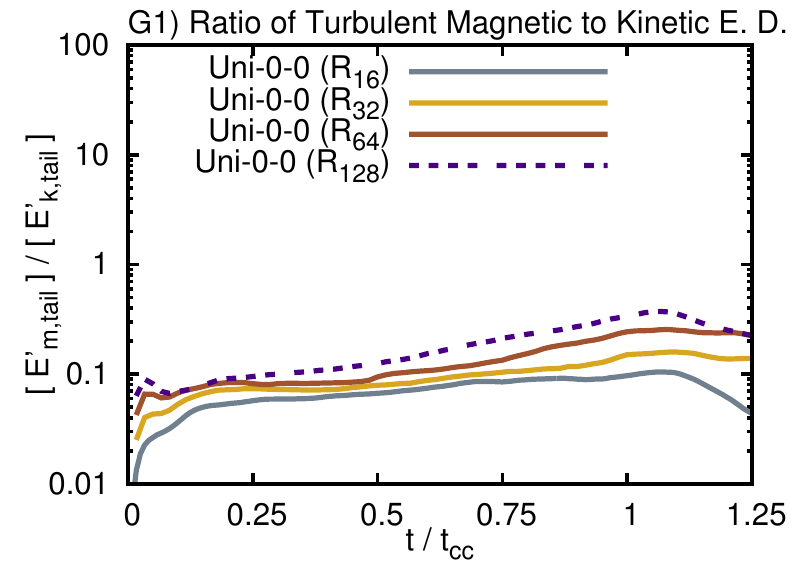}} & \hspace{-0.7cm}\resizebox{60mm}{!}{\includegraphics{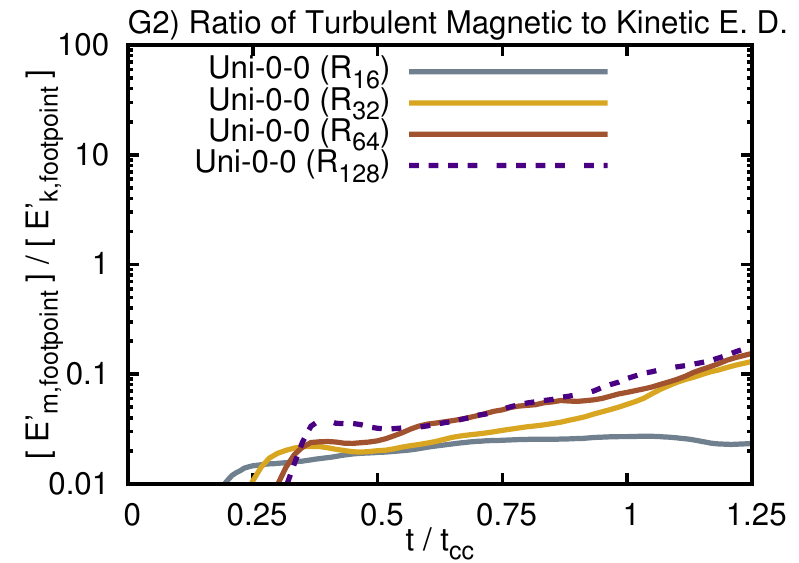}}\\
 \resizebox{60mm}{!}{\includegraphics{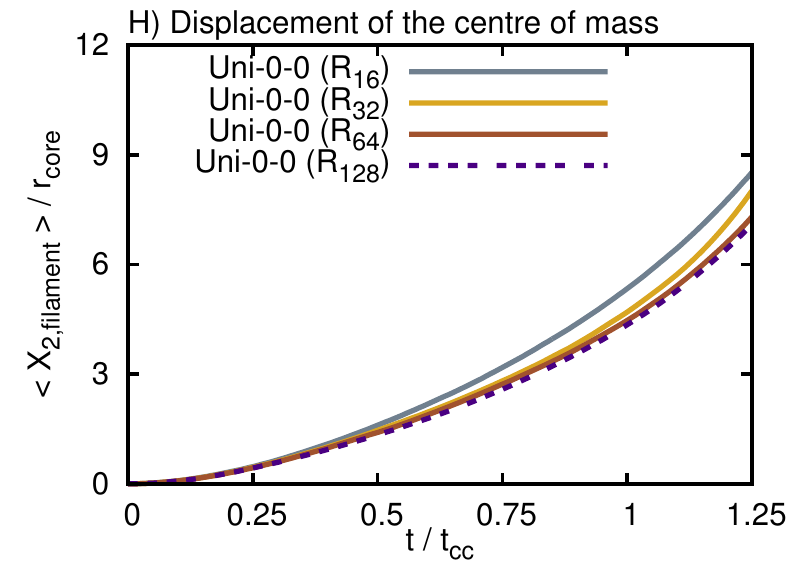}} &  \hspace{-0.7cm}\resizebox{60mm}{!}{\includegraphics{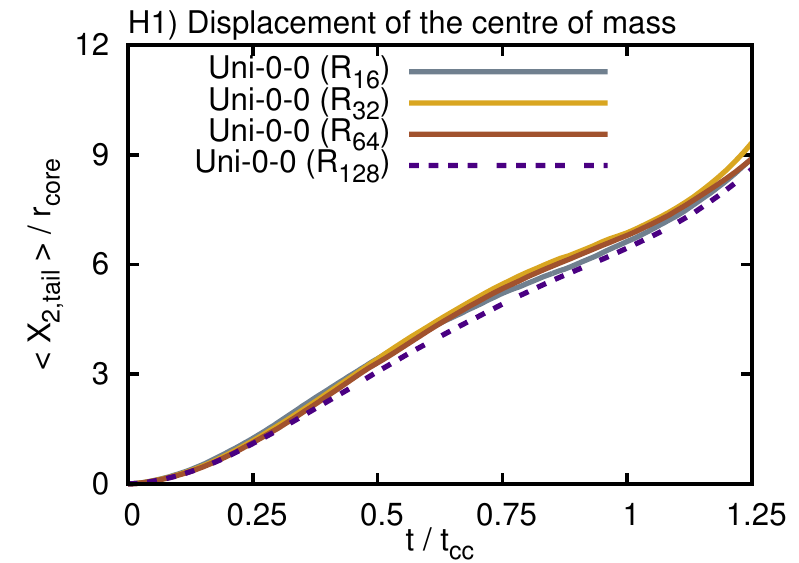}} & \hspace{-0.7cm}\resizebox{60mm}{!}{\includegraphics{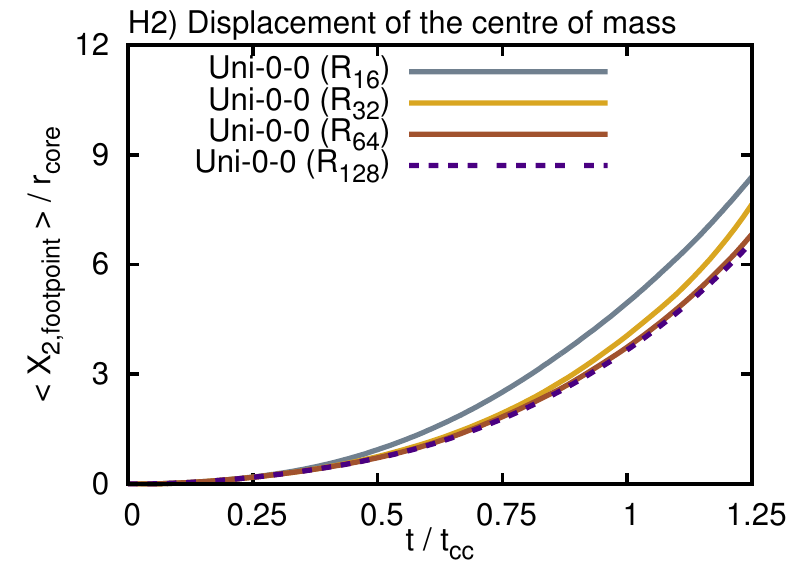}}\\
 \resizebox{60mm}{!}{\includegraphics{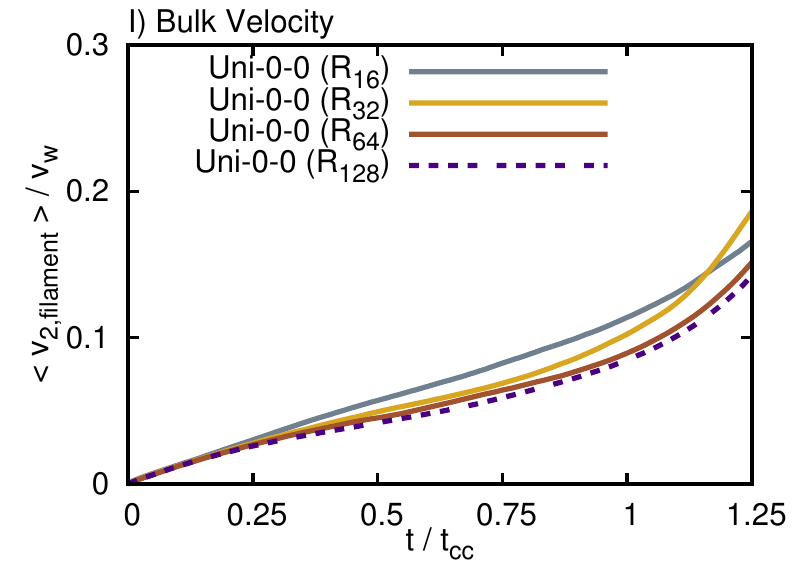}} &  \hspace{-0.7cm}\resizebox{60mm}{!}{\includegraphics{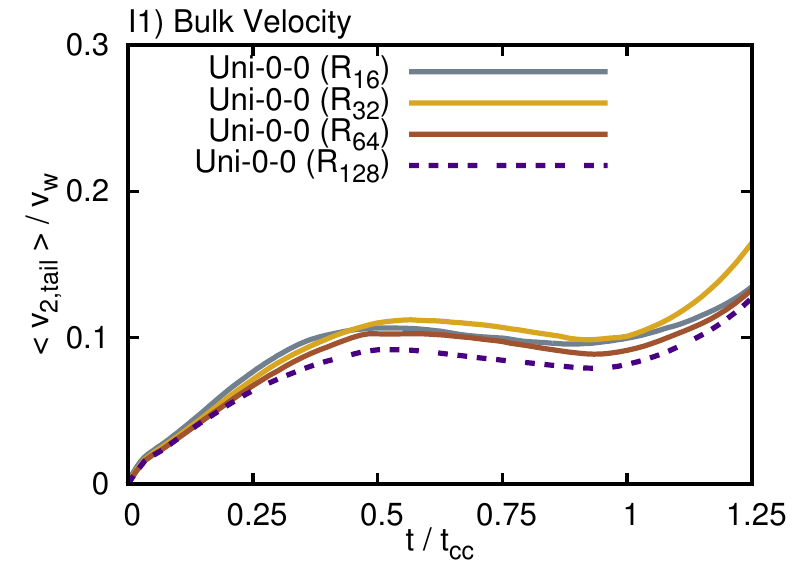}} & \hspace{-0.7cm}\resizebox{60mm}{!}{\includegraphics{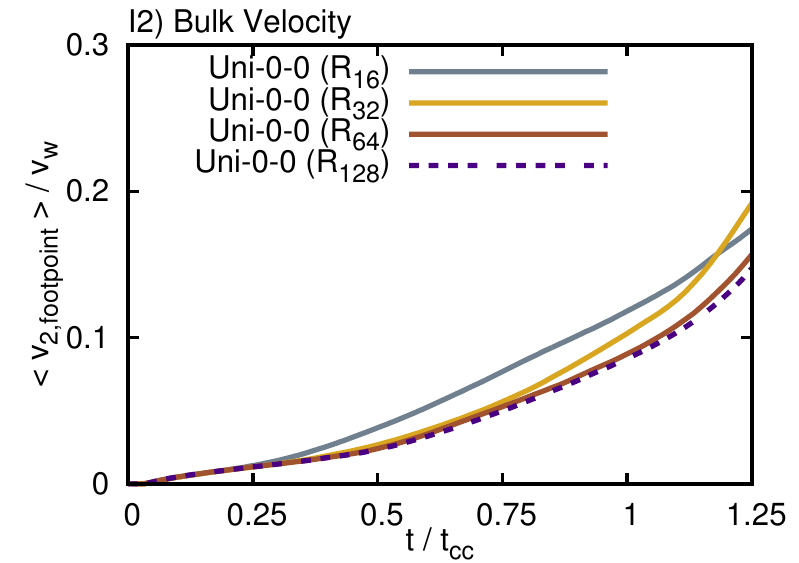}}\\
  \end{tabular}
  \caption{Time evolution of the diagnostics reported in Figures \ref{Figure6}, \ref{Figure7}, \ref{Figure10}, and \ref{Figure11} in model Uni-0-0 at different resolutions ($R_{16-128}$).}
  \label{FigureC2}
\end{center}
\end{figure*}

\begin{figure*}
\begin{center}
  \begin{tabular}{c c c}
 \textbf{Filament (Cloud)} &  \textbf{Filament Tail (Cloud Envelope)} & \textbf{Filament Footpoint (Cloud Core)}\\
  \resizebox{60mm}{!}{\includegraphics{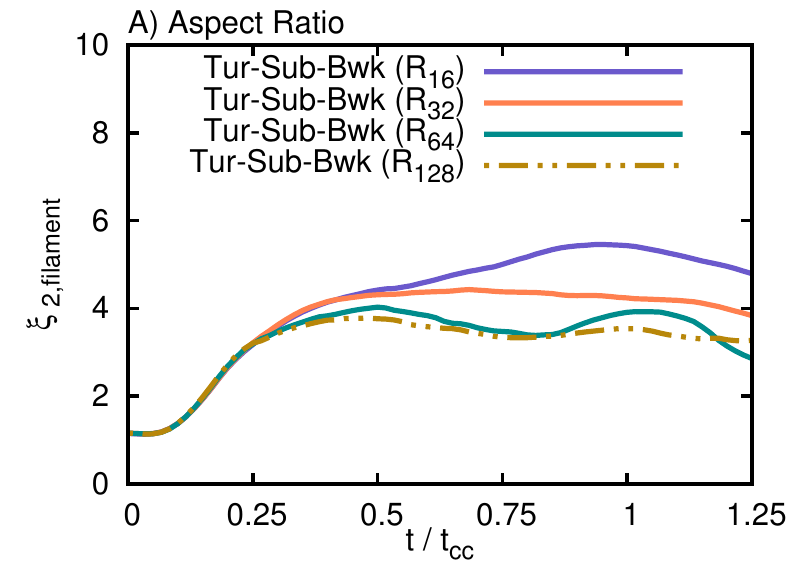}} & \hspace{-0.7cm}\resizebox{60mm}{!}{\includegraphics{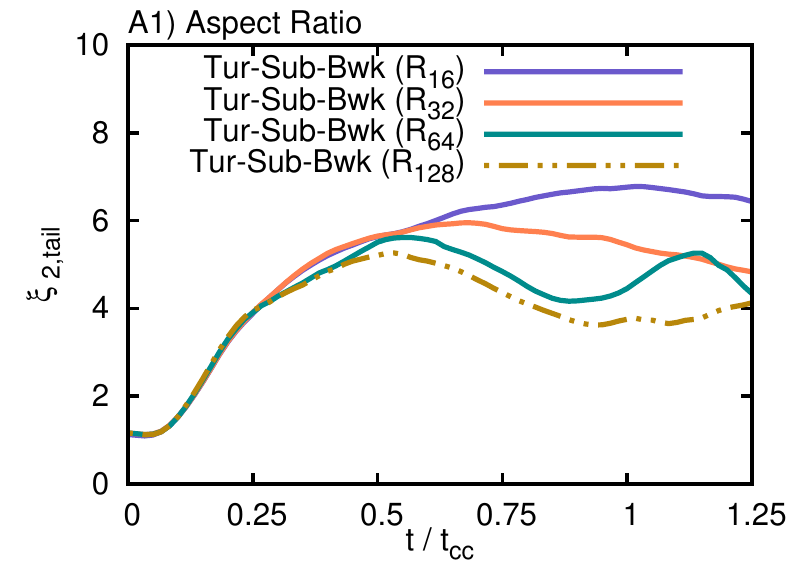}} & \hspace{-0.7cm}\resizebox{60mm}{!}{\includegraphics{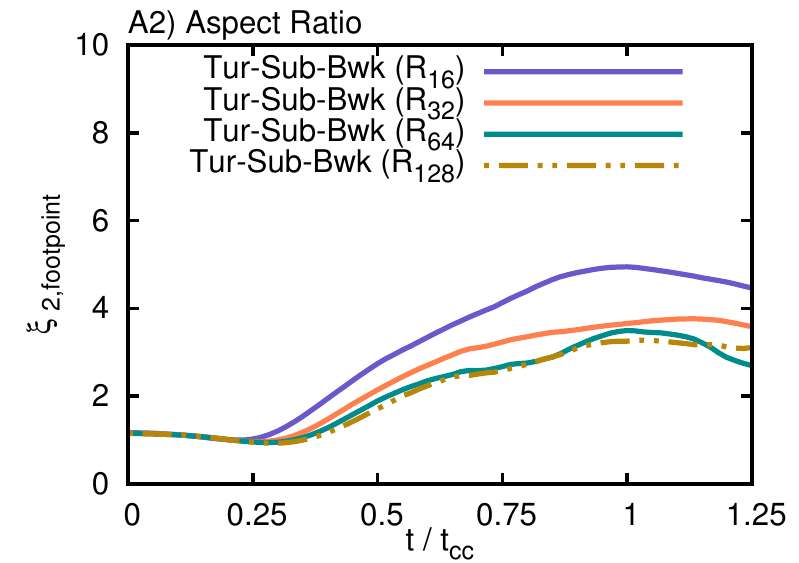}}\\
 \resizebox{60mm}{!}{\includegraphics{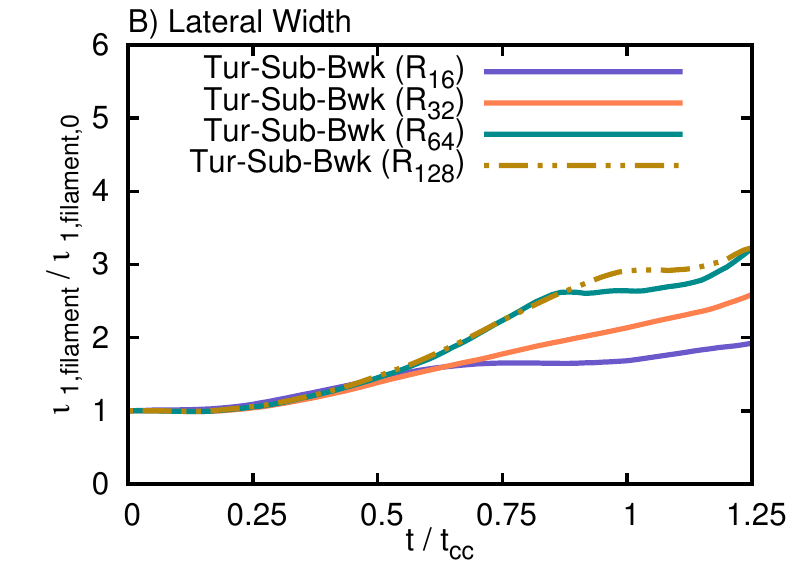}} & \hspace{-0.7cm}\resizebox{60mm}{!}{\includegraphics{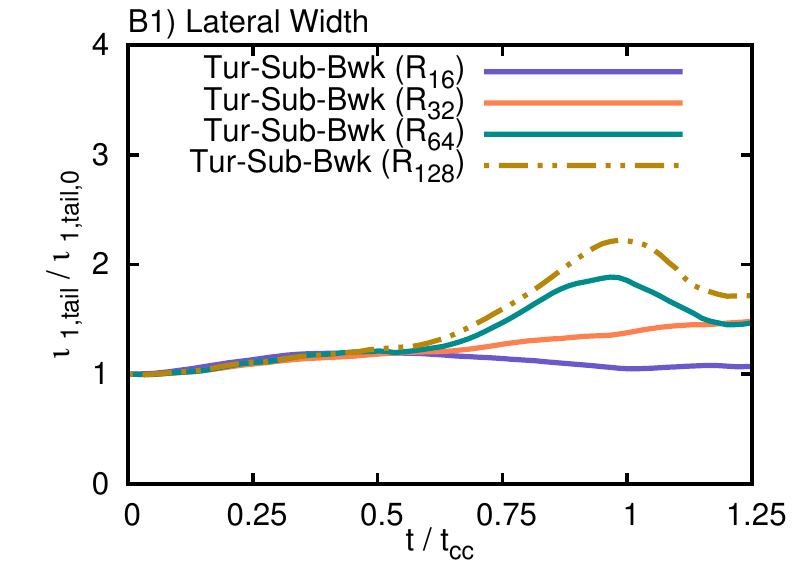}} & \hspace{-0.7cm}\resizebox{60mm}{!}{\includegraphics{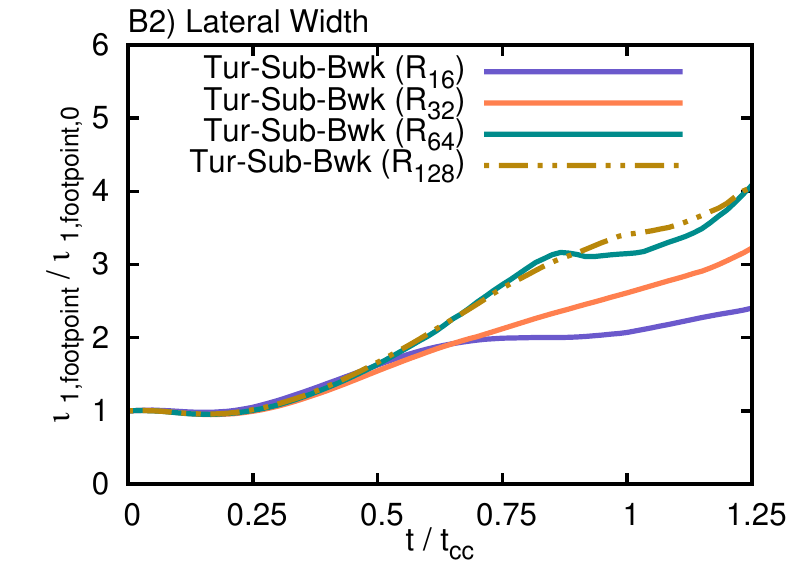}}\\
 \resizebox{60mm}{!}{\includegraphics{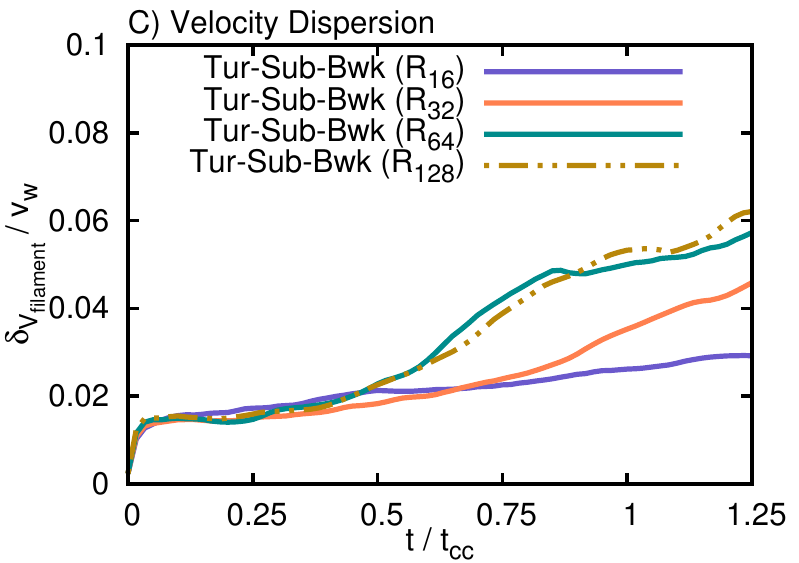}} &  \hspace{-0.7cm}\resizebox{60mm}{!}{\includegraphics{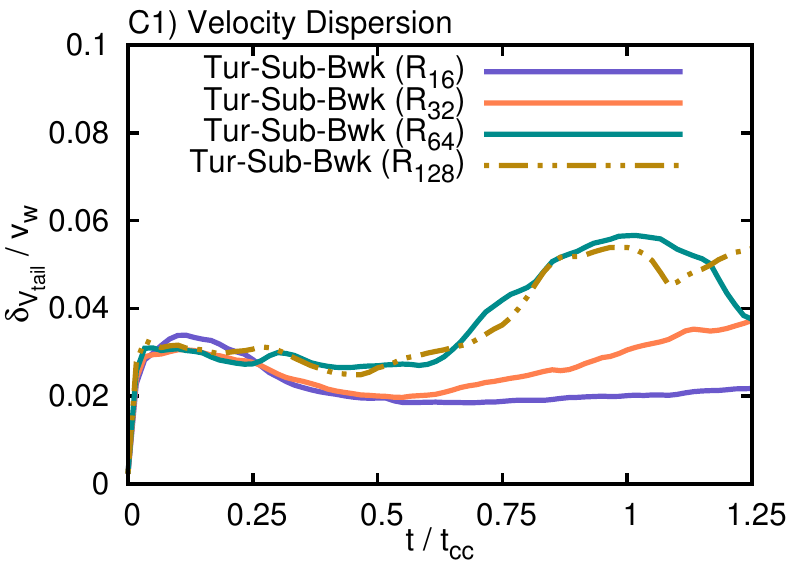}} & \hspace{-0.7cm}\resizebox{60mm}{!}{\includegraphics{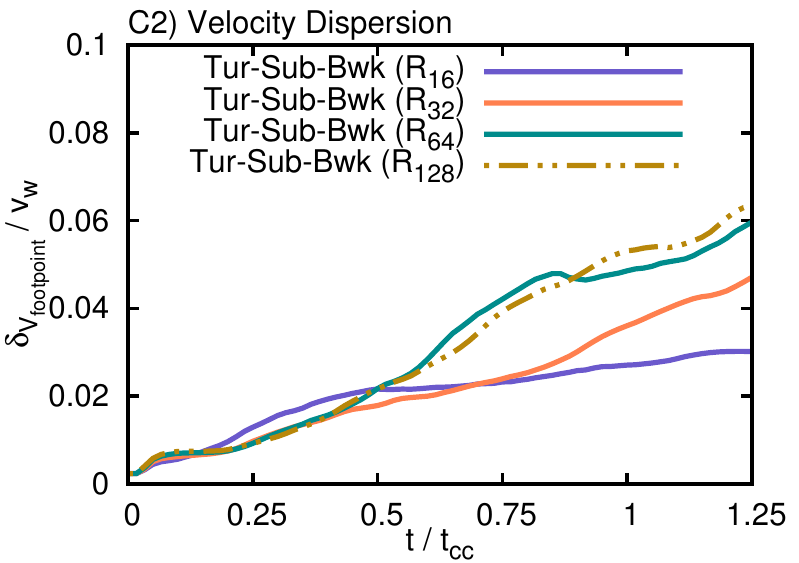}}\\
 \resizebox{60mm}{!}{\includegraphics{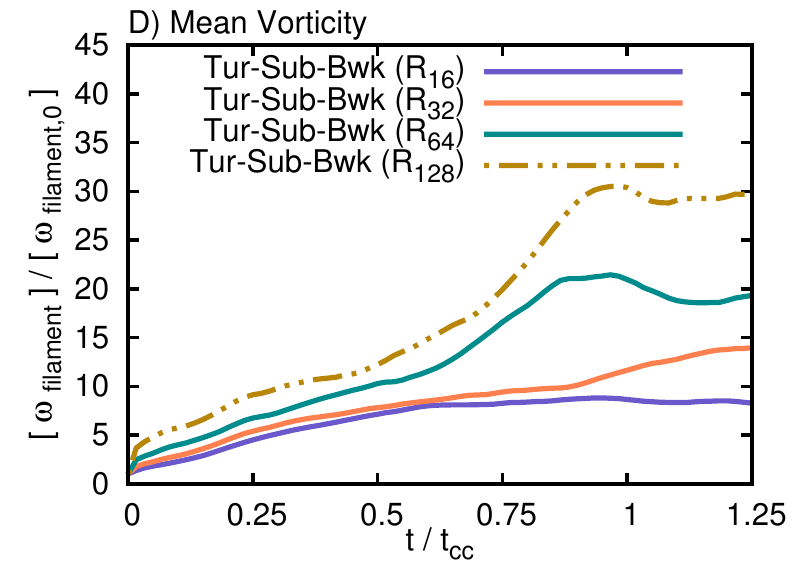}} &  \hspace{-0.7cm}\resizebox{60mm}{!}{\includegraphics{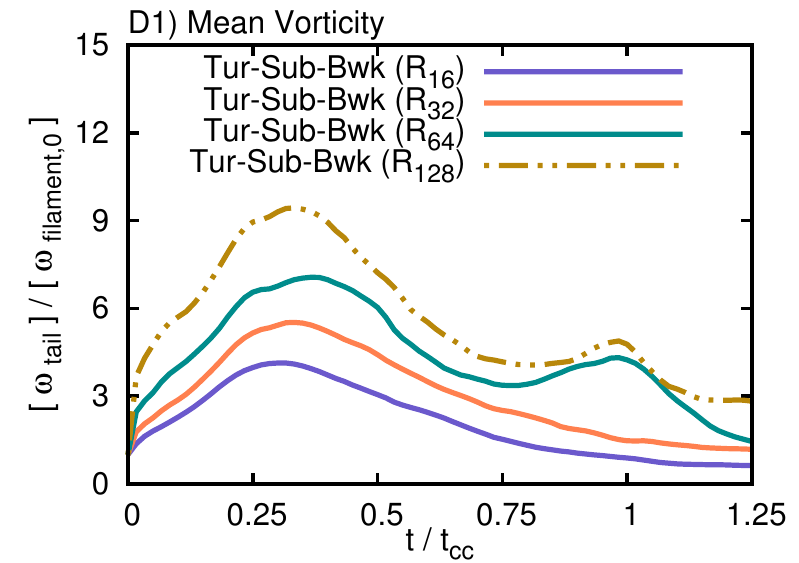}} & \hspace{-0.7cm}\resizebox{60mm}{!}{\includegraphics{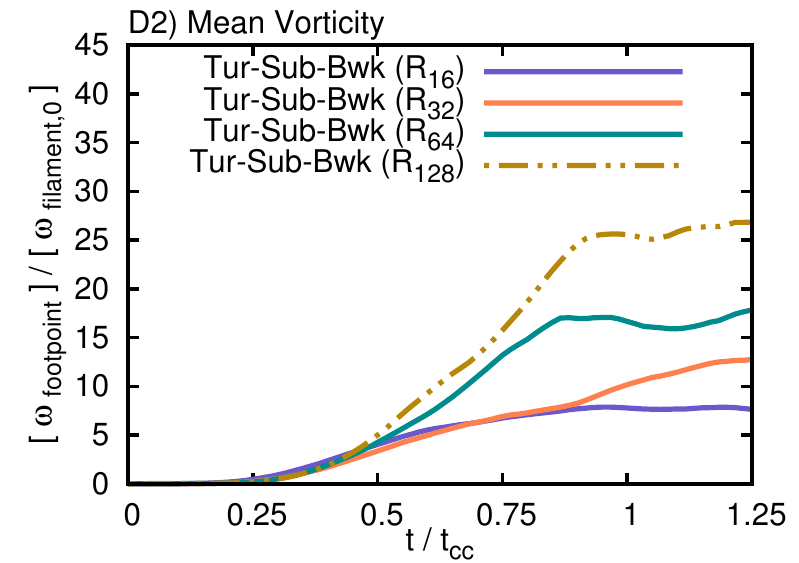}}\\
  \end{tabular}
  \caption{Time evolution of the diagnostics reported in Figures \ref{Figure5} and \ref{Figure9} in model Tur-Sub-Bwk at different resolutions ($R_{16-128}$).}
  \label{FigureC3}
\end{center}
\end{figure*}

\begin{figure*}
\begin{center}
  \begin{tabular}{c c c}
 \textbf{Filament (Cloud)} &  \textbf{Filament Tail (Cloud Envelope)} & \textbf{Filament Footpoint (Cloud Core)}\\
  \resizebox{60mm}{!}{\includegraphics{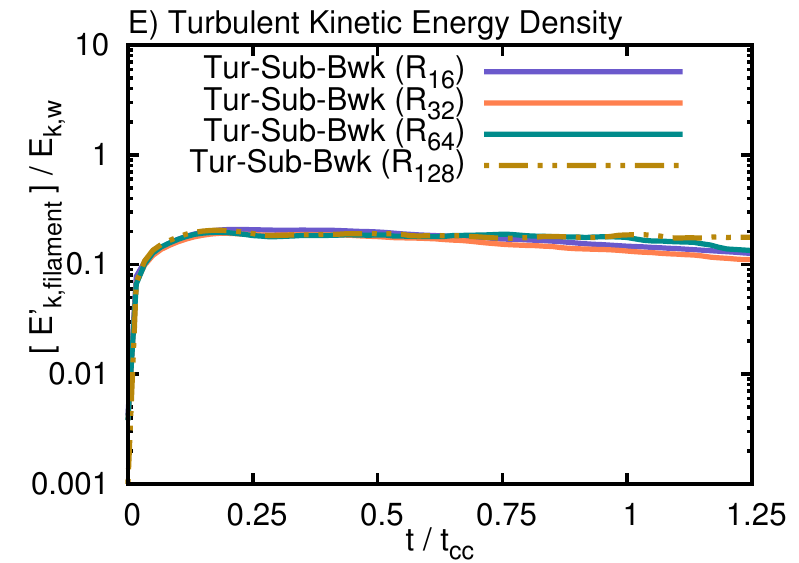}} & \hspace{-0.7cm}\resizebox{60mm}{!}{\includegraphics{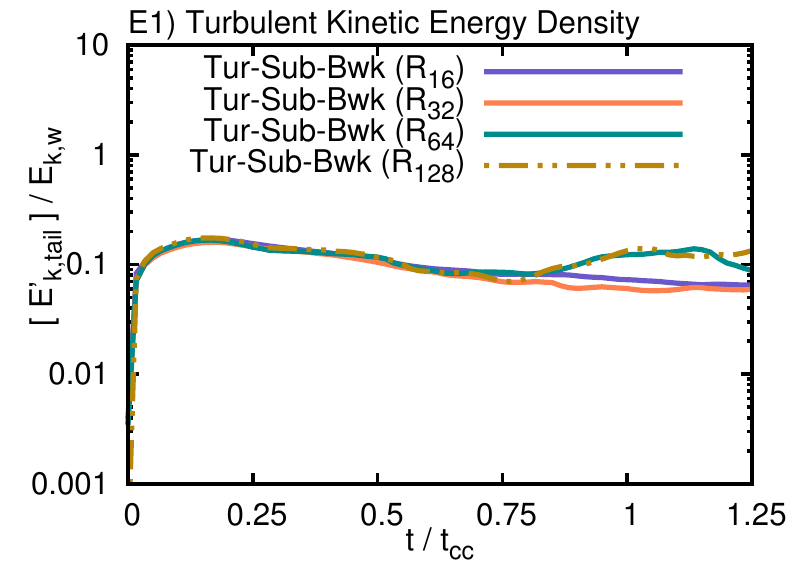}} & \hspace{-0.7cm}\resizebox{60mm}{!}{\includegraphics{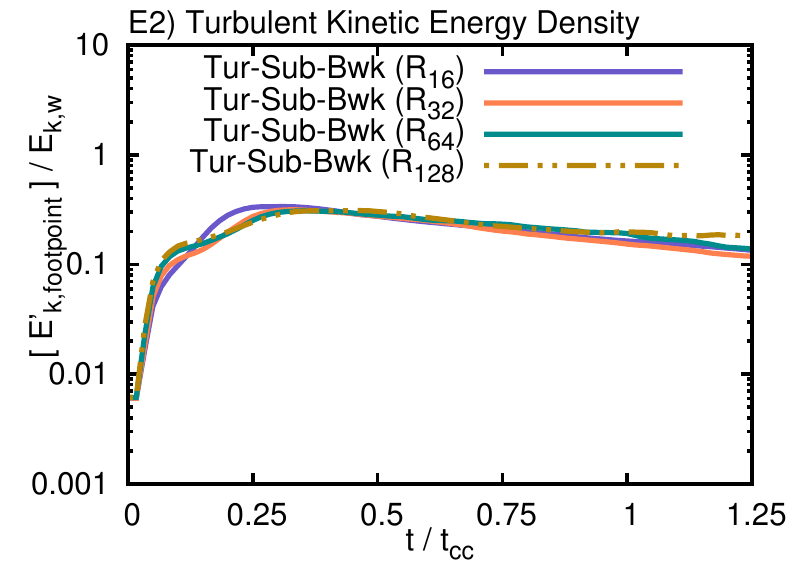}}\\
 \resizebox{60mm}{!}{\includegraphics{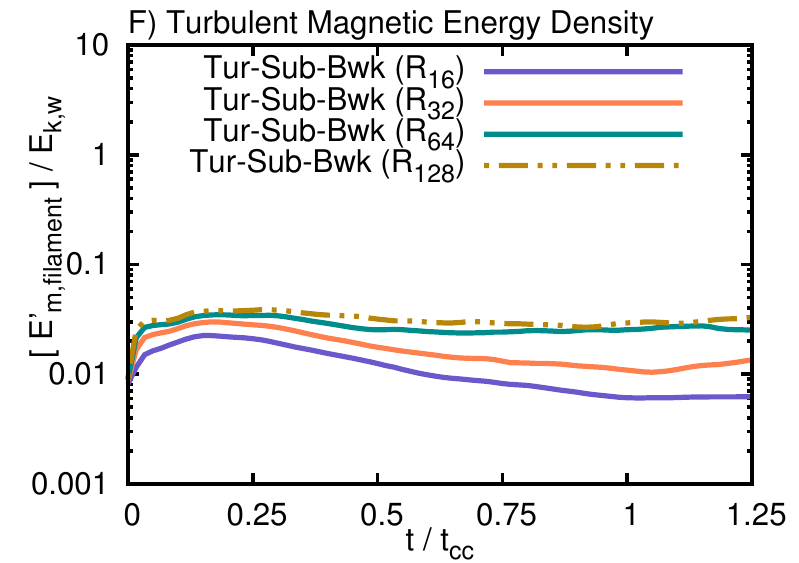}} & \hspace{-0.7cm}\resizebox{60mm}{!}{\includegraphics{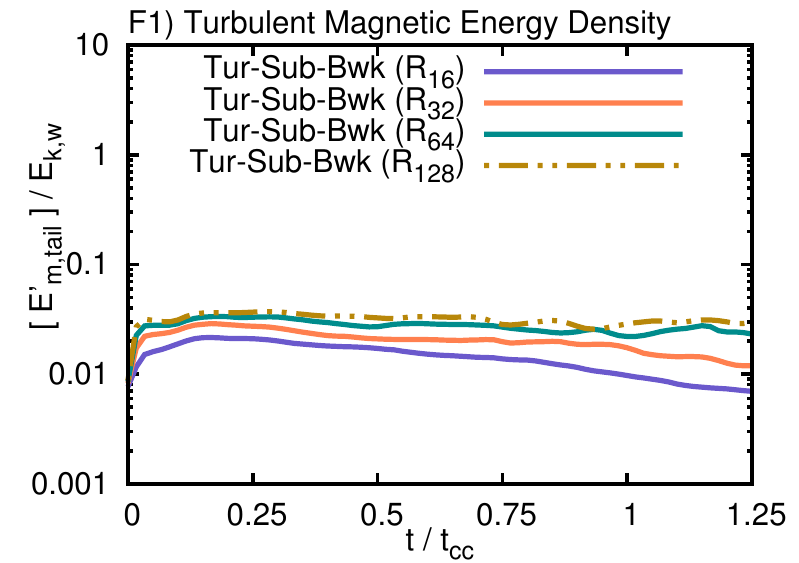}} & \hspace{-0.7cm}\resizebox{60mm}{!}{\includegraphics{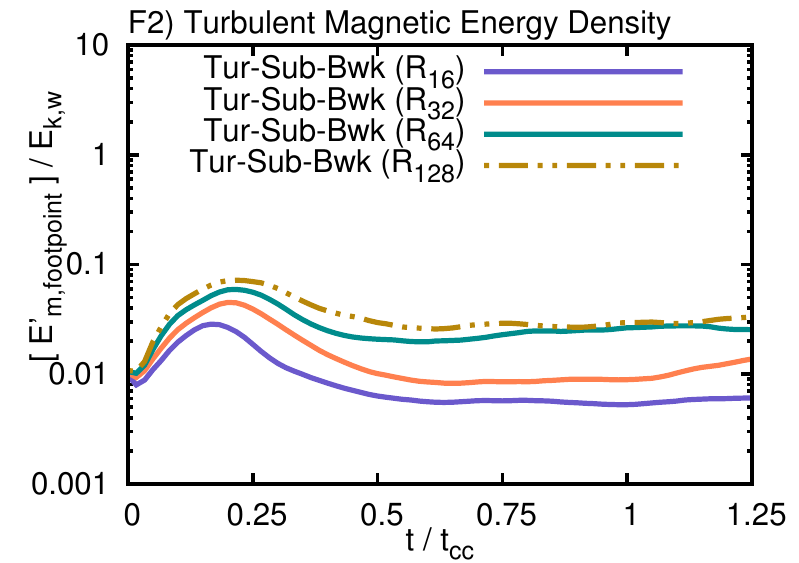}}\\
 \resizebox{60mm}{!}{\includegraphics{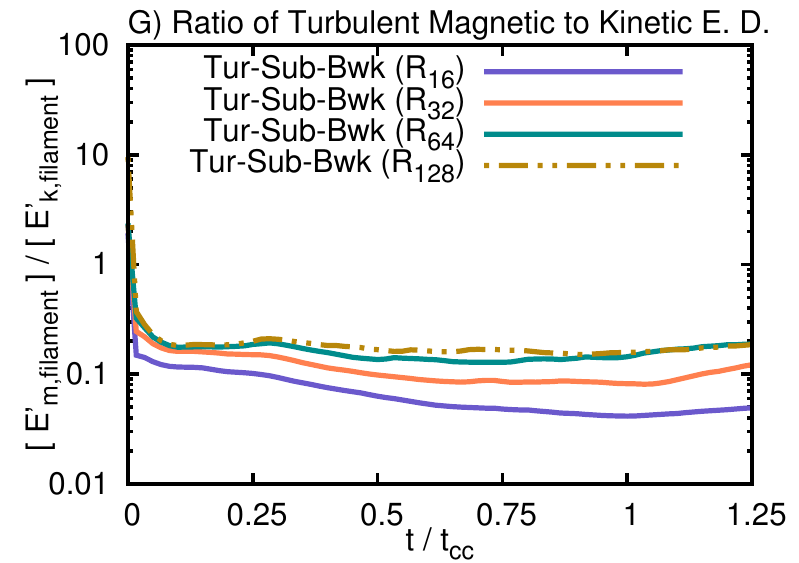}} &  \hspace{-0.7cm}\resizebox{60mm}{!}{\includegraphics{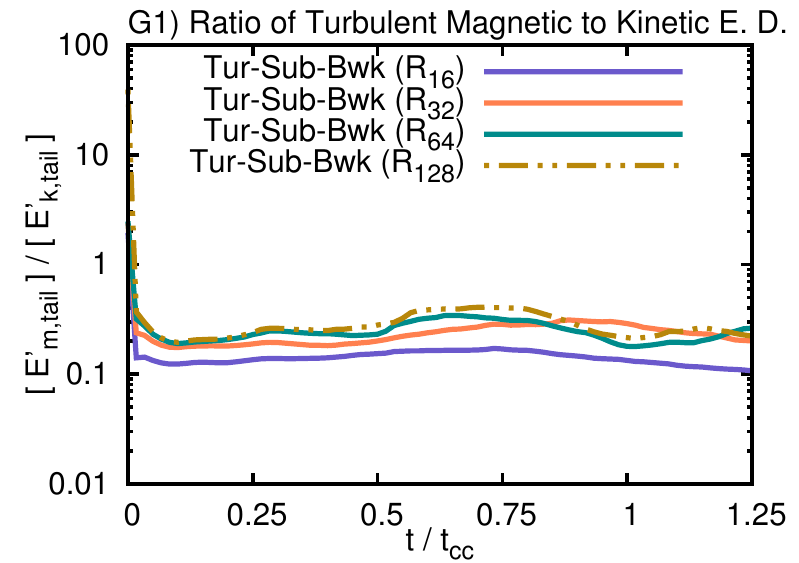}} & \hspace{-0.7cm}\resizebox{60mm}{!}{\includegraphics{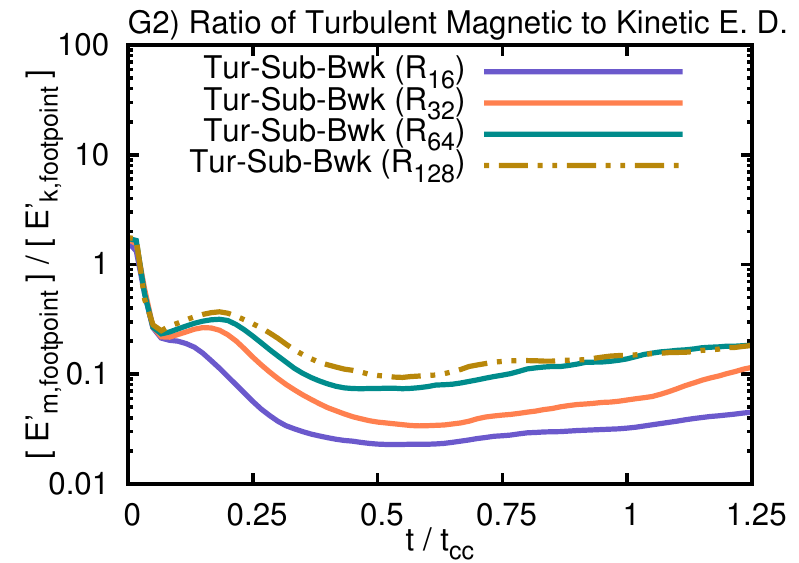}}\\
 \resizebox{60mm}{!}{\includegraphics{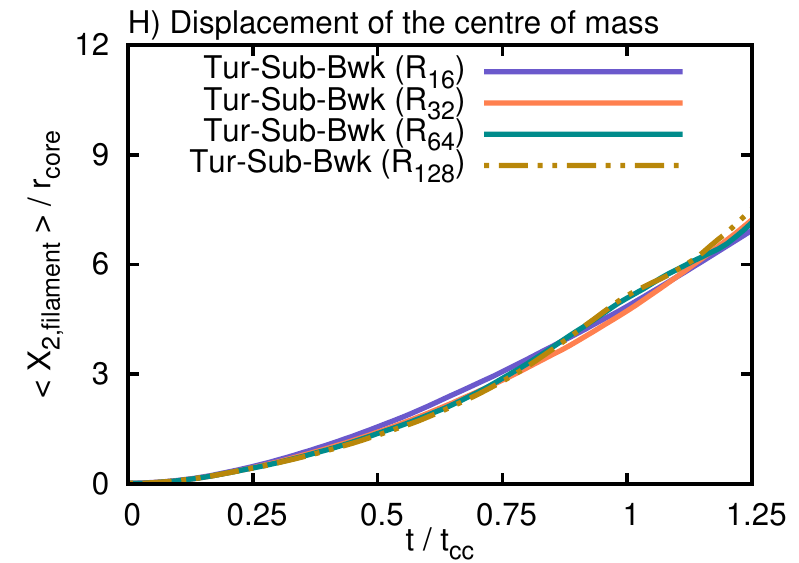}} &  \hspace{-0.7cm}\resizebox{60mm}{!}{\includegraphics{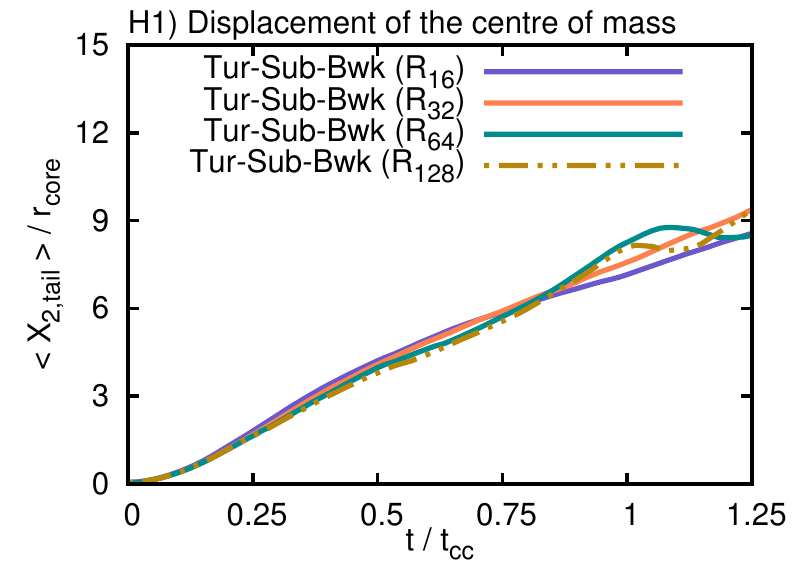}} & \hspace{-0.7cm}\resizebox{60mm}{!}{\includegraphics{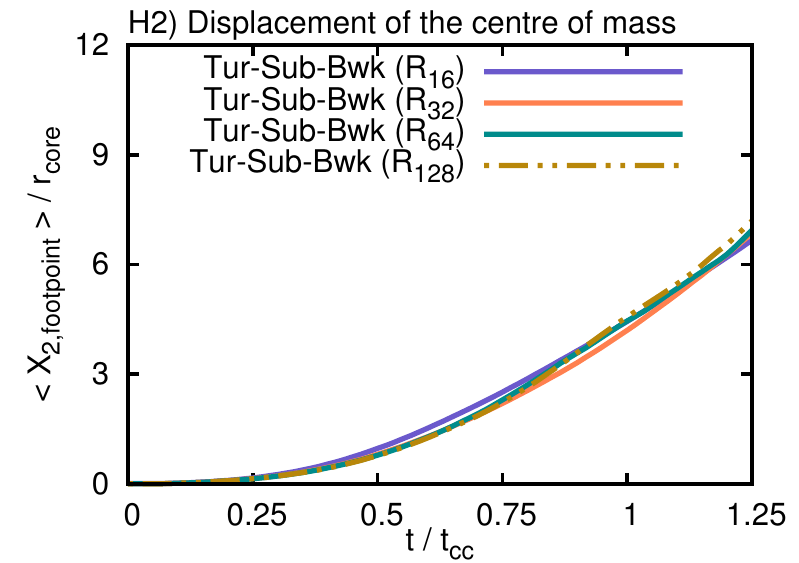}}\\
 \resizebox{60mm}{!}{\includegraphics{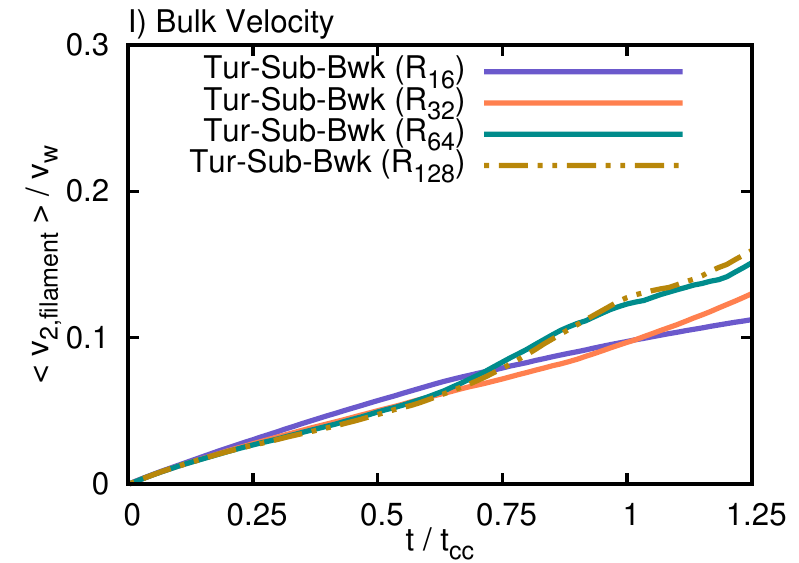}} &  \hspace{-0.7cm}\resizebox{60mm}{!}{\includegraphics{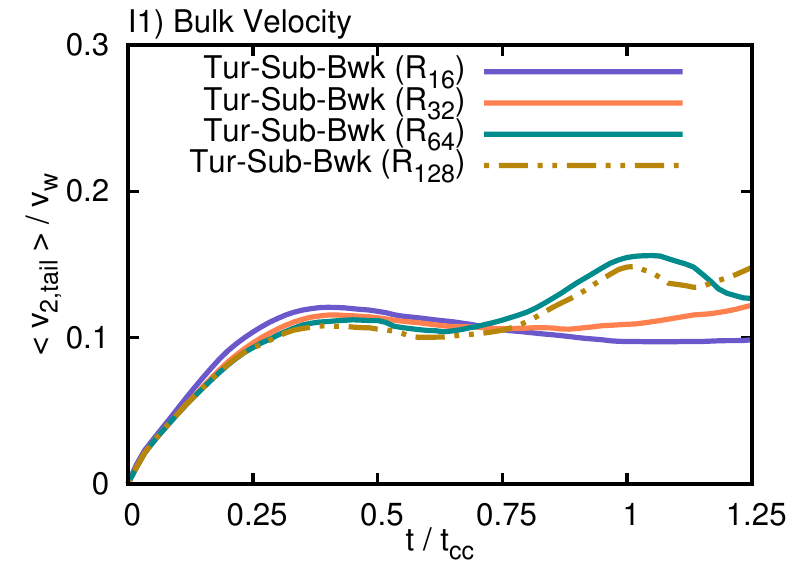}} & \hspace{-0.7cm}\resizebox{60mm}{!}{\includegraphics{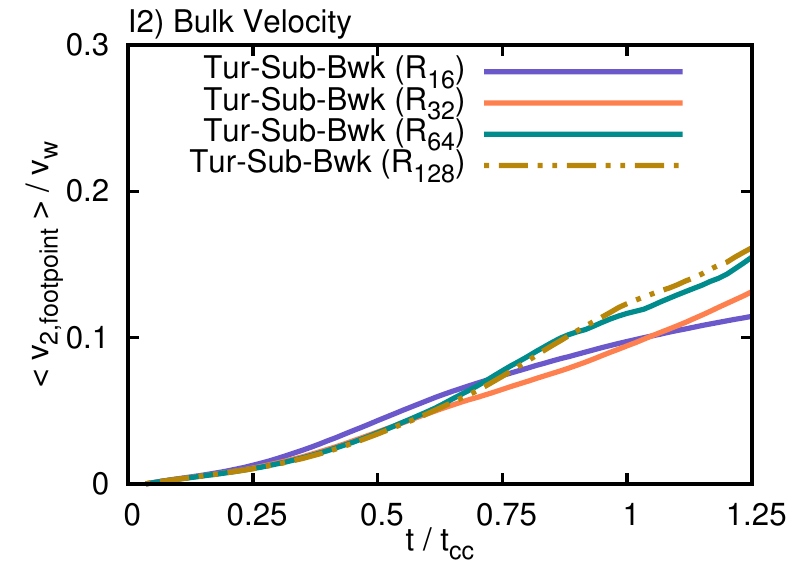}}\\
  \end{tabular}
  \caption{Time evolution of the diagnostics reported in Figures \ref{Figure6}, \ref{Figure7}, \ref{Figure10}, and \ref{Figure11} in model Tur-Sub-Bwk at different resolutions ($R_{16-128}$).}
  \label{FigureC4}
\end{center}
\end{figure*}

An important aspect to be investigated when studying wind-cloud systems numerically is the influence of the grid resolution on the results. A resolution study allows to ascertain if the results are trustworthy and determine how affected they are by numerical diffusion (e.g., see \citealt{1994ApJ...420..213K,1995ApJ...454..172X,2006ApJS..164..477N,2007PhDT........54N,2010ApJ...722..412Y,2016MNRAS.457.4470P} and Sections 4.5 and 5.4 in \citealt{2016PhDT.......154B} for previous discussions on the effects of resolution upon wind/shock-cloud/bubble systems). In this Appendix we investigate if the quantitative results presented in both \citetalias{2016MNRAS.455.1309B} and this paper hold for different numerical resolutions. This study is important because even when sophisticated solvers are utilised, capturing the physics of wind-cloud interactions greatly depends on the choice of numerical resolution (mesh spacing). In the models presented in our papers, care should be taken when selecting the resolution as the disruption of clouds occurs as a result of the growth of dynamically-unstable perturbations (i.e., KH and RT instabilities). These perturbations grow at different length scales, so the selected numerical resolution (i.e., the number of grid cells per cloud radius) for a particular simulation should ensure that the range of wavelengths at which these instabilities occur is sufficiently well resolved (see also \citealt{2016MNRAS.457.4470P}).\par

To investigate the effects of the numerical resolution on the results presented in this series of papers, we perform two sets of numerical simulations: one for a model with a uniform cloud embedded in an oblique magnetic field (relevant for model MHD-Ob in \citetalias{2016MNRAS.455.1309B} and models UNI-0-0 and Uni-0-0 in this paper), and one for a model with a turbulent cloud that has a log-normal density distribution, a subsonic velocity field (${\cal M}_{\rm w}=0.33$), and a turbulent magnetic field with $[~\beta_{\rm tu}~]=4$ (relevant for model Tur-Sub-Bwk). Figures \ref{FigureC1} and \ref{FigureC2} show the evolution of the diagnostics presented in Figures \ref{Figure5} and \ref{Figure9}, and Figures \ref{Figure6}, \ref{Figure7}, \ref{Figure10}, and \ref{Figure11}, respectively, for filament (left-hand side column), tail (middle column), and footpoint (right-hand side column) material in model Uni-0-0. The plots in Figure \ref{FigureC1} indicate that all resolutions ($R_{16-128}$) capture the overall evolution of the morphological properties of filaments, except for the mean vorticity enhancement where the lowest resolution of $R_{16}$ fails to properly capture small-scale vorticity in the interior of the cloud. Similarly, the plots in Figure \ref{FigureC2} indicate that resolutions $R_{\geq 32}$ capture the magnetic and kinematic properties of the uniform cloud. The transverse velocity dispersion, mean vorticity enhancement, and turbulent magnetic energy density exhibit the largest differences in this set of simulations (over different resolutions) as they depend on small-scale vorticity production. However, the differences between diagnostics at resolutions of $R_{64}$ and $R_{128}$ remain within $\sim10\,\%$ up to $t/t_{\rm cc}=1.0$, without trends with increasing resolution. Thus, if we consider all the diagnostics in Figures \ref{FigureC1} and \ref{FigureC2}, we find that convergence occurs at resolutions of $64$ cells per cloud radius, i.e., $R_{64}$, for this particular setup.\par

Figures \ref{FigureC3} and \ref{FigureC4} show the evolution of the diagnostics presented in Figures \ref{Figure5} and \ref{Figure9}, and Figures \ref{Figure6}, \ref{Figure7}, \ref{Figure10}, and \ref{Figure11}, respectively, for filament (left-hand side column), tail (middle column), and footpoint (right-hand side column) material in the turbulent model, Tur-Sub-Bwk. Similarly to the above case, the plots of Figure \ref{FigureC3} show that convergence of the morphological properties in this model is achieved at a resolution of $R_{64}$, except for the mean vorticity enhancement, which grows with increasing resolution. The reason for this behaviour is that low resolution setups do not capture properly the turbulent density, velocity, and magnetic fields in the initial cloud. Thus, as we increase the grid resolution, we also capture more details of the original turbulent distributions and diagnostics that strongly depend on them, such as the vorticity. The plots of Figure \ref{FigureC4} show similar results for $R_{64}$ and $R_{128}$, with the magnetic and kinematic properties converging to within $10\,\%$ up to $t/t_{\rm cc}=1.0$, with the exception of the turbulent magnetic energy, which also shows a mild growth trend with increasing resolutions. Overall, if we consider all the diagnostics in Figures \ref{FigureC3} and \ref{FigureC4}, we find that convergence in turbulent cloud models also occurs at resolutions of $64$ cells per cloud radius, i.e., $R_{64}$, except for the mean vorticity enhancement and the turbulent magnetic energy density. Thus, we conclude that the resolutions utilised for the models presented in both \citetalias{2016MNRAS.455.1309B} and this paper, i.e., $R_{64}$ and $R_{128}$ for setups with M and S configurations, respectively, are adequate to capture the morphological properties, energetics, and dynamics of wind-swept uniform and turbulent clouds.\par

\section{Larger-domain simulations}
\label{sec:Appendix4}

\begin{figure*}
\begin{center}
  \begin{tabular}{c c c}
 \textbf{Filament (Cloud)} &  \textbf{Filament Tail (Cloud Envelope)} & \textbf{Filament Footpoint (Cloud Core)}\\
  \resizebox{60mm}{!}{\includegraphics{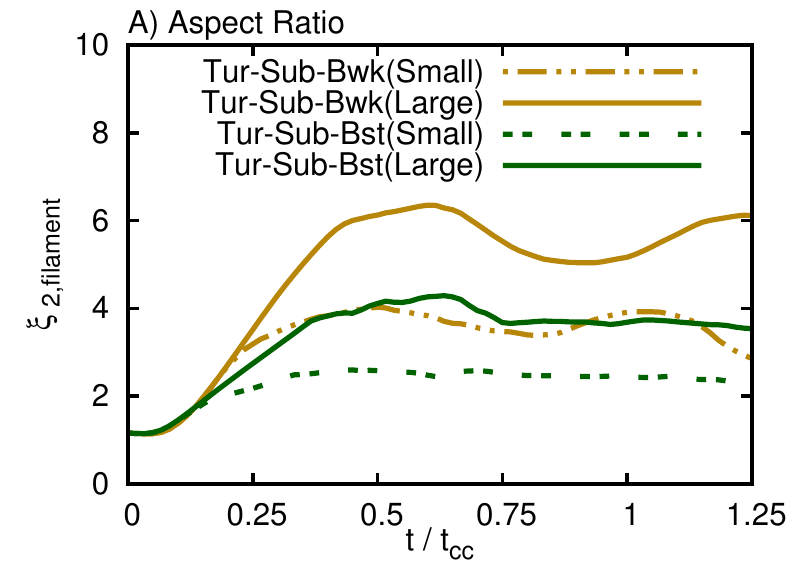}} & \hspace{-0.7cm}\resizebox{60mm}{!}{\includegraphics{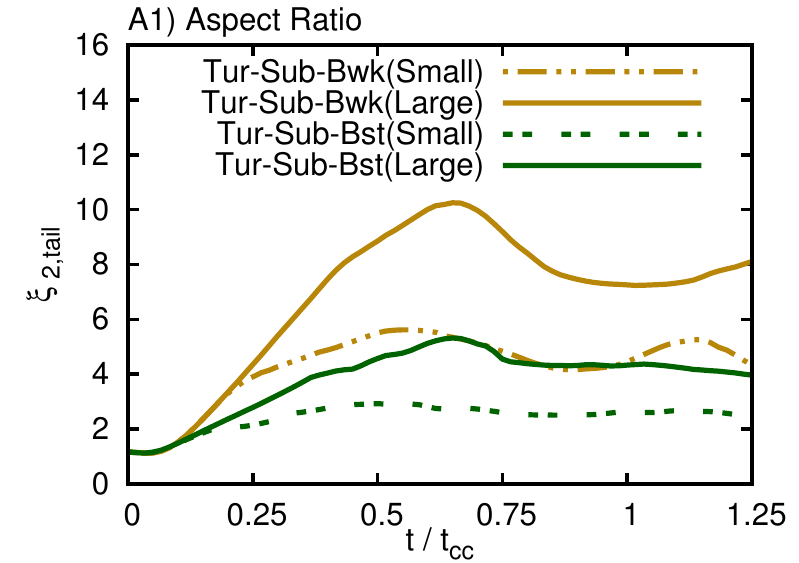}} & \hspace{-0.7cm}\resizebox{60mm}{!}{\includegraphics{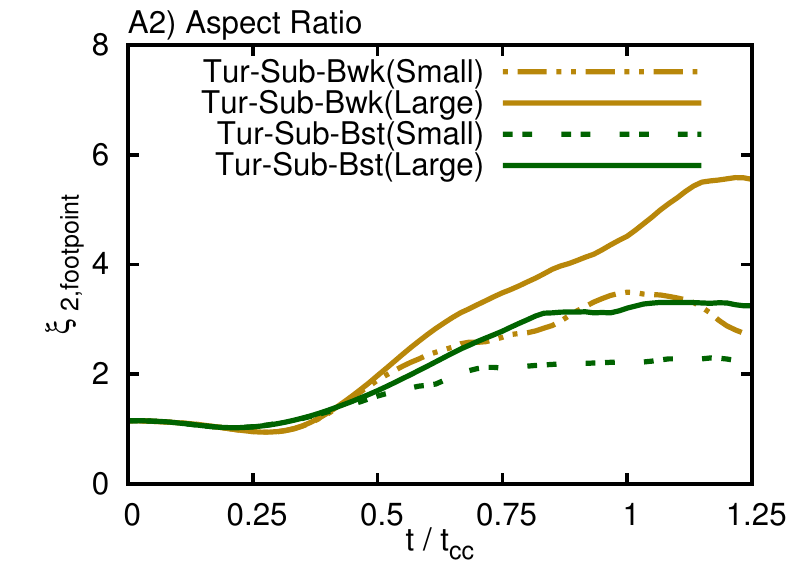}}\\
 \resizebox{60mm}{!}{\includegraphics{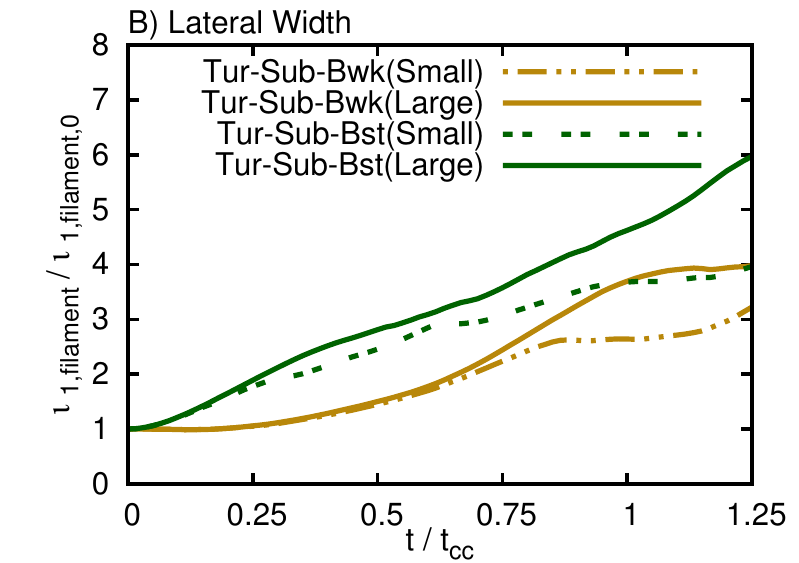}} & \hspace{-0.7cm}\resizebox{60mm}{!}{\includegraphics{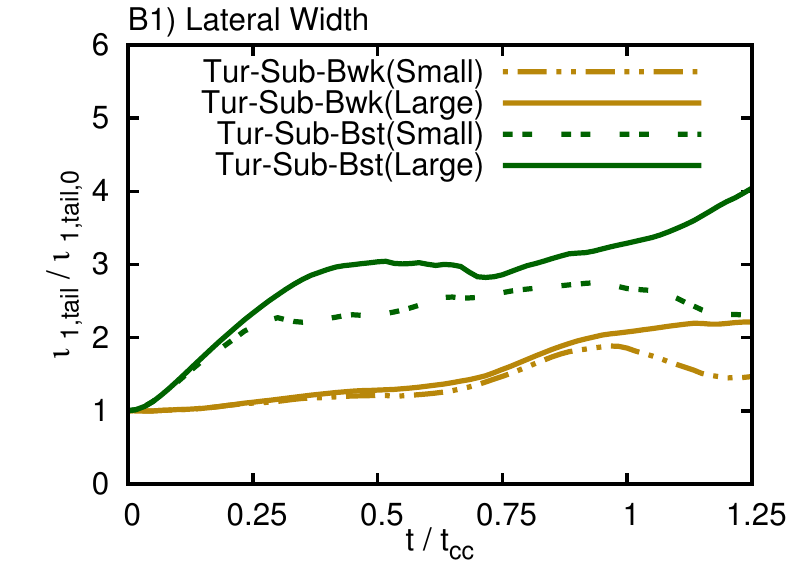}} & \hspace{-0.7cm}\resizebox{60mm}{!}{\includegraphics{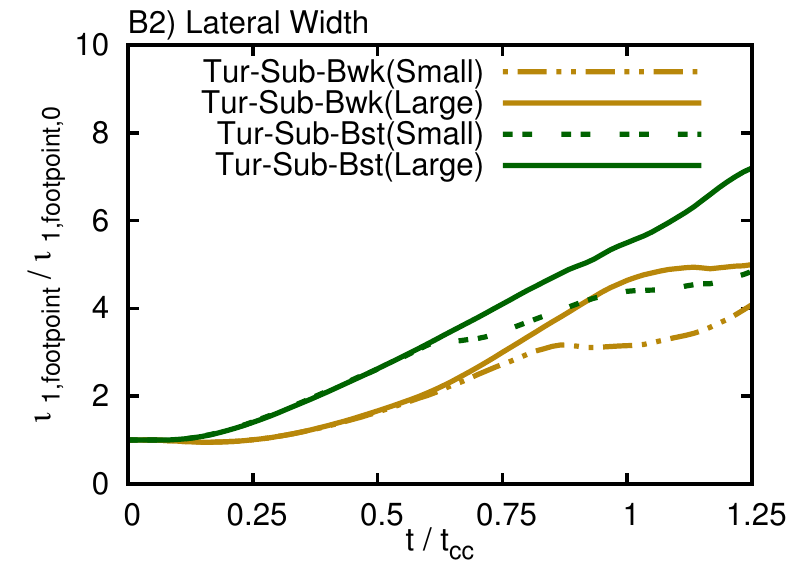}}\\
 \resizebox{60mm}{!}{\includegraphics{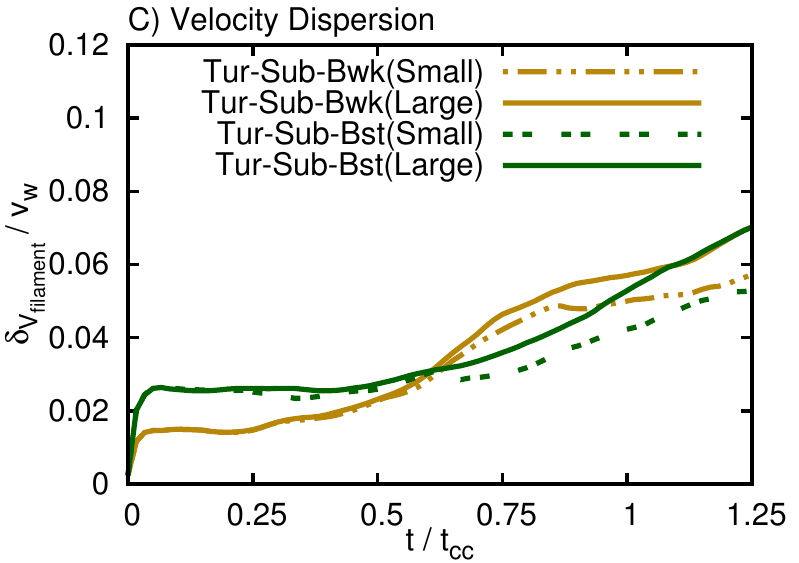}} &  \hspace{-0.7cm}\resizebox{60mm}{!}{\includegraphics{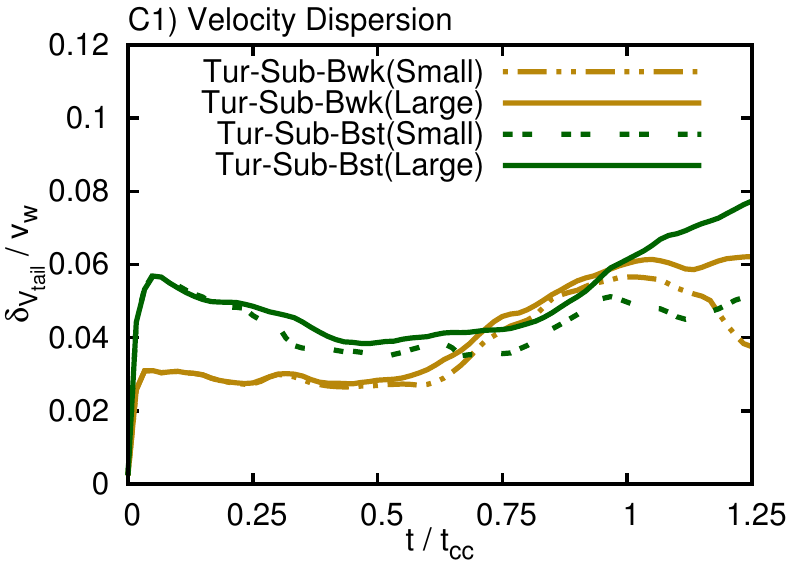}} & \hspace{-0.7cm}\resizebox{60mm}{!}{\includegraphics{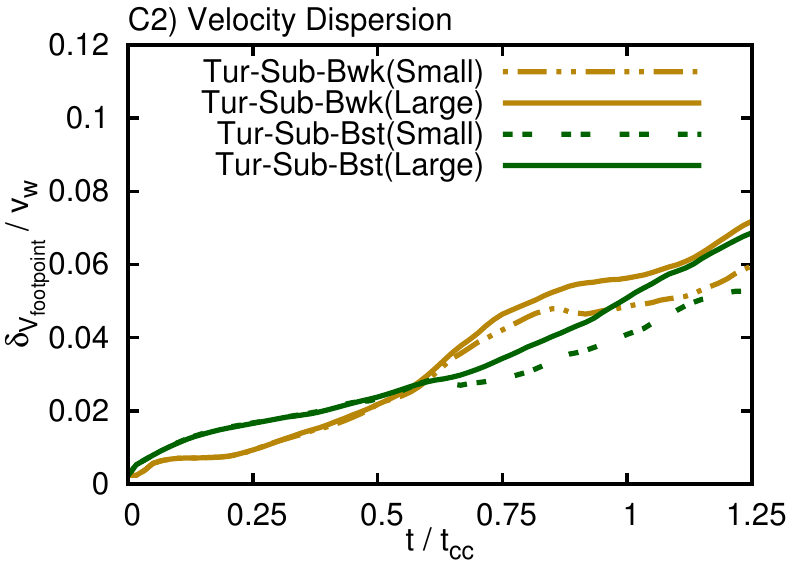}}\\
\resizebox{60mm}{!}{\includegraphics{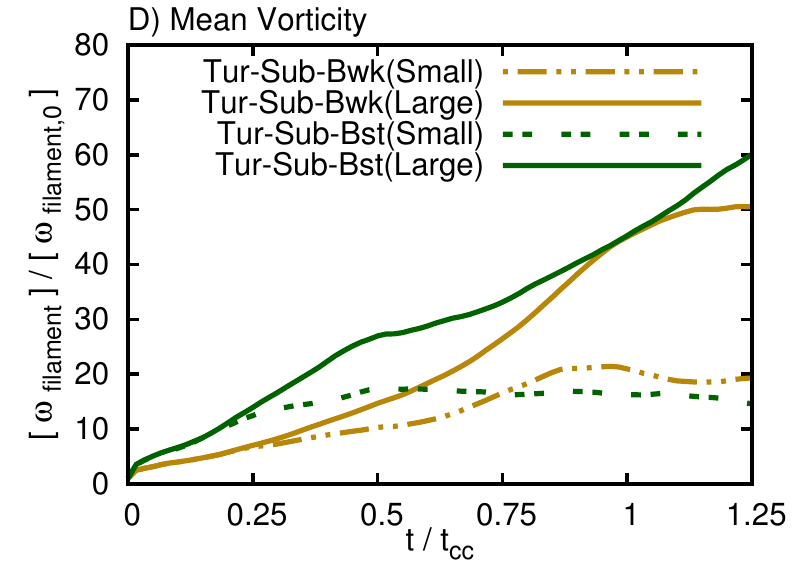}} & \hspace{-0.7cm} \resizebox{60mm}{!}{\includegraphics{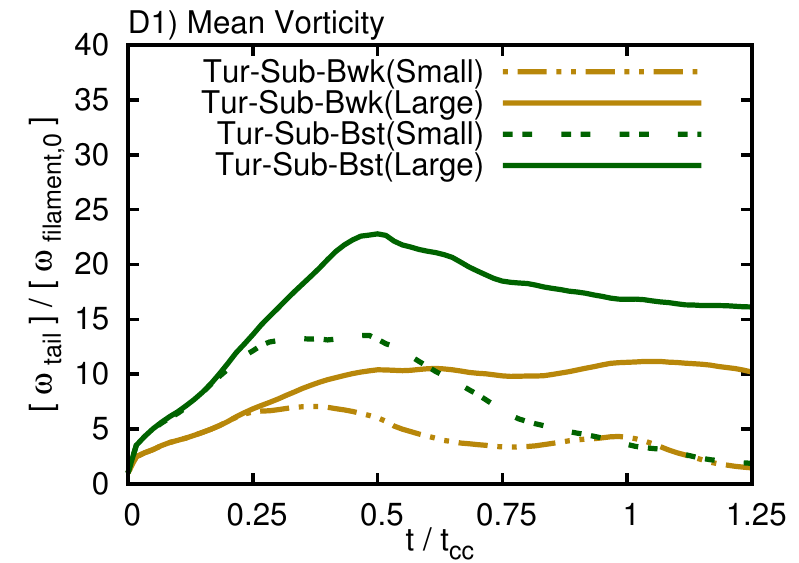}} & \hspace{-0.7cm}\resizebox{60mm}{!}{\includegraphics{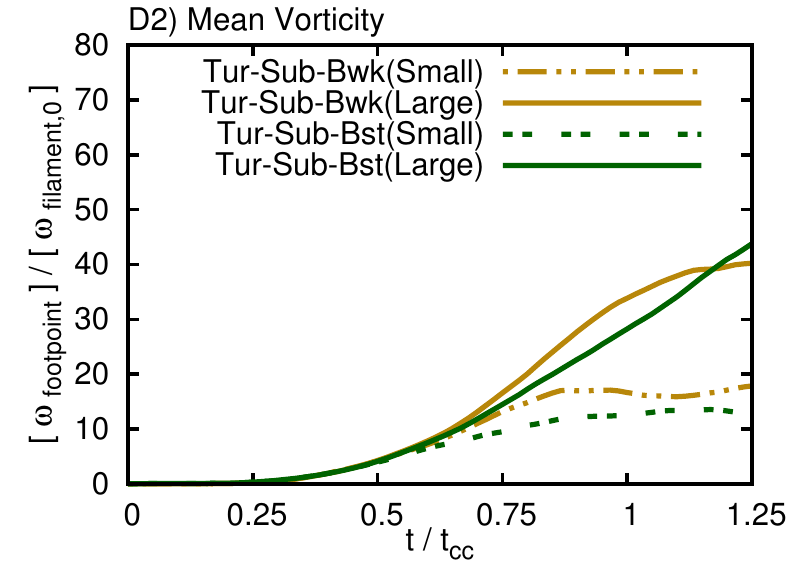}}\\
  \end{tabular}
  \caption{Time evolution of the diagnostics reported in Figures \ref{Figure5} and \ref{Figure9} in two models, Tur-Sub-Bwk and Tur-Sub-Bst, at the same resolution ($R_{64}$), but different computational domain sizes (S and L configurations).}
  \label{FigureD1}
\end{center}
\end{figure*}

\begin{figure*}
\begin{center}
  \begin{tabular}{c c c}
 \textbf{Filament (Cloud)} &  \textbf{Filament Tail (Cloud Envelope)} & \textbf{Filament Footpoint (Cloud Core)}\\
  \resizebox{60mm}{!}{\includegraphics{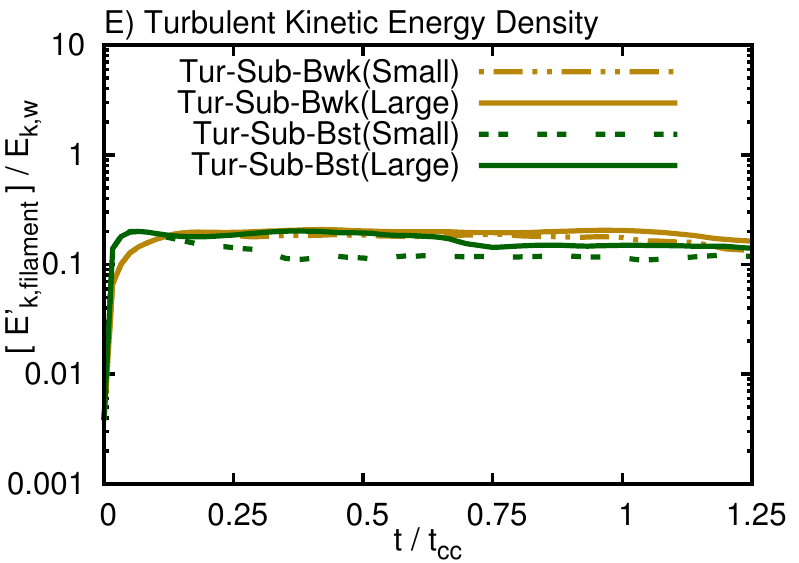}} & \hspace{-0.7cm}\resizebox{60mm}{!}{\includegraphics{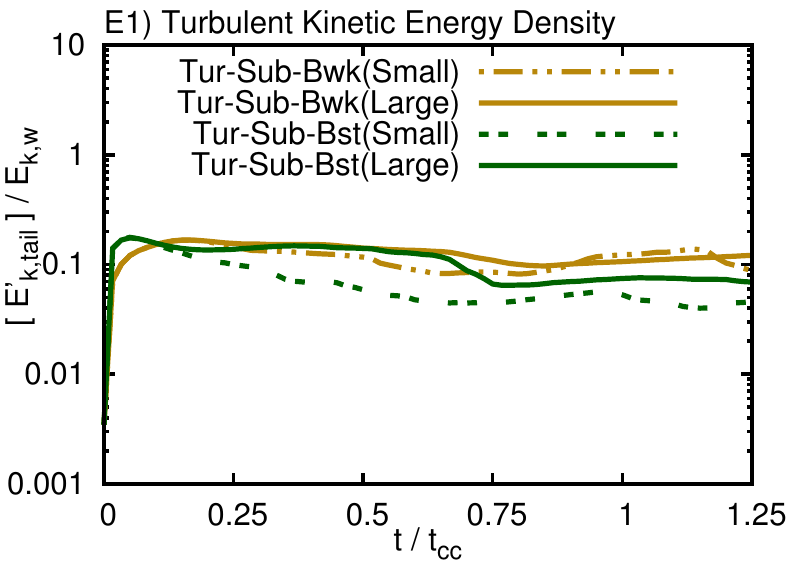}} & \hspace{-0.7cm}\resizebox{60mm}{!}{\includegraphics{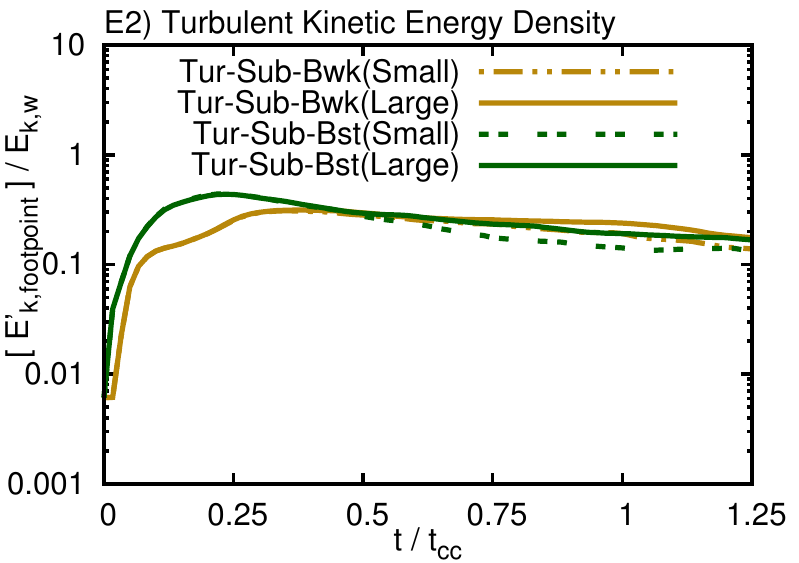}}\\
 \resizebox{60mm}{!}{\includegraphics{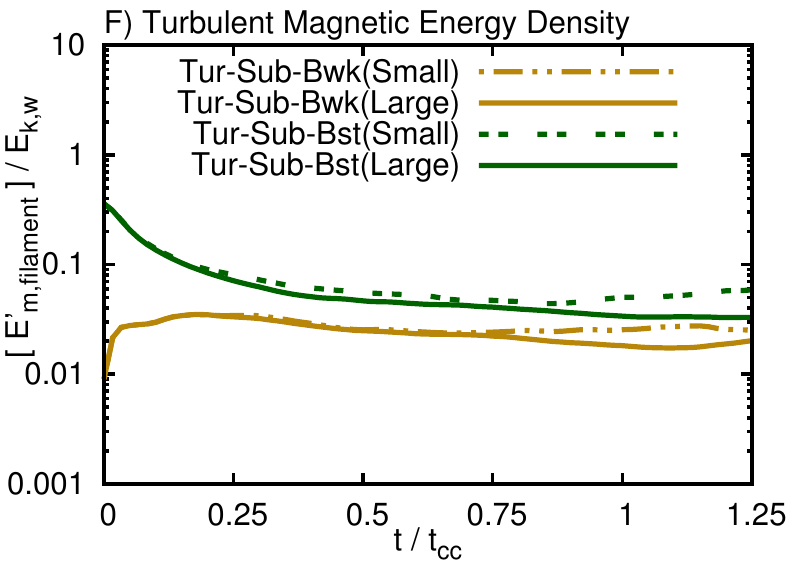}} & \hspace{-0.7cm}\resizebox{60mm}{!}{\includegraphics{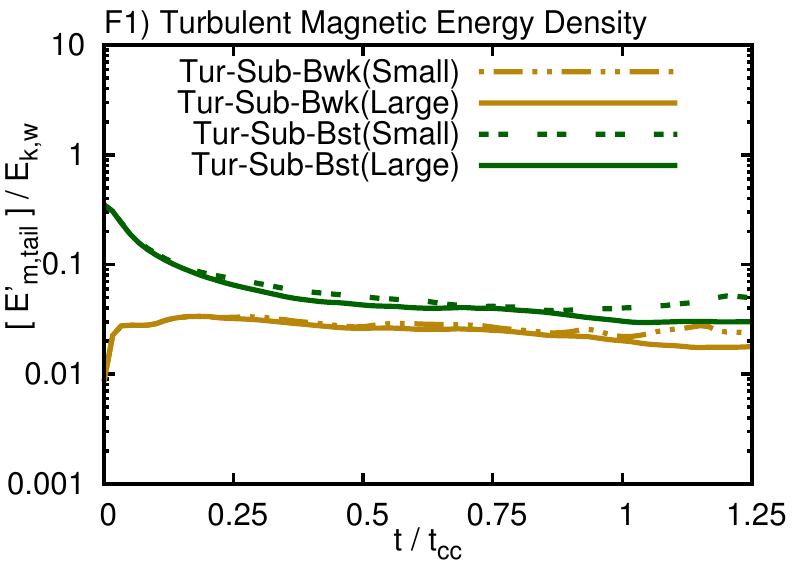}} & \hspace{-0.7cm}\resizebox{60mm}{!}{\includegraphics{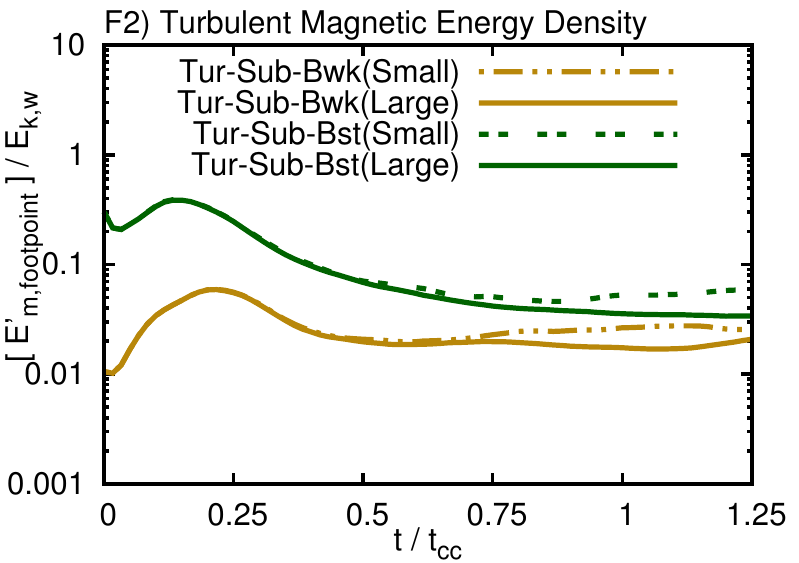}}\\
 \resizebox{60mm}{!}{\includegraphics{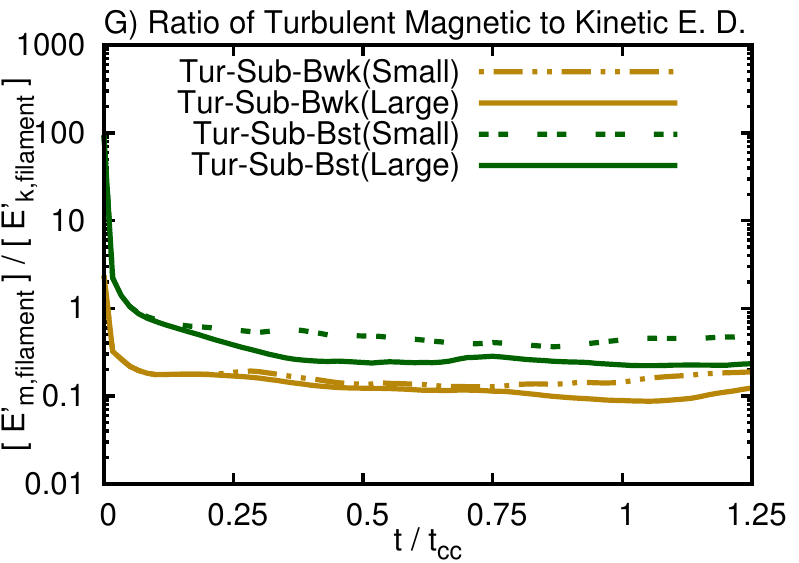}} &  \hspace{-0.7cm}\resizebox{60mm}{!}{\includegraphics{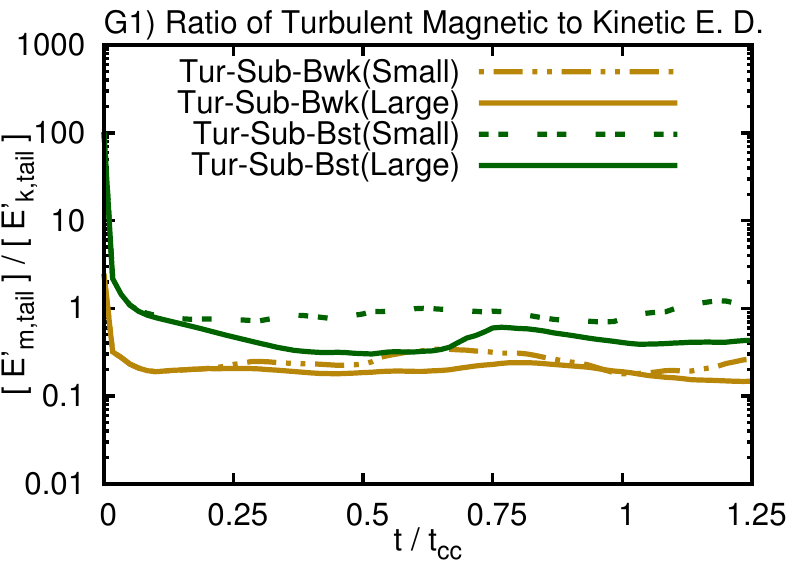}} & \hspace{-0.7cm}\resizebox{60mm}{!}{\includegraphics{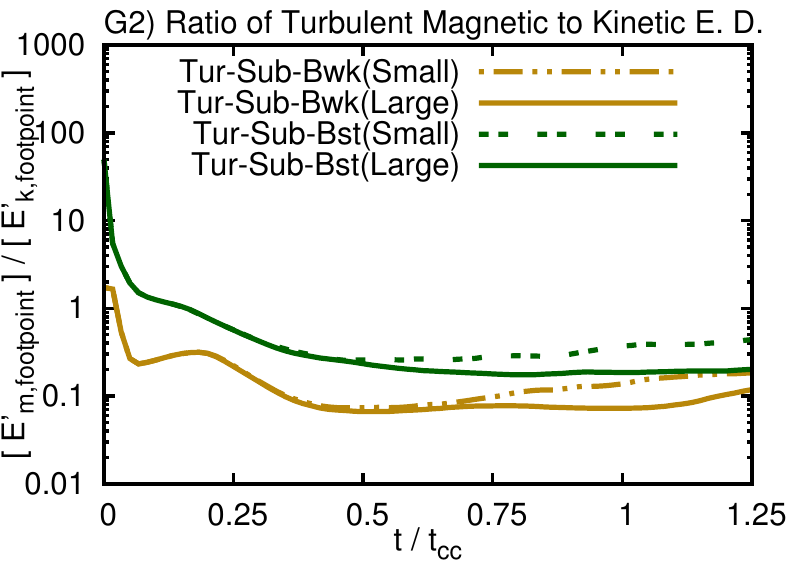}}\\
 \resizebox{60mm}{!}{\includegraphics{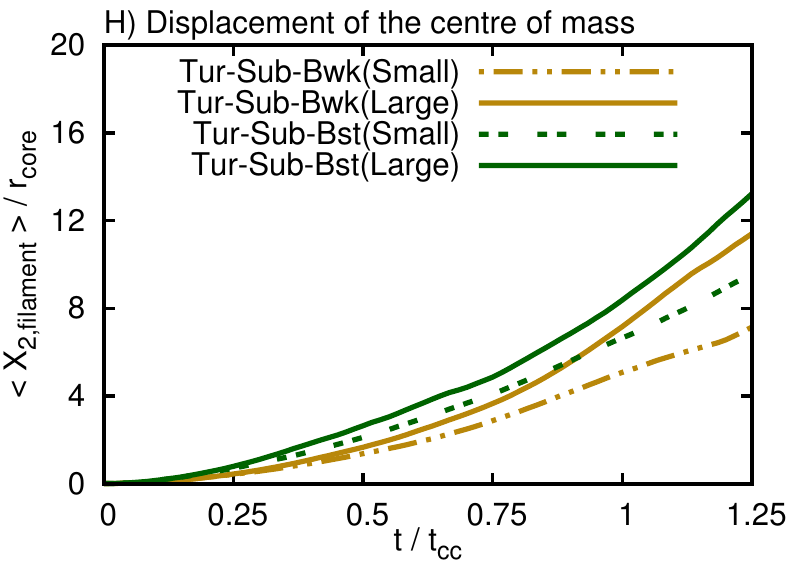}} &  \hspace{-0.7cm}\resizebox{60mm}{!}{\includegraphics{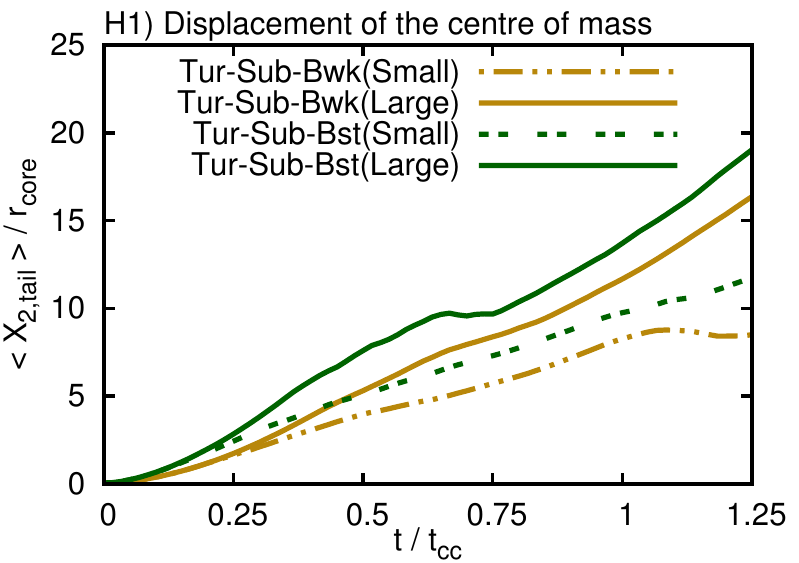}} & \hspace{-0.7cm}\resizebox{60mm}{!}{\includegraphics{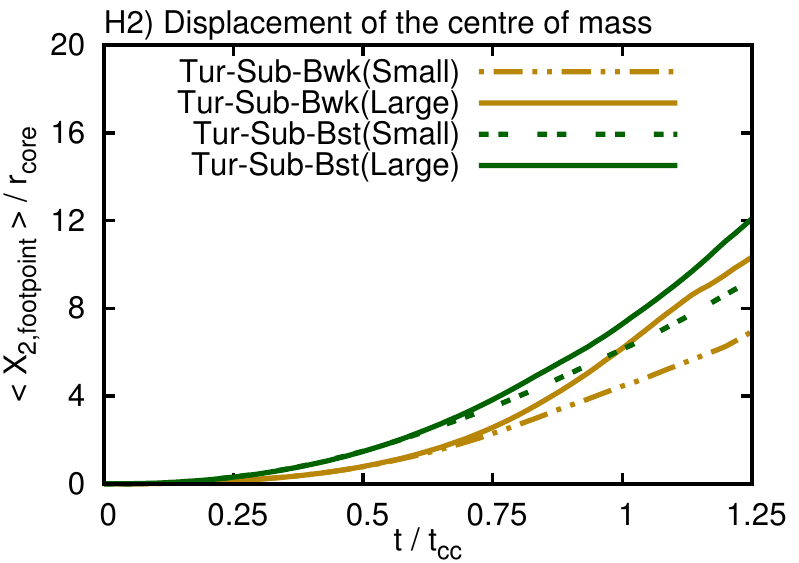}}\\
 \resizebox{60mm}{!}{\includegraphics{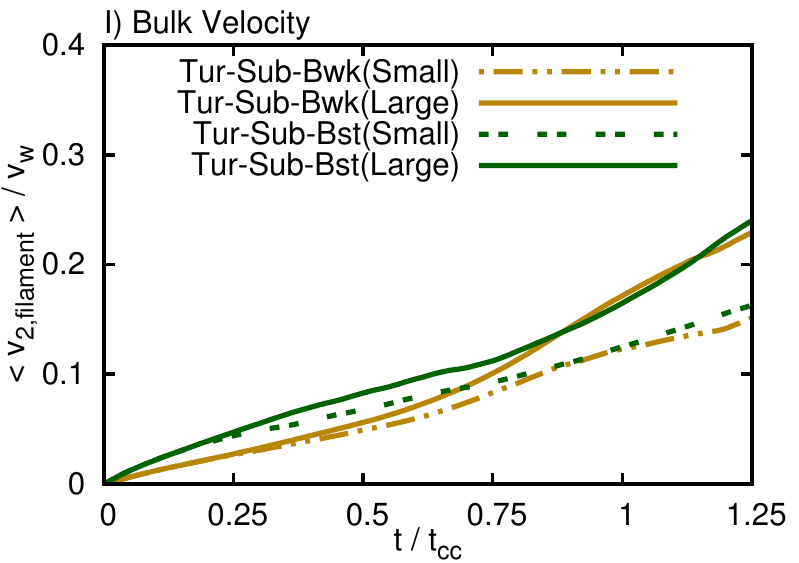}} &  \hspace{-0.7cm}\resizebox{60mm}{!}{\includegraphics{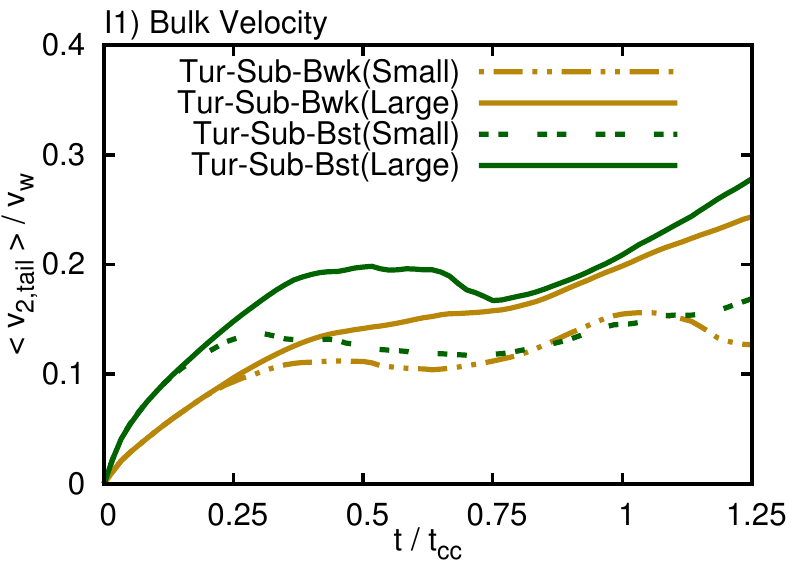}} & \hspace{-0.7cm}\resizebox{60mm}{!}{\includegraphics{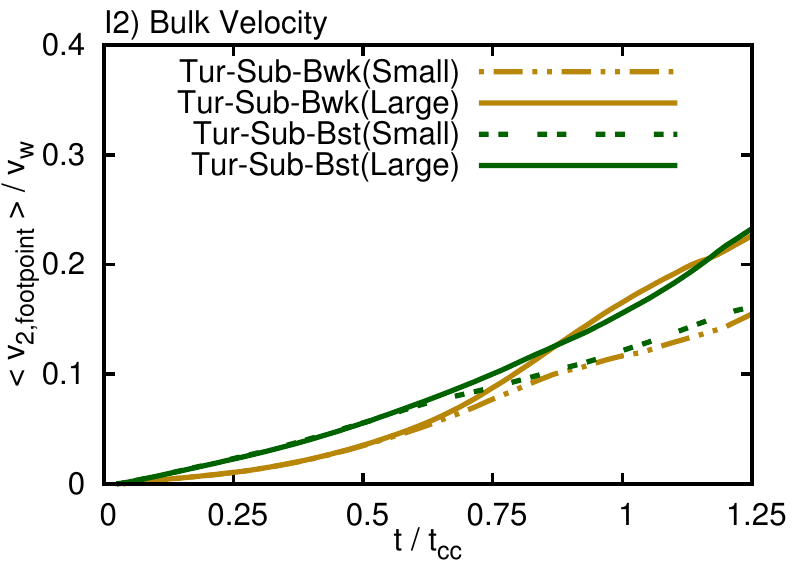}}\\
  \end{tabular}
  \caption{Time evolution of the diagnostics reported in Figures \ref{Figure6}, \ref{Figure7}, \ref{Figure10}, and \ref{Figure11} in two models, Tur-Sub-Bwk and Tur-Sub-Bst, at the same resolution ($R_{64}$), but different computational domain sizes (S and L configurations).}
  \label{FigureD2}
\end{center}
\end{figure*}

In this Appendix we discuss the effects of the simulation domain size on the diagnostics presented through this manuscript. In a similar manner to \citetalias{2016MNRAS.455.1309B}, we compare two sets of simulations with the same initial conditions (to those of models Tur-Sub-Bwk and Tur-Sub-Bst), the same numerical resolutions of $64$ cells per cloud radius (i.e., $R_{64}$), and different domain sizes (large or L configuration and small or S configuration). The computational domain in models Tur-Sub-Bwk(Large) and Tur-Sub-Bst(Large) is twice the size of the domain in models Tur-Sub-Bwk(Small) and Tur-Sub-Bst(Small), and it covers the spatial range: $-4\,r_{\rm c}\leq X_1\leq 4\,r_{\rm c}$, $-2\,r_{\rm c}\leq X_2\leq 22\,r_{\rm c}$, and $-4\,r_{\rm c}\leq X_3\leq 4\,r_{\rm c}$, where $r_{\rm c}$ is the radius of the cloud.\par

Figure \ref{FigureD1} shows the evolution of the parameters presented in Figures \ref{Figure5} and \ref{Figure9} in Section \ref{sec:Results} for filament (left-hand side column), tail (middle column), and footpoint (right-hand side column) material, in both models, Tur-Sub-Bwk and Tur-Sub-Bst. The panels of Figure \ref{FigureD1} indicate that the aspect ratio and the mean vorticity enhancement are underestimated in the small-domain simulation, with differences being as large as $\sim80\,\%$ and $\sim150\,\%$, respectively. The curves in these plots start to diverge when material starts to flow out of the smaller simulation grid, either through the back or the sides of the computational domain. Thus, tail material is affected the most by the choice of domain size as it is the first to be stripped off the cloud by the wind ram pressure and dynamical instabilities. Indeed, divergence of the aspect ratio and the mean vorticity enhancement in tail material starts at $t/t_{\rm cc}\sim0.25$ in both models, mainly due to fast, low-density material leaving the domain through the back surface of the domain. In the case of footpoint material, on the other hand, divergence of the aspect ratio and the mean vorticity enhancement only starts to occur at $t/t_{\rm cc}\sim0.5-0.6$ in both models, due to material leaving through both the back (at $t/t_{\rm cc}\sim0.5$) and the sides of the domain (at $t/t_{\rm cc}\sim0.6$). The other two parameters in Figure \ref{FigureD1}, i.e., the lateral width and velocity dispersion, are not as affected and errors remain below $\sim25\,\%$ in both models until $t/t_{\rm cc}\sim1.0$.\par

Figure \ref{FigureD2} shows the evolution of the parameters presented in Figures \ref{Figure6}, \ref{Figure7}, \ref{Figure10}, and \ref{Figure11} in Section \ref{sec:Results} for filament (left-hand side column), tail (middle column), and footpoint (right-hand side column) material, in models Tur-Sub-Bwk and Tur-Sub-Bst. Overall, the trends of the curves presented in this figures are the same, implying that our conclusions with respect to the energetics and dynamics of wind-swept clouds are independent of the computational simulation size. Tail diagnostics are again more affected by the domain size than footpoint diagnostics because tail material leaves the smaller simulation domain earlier. The tail parameters that are most affected by the size of the simulation domain are those related to the kinetic and kinematic properties of the filaments, such as the turbulent kinetic energy density, the displacement of the centre of mass, and the bulk speeds. Differences between large- and small-domain simulations are as large as $\sim60\,\%$, $\sim 40\,\%$, and $50\,\%$ in both models for each of these parameters, respectively, up to $t/t_{\rm cc}=1.0$. In the case of footpoint material, errors are lower and remain below $\sim 25\,\%$ for all parameters in both models over the same time-scale.\par

Another aspect that we highlight here is that smaller-domain simulations bias the results towards gas with lower kinetic energy densities and slightly higher magnetic energy densities than larger-domain simulations, thus favouring higher ratios of the two turbulent energy densities. This indicates that the actual ratios of turbulent magnetic to kinetic energy densities in the wind-cloud systems studied in this paper are more likely to be on the lower end of the range reported in Section \ref{subsubsec:EnergyDensities}, i.e., $[~E'_{\rm m,filament}~]/[~E'_{\rm k,filament}~]\sim 0.1-0.4$. In a similar manner, we observe that both the displacement of the centre of mass and the bulk speeds are underestimated in smaller-domain simulations. Consequently, the values reported in Section \ref{subsubsec:BulkSpeedandDistance} for these parameters, at $t/t_{\rm cc}=1.0$, should be regarded as reference lower limits of the travelled distances and bulk velocities of filaments, which are more accurately measured in the larger-domain simulations discussed in Section \ref{subsubsec:Entrainment}. Therefore, if more precise measurements of these global diagnostics are required (especially after the break-up time), future numerical work on wind-cloud systems should consider computational domains with sizes similar to the M or L configurations discussed in this paper.\par

Finally, an important remark to make based on the analysis presented in this Appendix is that even though we find that the choice of computational domain sizes does have an effect on the evolution of our diagnostics, when we compare different models with the same resolution and domain configuration, we find the same qualitative and quantitative results previously discussed in Section \ref{sec:Results}. Moreover, the errors are expected to be much lower when the M configuration is utilised as in our self-consistent models UNI-0-0, TUR-SUP-BST, and TUR-SUP-BST-ISO. Thus, we conclude that our domain configurations M and S are adequate for the purpose of comparing different models to one another and also for extracting reliable qualitative and quantitative information from the simulations.\par

\section{PDF of turbulent densities}
\label{sec:Appendix1}

\begin{figure}\centering
\includegraphics[scale=1]{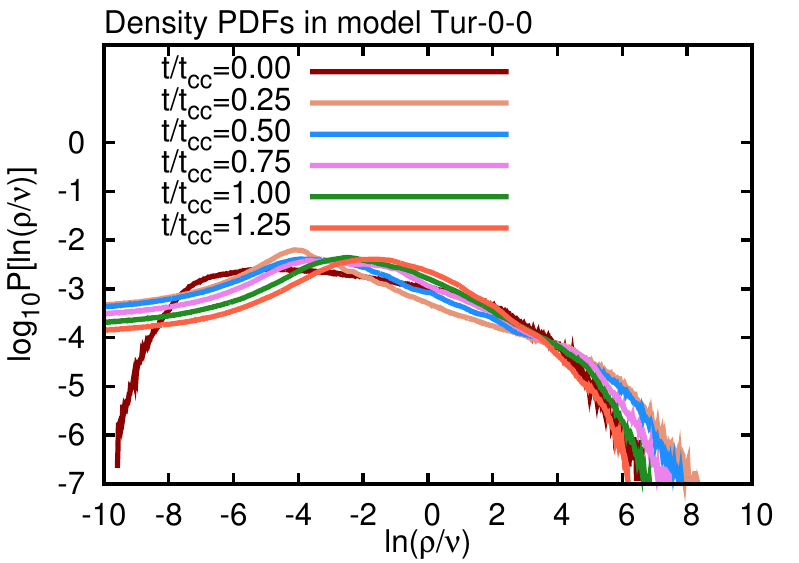}
\caption{Evolution of the gas density PDF in the turbulent cloud of model Tur-0-0 normalised with respect to the mean density, for $0\leq t/t_{\rm cc}\leq1.25$. The initial density field ($t/t_{\rm cc}=0$) is taken from \citealt{2012ApJ...761..156F}, representing typical physical conditions of ISM turbulent clouds. An initial compression phase ($t/t_{\rm cc}\sim0.25$) makes the cloud gas denser at the beginning of the interaction, but mass stripping ($t/t_{\rm cc}\geq0.5$) and mixing of cloud material with ambient gas make the density decrease and create a heavy and flat tail on the low-density end of the PDF.}
\label{FigureA1}
\end{figure}

In this Appendix we present the evolution of the probability density function (PDF) of the mass density of the cloud/filament in model Tur-0-0 (see Figure \ref{FigureA1}). The initial density distribution of the turbulent cloud in this model (and in all the other models with turbulent clouds, i.e., TUR-SUP-BST, TUR-SUP-BST-ISO, Tur-Sub-0, Tur-Sup-0, Tur-Sub-Bwk, and Tur-Sub-Bst) has the following mathematical form:

\begin{equation}
{\cal P}(\rho)=\frac{1}{s\sqrt{2\pi}}{\rm e}^{-\frac{[\ln(\rho)-m]^2}{2s^2}},
\label{eq:PDF}
\end{equation}

\noindent where $\rho$ is the mass density, $m$ and $s$ are the mean and standard deviation of the natural logarithm of the ISM density field. Accordingly, the mean and variance of the ISM density field are

\begin{equation}
\nu={\rm e}^{(m+\frac{s^2}{2})}, {\rm and}
\label{eq:PDFMean}
\end{equation}

\begin{equation}
\sigma^2=\nu^2({\rm e}^{s^2}-1), 
\label{eq:PDFVariance}
\end{equation}

\noindent respectively (see Appendices A and B in \citealt{2007ApJS..173...37S}). Figure \ref{FigureA1} shows the density PDF of cloud gas taken from snapshots of model Tur-0-0 at different instances of its evolution (i.e., for $0\leq t/t_{\rm cc}\leq 1.25$). Three effects are seen in these curves: i) at the beginning of the interaction the cloud is compressed by wind-driven shocks, thus increasing the variance of the density PDF (see the curve at $t/t_{\rm cc}=0.25$). Then, ii) the formation of the filamentary tail by mass stripping, causes cloud and wind gas to mix, thereby causing the cloud density to steadily decrease and the PDF to become skewed towards low-density values (see the high-density and low-density ends of curves, respectively, for $t/t_{\rm cc}\geq0.5$). As time progresses, iii) the low-density tail in the PDF also starts to flatten as cloud gas is continuously mixed with ambient gas and as the filament is eroded by dynamical instabilities. A complete analysis of the evolution of PDFs of wind-swept turbulent clouds will be the topic of a future paper.

\section{Divergence cleaning algorithm}
\label{sec:Appendix2}
\begin{figure}\centering
\includegraphics[scale=1]{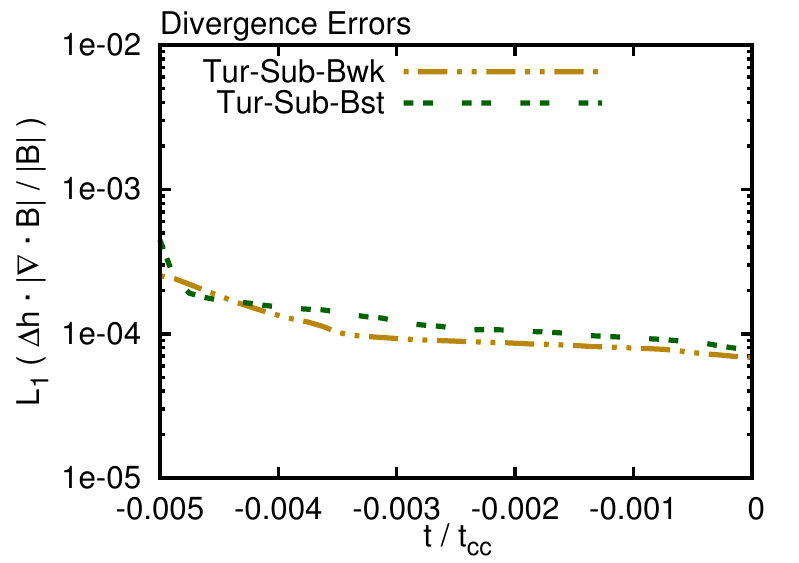}
\caption{Cleaning of divergence errors in the initial magnetic fields for models Tur-Sub-Bwk (double-dot-dashed line) and Tur-Sub-Bst (two-dashed line; same as for models TUR-SUP-BST and TUR-SUP-BST-ISO). The time on the $X$ axis is in units of the cloud-crushing time, $t_{\rm cc}$, and it is negative as it corresponds to the pre-processing phase of the magnetic fields, i.e., before the simulations of these models are started and updated with the MHD formulation of Section \ref{sec:Method}.}
\label{FigureB1}
\end{figure}

In this Appendix we provide additional details on the method utilised to pre-process the initial, turbulent magnetic fields in models TUR-SUP-BST, TUR-SUP-BST-ISO, Tur-Sub-Bwk, and Tur-Sub-Bst, so that they comply with the solenoidal condition. As stated in Section \ref{subsec:Initial and Boundary Conditions}, a direct interpolation of the magnetic field components taken from the simulation of turbulent flows (reported in \citealt{2012ApJ...761..156F}) into our simulation grids is not possible because the interpolated fields would contain magnetic monopoles at the cloud boundaries. To solve this issue and initialise models TUR-SUP-BST, TUR-SUP-BST-ISO, Tur-Sub-Bwk, and Tur-Sub-Bst with solenoidal magnetic fields, we first clean the divergence errors in them by following the steps below:

\begin{enumerate}
  \item The magnetic field components are extracted from a data cube of simulation 21 in \cite{2012ApJ...761..156F}.
  \item These components are then scaled to the desired turbulent plasma beta, $[~\beta_{\rm tu}~]$, for each model and subsequently interpolated into simulation grids with the same resolution of the grids described in Section \ref{sec:Method}. The resulting magnetic field in each model is then the sum of a uniform oblique field, $\bm{B_{\rm ob}}$, plus the turbulent field, $\bm{B_{\rm tu}}$ (see Section \ref{sec:Method} for further details).
  \item The interpolated magnetic fields violate the free-divergence constraint at the boundaries of the cloud, so we clean the divergence errors by using a mixed hyperbolic/parabolic correction technique by \cite{2002JCoPh.175..645D} that introduces a generalised Lagrange multiplier ($\psi$; hereafter GLM) to couple the divergence constraint, $\bm{\nabla\cdot}\bm{B}$, with the MHD conservation laws (see Equations \ref{eq:MassConservation}--\ref{eq:induction}). The induction equation (see Equation \ref{eq:induction}) is then replaced by:
  
\begin{equation}
\frac{\partial \bm{B}}{\partial t}+\bm{\nabla\cdot}\left(\bm{v}\bm{B}-\bm{B}\bm{v}\right)+\nabla\psi=0,
\end{equation}

\noindent while the solenoidal constraint reads:

\begin{equation}
\frac{\partial \psi}{\partial t}+c_{\rm h}^2\bm{\nabla\cdot}\bm{B}=-\frac{c_{\rm h}^2}{c_{\rm p}^2}\psi,
\end{equation}    

\noindent where $c_{\rm h}=C_{\rm a}\Delta h/\Delta t^n$ is the maximum admissible speed at which divergence errors propagate given the time-step restriction, $c^2_{\rm p}=\Delta hc_{\rm h}/\alpha$ is a diffusion coefficient, $\Delta h=\min(\Delta X_{\rm j})$ is the minimum cell length, $\Delta t^n$ is the time increment, and $\alpha=0.2$ is a dimensionless parameter controlling the optimal rate at which monopoles are damped (see \citealt{2010JCoPh.229.2117M} and \citealt*{2010JCoPh.229.5896M} for details of the implementation in Godunov-type schemes and high-order schemes, respectively).

  \item This formulation reconfigures the original magnetic field at wind-cloud interfaces and enforces the zero-divergence condition by transporting the divergence errors towards the domain boundaries and damping the existing magnetic monopoles \citep{2002JCoPh.175..645D}.
  \item We save the components of the divergence-free magnetic fields once the solutions are stable and the divergence errors (\citealt{2010JCoPh.229.2117M}),
\begin{equation}
L_1(Q)=\frac{1}{N_{\rm X_{1}}N_{\rm X_{2}}N_{\rm X_{3}}}\Sigma_{\rm i,j,k}|Q_{\rm i,j,k}-Q^{\rm ref}_{\rm i,j,k}|,
\label{eq:errordiv}
\end{equation}

\noindent where $Q_{\rm i,j,k}=\Delta h|\bm{\nabla\cdot}\bm{B}|_{\rm i,j,k}/|\bm{B}|_{\rm i,j,k}$, are below $\sim10^{-4}$ in models TUR-SUP-BST, TUR-SUP-BST-ISO, Tur-Sub-Bwk, and Tur-Sub-Bst (see e.g., Figure \ref{FigureB1}). The divergence errors at these stages are sufficiently small not to cause disturbances in the density and magnetic fields of the simulations. It is necessary to add a parameter $\eta=0.1$ of the maximum magnetic field ($|\bm{B_{\rm max}}|$) in the denominator of Equation \ref{eq:errordiv} to avoid undefined divisions by zero or the appearance of artificially high values when $|\bm{B}|_{\rm i,j,k}$ is small. \cite{2012JCoPh.231.7214T} followed a similar approach when testing a divergence cleaning algorithm in smoothed-particle magnetohydrodynamics (SPMHD). The main advantage of following this divergence-cleaning process on the magnetic field of the cloud is that the initial magnetic configuration does not contain sheaths of high magnetic tension around the cloud as it would in cases where the cloud magnetic field is merely truncated at the edge of it. This assures a smooth transition from the ambient magnetic field into the cloud magnetic field in the initial conditions (see e.g., the 3D streamline plot at $t/t_{\rm cc}=0$ of the magnetic field of model MHD-Tu-S in Panel C of Figure 7.6 in \citealt{2016PhDT.......154B}).
  \item The new solenoidal magnetic fields are interpolated into the grids of models TUR-SUP-BST, TUR-SUP-BST-ISO, Tur-Sub-Bwk, and Tur-Sub-Bst, and then the MHD wind-cloud simulations of these models are initialised with the constrained transport formulation described in Section \ref{subsec:SimulationCode}.
\end{enumerate}

\section{Magnetic Energy Enhancement}
\label{sec:Appendix5}

\begin{figure}
\begin{center}
  \begin{tabular}{c}
  \resizebox{80mm}{!}{\includegraphics{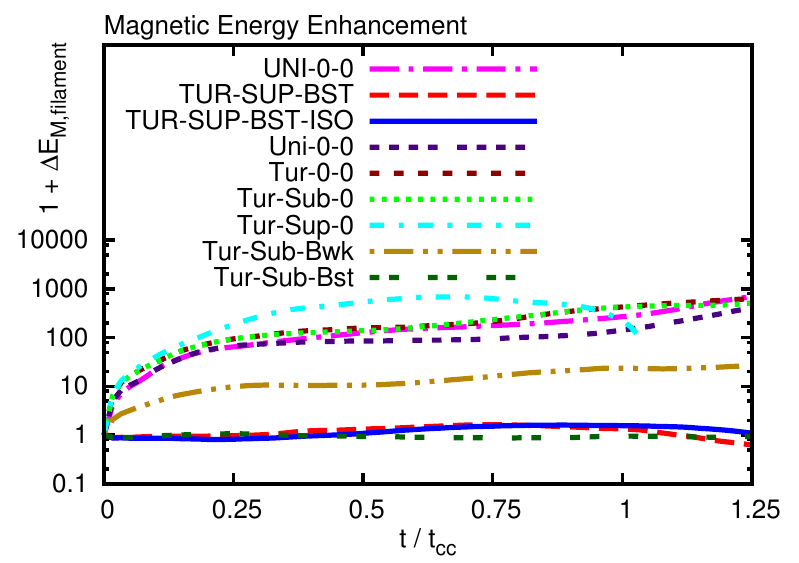}}\\
  \end{tabular}
  \caption{Time evolution of the magnetic energy enhancement in the filaments of nine models with the M and S configurations, namely models UNI-0-0 (dash-dotted line), TUR-SUP-BST (dashed line), TUR-SUP-BST-ISO (solid line), Uni-0-0 (four-dashed line), Tur-0-0 (long-dashed line), Tur-Sub-0 (dotted line), Tur-Sup-0 (short dash-dotted line), Tur-Sub-Bwk (long-dash-two-dotted line), and Tur-Sub-Bst (two-dashed line).}
  \label{FigureE1}
\end{center}
\end{figure}

In this Appendix we show the evolution of the magnetic energy enhancement in nine models with the M and S configurations discussed in Sections \ref{subsec:FilamentFormation} and \ref{subsec:Turbulence} (see Figure \ref{FigureE1}). The ratio of filament to initial cloud magnetic energy or the variation of the magnetic energy contained in filament material at a specific time, $t$, is calculated by the following equation

\begin{equation}
\Delta E_{M_{\alpha}}=\frac{~E_{M_{\alpha}}~-~E_{M_{\alpha,0}}~}{~E_{M_{\alpha,0}}~},
\label{eq:MEenhancement}
\end{equation}

\noindent where $~E_{M_{\alpha}}~=\int \frac{1}{2}\,C_{\alpha}\,|\bm{B}|^2dV$ is the total magnetic energy in cloud/filament material, $~E_{M_{\alpha,0}}~$ is the initial magnetic energy in the cloud, and $\alpha$ refers to filament, footpoint, or envelope material (see Section \ref{sec:Method} for further details). The evolution of this parameter in Figure \ref{FigureE1} shows two effects. First, the stronger the initial magnetic field in the cloud, the faster its growth reaches saturation. In fact, in models without turbulent magnetic fields (UNI-0-0, Uni-0-0, Tur-0-0, Tur-Sub-0, and Tur-Sup-0), the magnetic energy grows by a factors of $\Delta E_{M_{\rm filament}}\sim100-1000$, while in the turbulent scenarios it only grows by factors of $\Delta E_{M_{\rm filament}}\sim10$ (in the weak-field case: Tur-Sub-Bwk) or it is already saturated from $t/t_{\rm cc}=0$ (in the strong-field cases: TUR-SUP-BST, TUR-SUP-BST-ISO, and Tur-Sub-Bst). Second, the ratio of magnetic energy (and magnetic field strength) in the filament to that in the initial cloud remains nearly constant $\sim 1$ throughout the simulations (provided that self-consistent, strong, turbulent magnetic fields are added to the clouds). This has important implications for astrophysical filaments as it indicates that filaments have the same magnetic field strength as their progenitor clouds.

\section{Mixing fraction}
\label{sec:Appendix6}

\begin{figure}
\begin{center}
  \begin{tabular}{c}
  \textbf{Filament Tail (Cloud Envelope)}\\
  \resizebox{80mm}{!}{\includegraphics{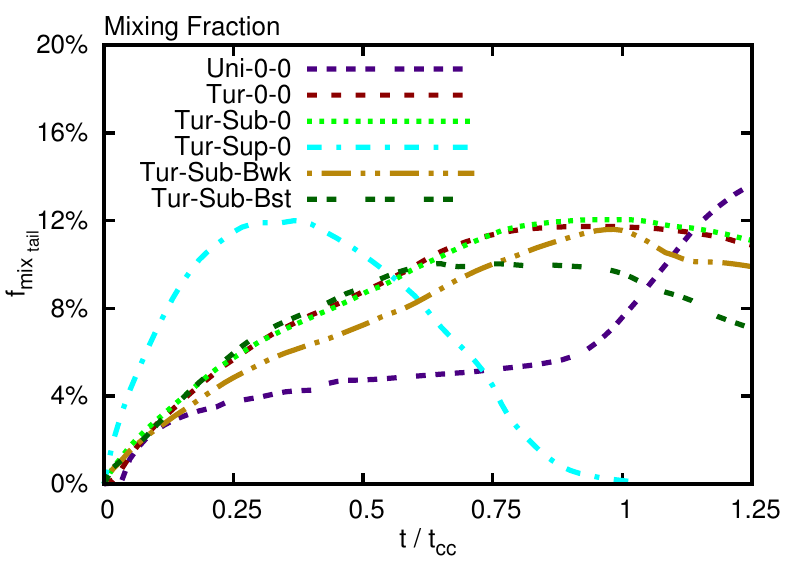}} \\
   \textbf{Filament Footpoint (Cloud Core)}\\
  \resizebox{80mm}{!}{\includegraphics{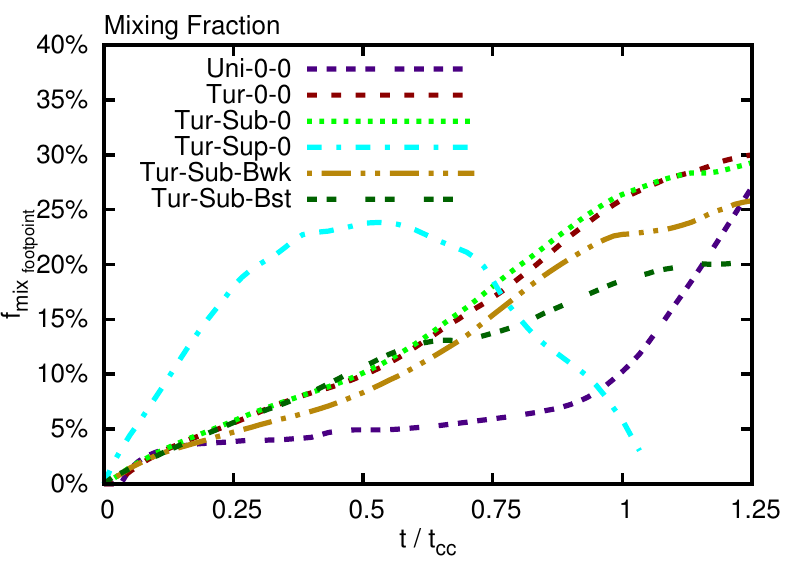}}\\
  \end{tabular}
  \caption{Time evolution of the mixing fractions in the tails (top panel) and footpoints (bottom panel) of filaments in six models with the S configuration, namely models Uni-0-0 (four-dashed line), Tur-0-0 (long-dashed line), Tur-Sub-0 (dotted line), Tur-Sup-0 (short dash-dotted line), Tur-Sub-Bwk (long-dash-two-dotted line), and Tur-Sub-Bst (two-dashed line).}
  \label{FigureF1}
\end{center}
\end{figure}

In this Appendix we show the evolution of the mixing fraction in the models with the S configuration discussed in Section \ref{subsec:Turbulence} (see Figure \ref{FigureF1}). The degree (percentage) of mixing between cloud and wind gas is calculated by the following equation

\begin{equation}
f_{{\rm mix}_{\alpha}}=\frac{\int \rho C^{*}_{\alpha}dV}{M_{{\rm cl},0}}\times 100\,\%,
\label{eq:MixingFraction}
\end{equation}

\noindent where the numerator is the mass of mixed gas, with $0.1\leq C^{*}_{\alpha}\leq 0.9$ tracking material in mixed cells, and $M_{{\rm cl},0}$ represents the mass of the cloud at time $t/t_{\rm cc}=0$. The evolution of this parameter in both the tail (top panel of Figure \ref{FigureF1}) and footpoint (bottom panel of Figure \ref{FigureF1}) of filaments shows the protective effects of strong, turbulent magnetic fields. Magnetic shielding grows as we increase the initial tension of the turbulent magnetic field and this prevents the mixing of cloud/filament and wind gas as it suppresses KH instabilities at fluid interfaces. As a result the mixing fractions in models Tur-Sub-Bwk and Tur-Sub-Bst become lower than in its counterparts as the simulations progress.

\bsp	
\label{lastpage}
\end{document}